# HABILITATION À DIRIGER DES RECHERCHES

Par **Philippe RAUFFET**

## Tools and methods for Human-Autonomy Teaming

Contributions to cognitive state monitoring and system adaptation

**Rapporteurs avant soutenance** :

| | |
|---|---|
| Frank Flemisch | Professeur, FKIE – RWTH Aachen |
| Catherine Da Cunha | Professeur des Universités, Centrale Nantes |
| Frédéric Vanderhaegen | Professeur des Universités, Université Polytechnique des Hauts-de-France |

**Composition du Jury :**

| | | |
|---|---|---|
| Président : | Benoît Le Blanc | Professeur des Universités, Institut Polytechnique de Bordeaux - ESC |
| Examinateurs : | Frank Flemisch | Professeur, FKIE – RWTH Aachen |
| | Catherine Da Cunha | Professeur des Universités, Centrale Nantes |
| | Frédéric Vanderhaegen | Professeur des Universités, UPHF |
| | Nadine Couture | Professeur des Universités, ESTIA |
| | Christine Chauvin | Professeur des Universités, Université Bretagne Sud |
| | Gilles Coppin | Professeur, IMT Atlantique |

**Résumé :** Le paradigme Human-Autonomy Teaming (HAT) a récemment émergé pour concevoir des équipes hybrides, dans lesquelles un opérateur humain coopère avec un agent artificiel autonome. Un défi majeur est de transformer cet agent autonome en un meilleur coéquipier, capable de plus d'interdépendance avec les humains. Les travaux présentés explorent deux axes, soutenus par des collaborations industrielles (dans les domaines du transport et des systèmes industriels), des partenariats académiques (notamment l'Australie Méridionale), et l'encadrement de doctorants.

Le premier axe porte sur le suivi des états cognitifs, afin de doter la machine d'une capacité de détection des difficultés rencontrées par l'humain. Pour répondre à cette question, une approche globale est proposée pour classifier la charge mentale de l'opérateur à partir de la fusion de données physiologiques et comportementales. Puis sont explorés les mécanismes de régulation mis en œuvre par les humains, en étudiant le concept de contrôle cognitif et sa relation avec la charge mentale.

Le deuxième axe traite des mécanismes d'adaptation de ces équipes, pour rendre la machine plus "compatible" avec l'homme. Deux pistes sont explorées. L'une porte sur l'amélioration hors ligne du savoir-coopérer des machines, à l'aide de la méthode CWA et des techniques IDM. L'autre s'intéresse à l'adaptation en ligne de la coopération homme-machine, où le système autonome peut être un coéquipier à l'intérieur de l'équipe, ou un coach au-dessus de l'équipe.

Enfin, des pistes de recherche sont ouvertes, soutenues par des initiatives en cours en France et à l'étranger. Elles concernent la consolidation d'une approche multi-niveau et neuro-ergonomique pour le suivi des états cognitifs, la construction d'un dialogue transparent entre l'humain et l'agent autonome, une prise en compte des situations transitoires et longitudinales, et le passage à l'échelle pour étudier des configurations plus complexes de ces équipes hybrides.

**Abstract:** The Human-Autonomy Teaming paradigm (HAT) has recently emerged to model and design hybrid teams, where a human operator must cooperate with an artificial agent, able to independently evolve in dynamic and uncertain situations. An important challenge in HAT is to transform autonomous systems into better teammates, capable of joining humans in highly interdependent activities. The presented works explore two main avenues, supported by industrial collaborations (in the domain of transportation and industrial systems), academic partnerships (especially with South Australian universities), and with the supervision PhD students.

The first axis deals with the monitoring of cognitive states, to equip the machine with an ability to detect when human face difficulties. To address this question, a global approach is proposed to classify operator's mental workload from the fusion of multi-sourced physiological and behavioral data.

The second axis focused on the mechanisms for adapting human-autonomy teaming, making machine more "compatible" with human. Two kinds of solution are explored. One focused on the offline enhancement of the know-how-to-cooperate of machines, with the aid of CWA method and MDE techniques. The other deals with online adaptation of human-machine cooperation, where autonomous system can be considered inside the team - as a teammate - or above the team - as a coach.

Finally, new research directions are opened, supported by ongoing initiatives in France and abroad. These perspectives relate to the consolidation of a multilevel approach for cognitive state monitoring, the building of a transparent dialogue between human and autonomy, a deeper consideration of transitional and longitudinal situations in HAT, and the scale-up challenge of studying HAT with human teams.

# Acknowledgements

There are many who helped me along the way on this journey. I cannot start this manuscript without expressing all my gratitude to all of them. First, I wish to acknowledge the members of the committee. I am very honoured that all my eminent colleagues have accepted to examine my research works. Many thanks to Pr. Frank Flemisch, Pr. Catherine Da Cunha and Pr. Frédéric Vanderhaegen to review my manuscript, as well as to Pr. Christine Chauvin, Pr. Nadine Couture, and Pr. Benoît Le Blanc to participate to the discussions around my scientific project.

I also would like to address a special thanks to Pr. Christine Chauvin, Pr. Pascal Berruet, and Pr. Gilles Coppin, from whom I learnt so much about human-machine cooperation, cognitive ergonomics, and sociotechnical system engineering. They have guided me during the first years of my career, through the multiple interactions we had, and we still have around challenging research projects and PhD supervisions. Of course, I have also big thoughts for my great colleagues in Lorient and in Brest: Gaël, Laurent, Clément, Natalie, Julie, David, Philippe, Sorin, Lama, Stéphane, and Tabatha. I have shared many interesting scientific discussions with them, but also friendly and funny moments through various activities, and especially music: Funk and Rock with Gaël and the Flying Peanuts, Jazz and Blues with Clement and the Fluppies and the Babass!

To me, research and teaching are a good way to have fun, by interacting with curious and open students, to explore new pathways. I am very glad to have met and worked with Alexandre, Sophie, Amine, Loïck, Yvan and Bailey, my PhD students!

Research is also a matter of collaboration: many thanks to the French academic and industrial partners who accepted to open their field of activity to the research, in various domain such as aviation, road and maritime transportation, or industry 4.0. I am very grateful to Jean-Pierre and Irène, Marie-Pierre and Damien, Jean-Samuel, Patrick and Lauren, Mathieu and Emmanuel. I have also enjoyed a lot to work with Anna, Lang, Siobhan, Charlotte, Stephanie, Assaf and Oren, my international colleagues from South Australia and Israel. It is also a rewarding experience to participate to the adventure of the new IRL CROSSING in Adelaide, and to share good times at IRL offices (and in Exeter, our second office) with jean-Philippe and Cedric.

Finally, all my love to Amelie, Gaëtan and Florian! The Australian experience was a professional project, but also a family project. I am very happy to discover Australia with them. It is nice to see how fast kids learn English and Australian rules football. A last and special dedication to Amélie, who again supported (or bore) me for a new adventure with Gertrude II!

# Table of contents

# Introduction and motivations

Associate professor since September 2012 at the Université Bretagne Sud, I joined the multidisciplinary team IHSEV (Human-Systems Interactions and Virtual Environments) of the Lab-STICC laboratory, bringing together researchers in automation, computer engineering and ergonomic psychology. With the restructuring of the laboratory in 2020, I am now part of the FHOOX team (Human and Organizational Factors, AutOmatics and compleX systems), a team I have been leading since 2020. This integration within the Lab-STICC follows a PhD carried out at Centrale Nantes, at the IRCCyN laboratory (now LS2N), on a subject at the crossroads between business modelling, information technology and management sciences.

My research is therefore multidisciplinary, and the transition between my thesis topic and my current work as an associated professor required a certain thematic adaptation. My work is mainly interested in the modelling and the management of socio-technical systems, and more particularly in the domain of human-machine cooperation and the management of capacities (individual, collective or organizational) within such systems.

In the various projects and topic that I have explored or that I am currently investigating, I have therefore sought to formalize models and methods coming from the human and social sciences, with technical and digital standards and approaches coming from engineering sciences. I thus used the contributions of system engineering, computer engineering and automation, to develop methods for designing information systems and piloting systems with adjustable autonomy (cockpit, supervision interface), where the cooperation between human and machine and the cognitive states of human operators must be modelled, evaluated, and managed.

This manuscript therefore aims to summarize the scientific questioning that has guided my research activities over the past ten years. It also presents the contributions that have been made through various collaborations, whether with doctoral students, master's students, postdoctoral researchers that I had the chance to supervise, with the many colleagues, in Brittany, in France or abroad, with whom I had the pleasure of discussing and building new proposals, and with industrial partners, who allowed me to mix theoretical frameworks with practical applications.

The first chapter summarizes the different activities I have conducted as associate professor at Université Bretagne Sud.

The second chapter provides the state of the science that guided me and helped define the two main research axes I have investigated around the question of Human-Autonomy Teaming.

The contributions to these research axes are then synthesized in two distinct parts: the first part deals with the monitoring of operator functional states (chapters 3 and 4), whereas the second part relates to the mechanisms of adaptation for human-autonomy teams (chapters 5 and 6).

Finally, the last section explains the scientific outlooks and the roadmap I have projected for the coming years, within the collaborative environment I had the opportunity to build this previous decade.

# Chapter 1.  Extended Curriculum Vitae

*"Do not be too timid and squeamish about your actions. All life is an experiment."*

— **R.W. Emerson**

This first chapter describes my educational and scientific activities since I arrived at the Université Bretagne Sud in 2012, by joining the Lab-STICC laboratory. After a brief overview of my background and key figures on my teaching and my research (section 1.1), I will detail more particularly the context and the achievements of my scientific activities (sections §1.2 and §1.3).

## 1.1. ACTIVITY OVERVIEW

### 1.1.1. Academic background

*Table 1: Curriculum Vitae*

| Positions and administrative responsibilities | | |
|---|---|---|
| **Since Feb. 2021** | Invited researcher at University of Adelaide (Sabbatical attributed by the French National University Council, 61st section) <br><br> CNRS delegation at IRL CROSSING since September 2021 | University of Adelaide, CNRS IRL CROSSING <br><br> Adelaide, Australia |
| **Since Sept. 2012** | Associate Professor in Automatics and Human Factor Engineering, <br><br> Member of the multidisciplinary IHSEV team, (Human System Interaction and Virtual Environment), then FHOOX team since 2018 (Human Factors, Organization, Automatics and complex systems) | Université Bretagne Sud, Lab-STICC UMR CNRS 6285 <br><br> Lorient, France |
| **Oct. 2010 - August 2012** | Temporary lecturer and research assistant | |
| **Jan. 2008 - Oct. 2010** | Research engineer in Enterprise Modelling, working on the French national project Pilot 2.0 | Centrale Nantes <br><br> France |
| **Nov. 2007 - Jan. 2008** | R&D Engineer in Enterprise Modelling | MNM Consulting <br><br> Paris, France |
| **April 2007 - Oct. 2007** | Research Engineer in Knowledge and Innovation Management | Indutech Pty Ltd & Stellenbosch University <br><br> Stellenbosch, South Africa |
| Training and Diploma | | |
| **2007 - 2010** | PhD in Industrial and Computer Engineering | Centrale Nantes <br><br> France |
| **2004 - 2007** | Engineer in Industrial Engineering <br><br> Master of Science in Applied Mechanics, <br><br> *option:* Product and Industrial System Design | Centrale Nantes <br><br> Nantes, France |
| **2002 - 2004** | Preparatory class for engineering school (Maths and Physics) | Lycée Louis Le Grand <br><br> Paris V$^e$, France |

### 1.1.2. Administrative responsibilities at UBS and Lab-STICC

**Leader of the research team FHOOX** (Human and Organizational Factors, autOmatics and complex systems) in Lab-STICC in 2020

**Director of a Technological Bachelor in Industrial Engineering and Quality Management** (LOGIQ: Management de la Logistique, de l'Organisation, de la Gestion Industrielle et de la Qualité) since 2015

**Member of the management board in IUT of Lorient** since 2015

**Member of the research committee in IUT of Lorient** since 2018

**Member of 3 recruitment committees** for a position in QLIO department in 2018, 2019 and 2020 (Postodoctoral fellowship and associate professor position)

### *1.1.3. Synthesis of teaching and scientific activities*

**Teaching at IUT of Lorient (Bachelor level, DUT QLIO and LOGIQ)**

*Lectures and practical training (230h a year from 2012 to 2018, 150h a year since 2018):* Reliability engineering, Operation management, Risk management, Maths and applied physics, Petri net simulation, Quality standards and metrology

*Pedagogical development:* Design of a SPOC (Small Private Online Course) in 2019, and responsible for the technological development of the production job in QLIO department (2012-2015), in relation with the research in Industrial Engineering, Automatics and Human Factors (purchasing with FEDER grant, implementation of AR glasses, 3D printer, Straton automata)

*Supervision:* Mentoring of student projects, supervision of apprentices and trainees in industry.

**Teaching at ENSIBS (Master level in Industrial Engineering)**

*Lectures and practical training (15h, since 2012):* Reliability engineering, Risk management

*Supervision:* Mentoring of several student projects on the topic of Industry 4.0 (mental workload, virtual reality, ecological interfaces).

**Teaching at Centrale Nantes (Master level in Industrial Engineering)**

*Lectures (15h, since 2012):* Enterprise modelling and management of organization capabilities (in collaboration with my colleagues from Centrale Nantes, within the international master "Industrial Engineering – Smart connected enterprise").

**Scientific production**

The full list of my publications is given in the beginning of the bibliography section. It should be noted that the references to my own publications are further referred between brackets with a letter and a number (like [J1] for the first paper published in a journal, or [C40] for the 40th paper accepted in a conference). For the other and external references, the APA norm was adopted.

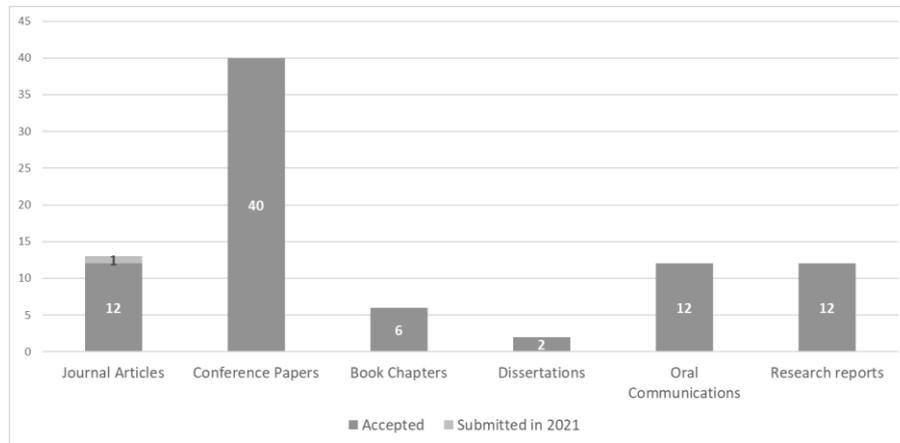

*Figure 1: Scientific production*

**Awards and invited presentations**

*Recipient of the Ph.D. and Research Supervising Bonus (Rank A),* attributed by the French National University Council (61st section) since 2018.

*Best paper Awards* at the 4th International Conference CHIRA in 2020 [C36].

*European Best Paper Awards* at the 13th International Conference IFAC HMS in 2016 [C24].

*Invited conferences:* Presentations of my research work at University of South Australia (UniSA) and University of Adelaide (UofA) [O5, O7, O11], as well as in TUM in Munich, Germany [O9].

**Supervision and mentoring**

*PhD students:* Three completed PhD (2 in 2017, 1 in 2021), 3 PhD currently supervised

*Master students and postdoctoral fellows:* Supervision and mentoring of several research fellows, especially with the hosting of 2 Australian researchers in Lorient between October 2018 and February 2019.

**Scientific responsibilities and involvement in research projects**

*Industrial projects:* Collaboration projects with AXA, IRT System X (Renault), Naval Group, SEGULA Technologies, Dassault Aviation, Thales, Airbus, DGA.

*National projects:* 3 ANR projects (PILOT 2.0, ASTRID TAPAS, HUMANISM), and 4 AMI ADEME projects (SOLENN, PASSION, AERONAV, SEANATIC).

## 1.2. CONTEXT OF RESEARCH WORKS

### 1.2.1. A multidisciplinary research with international and industrial collaborations

Since I was appointed as Associate Professor at Université Bretagne Sud, I have mainly worked on the questions of human-machine cooperation and cognitive states monitoring, by mixing models coming from cognitive ergonomics (naturalistic decision making, teamwork, cognitive control, shared situation awareness, etc.) and methods and tools from computer engineering, signal processing and automatics (physiological signal processing, data fusion, supervised classification, simulation, etc.).

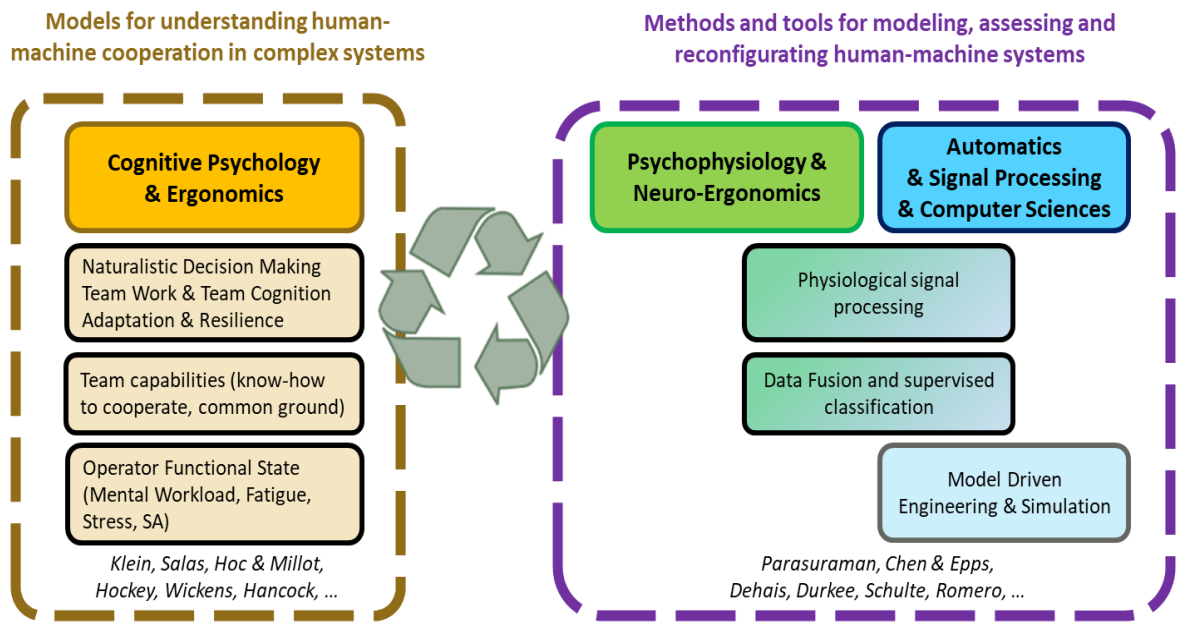

*Figure 2: A multidisciplinary research*

My work is related to various fields of application, ranging from transport (civil and military, as in the ANR ASTRID TAPAS project) to energy (AMI ADEME Solenn), including manufacturing systems (with a recent application on the topic of the Factory of the Future, in particular with the ANR Humanism project and the ADEME Seanatic project, cf. section 1.3), with a desire to make a link between the different facets of the profession of associate professor: on the one hand by trying to combine research and teaching, by participating in the setting up of the various technical platforms of the Université Bretagne Sud, and on the other hand by anchoring research in collaboration with several industrial partners (Dassault, Thales, AIRBUS, ERDF, Naval Group, ENSM, Renault, AXA, cf. Research Reports section in personal publications).

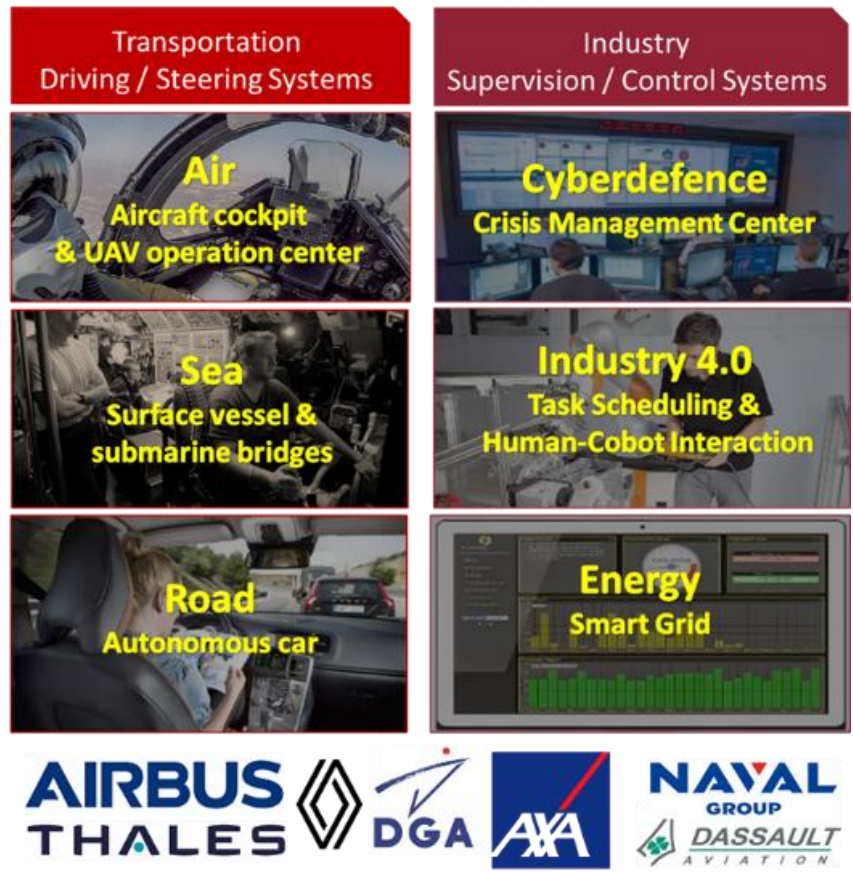

*Figure 3: Application domains and industrial partners*

In addition, since the beginning of my scientific career, I have sought to place my activity in a more international context, by developing a partnership with foreign colleagues. I thus built a collaboration with South Africa in 2007 and 2008, and I have conducted numerous exchanges with Israel since 2016 and with Australia since 2017 on the question of the real-time evaluation of operating states (such as mental workload, attention or fatigue).

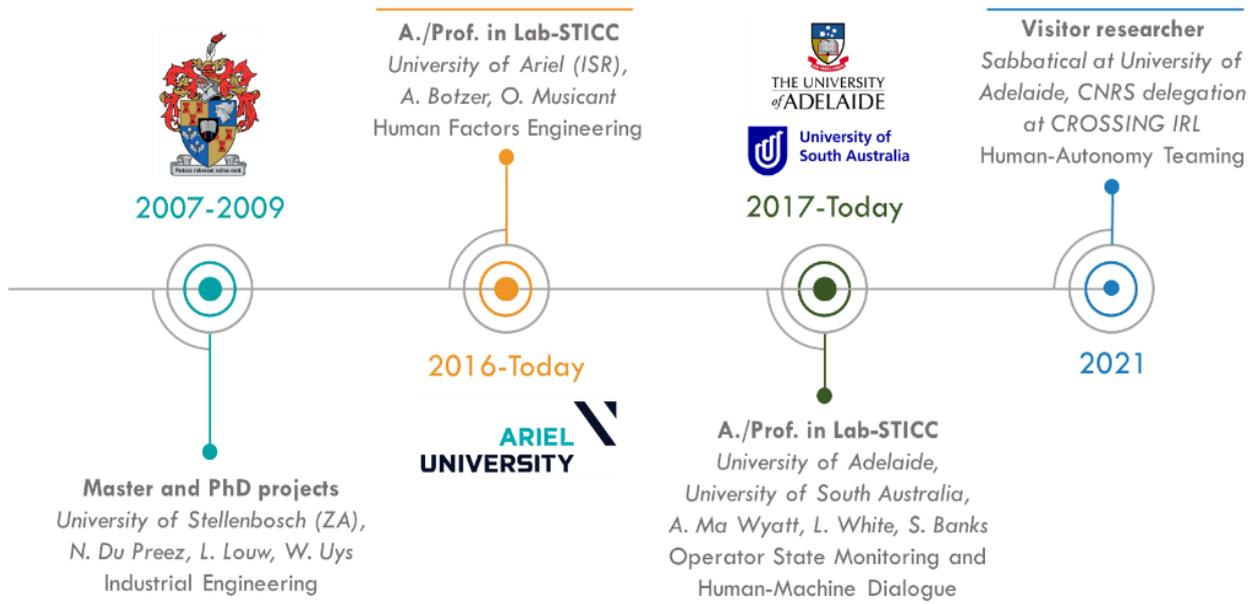

*Figure 4: International collaborations*

### *1.2.2. The Australian experience: one year in Adelaide*

**From the collaboration with South Australian universities…**

In 2017, I had the opportunity to participate to an initiative aiming at boosting the collaboration between South Australia and Brittany, in line with the launch of WASAA consortium in 2016 (Western Alliance for Scientific Actions with Australia). Different actions were conducted to strengthen this collaboration:

- I presented my research works in three workshop, as invited researcher by University of South Australia (UniSA, during two workshops organized by Pr. Siobhan Banks in 2017 and 2019), and by University of Adelaide (UofA, a workshop organized by Pr. Anna Ma-Wyatt in 2018).
- I hosted several Australian researchers in Lorient:
  - two invited professors came in Lorient for a week (Pr. Siobhan Banks in November 2018, Pr. Anna Ma-Wyatt in July 2019),
  - two young researchers also visited Lab-STICC in Lorient for three months in 2018 (a PhD student, Charlotte Gupta, and a postdoctoral fellow, Stephanie Centofanti).
- A shared communication was presented at the international IFAC conference on Human-Machine Systems (IFAC HMS), in September 2019.
- Different proposals, for PhD or postdoctoral positions, were jointly written on the topics of human attention modelling, human-machine dialogue and fatigue management in teams.
- Three applications were also jointly submitted to different Australian calls for project, with Pr. Anna Ma-Wyatt and Pr. Langford White from University of Adelaide:
  - Two projects in the framework of the programs HPR-NET and RN-UDS from Defence Science Technology Group,
  - One in the framework of the program FSP from Lockeed Martin.

These proposals aimed at reinforced the capabilities for monitoring human and collective performance in submarines, and to develop countermeasures able to tackle with a performance decrement.

### …to the creation of the French-Australian laboratory CROSSING

In 2018 also emerged the idea to build a CNRS international research laboratory to specifically address the question of human-autonomy teaming, on the initiative of Pr. Gilles Coppin. This laboratory aimed at gathering French and Australian experts in Artificial Intelligence, Autonomous Systems, Human Factors, and at boosting the collaboration between France and the three South Australian Universities (University of Adelaide, University of South Australia, and Flinders University). It also involves Naval Group as industrial and research partner.

I especially participated to definition of two research axes in this project:

- A first axis focused on the proposal of new models to understand and anticipate human behaviour, based on the capture of human physiological, cognitive, and emotional state to augment individual and team decision and predict human performance.
- A second axis dealing with new solutions for managing hybrid teams, based on the elaboration of new models for human-intelligent machine cooperation in naturalistic and realistic settings.

It is in this dynamic and enabling context that I have undertaken to spend one year in Adelaide in 2021.

## A one-year visit in Adelaide to consolidate collaborations

This one-year visit in Adelaide took place between February 2021 and February 2022. The path to this destination was not really a highway to heaven, especially due to the COVID situation and the resulting restrictions in Australia and in France. But finally, thanks to the friendly and invaluable of my Australian colleagues, and especially Pr. Anna Ma-Wyatt, I was able to get the visas and the exemption to travel to South Australia with my family.

Supported by a research sabbatical attributed by the 61$^{st}$ section of French National University Council, then by a CNRS delegation at IRL CROSSING since September 2021, I was able to participate to the first steps of the CROSSING (started in February 2021) with my French colleagues Jean-Philippe Diguet and Cédric Buche, and to continue the ongoing works with Professors Anna Ma-Wyatt (on attention modelling), Langford White (on agent transparency and human-machine dialogue), and Siobhan Banks. Especially, I co-supervised the PhD of Bailey Hadlum at UniSA on the topic "Fatigue and teamwork" (granted by Naval Group), as well as a postdoc on real-time human performance monitoring.

## 1.3. DETAILED SCIENTIFIC ACTIVITIES

### 1.3.1. *Supervision of PhD student and postdoctoral fellows*

I have been and I am still lucky to work with motivated and nice graduate and post-graduate students, to explore challenging topics and have fun in research.

| |
|---|
| **PHD SUPERVISION (3 COMPLETED, 3 IN PROGRESS)** |
| BURGUIN Yvan *(Co-Supervisor 25%)* <br> **Topic**: Automated reconfiguration of cyberdefence training scenarios from machine learning applied to operators' physio-psychological data [TH6] <br> **Context**: PhD in collaboration with IRIS team from LabSTICC (UBO), Breizh Cyber Valley grant <br> **Start date:** October 2021 **End date:** October 2024 |
| HADLUM Bailey *(Co-Supervisor 50%)* <br> **Topic**: Fatigue and team performance in maritime environment [TH5] <br> **Context**: PhD in collaboration with University of South Australia (UniSA), funded by Naval Group <br> **Start date:** July 2021 **End date:** January 2025 |
| SIMON Loïck *(Principal Supervisor 50%)* <br> **Topic**: Development of design rules and mechanisms for adapting the human-machine dialogue [TH4] <br> **Context**: PhD funded by the French Environment Agency (ADEME), SEANATIC project <br> **Start date:** July 2020 **End date:** July 2023 |
| LAOUAR Amine *(Co-supervisor 50%)* <br> **Topic**: Definition of principles for decision-making among crews in a cockpit of the future [TH3] <br> **Context**: CIFRE collaboration with AIRBUS (Aircraft manufacturer) <br> **Start date:** January 2017 **End date:** March 2021 |
| PRAT Sophie *(Co-supervisor 50%)* **Current position:** R&D Engineer at SEGULA Toulouse <br> **Topic**: Verification by simulation in an automated control-command design process [TH2] <br> **Context**: CIFRE collaboration with SEGULA Technologies (Technological consulting group) <br> **Start date:** February 2014 **End date:** December 2017 |
| KOSTENKO Alexandre *(Co-supervisor 50%)* **Current position:** R&D Engineer at ALTRAN Toulouse <br> **Topic**: Multidimensional and dynamic evaluation of the control of the situation by the operator: creation of a real-time mental load indicator for drone supervision activity [TH1] <br> **Context:** Funding from Brittany region and DGA <br> **Start date:** October 2013 **End date:** April 2017 |

| **SUPERVISION OF MSC INTERNSHIPS (3 COMPLETED)** |
|---|
| LEGENDRE Mathieu *(Co-supervisor)*<br>**Topic**: Discomfort and discrepancies between driving style and car behaviour [MA3]<br>**Context**: DAAV project funded by German French Alliance for Industry of the Future<br>**Start date:** February 2020　　　**End date:** October 2019 |
| THIEBAUT-RIZZONI Tabatha *(Principal supervisor)*<br>**Topic**: Design principles for the future intelligent vessel bridges [MA2]<br>**Context**: project ADEME PASSION, collaboration with ENSM, the Merchant Navy School<br>**Start date:** February 2019　　　**End date:** October 2019 |
| GUPTA Charlotte *(Principal supervisor)*<br>**Topic:** Combining chronobiology and neuro-ergonomics: metrics and experimental platforms [MA1]<br>**Context**: Nicolas Baudin Programme (funded by French Embassy in Australia)<br>**Start date:** October 2018　　　**End date:** February 2019 |

| **SUPERVISION OF POSTDOCTORAL FELLOWS (4 COMPLETED)** |
|---|
| DIALLO Lama *(Co-supervisor)*<br>**Topic**: Classification of cognitive operator states and control modes<br>**Context**: Project PRECOGS, funded by DGA in the programme Man-Machine Teaming<br>**Start date:** February 2020　　　**End date:** October 2020 |
| KOSTENKO Alexandre *(Principal supervisor)*<br>**Topic**: Classification of cognitive operator states and control modes<br>**Context**: Project PRECOGS, funded by DGA in the programme Man-Machine Teaming<br>**Start date:** February 2019　　　**End date:** February 2020 |
| CENTOFANTI Stephanie *(Principal supervisor)*<br>**Topic**: Fatigue and teamwork: theoretical background and methodological framework<br>**Context**: Australian researcher hosted in Lorient (UniSA funding)<br>**Start date:** October 2018　　　**End date:** December 2018 |
| KOSTENKO Alexandre *(Principal supervisor)*<br>**Topic**: Supervised learning of cognitive states from physiological data<br>**Context**: Project TRM 104, funded and in collaboration with Dassault Aviation (fighter jet manufacturer)<br>**Start date:** October 2016　　　**End date:** May 2017 |

### 1.3.2.   Responsibilities and involvement in scientific projects

I have also learnt a lot and gain much experience from local and national projects with industrial partners and scientific colleagues.

**Industrial collaborations**

- *Thales and Dassault Aviation (2019-2021):* Co-Responsible for PRECOGS project (poly-recognition of cognitive states), within the PEA Man-Machine Teaming Batch 1 (140 k€)
- *AXA (2016):* Responsible for a scientific collaboration contract with AXA France, on the monitoring of vigilance in car with connected objects (15k€)
- *Dassault Aviation (2016):* Co-Responsible of TRM104 project, a collaboration on the supervised machine learning of physiological data to classify operator functional states and reconfigure the human-machine dialogue (90 k€)
- *Airbus (2017-2020):* Co-Responsible of the industrial collaboration around the CIFRE PhD of Amine Laouar (30 k€)
- *Renault / IRT System X (2016):* Study on the analysis of driver behaviour and the effect of different Augmented Reality cognitive aids, in automated driving situations (10k€)

- *Ségula Technologies (2014-2017):* Co-Responsible of the industrial collaboration around the CIFRE PhD of Sophie Prat (30 k€)
- *Naval Group (2012-2015):* Contributor to a study on the evaluation of an ecological HMI for submarine steering, and on the development of a method for dynamic function allocation

**National projects**

- *AMI ADEME SEANATIC (2020-2023):* Responsible for the work package « Adaptive Interface » in Seanatic Project. The project aims at using IoT sensors an machine learning to develop new services for smart and connect vessels, especially predictive maintenance for seamen, ship owners and ship builders. This project results from a consortium with IoT.bzh, Piriou, Thalos and Azimut (500k€)
- *ANR HUMANISM (2017-2020):* Contributor to the work packages « state of the art about Operator 4.0 », and « Experimental validation », on a project studying the human-machine cooperation within "smart factories"
- *AMI ADEME PASSION (2017-2020):* Industrial collaboration with ENSM and IX Blue for the design of a new intelligent vessel bridge
- *AMI ADEME AERONAV (2016-2017):* Industrial collaboration with ENSM on the analysis of the piloting of High-Speed Crafts
- *AMI ADEME SOLENN (2014-2017):* Industrial collaboration with ERDF, Vity and Delta Dore on the implementation of smart grids and their adoption by users.
- *ANR ASTRID TAPAS (2011-2015):* Scientific collaboration with Dassault Aviation and the naval air base of Landivisiau. The project dealt with the development of human-human cooperation and human-machine teaming within fighter jet squadron.

**Internal projects in University of South Brittany**

- *Cyberdefense UBS:* Contribution to different studies on the evaluation of stress and teamwork in cyberdefense teams (experiments in 2016 and 2017 with students in cybersecurity center C4 in Vannes)
- *Smart Factory - SCAP 4.0:* Contribution to the development of the platform SCAP 4.0 for Industry 4.0 (supervision of projects with students in industrial engineering, development of scenario and experiments)

### *1.3.3.  Other activities*

Finally, I have been also participated to other scientific actions, like peer-reviewing and dissemination towards students and general audience.

**Editorial activities**

*Member of the scientific committee of an international conference (4):* Conference H-Workload (Dublin, June 2017, Amsterdam, 2018, Roma, 2019, Dublin 2021)

*Reviewer for international journals (15)*
- IEEE Transactions on Human-Machine Systems (2 article in 2018 and 2021)
- Fire Safety Journal (1 article in 2018)
- Frontiers in Psychology (1 article in 2020)

- Cognition Technology and Work (3 articles in 2017, 2019 and 2020)
- Reliability Engineering and System Safety (1 article in 2017)
- Applied Ergonomics (2 articles in 2016, 1 in 2018)
- International Journal of Product Life Management (2 articles in 2015 and 2016)
- International Journal of Transactions on Engineering Management (2 articles, in 2014 and 2015)

**Scientific committees for PhD students**

Member of the PhD follow-up committees since 2020 for Chloé Lourdais, PhD student in Centrale Nantes, on the topic "Optimization of an HMI for remote online medical consultation reducing patient mental workload", supervised by Pr. Emilie Poirson and Dr. Liang Ma.

External examiner of the PhD dissertation defended by Lineth Rodriguez in 2019, on the topic "A qualitative analysis to investigate the enablers of big data analytics that impacts sustainable supply chain", supervised by Pr. Catherine Da Cunha.

**Links between research and teaching**

*Science festival:* Responsible for Lab-STICC workshops during Science Festival 2014

*SPOC/MOOC:* Design of an online lecture in English in 2020, based on my research works on cognitive states, in the programme "Human and Organizational Factors in Maritime Domain"

*Industry 4.0 for teaching and research:* Responsible of the technologic platform of the Bachelor department (supervision of engineer student projects, investments in new equipment), responsible for the investment in sensors and softwares for a human factor analysis platform (CPER Smart Factory 2021-2026, CPER PFT Compositic 2014).

# Chapter 2. Human-Autonomy teaming: literature review and research axes

*"To you, a robot is a robot. Gears and metal; electricity and positrons. Mind and iron! Human-made! If necessary, human-destroyed! But you haven't worked with them, so you don't know them. They're a cleaner, better breed than we are." – I. Asimov*

This chapter aims to explain and position the research axes that I have developed over the past ten years. To do that, the following paragraphs first recall the industrial needs and the scientific issues (the "why" of my work), then sets out the analysis units of my research (the "what"), as well as the main theoretical, methodological, and practical levers that could be used to solve these issues (the "how").

## 2.1. CHALLENGES: HUMANS AND THE RISE OF AUTONOMOUS AND INTELLIGENT SYSTEMS

Advances in artificial intelligence, machine learning and cognitive modelling lead to rethinking decision-making processes and the distribution of work between humans and machines. Whatever the field (Industry 4.0, autonomous transport systems, etc.), we observe an increased use of new technologies (digital twin, cobotics, virtual and augmented realities, machine learning) to optimize, customize and make flexible the processes of execution, supervision or planning (Hozdic, 2015, Longo et al., 2017). The introduction of these technological revolutions allows to manage highly complex processes, thanks to autonomous and intelligent cyber-physical systems. This technological change is also accompanied by a societal transformation: with the disappearance of low-value-added activities, we are witnessing the emergence of "expert" professions. Thus, human operators will increasingly be entrusted with supervision and control activities, to the detriment of operational tasks (Hirsch-kreisner, 2014, Rüssman et al., 2015).

The role of the operator 4.0 (Romero et al., 2017) therefore needs to be redefined, whether human is the driver or the pilot of a transportation system (autonomous car, drone swarm, fighter plane) or the supervisor of an industrial system (smart grid, production line, ship maintenance system). Particularly, we must be concerned about the new interferences that will appear between human and these cyber-physical systems with increased capabilities. The joint activity of these two types of agents (human and autonomous machine) can indeed influence the performance of the process, but also the interactions between these two agents. Moreover, in addition to the question of the coordination of this co-activity, the structure of cooperation or authority between the two agents should be addressed.

Thus, many questions remain to be clarified to improve Human-Systems integration in the era of autonomous agents and artificial intelligence. It is indeed necessary to keep "human in the loop", by allowing him to maintain a mental representation of situations, and by improving his ability to supervise autonomous agents. The question of optimizing the cooperation of humans with cyber-physical systems (CPS) therefore becomes a critical issue.

### 2.1.1. The Human-Autonomy Teaming paradigm

The Human-Autonomy Teaming paradigm has recently emerged to characterize the evolution of these hybrid human-system teams (also acronymized HAT, for Human Autonomy Teams), and to develop new ways of thinking about the mechanisms of cooperation within these entities (O'Neill et al., 2020). This paradigm is used to describe the association of human operators with intelligent and autonomous agents, with each team member working interdependently towards a common goal (Chen et al., 2016, McNeese et al., 2018). The concept is not recent: Woods (1996) already suggested automation would tend more towards the empowerment of machines, and the introduction of these intelligent agents in larger systems would lead to rethinking the composition of the team. However, this concept has been refined during the last decade, to specify the specific attributes of these "human-autonomy teams".

According to Demir et al. (2016) and Myers et al. (2019), the autonomous agent in this type of team must be a digital or cyber-physical entity, with its own capabilities, full or partial, for self-governance, decision-making, adaptation, and communication. This "power of action" (also named "agentic capability", Larson and DeChurch, 2020) allows the intelligent machine to dynamically manage many functions related to the execution of tasks (the part of the activities which correspond to the "taskwork", what contributes to the objective performance in relation to the missions to be fulfilled and the goals to be achieved).

Nevertheless, bringing together an autonomous agent and a human operator does not necessarily constitute a "human-autonomy team". Indeed, to observe the effective emergence of a team, the autonomous agent must possess another important property: it must be perceived and accepted by human operators as a full member of the team. Larson and Dechurch (2020) indicate that the intelligent agent must fulfil a very distinctive role and must make a clearly identifiable contribution within the team, as it could be the case for a human teammate. The autonomous agent should therefore be considered as a legitimate team member, and no longer as a simple support tool (Groom & Nass, 2007; Grimm et al., 2018; Lyons et al., 2018). This distinction between "teammate" and "tool" is a discriminating element, isolating HAT configuration from the configurations previously explored around HAI (Human-Automation Interaction) and LOA (Levels of Automation). Hence, in a HAT, the autonomous system must also be able to participate in activities that correspond to "teamwork" (Seeber et al., 2020), such as coordination or reallocation of the activity.

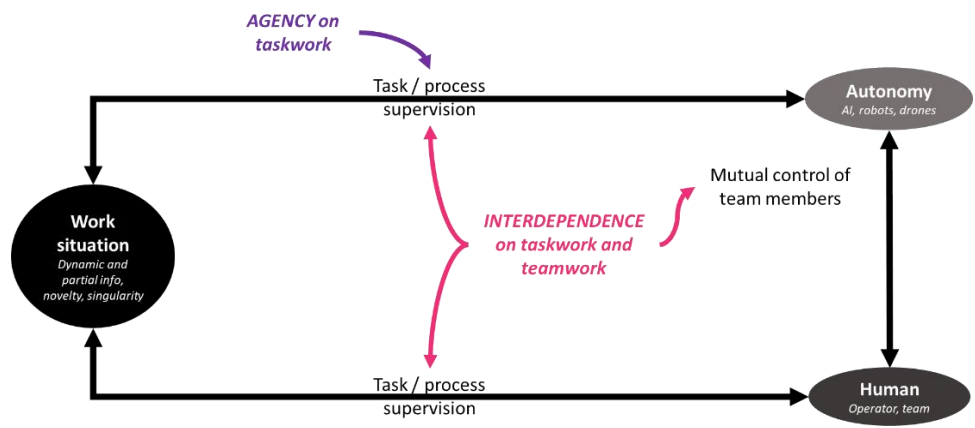

*Figure 5: The HAT paradigm considered with a triadic relationship between work situation, autonomous agents and human operators*

To sum up, a human-autonomy team therefore emerges from a triadic relationship between the work situation, the autonomous agent and the human operator. Each team member must contribute to the work on the task ("taskwork"), thanks to their abilities to perceive, analyse or act on the work situation. These abilities can be developed by each team member, to increase their "power of action", their "agentic capability" (Wynnes and Lyons, 2018). In addition, each team member must also be able to participate in teamwork and bring out a certain "interdependence" (Walliser and al., 2017), by monitoring the states or actions of other team members, by understanding each other's role, and by identifying potential interferences between the actions of the team, or possible needs from teammates.

As Johnson et al. (2011) underlined, "increasing the effectiveness of HAT teams is not based solely on work seeking to make the machines even more independent. An important challenge is also to find solutions to transform autonomous systems into better teammates, capable of joining human operators in highly interdependent and sophisticated cooperation activities".

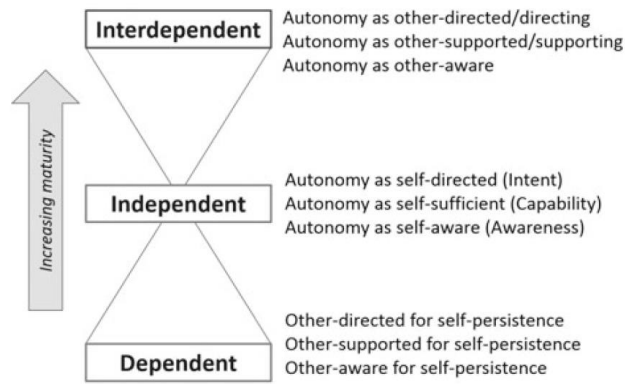

*Figure 6: Maturity of HAT: from dependence to interdependence*

This perspective is again emphasized by Abbass et al. (2018), who explain that autonomous systems must mature to become trusted teammates. Making the analogy with models of development and emancipation of children becoming adults, these authors explain that machines have already passed from the stage of dependence (where humans are always present to control the behaviour of machines, to switch them from one state or one mode to another) to the stage of independence (where machines, aware of the environment thanks to their sensors and their data analysis capabilities, can perform many tasks and adapt to different situations without human intervention). However, this stage of independence is limited and necessarily transitory: the machine, like Icarus freeing himself from the tutelage of his father Daedalus, cannot continue to progress if it does not experience social interdependence. Indeed, it is by interacting with the others that the machine could adapt its behaviour, but also influence the behaviour of others and improve collective performance. The pathway to interdependence, however, is a tortuous one: the trap of co-dependence must be avoided, where machines and humans cannot do without each other, and where they both risk losing their autonomy. As Gaspard Koenig (2019) warns, intelligent systems, assisting humans and making them lazy, could put an end to the idea of an autonomous and responsible human, who could no longer in turn challenge and train the machine to solve new problems.

### 2.1.2. Domain-dependent pathways for HAT interdependence

The introduction of autonomy in transportation and manufacturing systems, as well as the definition of its relationships with the human operator, have given rise to different taxonomies of human-machine symbiosis. Three of these taxonomies are reviewed in the following paragraphs, to illustrate the efforts to conceptualize the Human-Autonomy Teaming paradigm and the notion of interdependence, specific to some work domains I have explored with industrial partners, in the domains of manufacturing, autonomous car driving and aircraft piloting.

**Manufacturing operations: from operator 1.0 to operator 4.0**

Romero, Bernus, Noran, Stahre and Fast Berglund (2016) explained that the place of the operator and the interactions with the physical and digital components in production systems can be summed up in a taxonomy resulting from the historical evolution of the industry (see Figure 7).

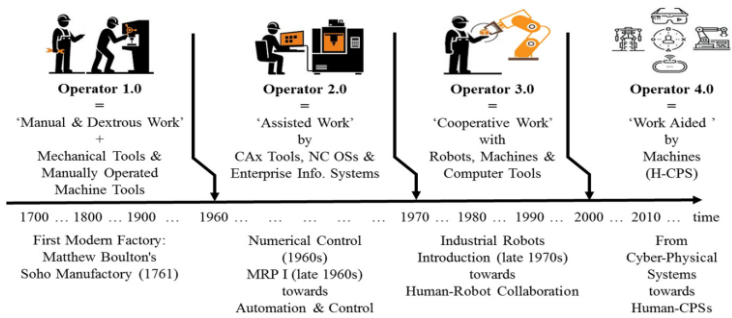

*Figure 7: (R)evolution of human-machine cooperation in Industry (Romero et al., 2016)*

These authors distinguish 4 categories - or 4 generations - of operators (this is an incremental improvement of the space of possibilities and the capacities of the operator over time, and not a discriminating vision and disruptive generations):

- *Operator 1.0* is a person with essentially manual know-how, assisted or not by mechanical machines, which are controlled manually; Machines are used as simple tool, the expertise in only possessed by the human.

- *Operator 2.0* is a human agent who performs his tasks using computer resources, ranging from computer-aided design tools to numerical control machines, including company information systems (ERP, PLM, etc.); Knowledge is partially externalized to the machines under the form of rules, but machines are still dependent from humans: machine are only useful if human has enough expertise to understand the rules, or to program and control the machines.

- *Operator 3.0* represents the generation of cooperative work between man and machine (in the broad sense, the machine can be a robot, a computer, an automaton, etc.); Machines begin to be independent, aware of their environment, able to self-reconfiguration.

- Finally, *Operator 4.0* can be defined as a flexible and adaptable worker or user, who carries out activities being assisted by intelligent machines and autonomous systems, which will provide the necessary information or the required actions only when the situation where the state of the team requires it. Machines are interdependent with humans, providing flexible and customized assistance to operators.

**Car driving: levels of autonomy and interdependence between human and machines**

In the domain of autonomous driving, interdependence between humans and machines has been defined, and even standardized (SAE, 2018), along six levels of autonomy (cf. Figure 8), from Level 0 (full manual driving) to Level 5 (full automation). These different levels describe the different system capabilities related to the perception of the environment and the system actions.

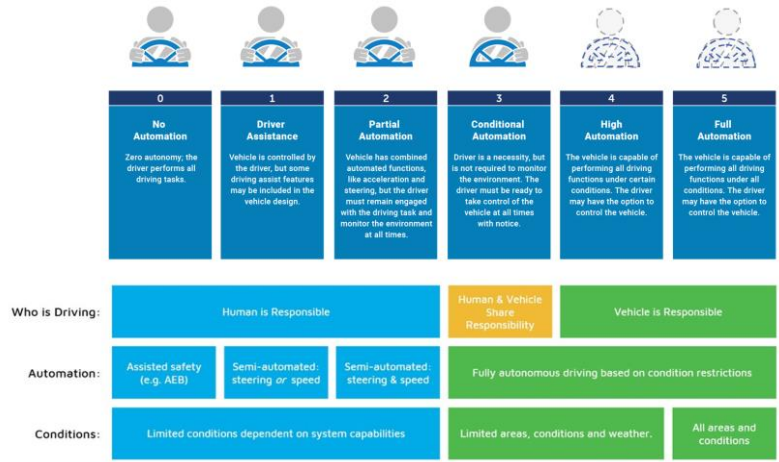

*Figure 8: Autonomous car, levels of autonomy and interdependence*

These different capabilities (cruise control, automated cruise control, forward collision avoidance, lane following or keeping assist, etc.) can be conditionally automated according to the road context, but also actionable by the drivers. The cooperation between human drivers and autonomous cars therefore raises the questions of authority and responsibility about the driving tasks. For instance, at SAE Level 3 of conditional automation, the vehicle most often operates automatically; it can overtake other vehicles and return to its lane, and it is able to pick routes according to pre-defined destinations. However, drivers are occasionally required to regain control within an acceptable time (about ten seconds) to be able to respond to road situations that the automation cannot negotiate, such as missing lane markings or heavy weather conditions (Diederichs et al., 2015; Zeeb, Buchner & Schrauf, 2016; Reilhac et al., 2017).

Hence, humans must be kept in the control loop since they still have to detect and react to a change in road conditions or in driving mode. However, high automation level pushes them to do non-driving tasks (e.g., media reading and playing, communication) to the detriment of driving tasks (i.e. the supervisory control of the automated driving), reducing their ability to regain control properly.

It is therefore a big challenge to better manage the interdependence between humans and cars, to avoid this irony of automation (Bainbridge, 1983), as well as to reduce some dangerous or discomfortable experiences resulting from automation surprise (Sarter, Woods & Billing, 1997). Indeed, the discrepancies between human driving styles and car driving modes, in terms of road perception, diagnosis and driving actions could lead to risky situations (difficulty to regain control), but also to unpleasant situations (where the drivers do not understand or trust the choices and the behaviour of the car, Hartwich et al., 2018).

**Aircraft piloting: autonomy and crew management inside and outside cockpit**

The gradual introduction of automation allowed to reduce the number of pilots in the cockpits. In the 50s, piloting a commercial aircraft required 5 persons (a pilot, a co-pilot, a flight engineer, a navigator, and a radio operator). Advances in voice communications have led to the elimination of the radio operator position. Then, the arrival of modern navigation equipment (for example inertial navigation units) led to the abolition of the position of navigator. And finally, the computerization of the cockpits has been accompanied by the development of engine monitoring equipment and on-board services (fuel circuit, hydraulic and electrical networks, air conditioning) of the aircraft, making the flight engineer position superfluous. Nowadays, the crew of an airliner is therefore made up of two pilots (a captain, more experienced, and a co-pilot or first officer), who essentially have roles in planning, supervision, and decision-making (see Figure 9, left).

The "Single Pilot Operations" configuration (SPO, see Figure 9, middle) is the logical next step in reducing the number of crew members in the cockpits. Its implementation should allow operators to reduce the budget allocated to flight personnel, while maintaining a level of safety comparable to the current one, or even improving it (Neis, Klingauf & Schiefele, 2018). However, the transition from two pilots to one is a new and huge challenge. Indeed, questions arise, such as the consequences of a heavy mental workload or an inability of the pilot to perform his tasks (Boy, 2014). In addition, simply developing automation will not suffice to ensure the desired level of security (Bilimoria et al., 2014, Neis et al., 2018). Indeed, compensating for the withdrawal of a pilot by increased automation can result in emerging problems, such as the out-of-the-loop syndrome or difficulties in making decisions (Neis et al., 2018), again bringing into play the notions of automation ironies and surprises.

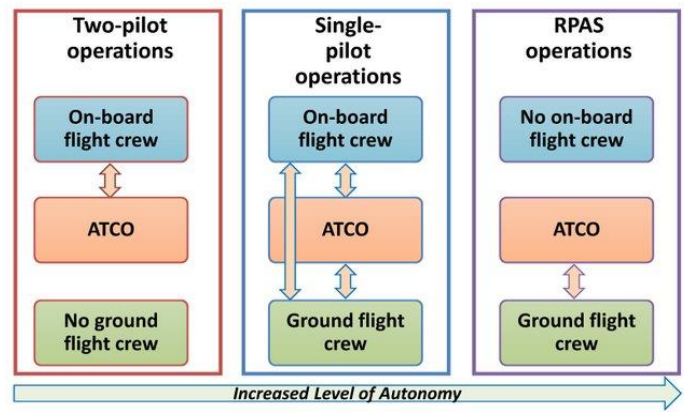

*Figure 9: Impact of autonomy increase on aircraft crew*

The literature therefore mentions certain new avenues for designing a safe cockpit. The current cockpits are certified so that the aircraft can be piloted by a crew reduced to n-1, i.e., a single pilot, in the event that the second should no longer be able to perform his duties. For an SPO cockpit, a crew reduced to n-1 entails the need to design an aircraft that can be completely autonomous, with a supporting ground assistant (RPAS for Remotely Piloted Aircraft System, see Figure 9, right). However, under nominal conditions, this total autonomy of the aircraft raises the question of the role of the human operator (Boy, 2014). In addition to helping ground operators, the solution the industry is moving towards is to design automation that can optimally assist the pilot (Young, Stanton & Harris, 2007). The tasks would then be dynamically distributed between the human operator and the automation systems, depending on the situation: we speak of adaptive automation (Neis et al., 2018). As a result, automated systems would no longer just perform assigned tasks; they would become full agents of a team formed with the human operator, each providing specific skills and collaborating with the other (Bilimoria et al., 2014). In other words, the automated system would become a co-pilot (Cummings, Stimpson, & Clamann, 2016) and the tasks of the three actors would be defined according to the pilot's state (normal or not) and flight conditions (normal or degraded). Thus, in the event of the incapacity of the pilot, the aircraft landing could be ensured by the ground operator who would interact with the automated system.

### 2.1.3. Multicriteria performance of interdependence in HAT

These three domain-specific taxonomies have underlined that human-machine cooperation activities may occur in dynamic situations, and that the management of the interdependence between autonomy and human are crucial for HAT performance. It is therefore important to characterize the main key performance indicators for the interdependence in these hybrid teams.

The interdependence of autonomous agent is assumed to increase the effectiveness of human-autonomy teams, especially by providing humans with better situation awareness, better trust in autonomy (about the rationality or the reliability of agent's behaviour), and finally better decision accuracy (making human able to better detect false positives or misses from autonomous agent diagnosis). Nevertheless, this agent interdependence can also overload human operator, giving them too much work to process to decide or to interact within a given time budget.

These 4 different criteria can be measured in different ways.

**Trust.** To measure trust in autonomy, authors used "direct" and subjective questionnaires or scales (like HRI trust scale or Trust in automation scale, see Lewis et al. (2018) for survey paper), or "indirect" and objective measures. Trust is mainly observed along two main dimensions, based on performance, or social relation, studied independently or together (see Law and Scheutz (2021) for a recent literature review).

Performance-based trust is based on the perceived reliability or explainability of the autonomous agent, and can be expressed in terms of:
- Task intervention (how human let the robot do some tasks, and how often human takes over doing the task from the robot),
- Task delegation (selection of an agent among different teammate, human or robot),
- Following advice (by accepting or not a recommendation, and sometimes when humans are not completely blind on the situation, changing their own decision by the robot's suggestion)

On the contrary, relation-based trust is more focused on how autonomy can be socially trusted by a human (human trust the robot not or not only because this one seems to be rationale, but rather because it seems to be sympathetic or acting as a teammate knowing how to socially cooperate with the human (Tulli et al., 2019; Kampik et al., 2019).

**Decision accuracy.** Accuracy of operators are often measured in using automated recommendations, namely hit rates and correct rejection rates. Hit rates refer to the proportion of time participants correctly accept the agent's recommendation (the inverse of automation disuse). Correct rejection rates refer to the proportion of time participants correctly reject the agent's recommendation when it is wrong (the inverse of automation misuse).

**Mental workload.** Response time can be used as a simple indicator to observe different levels of mental workload (Akash et al., 2020). Another means is to use secondary tedious task, and to assess how the performance on this task changes when the task demand varies. Different subjective scales can also be used for an instantaneous assessment (like ISA or SMEQ scales). Chapter 3 will particularly focus on the contributions I made on the monitoring of this indicator.

**Situation Awareness.** Situation awareness has been mainly studied with objective freeze probe questionnaire (like SAGAT, SPAM, etc., questioning the human about its awareness of different components of the situation), or with subjective measures, composed of Likert scales on different situation dimensions (like SART, see Endsley, 2020 for survey paper). In addition of these questionnaire techniques (imposing to freeze the course of action from the operator), other works also try to propose some eye related situation awareness metrics (Bhavsar et al., 2017, Schwerd & Schulte, 2020), for instance based on entropy measures on gaze transitions or dwell times between different task-related areas of interests.

**HAT effectiveness.** To guarantee effective teamwork (see Fig. 8), it is therefore necessary to balance and optimize these 4 criteria, preventing the human-autonomy team from becoming a "failing" team:

- **A "blind" team** where robot does not share enough information to reach a satisfying shared SA. There is a risk of not detecting a new and singular situation, with a quick decrement of decision accuracy when situation changes.
- **A "rogue" team** where the distribution of work is not very well managed (the interdependence of agent is too intrusive), potentially leading to a high workload for human. There is a risk of decision failure when time constraints increase.
- **An "incompetent" team** where the human decision accuracy (given the autonomy's reliability) is bad. Here, human can enter a complacency loop, by over-trusting the autonomous system (loss of the Reagan's motto "trust but verify").
- **A "pseudo" team** where human does not trust autonomy. Here human workload may increase when there is too much information to analyse for checking agent reliability.

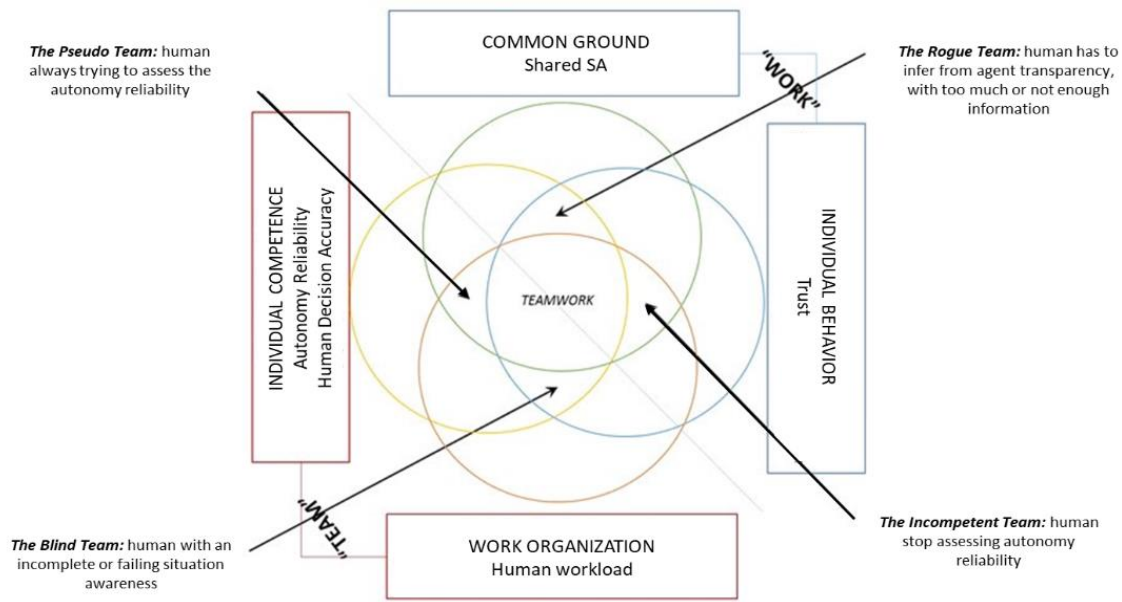

*Figure 10 : Dimensions for effective HAT (adapted from d'Angelo et al., 2019)*

## 2.2. UNITS OF ANALYSIS: HUMAN-MACHINE COOPERATION AND KNOW-HOW-TO-COOPERATE

The previous paragraphs have shown that the taxonomies proposed by the industry (intelligent manufacturing, autonomous car, modern aircraft) are very "Domain Specific Modelling" (DSM). However, these different approaches are based on common points, and on more generic human-machine cooperation models. In the next part, we will therefore adopt a more "Generic-Purpose Modelling" (GPM) point of view, to structure the invariants of this human-autonomy symbiosis.

### 2.2.1. Joint Cognitive Systems and Human-Machine Cooperation

**Towards joint cognitive systems**

According to Romero et al. (2016), the emergence of these Human-Autonomy Teams, where human operators are in symbiosis with cyber-physical systems and autonomous agents, is part of a new approach to design adaptive systems. Automation is seen here to increase human capacities, at the physical, sensory, and cognitive levels. "Symbiosis" between the human and the cyber-physical system should be achieved by creating a new hybrid (human and artificial) agent. This vision proposed by Romero et al. (2016) is in line with the notion of joint cognitive systems (Hollnagel & Woods, 2005), emphasizing the idea of co-agency where human and machine are considered together, rather than as separate entities linked by human-machine interactions.

This approach follows and extends the concept of "Balanced Automation System" (Romero et al., 2016, Fantini et al., 2018): it is no longer just a question of finding a balance between manual tasks and automated tasks, but to design an adaptive and dynamic automation, based on a dynamic allocation of functions between man and machine. According to Fantini et al. (2018), autonomous agents and algorithms do not therefore aim to replace humans, but they are on the contrary intended to become assistants or teammates, which allow the operator to continue working, despite age, handicaps, or inexperience, which help to keep people "in the control loop", and which improve the performance and comfort of workers.

**Human-Machine Cooperation: structural and functional approaches**

Cooperation has been the subject of many works in the literature. In the context of human-machine cooperation, cooperation can be understood through different models, according to a structural point of view (Millot and Lemoine, 1998) which defines the structure of relations between cooperating agents (human or machine), or to a functional point of view which describes the cooperative activities that are developed between these agents (Schmidt, 1991, Hoc, 2001).

With the structural approach, Millot and Lemoine (1998) distinguished two levels which allow to characterize the relations between cooperating agents as well as the properties of these agents:

- *Vertical cooperation* involves a hierarchical relationship between human and technical agents in which an agent at a higher hierarchical level supervises and controls the process. The agent at the higher level has authority and is responsible for the decisions. He can ask for help to the lower-level agent, which can give advice and help the supervisor decide. This link between the hierarchical structure and the notions of authority and responsibility is underlined by Flemisch et al. (2012), which indicates, for example, that it would be desirable for an agent to have authority if this agent is supposed to be responsible.

- *Horizontal or heterarchical cooperation* allows for the sharing of operational tasks between agents which act on the process at the same hierarchical level, without a relation of authority between agents. These tasks are not always independent, and the co-activity can then generate interference.

Whatever the cooperation structure, the two agents will have to manage the interferences linked to the co-activity, through different cooperation activities. This is what Schmidt (1991, 1994) calls the forms of cooperation, of which he proposes a typology:

- *Augmentative cooperation* aims to increase physical or intellectual capacities. Agents have the same know-how and cooperate on a common task which can be divided into similar sub-tasks. This form of cooperation helps reduce the workload when it increases and cannot be managed by a single agent anymore.

- *Debative cooperation* aims to compare the various points of view between agents who have the same know-how and who carry out the same task. The cooperating agents can then discuss how to carry out the task, the results of this task, etc. This form of cooperation allows for different perspectives on the same task, and for exercising mutual control within the team to improve performance and reduce errors.

- *Integrative cooperation* is based on the complementarity of agents' know-how. They will thus perform different sub-tasks to perform a common task, by using their different skills.

Finally, according to the functional approach of Hoc (2001), two agents are in a cooperative situation if they meet two minimum conditions. First, each agent pursues goals, and each can interfere with the other agent's activity. Interferences can occur between goals, shared resources, or procedures involving a coactivity. Second, each agent strives to deal with these interferences to facilitate the accomplishment of individual activities, or the accomplishment of the common task if it exists. This interference management can be observed at three levels of cooperation:

- *N1: cooperation in action (or execution).* This "micro" level, the closest to action, distinguishes between different operational cooperative activities of managing goals and procedures in real time and in the short term: local interference creation (e.g., disagreement), local interference detection (e.g., redundancy), anticipation and resolution of interference.

- *N2: cooperation in planning.* This "meso" level is characterized by cooperative activities of developing or maintaining a common frame of reference: maintaining and developing a common objective, a common plan, a distribution of functions.

- *N3: meta-cooperation.* This "macro" level incorporates all the activities that facilitate the two previous levels by developing a common code of communication and models of oneself and of the teammates. This level relies on a longer experience of cooperation, based on some structures of knowledge that emerged during the lower-level cooperative activities, and that will further ease the building of a common frame.

As the models from the literature outlined, human-machine cooperation involves many different processes that boost human-autonomy interdependence. To be implemented, these processes require specific skills from human and autonomous teammates.

### 2.2.2. Know-How-To-Cooperate

Millot and Lemoine (1998) developed the concept of Know-How-to-Cooperate to explain the ability of each teammate contributing to the team performance. They considered that humans and autonomous agents have two different kinds of skill, related to taskwork and teamwork:

- Agents have Know-How (KH) in controlling processes or carrying out tasks.
- Agents have Know-How-to-Cooperate (KHC) with other agents.

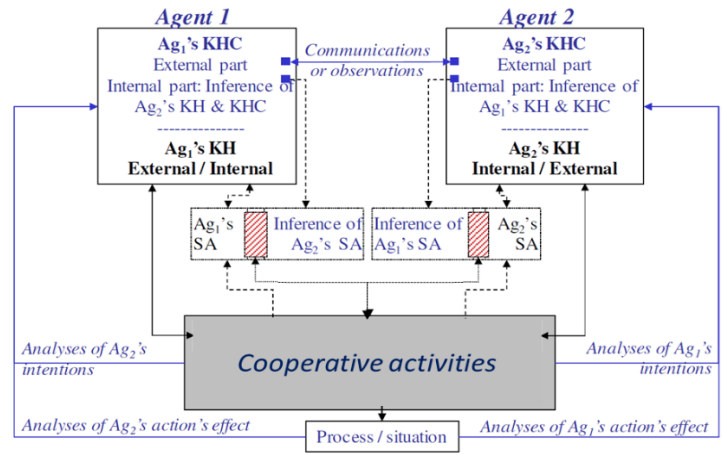

*Figure 11: Agents' KHC in cooperation activities (adapted from Pacaux-Lemoine, 2020)*

As summed by Pacaux-Lemoine (2020), this KHC allows one agent to take advantage of the complementary KH of other agents. This KHC can be considered with internal or external points of view (cf. Figure 11).

- Internal KHC is related to the analytical ability to build a model or a representation of other agents, to better understand and infer other teammates' concerns, expectations, and intentions.
- External KHC deals with the ability to communicate and observe the other agents. It is therefore more related to the perception and the interaction with the environment and the teammates, such as detecting changes in the behaviour of others (mimics, emotions, etc.), detecting changes in the environment or the supervised process resulting from teammates' action, or communicating verbally or through mediating tools.

This KHC must especially supports the cooperative activities described in the previous section, especially by providing agents with interference management capabilities (through coordination and conflict solving), but also with benevolent behaviour facilitating the achievement of others' goals (Millot & Lemoine, 1998).

### 2.2.3. Social and teamwork perspectives for HAT

To succeed in implementing "symbiotic" empowerment of the operator and developing the know-how to cooperate with autonomous systems, we can also approach human-machine cooperation with a perspective of social interactions (Fantini et al., 2018).

**Social perspectives and emergence of broker and interface agents**

In this sense, Romero, Wuest, Stahre and Gorecky (2017) refine their vision of the operator 4.0, by placing it at the centre of a social network, made up of other social human operators, but also of "social" machines and software. This perspective makes it possible, according to these authors, to communicate and exchange information to:

- Improve the representation of the situation by each actor.
- Align or modify the different activities of human and artificial agents (in terms of distribution of tasks and sharing of authority).
- Keep the man in the loop as much as possible, without compromising the objectives of the system.
- Follow and memorize the evolution of technical and human agents, to dynamically determine the capacities of each. These abilities are improved over time through practice and learning through the implemented activities.

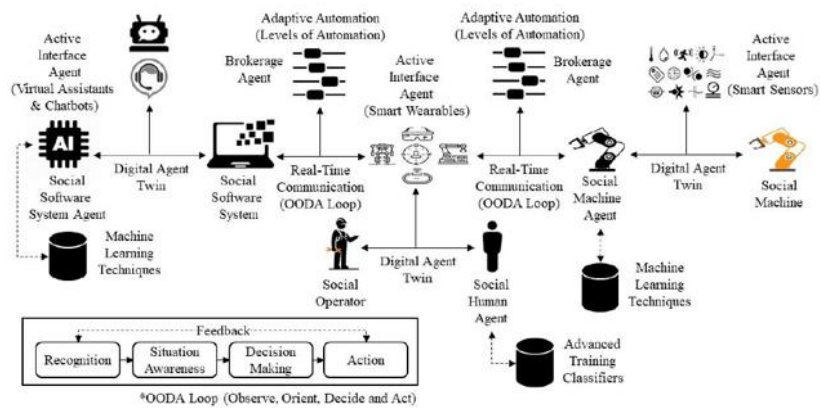

*Figure 12 – Architecture for a social plant (Romero et al., 2017)*

In the field of Industry 4.0, a multi-agent architecture is for example proposed by Romero et al. (2017) to better formalize these social interactions between humans and autonomous agents (cf. Figure 12). This one is based on the introduction of two types of agents:

- *Interface Agents* correspond to a set of rules and conditions supporting interactions between humans, technical agents, and the rest of the system. These interface agents are qualified as active, in the sense that they evolve continuously, by learning (based on observation, imitation, feedback from other agents or programming). They thus make it possible to personalize the assistance provided to the agent, making it dynamic and adaptive, and thus help to keep the social operator or the social machine in the loop when they face difficulties.
- *Broker Agents* correspond to the rules for the allocation of functions and the sharing or delegation of authority. They thus make it possible to adapt the level of automation to optimize human-machine cooperation.

**Teamwork considerations for Human-Autonomy Teaming**

In addition to this social architecture, it is also interesting to revisit the research works on teamwork developed for human teams. For instance, the well-known model of Salas et al. (2005), named the "big five" (that could be a South African name!), insists on some key factors contributing to team adaptability and therefore team effectiveness. They especially pointed out the crucial role of processes such as mutual performance monitoring and back-up behaviour, supported by dynamic knowledge structures, the shared mental models (cf. Figure 13).

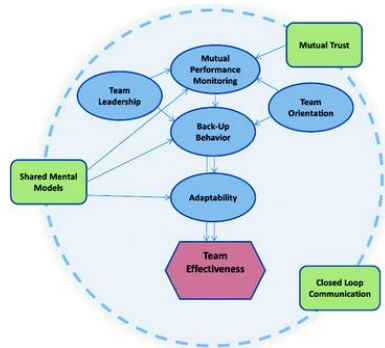

*Figure 13: Factors of teamwork effectiveness (Salas et al., 2005)*

For Salas et al. (2005), Mutual performance monitoring can affect team adaptability and effectiveness by identifying errors or lapses. Indeed, team member feedback can help individuals become more cognizant of their current performance and their potential deficiencies. This process is particularly important when team is engaging in stressful tasks, where team members are more likely to make errors due to mental overload.

*In Human-Autonomy Teams, this mutual performance monitoring is also an interesting concept to develop, especially to detect if human operators are still "functional" on the taskwork.*

Moreover, these authors also mentioned the importance of back-up behaviour, defined as a discretionary provision of resources and task-related effort towards another teammate, when a problem of workload distribution is perceived within the team. Three means of supplying back-up behaviours are distinguished: (a) to provide feedback and coaching to improve performance; (b) to assist the teammate in performing a task; and (c) to complete a task for the team member when an overload is detected.

*In Human-Autonomy Teams, these back-up mechanisms could be instantiated with some feedbacks from the machine to the human aiming at recovering situation awareness or state awareness (on the system or even on their own human state), or with the adaptation of work sharing between human and autonomous system.*

To sum up, attempts to model Human-Autonomy Teams and their internal "social" interactions highlight the emergence of a joint cognitive system. Obtaining such a "symbiotic" system must go through an optimization of human-machine cooperation, relying on specific levers, which are further developed in the following section.

## 2.3. LEVERS FOR HUMAN-AUTONOMY TEAMING

### 2.3.1. Lever 1: Monitoring Operator Functional State

To instantiate a mutual performance monitoring in human-autonomy teams, autonomy must be able to detect if an operator is still functional or not on the task. Indeed, many complex activities carried out by human-autonomy teams (such as operating a swarm of drones, driving an autonomous, or piloting a fighter jet or a commercial aircraft) are often characterized by high-risk situations and strong time pressure. That can raise problems in human performance, mental workload, or attentional tunnelling, that can ultimately lead to errors and getting human out of the loop. It thus becomes crucial to assess online the operator's ability to keep performing the mission, to anticipate potential performance impairments, as well as to activate appropriate countermeasures in time (change in transparency level, dynamic function allocation, etc.). To encapsulate the different elements contributing to a potential degradation of performance from an operator, Hockey (2001) proposes the notion of OFS, namely Operator Functional State. This concept is defined as "the variable capacity of the operator for effective task performance in response to task and environmental demands, and under the constraints imposed by cognitive and physiological processes that control and energise behaviour". This definition first underlines a strong relationship between the operator functional state and his/her performance on the tasks, leading to OFS classification using categories like "Capable / Incapable", "Low risk / Very risky", or more generally classes expressing a gap to expected performance.

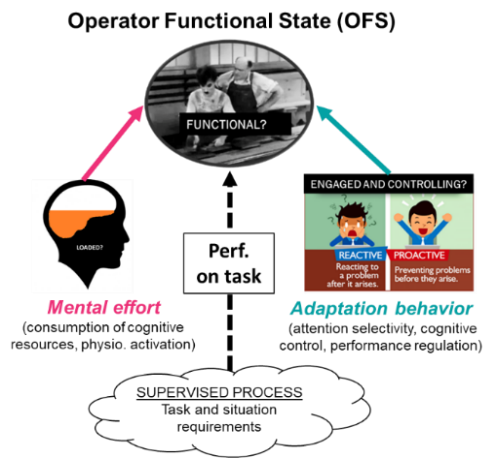

*Figure 14: Different components to study Operator Functional State*

However, it is often difficult to predict a performance collapse of an operator solely based on the analysis of the results of his/her activity. This difficulty is particularly salient for experienced operators: the observable degradations of their performance are indeed only slight and gradual (before a stall) since these operators have regulatory strategies to maintain during a certain time the effectiveness of the main tasks. Therefore, the time of reaction and adaptation of the system to a performance collapse may be too long for the situation and may cause irreversible effects. Thus, the recent research works on OFS (Yang et al., 2014) aims at coping with these difficulties for anticipating a decrease in operator performative capacity, *by combining* (cf. Figure 14), in addition to task performance monitoring, *some metrics related to mental effort* of human operators (i.e. the physiological activation, the consumption of cognitive resources), and *others related to their adaptive behaviour* (i.e. the regulations implemented by operators to efficiently adapt to the situation, according to the trade-off between performance and cognitive cost).

### 2.3.2. Lever 2: Sharing work between humans and agents

Function allocation between human and machine is a major problem in the design of human-machine systems (Chapanis, 1965), because it structures the way system, machines and personnel interact (Scallen & Hancock, 2001). Two approaches address this function allocation problem. Both are based on breaking down a task into several basic cognitive functions: information gathering, information analysis, decision-making, and action implementation.

The first approach considers the allocation as "static", i.e. the distribution of functions is defined and fixed at the time of the design of the system. Initially based on simple MABA-MABA lists (Man-Is-Better-At, Machine-Is-Better-At, Fitts, 1951), this static approach then relied on less binary and more scaled approaches (Sheridan & Verplank, 1979; Endsley & Kaber, 1999). Different levels of automation (LOA) are thus distinguished, to measure the impact on the overall performance of the system but also on operators' workload, as well as on Situation Awareness (or SA). Full automation may reduce SA and can be at the origin of the complacency phenomenon (Parasuraman, Sheridan, & Wickens, 2000). Conversely, the intermediate levels of LOA allow for greater human involvement in the control during normal operations and increase the operator's SA, thus facilitating rapid resolution of the problem when it occurs. However, regardless of the level of LOA chosen, this approach has a major limitation: it does not allow the system to adapt to variations in the situation or, at least, to variations that were not anticipated.

A second approach has appeared more recently. It offers a "dynamic" allocation (DFA: Dynamic Function Allocation), aiming to allocate different functions to man or machine according to the requirements of the situation and, above all, according to the mental load of the operator (Dearden, Harrison & Wright, 2000; Grote et al, 2000; Inagaki, 2003). The question thus moves "who does what" to "who does what under what conditions". While this dynamic approach opens interesting perspectives in terms of human-autonomy adaptability, it also brings other challenges, in particular to determine which one, between machine or human, will have the authority and the responsibility. It also joins the notion of reconfiguration of socio-technical systems (Berruet, 1998).

### 2.3.3. Lever 3: Maintaining and updating the Common Ground

**COFOR as a key element of cooperation, and its maintenance mediated by CWS**

In the literature on human-machine cooperation, several authors have underlined the central role of a structure of knowledge and representations shared by agents, called the common ground (Clark, 1991, Klein et al., 2005) or the COmmon Frame Of Reference (COFOR, Hoc, 2000). COFOR is built and maintained during cooperation situations and makes it possible to manage interference between agents to take it into account in decisions. Close to the concept of shared situation awareness (Salas et al., 1995), COFOR includes a shared representation of the situation, the team and other agents considered as resources. For Hoc (2000), the development and maintenance of a COFOR is therefore an important activity in cooperative situations by playing a central role in the understanding of communications. This shared representation is facilitated by a clear human-machine dialogue, the humans being able to communicate their intentions to the machine based on their decisions or actions, and the machines being able to communicate to the human their status or the activities they carried out.

It is necessary to develop tools to facilitate human-machine dialogue, making it possible to mediate the exchange of information to maintain or adjust COFOR (Millot & Pacaux-Lemoine, 2013). Several works (Hoc & Lemoine, 1998, Pacaux-Lemoine & Debernard, 2002) have shown the role of spaces for dialogue to support COFOR. Millot and Pacaux-Lemoine (2013) call Common Work Space (CWS) these systems aiming at assisting external representation and cooperation. CWS has been studied in different fields, in air traffic control (Millot and Pacaux-Lemoine, 2013), driving (Pacaux-Lemoine and Simon, 2015), or manufacturing (Pacaux-Lemoine et al., 2017). The level of detail of the CWS can be adapted, for example for an industrial system, from a global view (strategic level) to a more detailed view (tactical or even operational level). Millot, Pacaux-Lemoine and Debernard (2011) notably showed how the different forms of cooperation of Schmidt (1991) could be supported by this dialogue, to achieve the different cognitive functions: information taking, analysis, decision making and action implementation (Parasuraman, Sheridan & Wickens, 2000).

*It is therefore through this notion of COFOR, reified and externalized thanks to CWS, that autonomous systems could be more interdependent with their human teammates and more adapted to the situation and the state of the operator.*

**Agent transparency in the management of human-machine dialogue**

The quality of the dialogue between cooperating agents depends on several criteria (Wilson & Sperber, 2002). *It is indeed necessary to have a message whose content will be useful and accurate, i.e., a message with sufficient explanations from the sending agent, leaving little room for uncertainty, and comprising all the elements enabling a good inference from the receiving agent.* Several studies have therefore explored this question of the quality of dialogue, in particular by investigating the issue of transparency. *Indeed, agent transparency would make it possible to refine and guide the dialogue between the different agents.*

Agent transparency is defined by Chen, Procci, Boyce, Wright, Garcia & Barnes (2014) "as the descriptive quality of an interface pertaining to its abilities to afford an operator's comprehension about an intelligent agent's intent, performance, future plans, and reasoning process".

As suggested in the two literature reviews on transparency proposed by Baskhara et al. (2020) or Rajabiyazdi and Jamieson (2020), two main approaches have been adopted in the literature to describe and operationalize the concept of agent transparency, considered in terms of levels or in terms of dimensions: Chen et al.'s (2014) situation awareness–based agent transparency, known as the SAT model, and the framework of transparency for human-robot interaction proposed by Lyons (2013). In these two approaches, transparency is unidirectional: autonomous agents are the senders and humans are the recipients.

*Transparency levels:* the SAT model proposed by Chen et al. (2014) is founded on the Situation Awareness theory of Endsley (1995), on the BDI (Beliefs, Desires, Intentions) Agent framework of Rao and Georgeff (1995), and on the PPP (Process, Purpose, and Performance) framework related to Trust (Lee & See, 2004). The SAT model incorporates three levels of agent transparency. At level 1 (hereinafter referred to as L1), the agent provides basic information about its current state, goals, intentions, plans, progress, current and proposed actions. At level 2 (hereinafter referred to as L2), the agent provides rationale that justify its action or decision. At this second level, the human operator is provided with information about the agent's reasoning, the agent's behavioral capabilities and about constraints that the agent considers. At level 3 (hereinafter referred to as L3), the agent provides a projection of future outcomes (e.g., success rate, estimated time of arrival). At this third level, human operators are provided with information about the agent's anticipation of the future state, consequences, and uncertainties.

*Transparency dimensions:* according to Lyons (2013), two dimensions of agent transparency can be considered to support effective HATs:

- The robot-TO-human transparency concerns information about the robot which is communicated to the human operator. Here, the agent can be transparent about its intent or purpose (the intentional model), about the current task or previous tasks conducted (the task model), about the processes performed that led to a decision or an action (the analytical model), or about aspects of the environment (the environmental model).
- The robot-OF-human transparency concerns the robot's awareness about "the others" which is communicated to the human operator. Here, the agent can be transparent about its perception of the operator's state (the operator model), or about task allocation (the teamwork model).

## 2.4. DEFINITION OF MY RESEARCH AXES IN THE FRAMEWORK OF HUMAN-AUTONOMY TEAMING

This literature review about human-autonomy teaming and human-machine cooperation models and levers, help define my research problem and the two main research axes I have explored, and I am still exploring. The research problem can be summed up as follows (cf. Figure 15):

**"How to improve the interdependence between human and autonomy, to transform machines into better teammates"?**

This problem can be then decomposed into two main research questions:
- Axis 1: "How to make autonomy more sensitive to human state"?
- Axis 2: "How to make autonomy adaptable to human"?

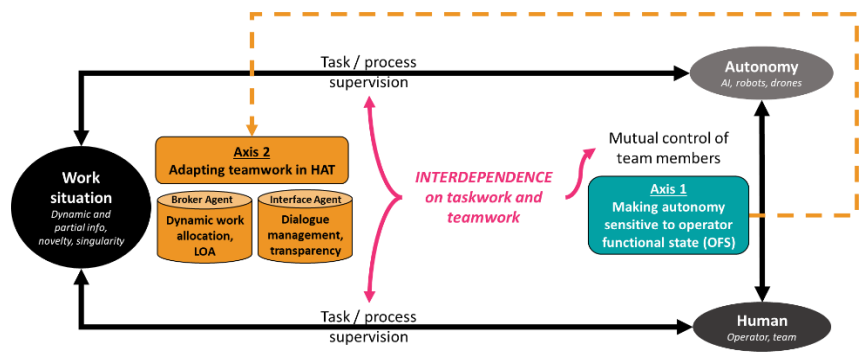

*Figure 15: Research axes on the question of human-autonomy interdependence*

### 2.4.1. Axis 1: Detect - making autonomy sensitive to human state

This first axis deals with the measurement, the classification and the monitoring of some cognitive states related to human performance, such as mental workload, mental fatigue, or stress. The idea is to equip the machine with an ability to detect and to predict when the human operator faces some difficulties to tackle with the dynamic situation or the autonomous system. This sensitivity to the human state could be seen as a first step to make the machine more interdependent with human activity. Indeed, the detection of a non-functional operator state could lead to some adaptations of the work sharing between human and machine, and that would also help humans be aware of their current states and their potential limitations, thanks to a reflexive feedback from the autonomous system. As the magic mirror provided to the Evil Queen in Snow White, the magic could answer to the questions asked by a human operator: "Magic mirror in my hand, am I still the most capable to manage the current situation?".

The contributions on this axis are developed in Part A. This part of the manuscript recalls the approach we have adopted and the main results we have obtained to address this question of the modelling and the monitoring of OFS. First, we proposed a global approach based on the concept of Mental Workload, to classify operator functional state from the fusion of multi-sourced physiological and behavioural data, and to detect some critical states, like mental overload (Chapter 3). Then we explored further the adaptation mechanisms and the regulations implemented by the human operators, by investigating the concept of cognitive control and its relationship with mental workload (Chapter 4).

### 2.4.2. Axis 2: Design offline/Adapt online - making autonomy compatible with human

The second axis focused more on the mechanisms for adapting human-autonomy teaming, making machine more "compatible" with human. The term of compatibility must be here understood at a more operational and short-term level that what Stuart Russell develop in his book "Human compatible", more focused on how an AI could continuously learn human preferences based on human behaviour, to avoid the risk of a kind of Terminator's Skynet AI misguided by rigid human-specified goals. This axis may be of course related to the first research question since the adaptation can be designed, or even dynamically triggered, thanks to the detection or the prediction of a non-functional operator state.

The contributions on this axis are synthesized in Part B, by especially distinguishing between two kinds of solution to adapt human-autonomy teaming. One focused on the proposal of methods for designing offline the know-how-to-cooperate of machines (Chapter 5). The other one deals with online monitoring and adaptation of human-machine cooperation (Chapter 6), where autonomous system can be considered inside the team - as a teammate of the human operator detecting and managing the interferences between their respective activities - or above the team - as a coach supervising interferences that occur with a team composed of human (cf. Figure 16).

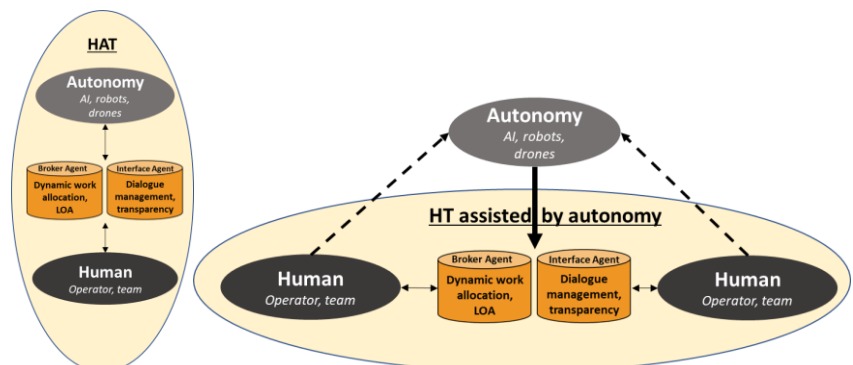

*Figure 16: Different configurations of human-autonomy teaming*

# Part A. Making autonomy sensitive to human: Contribution to the monitoring of operator functional state

# Chapter 3. Online and objective monitoring of Mental Workload

*"Stark: Check the heart, check the, check the, is it the brain? / J.A.R.V.I.S: No sign of cardiac anomaly or unusual brain activity. / Stark: Okay, so I was poisoned? / J.A.R.V.I.S: My diagnosis is that you've experienced a severe anxiety attack. / Stark: Me?"* − **Iron Man**

This chapter synthesizes the different works we have done on the evaluation of mental workload, as a first-order estimator to classify operator functional state, and to further trigger some system adaptation when a "non-functional operator state" is detected, like mental underload or overload (cf. §6.1).

These propositions mainly result from a PhD funded by DGA between 2013 and 2017, carried out by Alexandre Kostenko (2017). This PhD was a starting point that allowed to open new industrial partnerships with Dassault Aviation (TRM104 project in 2017), AXA (2018) and Thales (PEA MMT PRECOGS between 2019 and 2021), as well as international collaborations, with colleagues from Israel and Australia. This axis also results in many scientific papers and reports (cf. [C20, C22, C24, C25, C28, C34, C35, O6, O7, O9, O12, R5, R6, R9, R10]).

## 3.1. ONE CONCEPT, MANY APPROACHES TO UNIFY

### 3.1.1. A polysemic concept, between theory and data

The concept of mental workload (MWL) is extensively used, but is also very controversial, giving rise to various theoretical and methodological models (Cegarra and Chevalier, 2008). At present, debate continues on the dimensions to be considered to represent and estimate the MWL, and the way in which these dimensions are connected (Longo, 2017). Following the ergonomics principles of standard DIN ISO 10075-1:2017 (2018), mental workload is a construct that is not directly defined, but rather explained according to its causes and its effects. Mental workload is therefore related to the notion of mental stress (i.e., the causes of mental workload, the task load imposing constraints upon operators) and to the notion of mental strain (i.e. the mental effort, the cognitive cost of the task for the operators).

Over the last four decades, MWL has been modelled and assessed according to different approaches, following this twofold explanation. Firstly, there were exogenous approaches based on the study of activity constraints (Knowles, 1963) and performance (De Waard, 1996). Then, research has also focused on endogenous estimators, which are based on the concepts of operator capacity and mental effort (Sheridan & Stassen, 1979), and new possibilities of measuring physiological activity (Kahneman, 1973). These two approaches are complementary. Indeed, as emphasized by Christopher Wickens, Karel Brookhuis, Luca Longo and Sarah Sharples in a panel discussion during the first symposium on human mental workload in 2017 (Wickens et al., 2017), it could be useful to take advantage from both theory-driven approaches (with multi-dimensional models) and data-driven approaches (using massive data and information fusion). Grouping them together led to the emergence of a holistic model, as proposed by Hart and Staveland (1988).

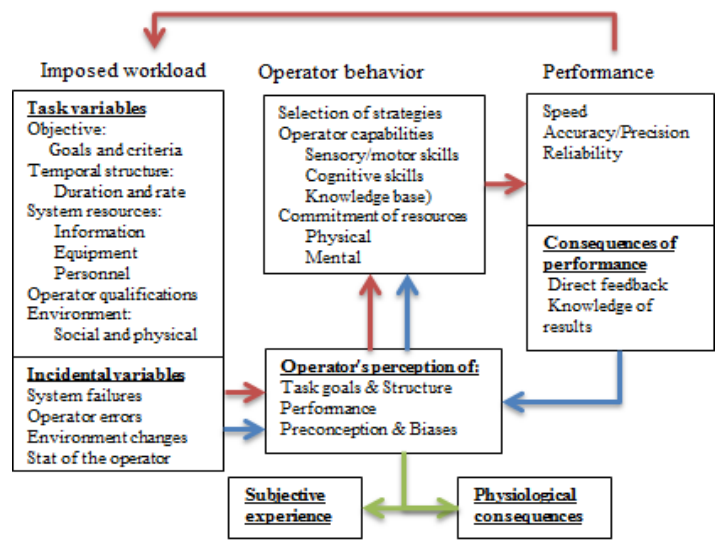

*Figure 17. MWL model, adapted from Hart and Staveland (1988)*

In this model (Figure 17), MWL is a multidimensional construct, relying on several key points:

- *MWL results primarily from information processing (in red):* The constraints (imposed workload) represent the input of the cognitive process. These constraints are thus perceived and interpreted by the operator, what guides him/her for choosing and implementing strategies (operator behaviour). The output of this process is represented by the results obtained from the actions of the operator (activity performance). It should be noted that the information processing is closed-looped: the achieved performance influences the variation of the constraints (for example, if the operator falls behind, the temporal constraint will increase). Constraints can be considered as "MWL drivers", and performance as a "MWL indicator"; as proposed by De Waard (1986), an underperformance is related to either an excessively high or an excessively low workload.

- *MWL can be estimated by its psychophysiological consequences (in green).* The information processing does not only result in the performance of an activity. The operator behaviour also generates effects on the operator himself/herself, which can be measured by objective physiological variations or subjective feelings. These two kinds of variables, introduced by the energetic approach and the concept of mental effort (Kahneman, 1973), can be therefore considered as "MWL indicators".

- *The central place of perception explains the dynamics of MWL (in blue).* Based on a perception of the situation and of his/her own activity (evaluation of constraints and performance), the operator interprets and anticipates the evolution of the situation. This understanding influences the choice of the strategies implemented. This continuous and reflexive assessment of the activity can therefore generate MWL variations, that can be regarded as another class of MWL drivers, that we refer to here as "MWL mediators".

The MWL can therefore be seen as a multidimensional construction, that can be estimated by indirect measures, since some variables are drivers or mediators (cause-based analysis of what happens on the task and the operator) or indicators (consequence-based analysis of what happens on the task and the operator, cf. Table 1).

*Table 1: Cause/Effect and Task/Operator dichotomies to characterize MWL*

| MWL | CAUSE RELATED TO | EFFECT ON |
|---|---|---|
| TASK | CONSTRAINTS | PERFORMANCE |
| OPERATOR | REGULATIONS | PSYCHO-PHYSIOLOGICAL STATE |

The perception process is central in the model; it guides the behaviour of the operator. However, this dynamic and reflexive evaluation of the situation is quite complex to measure. To better understand this black box "perception", it is important to investigate the regulation loops.

### 3.1.2. Feedback control of MWL with various regulation loops

In his studies on air traffic control, Sperandio (1971) identified two regulation loops.

- *Loop 1:* the feedback of the MWL, resulting from the implemented strategy, has a regulating effect on the future strategy. Indeed, if the implemented strategy imposes a high cognitive level, the operator could change strategy to obtain a lower cognitive level.

- *Loop 2:* the strategy implemented in response to the perception of constraint allows a regulation of the future level of constraint.

To these two loops, we can add a third regulation loop, presented by Leplat (2006):

- *Loop 3:* The difference between the actual and the expected level of performance will have an incidence on the selection of future strategies.

Hence, the dynamic behaviour of an operator is guided by the perception of the situation that can be modelled by the three different loops proposed by Sperandio and Leplat. These regulation loops depend on the assessment of variables already existing in the multi-dimensional model of Hart and Staveland.

**A feedback control system to regulate MWL**

To synthesize the consideration of regulation loops into MWL model, we proposed a feedback loop model (cf. Figure 18). This model, proposed by Alexandre Kostenko ([C24], Kostenko, 2017), illustrates the causal and temporal structure of MWL, where the dynamic constraints (the task or the situation demand) at given instant t-1 may generate some psychophysiological consequences at t, and some performance variations at t+1.

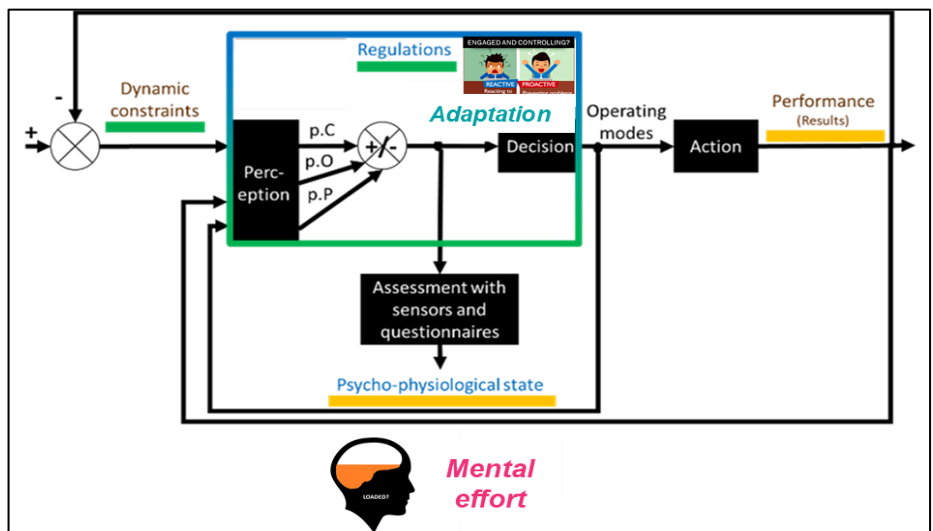

*Figure 18: MWL dynamic closed looped and multidimensional model*

That converges with the idea developed by De Waard (1996), about the use of psycho-physiological state monitoring for better assessing mental workload and for anticipating performance decrement, instead of only studying the well-known U-curve relating task demand and performance to estimate mental workload. In this sense, the variations of psycho-physiological indicators related to the notion of mental effort appear very central to the model. They allow to better understand the mobilization of cognitive resources between the perception of the situation difficulty and the decision-making of the operator.

Furthermore, this model also insists on the key place of regulation loops. Mental workload is therefore seen as a dynamic state, resulting from the adaptation mechanisms implemented by the operators. These mechanisms especially depend on some individual characteristics (such as operator's experience or his/her level of engagement in the task at a given moment), but they also rely on the self-perception and the self-assessment of some discrepancies between current task demand, task performance and adopted cognitive behaviour.

These are these adaptation mechanisms, often considered within a black box, we tried to better define as regulation loops.

**Integration of regulation loops in the modelling of Mental Workload**

According to the model of Hart and Staveland, the operator performs the information processing, and then chooses and implements a strategy primarily in response to the perceived constraints. Moreover, according to the models of Sperandio and Leplat, the perceived operator behaviour, task constraints, and performance have an influence on the choice of strategy, in a dynamic and reflexive assessment of the situation. The choice of strategy is therefore based on the perception of three criteria: the constraint, the behaviour, and the performance. It can be assumed that the pairwise comparison of these three criteria would allow the operator to select a strategy and regulate his or her activity. From this comparison, we have identified and integrate several regulation loops (cf. Figure 19, see [C24]):

- *The effectiveness loop (Performance-Based Regulation: PBR).* This loop corresponds to the comparison between the prescribed objectives (task variables partially defining the constraints) and the performance. If there is a great difference between the actual and the expected performance, this means that the behaviour is not effective. In such a case, the operator should implement more effective strategies.
- *The relevance loops* correspond to the comparison between the constraint and the behaviour. We can identify two comparisons: the prescribed strategies vs. the implemented strategies, and the dynamic constraint vs. the cognitive cost induced by the complexity of implemented strategies.
  - ○ *Compliance-based regulation (CBR):* in the first case, the operator is aware that the implemented strategies do not comply with the strategies prescribed for a targeted better performance (which form part of the constraints as variables characterizing how the tasks must be performed). The operator will regulate his/her activity by adopting a more pertinent behaviour, and by implementing the prescribed strategies.
  - ○ *Priority-based regulation (PRBR):* in the second case, if the dynamic constraints are high and the strategies implemented are the costliest, the behaviour can be also irrelevant. The operator should regulate his/her activity by defining priorities and implementing some of the less costly strategies. For instance, in studies of air traffic control, Sperandio (1988) observed that, when there are many airplanes, the controller focuses only on safety, and no longer takes account of fuel economy or time of transit, etc.
- *The efficiency loop (cost-based regulation, COBR),* which corresponds to the comparison between performance and behaviour. If the performance is good and the cost is very high, the behaviour is not the most efficient. The operator would regulate his or her activity by implementing less costly strategies.

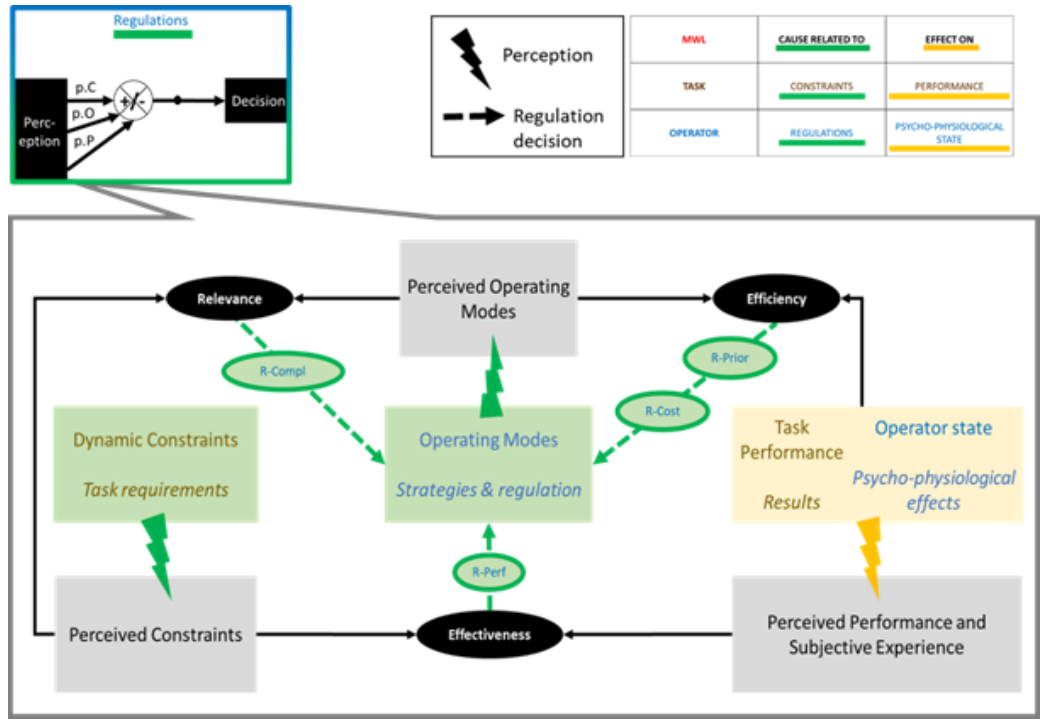

*Figure 19: Objective coding regulations based on pairwise comparisons of MWL dimensions*

### 3.1.3. A multi-sourced evaluation of mental workload

There are three categories of basic parameters commonly used in ergonomics research to assess MWL (Young et al., 2015): measures of task performance in the primary and/or the secondary task coupled with activity performance assessment, subjective reports and physiological measures. Besides these criteria, Sperandio (1971) proposed to evaluate the MWL by studying the variation of the implemented strategies (i.e., regulations). We briefly describe these types of measures, by regrouping them into the three classes of MWL variables.

**MWL Drivers (activity-based estimators: task load, constraints).** These measures involve evaluating the prescription (objective, strategies, etc.) and the quantity of information to be handled, etc. For example, in the traffic controller domain, many constraint indicators are used for evaluating the workload: the number of aircrafts to control (Arad, 1964), the number of conflicts to manage (Hilburn, 2004), etc.

**MWL Mediators (operator-based estimators: Variation of implemented strategies, regulations).** This measure consists of observing changes of strategy during the activity. The operator regulates the activity based on his/her evaluation of the situation. Table 1 summarizes how the regulation loops can be externally observed and qualified according to the objective states of the exogenous MWL variables (activity constraints, performance and behaviour).

*Table 2. Regulation loops of MWL*

| Objective evaluation of the perceived situation Comparison between 2 criteria | | Diagnosis | Regulation |
|---|---|---|---|
| Constraint (objectives) | Performance (results) | The results do not reach the objectives | **PBR:** The most effective strategies are implemented |

| Constraint (Prescribed strategies) | Behavior (implemented strategies) | The implemented strategies do not comply with the prescription | **CBR:** The prescribed strategies are implemented |
|---|---|---|---|
| Constraint (Dynamic demand and changes) | Behavior (Cognitive cost) | The behavior is too costly | **PRBR:** The least costly strategies are implemented |
| Performance (results) | Behavior (Cognitive cost) | The behavior could be less costly | **COBR:** The least costly strategies are implemented |

**MWL Indicators (activity or operator-based estimators).**

- <u>Performance:</u> The measure consists of studying, in an objective way, the success or failure in reaching the objectives, the duration of the tasks, etc. When the MWL is very high, the performance degrades (De Waard, 1996).

- <u>Subjective experience:</u> This measure involves direct questioning of the operators about the activity, to find out how they feel. Several questionnaires or standardized scales are typically used in ergonomics research and in the domain of "Human factors". The most commonly used scales are the NASA-TLX (Hart & Staveland, 1988), or the ISA (Jordan, 1992).

- <u>Physiological consequences:</u> This measure consists of observing the variation of physiological state (heart rate, etc.) and ocular activity (pupil diameter, etc.). When the MWL increases, there is also an increase in heart rate (Kramer, 1991) and average pupillary diameter (Beatty, 1982).

For Cegarra and Chevalier (2008) as well as for Hancock (2019), it is necessary to combine these different measures, because each has advantages and disadvantages, and none can claim to perfectly measure mental workload. Data fusion of multi-sourced information can thus improve the quality of the mental workload evaluation. This would contribute to a gain in sensitivity, as the measurements can complement each other, one measuring aspects that the other cannot address. This combination would also make it possible to improve the selectivity, since certain measurements (in particular physiological data) are difficult to control. Finally, it would facilitate interpretation and reinforce the diagnostic character (especially by crossing physiological measurements, subjective questionnaires, and elements of contextualization on the activity and the requirements of the task).

### 3.1.4. Regulation loops detection for online MWL monitoring

To integrate the regulation loops in the online monitoring of MWL, it is important to be able to code and to detect them from objective metrics, especially by studying the changes in operator behaviour and by understanding the rationales of these behavioural variations. To this purpose, we first proposed to detect the regulations according to the variations of operator compliance with the different tasks occurring at a given moment (a kind of multitasking performance). Then we classified the regulations into different types according to the cognitive trade-off, to understand if the regulation implemented by the human is more oriented by the research of a better performance, or the research of a lower cognitive effort. It must be noticed we considered here the task according to the nature of what the operator must achieve (cf. Figure 21), and independently from the number of objects to be processed in this task. Hence, a task becomes active when there is at least one object requiring action from the operator, and it gets inactive when there is no more object to treat. The detection and the classification of the regulation loops follow the rules presented above, as illustrated in the decision tree of Figure 20. Let be the following variables:

- **Ti** a task, with i between 1 and n the number of different tasks in a multitasking environment. Each Ti is associated with a prescribed strategy **PSi** (it can be an optimal operating mode between many, or simply doing the task vs not doing it)
- **ATi(t)** the activation of the task Ti at a given instant t (0 if Ti is inactive, 1 if Ti is active).
- **OTi(t)** the achievement status of the task Ti by an operator at a given instant t. Each task Ti is given a time budget. This time budget is used for each new object requiring the task Ti. If the operator does not achieve the task Ti with the prescribed strategy PSi for a given object, Oti(t) equals 0, until another object is processed in time (OTi(t) comes back to 1).
- **NPS(t)** the number of Prescribed Strategies to comply with. $NPS(t) = \sum_{i=1}^{n} ATi(t)$
- **CPS(t)** the number of Prescribed Strategies that an operator carries out.
  $CPS(t) = \sum_{i=1}^{n} \left( ATi(t) - OTi(t) \right) = NPS(t) - \sum_{i=1}^{n} OTi(t)$
- **Perf(t)** the global performance of the operator on the activity

We finally pose **ΔCPS(t)**=CPS(t)-CPS(t-1) and **ΔNPS(t)**=NPS(t)-NPS(t-1), the temporal variation of the variables between two consecutive instants.

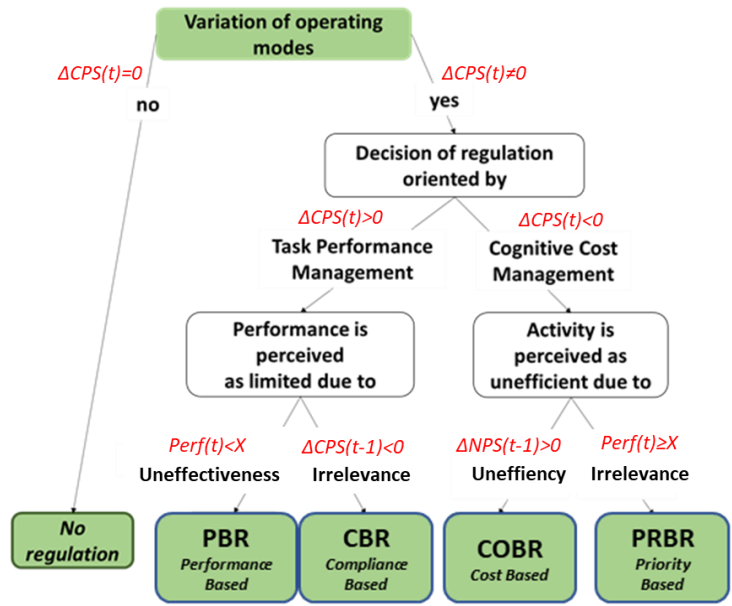

*Figure 20: Decision tree for coding regulation loops*

**Detection of a regulation loop**.

When ΔCPS(t)≠0, a regulation is detected.

**Classification of regulation loops based on the cognitive trade-off**

*Task performance management: PBR and CBR loops*.

If ΔCPS(t)>0, the regulation is considered with a performance orientation (the operator complies more with prescribed strategies).

- If Perf(t)<X, with X the threshold for acceptable performance, this is a PBR loop (the operator reacts to an ineffective situation, guided by the perception of performance)
- If ΔCPS(t-1)<0, this is a CBR loop (the operator reacts to the discrepancies between what must be done and what is actually done, guided by the compliance with given strategies)

*Cognitive cost management: COBR and PRBR loops.*

If $\Delta CPS(t) < 0$, the regulation is considered with a cognitive cost orientation (the operator does less, to reduce the strain).

- If $\Delta NPS(t-1) > 0$, this is a COBR loop, the activity raises a too high cognitive cost.

- If $\Delta NPS(t-1) \leq 0$, this is a PRBR loop, the operator prioritize to be more efficient with equivalent performance.

## 3.2. A USE CASE FOR MONITORING OPERATOR MENTAL WORKLOAD

To improve our understanding of the relations between these classes of variables and the related measures, and to validate our approach in (especially on the way to take regulation into account in the multidimensional model of MWL), a set of experiments was designed in an environment involving complex and naturalistic decision-making. The experiments were carried out in the specific domain of aerial surveillance, where a human operator must manage a swarm of drones to detect potential threats on a given location. A simulator, developed at IMT Atlantique, was used to define the scenario, and to collect the data (Coppin & Legras, 2012).

### 3.2.1. Surveillance activity with SUSIE simulator

The simulation consists in securing a given area by piloting a swarm of drones. The aim of the human activity is to find, identify and neutralize different mobile targets, that are hidden at the beginning of the simulation. This simulator was chosen because it presents several advantages for our study: it is adapted to the emergent multitasking activity of and operator managing several drones; it is ecological and complex enough to consider the future implementation of proposed MWL monitoring on real system ; and it however constitutes a microworld on which the experimenter can control the different simulation parameters (task difficulty, target location and appearance time, etc.) and can record behavioural (actions achieved by participants with mouse clicks) and performance indicators (e.g. reaction time for the different target states, or number of processed targets over time).

Susie is a Java based software that allows participants to interact with and to supervise a swarm of drones using a mouse-screen system. Only one operator is required, but some tasks can be or are achieved by an artificial agent. The system provides different information to the operator from two sources: a dynamic map and a message banner on the right (Figure 21, top image). The dynamic map gives information about the areas that the drones control such as the vehicles in these areas and their state. The message banner indicates the coordinates and direction of a vehicle that the operators need to assign high priority to its neutralization. The main task is to detect and neutralize the threats (i.e., hostile vehicles) on the map. When a vehicle is generated by the software, it is hidden, i.e., it is present on the map but invisible (it must be detected by drones sent by the participant). Before it is neutralized, the status of the vehicle changes several times (Figure 21, middle image). To advance from one status to the next, operators need to complete many sub-tasks, summarized as follows in the table at the bottom of Figure 21.

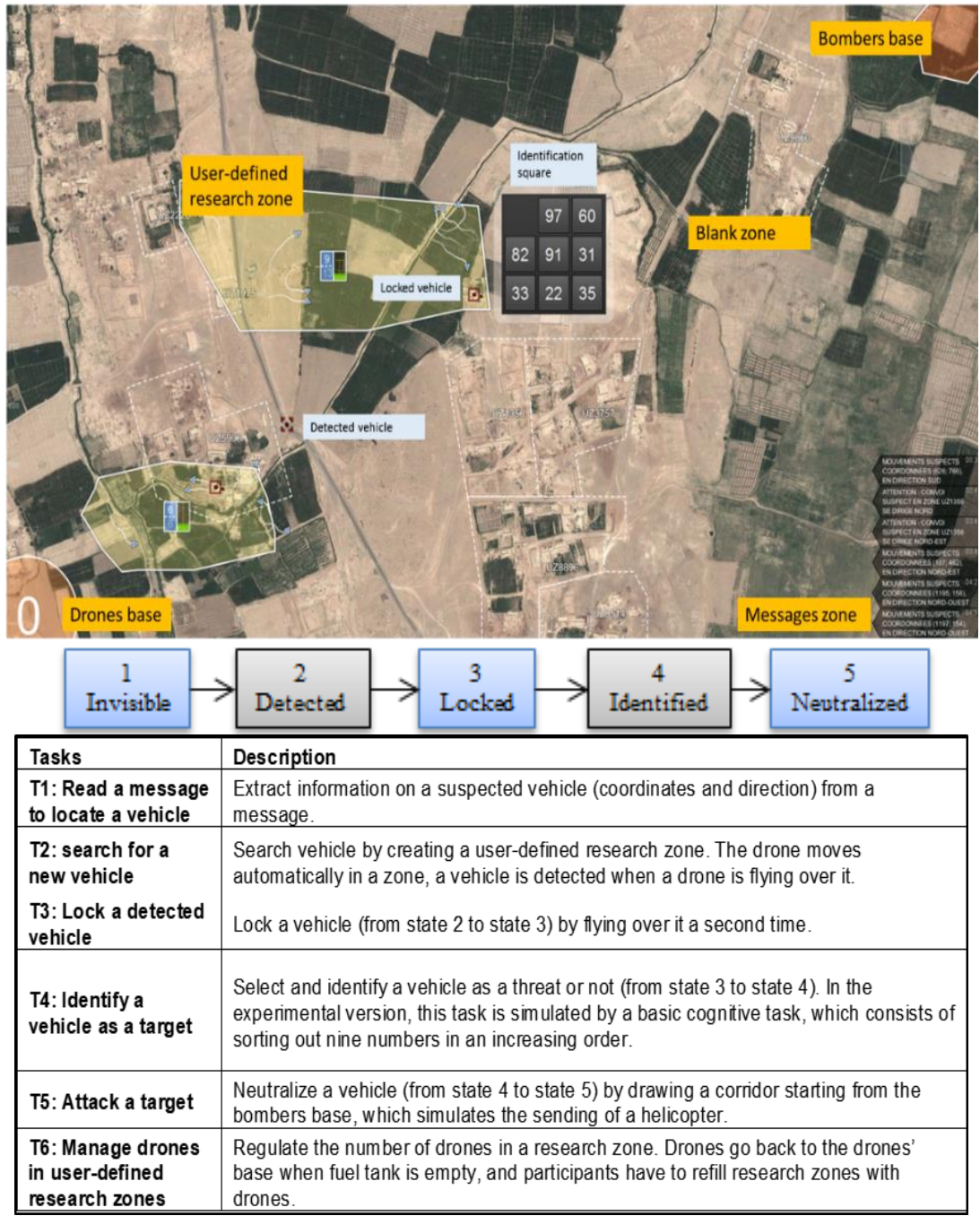

| Tasks | Description |
|---|---|
| **T1: Read a message to locate a vehicle** | Extract information on a suspected vehicle (coordinates and direction) from a message. |
| **T2: search for a new vehicle** | Search vehicle by creating a user-defined research zone. The drone moves automatically in a zone, a vehicle is detected when a drone is flying over it. |
| **T3: Lock a detected vehicle** | Lock a vehicle (from state 2 to state 3) by flying over it a second time. |
| **T4: Identify a vehicle as a target** | Select and identify a vehicle as a threat or not (from state 3 to state 4). In the experimental version, this task is simulated by a basic cognitive task, which consists of sorting out nine numbers in an increasing order. |
| **T5: Attack a target** | Neutralize a vehicle (from state 4 to state 5) by drawing a corridor starting from the bombers base, which simulates the sending of a helicopter. |
| **T6: Manage drones in user-defined research zones** | Regulate the number of drones in a research zone. Drones go back to the drones' base when fuel tank is empty, and participants have to refill research zones with drones. |

*Figure 21. SUSIE environment, sequential target states and related tasks*

### 3.2.2. Participants, scenario, and training

22 participants, aged from 22 to 30 (mean: 23, standard deviation: 2,4) took part in the experiment. For reasons of homogeneity, all were men from the same scientific diploma, and had good experience with video gaming (this point was controlled by a short questionnaire). Since there is no expert operator for the monitoring activity of swarm drones (the system is indeed experimental), we have considered that this population, with these scientific and practical skills, could represent the future drone operators.

The scenario lasted 20 minutes and was composed of two phases: phase A of low difficulty (10 minutes) and phase B of high difficulty (10 minutes). The difficulty level was modified by changing the distribution rate of new vehicles and new messages. To limit the effects due to the order in which the phases are carried out (learning effect, etc.) two similarly sized groups of participants were created. The first group carried out phase A followed by phase B, while the second group carried out phase B followed by phase A. Both theoretical training (10 minutes) and practical training (20 minutes) were carried out. The theoretical training consisted of presenting the system, while also providing the objectives and the prescribed strategies (or Operating Modes). Two objectives were set: i) Neutralize 25 vehicles in 20 minutes, ii) for all messages, draw a zone in relation to the message within a delay of two minutes. Six prescribed strategies were provided, one for each task (cf. Figure 21). The practical training consisted of taking the system in hand. During this phase, the participants were frequently reminded about the strategies.

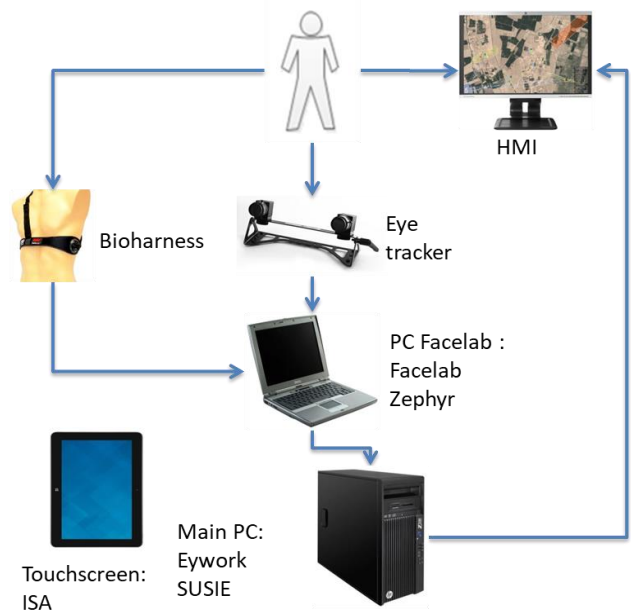

*Figure 22: Experimental set-up*

Participants performed the experiment in a room where the brightness is controlled and constant, to avoid effects on the variations in the pupillary diameter. In this room a space has been created for running the scenario on the simulator, with fixed desk and chair (to avoid parasitical movements of participants). The system supporting SUSIE software is composed of a 24" screen and a mouse connected to a PC. Data acquisition was carried out using a Seeing Machines FaceLAB5© eye tracker for pupillary response, a Zephyr Bioharness© heart rate belt, and a log (text file) of scenario events (vehicle appearance, message) and operator's mouse actions recorded on SUSIE. Subjective experience was also collected with ISA scale.

### 3.2.3. Data selection and processing

We define the different measures chosen in our experiment, according to the three classes of variables presented in §3.1.3.

**Drivers.** Two types of constraint were considered: informational constraint is related to the distribution rate of new vehicles and new messages fixed by a software parameter (in this experiment, it is low in phase A and high in phase B). Multi-tasking constraint correspond to the number of tasks an operator must carry out at a given time.

**Mediators.** To detect the occurrence of each of the four regulation loops (cf. Table 2) during the experiment, we measured the compliance between the prescribed strategies (or Operating Modes) and the implemented strategies at every second, for the six different tasks involved in this surveillance activity (cf. Figure 21). Then we looked at the objective evaluation of the situation, in terms of constraints, performance and cognitive cost, as explained in §3.1.4.

**Indicators.** This concerns performance as well as subjective, physiological and ocular measures.

- Performance measures. The performance value is calculated from two criteria: i) neutralization time (from the first detection to neutralization), ii) the observation that a zone relative to a message was drawn in a delay of two minutes or not.

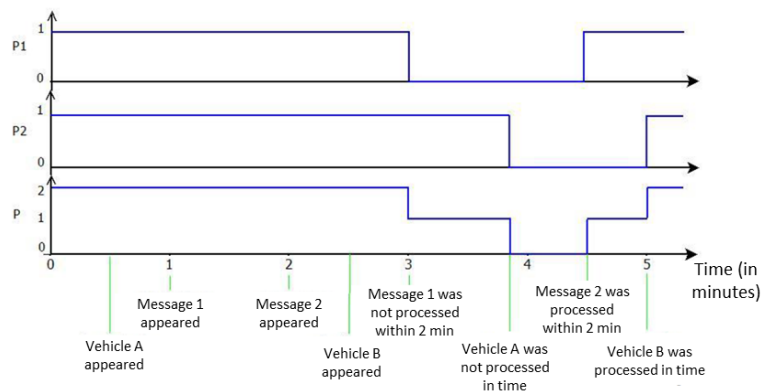

*Figure 23: Calculation of overall performance indicator*

- Physiological and ocular measures. The physiological data (Heart Rate Variability or HRV, and Pupillary Diameter or PD) were collected with the aid of the equipment described above. HRV was computed from the standard deviation of the NN interval on the last 100 heartbeats (SDNN method), following the formula $sqrt\left(\frac{\sum_{i=1}^{N}(RRi - mRR)^2}{N-1}\right)$. Pupil diameter was cleansed (diameter smaller than 2mm and larger than 8mm were excluded). All the physiological data were z-normalized to remove the interindividual differences.

- Subjective measures were recorded by using an ISA scale during the experiment, popping up every 90s on the screen.

## 3.3. A SYNTHETIC AND ONLINE INDICATOR OF MWL

### 3.3.1. Validation of the regulation loops

Two main hypotheses were studied to validate our approach on the regulation loops coding.

First, it was assumed that task constraints influence the implementation of the regulation loops. We expect that, in phase B, the operator complies less with the prescribed strategies and uses less costly strategies than in phase A. Second, the relationship between regulations loops and performance and physiological indicators was investigated. For each variable, we tested the data distribution using the Kolmogorov-Smirnov and Lilliefors tests. These tests allow us to choose between a t-test (parametric) and Mann-Whitney U test (nonparametric).

**Effect of constraints on regulation loops.** The first analysis consists of dividing the scenario according to the informational constraint level (i.e., phases A and B) and studying the effect of dynamic constraint on the implemented regulation loops. As depicted on Figure 24, the results show that the operator significantly implements PBR (performance-based), CBR (Compliance-based) and PRBR (priority-based) more often in phase B than in phase A (respectively, $t(21) = -3.15$ and $p < 0.01$; $Z(21) = -2.60$ and $p < 0.01$; $t(21) = -4.32$ and $p < 0.001$). On average, in phase A, a regulation loop is implemented 14 times, as against 19 times in phase B for PBR, 8 times as against 14 times in phase B for CBR and not at all as against 19 times in phase B for PRBR. No difference was found for COBR (cost based). Statistics showed operators' compliance with prescribed strategies is lower in phase B than in phase A ($t(21) = 5.81$ and $p < 0.001$).

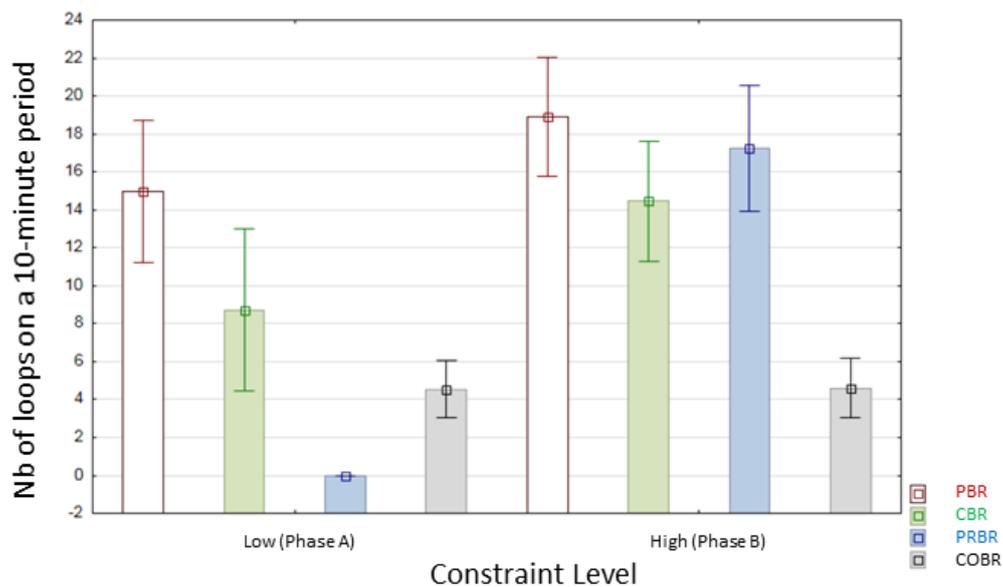

*Figure 24: Effect of constraint level on regulation implementation*

**Effect of regulation loops on performance.** The second analysis consists of dividing the scenario according to whether the operator regulates his own activity, for each loop. The objective is to analyse the effect of regulation on the indicators. This analysis is performed at constant constraint (only phase B was considered here). Statistical analyses revealed that implementation of the PBR loop implies a significant increase of performance ($Z(21) = -3.23$ and $p < 0.01$). On the contrary, the implementation of the PRBR loop results in a significant decrease of performance ($t(21) = -4.30$ and $p < 0.001$). Finally, no significant difference was observed physiological mental effort for all loops, nor for performance with the implementation of CBR and COBR loops.

**Findings.** These two analyses therefore showed that operators could implement different types of regulation to manage the dynamic constraints of the work situation, and that resulted in variation in performance and mental effort. Moreover, our proposed coding of the regulation loops (§3.1.4 and §3.2.3) seems consistent with what it can be expected in terms of regulation causes (constraint level) and consequences (variations in performance and physiological effort).

### 3.3.2. Design and validation of a synthetic online MWL indicator

**Approach to build an online MWL indicator, based on objective measures**

We first selected the different performance, behavioural and physiological metrics that were sensitive to the variations in task load. Hence, we statistically verified that the two performance indicators were significantly correlated to change in constraint. Moreover, we observed that only the pupillary response was sensitive to the variations in constraints or in regulations, but not HRV metrics. Finally, only PBR, PRBR and CBR regulation loops significantly varied with constraints. Therefore, HRV indicators and COBR loop were discarded from our data fusion.

To obtain a complete and robust measurement, we have carried out a data fusion from the selected indicators. To do this, we have selected probability theory (Bayesian network or BN) and fuzzy set theory. Bayesian networks are proving to be a relevant choice in our case. Indeed, they make possible a measure that manages the heterogeneity and uncertainty of low-level information. Ji, Lan and Looney (2006) use Bayesian networks to estimate operator fatigue in real time. Likewise, Rachedi (2015) implements a BN to characterize the state of the operator in the field of railway driving. Moreover, our choice to use fuzzy logic converges with the propositions of Yang et al. (2009) and is explained by the need to discretize continuous low-level information to feed Bayesian Networks with precise and uncertain information.

We first treated low-level information with fuzzy logic (cf. Figure 25), then we merged them by Bayesian networks.

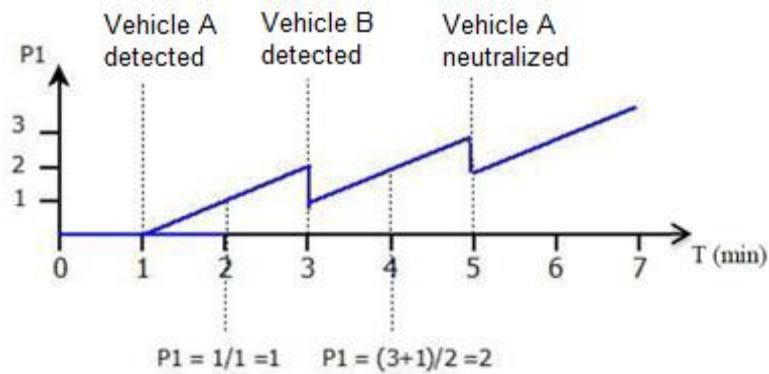

*Figure 25: Fuzzy logic for discretizing continuous data*

To build this Bayesian network, we first created four indicators representing the dimensions of mental workload (cf. Figure 26), one for constraint, one for behaviour, one for performance and one for physiological activity. Conditional Probability Tables were based on the statistical analysis of the collected data during the experiment.

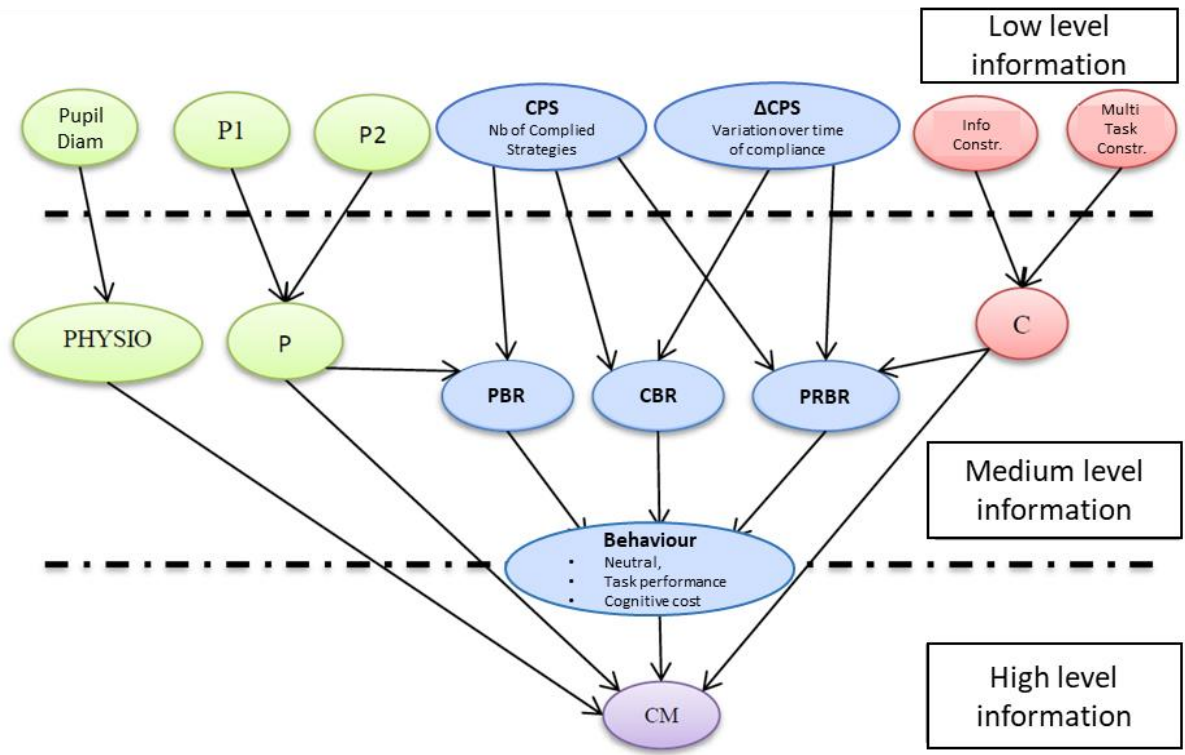

*Figure 26: Bayesian Network for the data fusion of MWL*

These four indicators were then merged to create an objective synthetic indicator of mental workload. This indicator was created with 5 levels, by analogy with the ISA subjective scale, which allows us to obtain a fine measurement.

This information fusion approach for mental workload converges with the current trend, in particular with the work of the AFRL 711th Human Performance Wing (Human Factors branch of the US Air Force), relating to the PACER - Performance Augmentation Computing Engine in Real time (Durkee et al., 2015), which contextualize physiological and performance data by measuring constraint level. There may also be links with the work of Schulte et al. (2015), which considered regulations to estimate mental workload in real time.

**Validation of the online objective MWL indicator**

To validate the obtained online MWL indicator resulting from the data fusion of objective metrics, we compared this computed this synthetic indicator with the ISA ratings, on the time stamps where ISA questionnaire appeared (cf. Figure 27, left side)

The synthetic and objective mental workload indicator and the subjective rating measured using the ISA scale are ordinal qualitative variables. A Spearman test was used to study the correlation between these two ordinal variables. This shows that the two measurements are positively correlated, r = 0.534, p <0.001 (cf. Figure 27, right side).

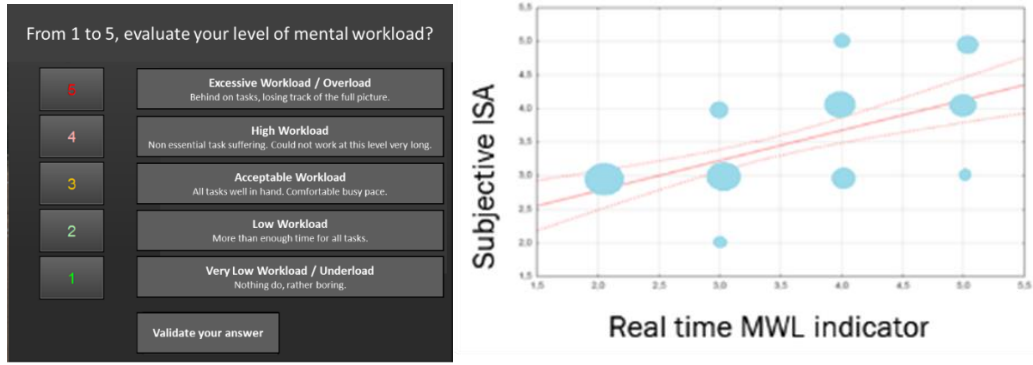

*Figure 27: Comparison between objective MWL indicators and subjective ISA scale*

### 3.3.3. Improving physiological data classification of mental effort with machine learning

The previous proposal for building an online MWL indicator relied on intuitive and simple approach for estimating physiological mental effort (Alexandre Kostenko's PhD). In a collaboration with Dassault Aviation (TRM104 project), we aimed at going further by applying supervised machine learning for classifying physiological data like pupil diameter and heart rate variability. To classify physiological and contextual data, different methods were proposed in the literature.

Kotsiantis (2007) summarizes the main methods of supervised machine learning. Some methods can only deal effectively with one type of data: discrete data for "Naïve Bayes Classifiers" and "Rule-Learners", and continuous data for neural networks (NN), k-Nearest Neighbours (kNN) and Support Vector Machines (SVM). Only decision trees can handle both types of data efficiently. Moreover, there are relatively recent references to the classification of physiological signals, which allow a slightly less generic analysis, but more centred on the use of learning techniques applied to the classification of Operator Functional States (cf. Table 6). Regarding these two syntheses, and since we seek to classify continuous data (like heart rate variability of pupillary diameter) with more than two classes of mental effort, 3 classification methods were considered here: Support Vector Machine (SVM), kNN (k-Nearest Neighbours) and Random Forest (RF). SVM is efficient for processing continuous data, and this technique is effective for managing multi-collinearity. Moreover, SVMs are more tolerant of missing data, better manage the phenomenon of over-learning and generally have better reliability. Furthermore, kNN and RF have the advantage of being of a multi-class nature.

**Classification and validation of task difficulty indicator**

Prior to classify the mental effort from physiological signals, we looked to determine accurate labels to supervise the machine learning. To do that, we first carried out a classification of the task difficulty level (TD), based on indicators of task constraint and performance.

*Creation of the task difficulty indicator.* The computation of the difficulty indicator TD is divided into two stages: first, raw data were processed by discretization into two or three categories: the threshold between these categories were set according to a preliminary study of the subjective experience of participants on different conditions of task difficulty. Moreover, the different discretized variables were initialized at the beginning of the scenario, by considering different constraint variables (cf. Table 3); then, a data fusion of the discretized indicators was achieved by aggregation, to obtain a global indicator of task difficulty with three categories (TD1=low, TD2=medium, TD3=high). Aggregation rules were defined as follows: if {N1d=High and N2d>=Medium and Entropy>=Medium} then TD=High.

*Validation of the task difficulty indicator.* To validate the difficulty indicator (TD), we studied the correlation between this indicator and the subjective assessment of the task difficulty, collected every 90 seconds with a Likert scale. As the difficulty indicator TD and the subjective assessment are ordinal qualitative variables, a Spearman test was therefore used to study the correlation. It showed that these two variables are positively correlated ($r = 0.711$, $p < 0.001$). The task difficulty indicator TD can therefore be used to supervise the classification of physiological signals.

*Table 3: Data processing for task difficulty indicator*

| Variables related to task difficulty | Raw data | Definition | State space | Discretized data | State space |
|---|---|---|---|---|---|
| **Constraint variables** | N1 | Represents the number of targets that must be processed | {0, 1, …} | N1d (targets) | **low**: N1 <= 5<br>**medium**: 5 < N1 <= 11<br>**high**: N1 > 11 |
| | N2 | Represents the number of messages that must be processed | {0, 1, …} | N2d (messages) | **low** : N2 <= 2<br>**high**: N2 > 2 |
| | Entropy | Represents the spatial entropy of the different targets on the whole monitored space | Continue | Entropyd | **low**: Entro <= 0.45<br>**medium**: 0.45 < Entro <= 1<br>**high**: Entro > 1 |

**Supervised classification of physiological states**

The aim of this stage is to classify the operator's mental effort into 3 categories, by applying a supervised learning of the physiological signals with the three different task levels defined in Stage 1. Two main physiological indicators (pupillary diameter and HRV) were selected for this classification. Figure 28 summarizes the principles of classification processes that were implemented. The reliability of the three supervised learning methods identified in the literature (SVM, kNN and Random Forest) will be evaluated by the cross-validation method, which consists of dividing the total sample into two subsamples, the first to be used for learning, and the second to test the learning. For each of these methods, we tested different settings. For SVM, we used several kernel methods (polynomial, RBF, sigmoid), under different parameters. We also tested several types of distance (Euclidean, Euclidean squared, Manhattan and Chebychev) and several K values for kNNs. RF algorithm was tested with different numbers of trees.

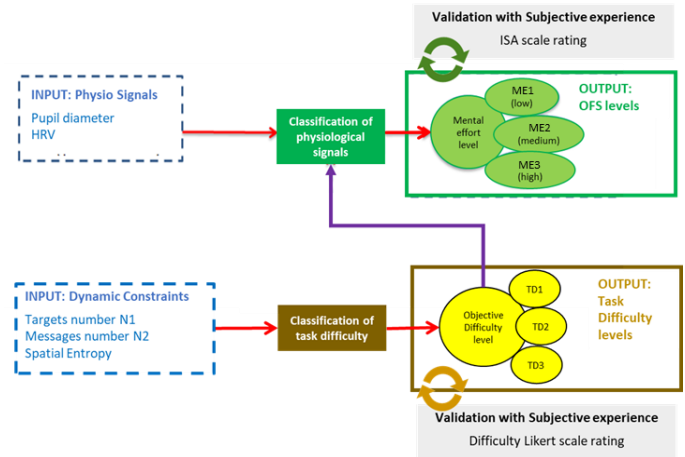

*Figure 28: Supervised learning of physiological data*

Data from 17 participants were collected to achieve the supervised learning of mental effort, within an experiment carried out on SUSIE simulator. After raw data cleansing and normalization, mean values of pupillary diameter and heart rate variability were calculated every second. The supervised learning was carried out at two different levels, one on all participants and one on each participant independently. For the first classification (all participants), the training was achieved from the data of 13 participants, and the data of the 4 left participants were used as a test sample to check the accuracy of the process. For the second one (for each participant), 75% of the data was used to form the training sample, and the resulting classification was then tested on the remaining 25% of the data. For both levels of learning, we made a classification into three and two classes.

The three different elected methods (SVM, kNN and RF) were tested with different parameters. As mentioned in Table 4, SVM and RF produced better classification on all participant data than kNN method, for both 3-classes (respectively 61,06%, 58,27% and 49,62%) and binary classification (respectively 83,12%, 79,70% and 77,43%).

*Table 4: Classification accuracy for all participants*

|  | Supervised learning algorithms | Best settings | Global accuracy | Accuracy of low class | Accuracy of medium class | Accuracy of high class |
|---|---|---|---|---|---|---|
| **3-classes classification** | SVM | kernel sigmoid (gamma = 0.5 & coeff = 0) | 61,06% | 55,61% | 13,81% | 100% |
|  | kNN | Chebychev Distance & k=1 | 49,62% | 76,51% | 47,06% | 46,98% |
|  | RF | 68 trees | 58,27% | 89,52 | 49,47% | 74,74% |
| **2-classes classification** | SVM | kernel sigmoid (gamma = 0.5 & coeff = 1) | 83,12% | 54,54% |  | 95,94% |
|  | kNN | Chebychev distance & k=18 | 77,43% | 49,46% |  | 91,83% |
|  | RF | 23 trees | 79,70% | 78,29% |  | 88,55% |

For the classification of individual data, Figure 29 shows that SVM and kNN are more accurate than RF method for producing 3-classes (respectively 78,29%, 74,94% and 44,06%) and binary classification (respectively 91,35%, 89,71% and 69,94%). Among all the tested methods, the SVM thus give the best results. In addition, the binary classification by the SVM also give better results than the classification in three classes (83,12% versus 61.06% on all participants, and 91,35% versus 78,29% at individual level).

| | Supervised learning algorithms | Best settings | Average global accuracy | Standard deviation | Minimum global accuracy | Maximum global accuracy |
|---|---|---|---|---|---|---|
| **3-classes classification** | SVM | kernel sigmoid (gamma = 0.5 & coefficient = 0) | 78,29% | 12,13% | 58% | 100% |
| | kNN | Distance Chebychev & k=1 | 74,94% | 14,79% | 55% | 100% |
| | RF | 68    Trees | 44,06% | 8,71% | 33% | 63% |
| **2-classes classification** | SVM | kernel sigmoid (gamma = 0.5 & coefficient = 1) | 91,35% | 9,08% | 63% | 100% |
| | kNN | distance Chebychev & k=18 | 89,71% | 9,87% | 63% | 100% |
| | RF | 23    Trees | 69,94% | 12,30% | 42% | 95% |

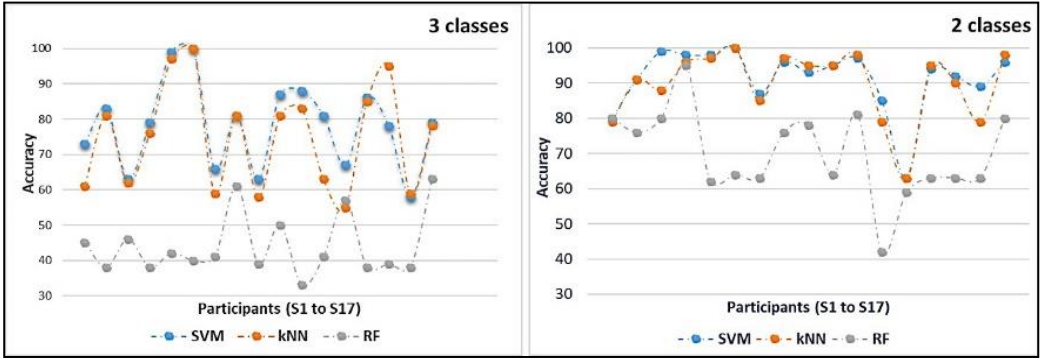

*Figure 29: Classification accuracy for each participant*

**Validation of the mental effort classification**

The classification was therefore *a posteriori* validated, by comparing the resulting mental effort classification values with the subjective values collected on ISA questionnaire, collected every 90 seconds during the scenario. The contingency table (cf. Table 5) between mental effort level and ISA level is given below. We analysed whether the mental effort was associated with ranking on the ISA questionnaire on these temporal points. Based on the results of the study, those with higher mental effort ranks were more likely to have scores that ranked higher on the ISA questionnaire, r=0.42, p<0.05.

*Table 5: Contingency table for mental effort and ISA*

| | | Mental Effort | | |
|---|---|---|---|---|
| | | Level 1 (Low) | Level 2 (Medium) | Level 3 (High) |
| ISA | ISA 1 | 52% | 16% | 1% |
| | ISA 2 | 33% | 30% | 8% |
| | ISA 3 | 11% | 23% | 23% |
| | ISA 4 | 4% | 22% | 44% |
| | ISA 5 | 0% | 10% | 24% |
| Total | | 27 | 195 | 104 |

To sum up, compared to many studies that arbitrarily and *a priori* set the level of task difficulty, the proposed classification method uses dynamic and objective task difficulty labels to supervise data learning, and these labels were cross validated by the subjective experience expressed by participants. Moreover, the results of supervised learning methods on HRV and pupil diameters to classify mental effort showed that the SVM method, and to a lesser extent the kNN method, produced robust classifications.

Table 6 shows that the results of the present study are similar, and sometimes better than the other works found in the literature.

*Table 6: Methods applied to physiological data classification*

| Authors | Data | Number of classes | Supervised learning method | Accuracy |
|---|---|---|---|---|
| Gagnon et al. (2016) | HR, BR, Gaze position, PD, BLKf, HRV | 2 | kNN | Individually: 91% All subjects: 89% |
| | | | SVM | Individually: 84% All subjects: 83% |
| Gilani, (2016) | HRV, EEG | 2 | kNN | All subjects: 81% |
| | | | SVM | All subjects: 56% |
| | | | RF | All subjects: 87% |
| Yin et Zhang (2011) | EEG, ECG, et EOG | 2 | LS-SVM | Individual: 93% |
| | | 3 | SVM (1 vs 1) | Individual: 72% |
| Zhang et al. (2015) | EEG, ECG, et EOG | 3 | SVM | Individual: 69% |
| **Current study** | **HRV, PD** | **2** | **SVM** | **Individually: 92% All subjects: 83%** |
| | | | **kNN** | **Individually: 90% All subjects: 77%** |
| | | | **RF** | **Individually: 70% All subjects: 80%** |
| | | **3** | **SVM** | **Individually: 78% All subjects: 61%** |
| | | | **kNN** | **Individually: 75% All subjects : 50%** |
| | | | **RF** | **Individually: 40% All subjects : 58%** |

## 3.4. DISCUSSION

OFS monitoring opens interesting prospects for HAT, especially as a trigger for adaptation mechanisms, such as dynamic function allocation or dialogue management. This can help reduce risk, increase system safety and performance. However, this OFS monitoring must also be considered cautiously to guarantee the evaluation of the OFS is not perverted (Wickens et al., 2017). Indeed, sometimes, the question of OFS is only used to reduce the size of the teams, guide only by the maintenance of an acceptable level of workload and performance, without increasing operator comfort or engagement level.

With the approaches we have developed in this chapter, we considered here mental workload as a first approximation of the OFS. One of the key points in the modelling and the monitoring of mental workload is the integration of mental effort and regulations. Understanding the behavioural regulation loops and the variation in physiological mental effort allows to anticipate the deleterious consequences of an inadequate mental workload, such as underload or overload (purple sections on 30, adapted from De Waard, 1996). Another important key point of our proposal is to produce a non-binary OFS classification, to be able to anticipate certain performance decrement, and react on time and in appropriate manner. It is why the proposed 5-level objective indicator of MWL can help to better trigger HAT adaptation (cf. §6.2).

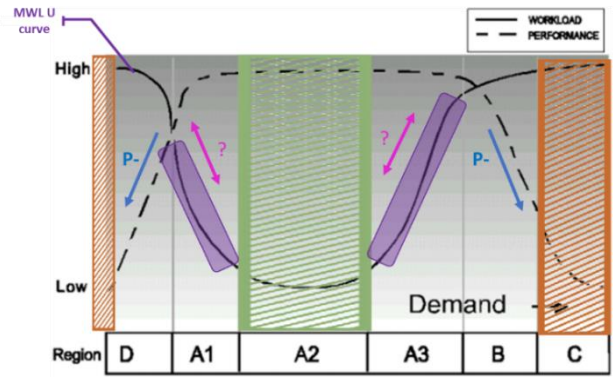

*Figure 30: OFS monitoring to anticipate later performance decrement*

Finally, it is also worth noting the importance of contextualizing objective indicators of mental workload. In particular, it is necessary to have a model of the task that help assess the level of constraint, and to relate it to the mental effort and the actions implemented by the operator. Indeed, with understanding the context and the relation with task demand, there is a risk to activate inappropriate countermeasures or alerts, for instance when the operator produces an important effort to overcome a problem or to find a strategy. It is why it is important to study mental workload over a long term rather instantaneously, and to understand better the dynamics of the regulation loops that expert can implement when facing complex situations. The next chapter explores these adaptation mechanisms, by investigating the question of cognitive control.

# Chapter 4.  Consideration of cognitive control in human state monitoring

*"It is not the strongest of the species that survives, nor the most intelligent. It is the one that is the most adaptable to change."* **– C. Darwin**

Monitoring and process control activities carried out in dynamic situations are characterized by the management of uncertainty and risk, the multiplicity of tasks, and the complexity of the controlled systems (Hoc, 1996). From a cognitive perspective, these activities call for diagnostic/prognostic and decision-making processes that use both internal data processing (i.e., mental models relating to the controlled systems or to environmental dynamics) and external data processing (i.e., information that is available in the environment or interfaces). Hence, in dynamic environments, the pilot of a mobile vehicle (e.g., aircraft, car or ship) is more or less proactive (when he/she relies on mental models to act) or more or less reactive (when his/her actions are mainly driven by external data).

This chapter illustrates the studies we conducted to better understand the adaption mechanisms implemented by human operators, by dynamically adopting different cognitive control modes. This research work especially relies on the CIFRE PhD of Amine Laouar (2021), funded by AIRBUS, between 2018 and 2021. The following paragraphs explain the approach and the findings of a laboratory experiment we did during this PhD, and that was also replicated, with convergent results in a simulator study with Airbus, as well as in the project PRECOGS with Thales and a collaboration with IRT System X (Renault). These works were disseminated through different papers [J10, J11, C31, C37, O12, R9, R10].

## 4.1. THE CONCEPT OF COGNITIVE CONTROL

### 4.1.1. Definition

Cognitive control is one of the key concepts in contemporary cognitive science. It refers to processes that allow information processing and behaviour to vary adaptively from moment to moment, depending on current goals, rather than remaining rigid and inflexible.

In the field of cognitive ergonomics, two models of cognitive control have been proposed to account for the behaviour of operators in dynamic situations: the SRK taxonomy of Rasmussen (1983) and the COCOM model proposed by Hollnagel (1993). As Hoc and Amalberti (2007) point out, these two models focus on two different aspects of cognitive control. The SRK taxonomy considers the level of abstraction of the data processed during supervision activities (sub-symbolic vs. symbolic data), whereas the COCOM model accounts for the more or less reactive or proactive nature of the observed behaviours. The taxonomy defined by Rasmussen distinguishes three different levels of control. The skill-based level results in the implementation, without conscious attention, of cognitive automatisms and automated and strongly integrated patterns of actions. At the rule-based level, behaviour is guided by known rules or procedures. The knowledge-based level is brought into play to solve new problems requiring the definition of new rules, innovation, and creativity.

The COCOM model puts the emphasis on temporality. It distinguishes among four main control modes (Hollnagel, 1993, 2002): strategic, tactical, opportunistic, and scrambled. The strategic mode is used only when there is considerable time available. It involves managing several goals simultaneously and using predefined or generated plans to address a situation. Hence, it requires considerable attentional resources. The tactical mode is based upon using known rules and is used to process a limited number of goals. When the available time is only just sufficient, operators are likely to use an opportunistic mode that focuses on managing one goal only. Hence, the resulting choice of action is determined by the most salient information. Finally, the scrambled control mode is used when the time available is extremely limited. In that case, planning is impossible, and the choice of action is random; consequently, the operators no longer control the situation.

### 4.1.2. Dynamics of cognitive control and mental workload

Many studies have already used the notion of control modes to account for operators' performance in dynamic situations (Stanton, Ashleigh, Roberts, & Xu, 2001; Eriksson & Stanton, 2017; Chauvin, Said, & Langlois, 2019). Stanton et al. (2001) have especially shown people move between cognitive control modes (coded with COCOM models) in a linear manner when the situation does not change too abruptly. They did not observe any disruptive transitions (for instance between scrambled and strategic modes), but instead slight variations towards neighbour modes (like from tactical mode to either strategic mode or opportunistic mode). Research conducted in cognitive neuroscience has also identified some relationships between control modes and neurophysiological activation (Braver, 2012). The proactive control mode is characterized by the maintenance of goal-relevant information in working memory, which optimizes attention, perception, and response preparation. It relies on a sustained activity of the dorsolateral prefrontal Cortex (dlPFC). In the reactive control mode, attention is mobilized as part of a late correction mechanism, and decision making is guided by stimuli (Mäki-Marttunen, Hagen, & Espeseth, 2019a). This mode of control is linked to a more transient activation of the dlPFC (Ryman, El Shaikh, Shaff, Hanlon, Dodd, Wertz, C. & Abrams, 2019). As summarized by Braver (2012, p. 106), "proactive control relies upon the anticipation and prevention of interference before it occurs, whereas reactive control relies upon the detection and resolution of interference after its onset". In the field of cognitive neuroscience, different experimental studies have shown a link between cognitive workload and cognitive control, as a heavy workload leads to the adoption of a reactive control mode (Mäki-Marttunen, Hagen, & Espeseth, 2019b). Furthermore, Cegarra, Baracat, Calmettes, Matton & Capa (2017) examined the relations between cognitive control modes and mental workload. In this previous study, the notion of cognitive control was viewed from the perspective of Rasmussen's Skills, Rules, Knowledge (SRK) taxonomy (Rasmussen, 1983). Hoc and Amalberti (2007) explained that the SRK taxonomy deals with an aspect of cognitive control that involves the level of abstraction of the data processed as part of monitoring activities (i.e., sub-symbolic or symbolic data). In the present study, another aspect of cognitive control is focused on, namely the source of the data being processed: proactive control involves internal data (i.e., mental models), whereas reactive control involves external data. Cegarra et al. (2017) have especially shown that the skill-based level is associated with a lower mental load than the rule-based level. However, to our knowledge, the relationship existing between the four modes of the COCOM model, and the mental workload of operators was not investigated yet.

Here, mental workload is viewed from both aspects of mental stress (i.e., the constraints imposed upon operators) and mental strain (i.e., the cognitive cost of the task for the operators), following the ergonomics principles of standard DIN ISO 10075-1:2017 (2018). This study aimed at:

- Investigating how cognitive control transits from one mode to another one (as defined by the COCOM model) when the mental stress varies between low and high constraints,
- Exploring if some physiological measures of mental effort could be sensitive and related to different control modes.

## 4.2. A MATB-II EXPERIMENTAL APPROACH FOR INVESTIGATING COGNITIVE CONTROL

The experiment used the MATB-II microworld Multi-Attribute Task Battery (Santiago et al., 2011), which has already been used in the studies of Cegarra et al. (2017) and Vanderhaegen et al. (2019).

The experiment entailed asking participants to execute a main task for which optimum performance would require adopting a strategic mode; the task involved managing the content of fuel tanks. The task was repeated three times, and its complexity (i.e., mental stress) increased each time.

### 4.2.1. Participants, experimental set-up, and scenario

Twenty participants in the 18-21 age group (M = 18.55; SD = 0.83), all male, were recruited from among the student population of Université Bretagne Sud. All showed normal hearing and normal vision (or corrected to normal vision). The participants were informed of their rights and provided written consent for their participation, in line with the Helsinki Declaration. Participants were asked to perform tasks in the MATB-II environment shown in Figure 31. MATB-II is a microworld that enables people to execute four tasks that are characteristic of flying an aircraft. In this experiment, one of these tasks was used as a "main task".

The main task, called "Resource management", simulated process monitoring. It involved the fuel management of a civil aircraft, using a set of six fuel tanks and the pumps that connect them. The instructions were to keep both upper tanks at a stable level of 2,500 units (symbolized by blue marks on the tanks), keeping in mind that the level was automatically reduced to simulate the fuel consumption of the engines. To do so, both tanks needed to be continuously supplied from the other tanks through the pumps that, however, could break down. The secondary tasks were as follows: a tracking task that involved keeping a target at the centre of a marker, and a system monitoring task that involved spotting anomalies in the position of markers (see the dial in the upper left corner of Figure 31).

A preliminary phase was used to explain the tasks to be executed during a 15-minute training session followed by a test session aimed at ensuring that participants had fully understood the instructions. The experimental phase was broken down into three 7-minute sequences (see Figure 32). The first sequence involved the main task only (referred to as "resman"), the second one involved the main task and the secondary system monitoring task (named "with track"), and the third one required participant to execute both the main task and the secondary tracking task (named "with sysmon").

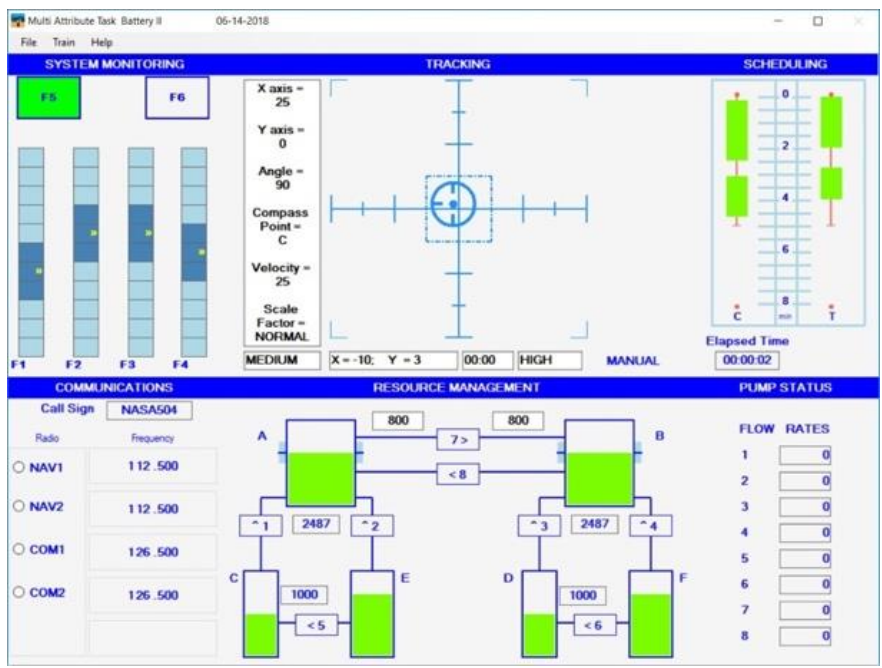

*Figure 31: Screen capture of the MATB-II window*

It should be noted that the main task ("resman") is a continuous task, since the participants must manage the levels of the 2 reservoirs which evolve every second, and whose dynamics (filling or emptying) can change with the occurrence of failures. Furthermore, as explained by Philips et al. (2007) and Gutzwiler and Wickens (2015), we can also distinguish the two different secondary tasks of our scenario: tracking is a continuous task, which requires permanent control of the trajectory, while the monitoring system is a discrete task, consisting of acknowledging alarms when they appear on the screen.

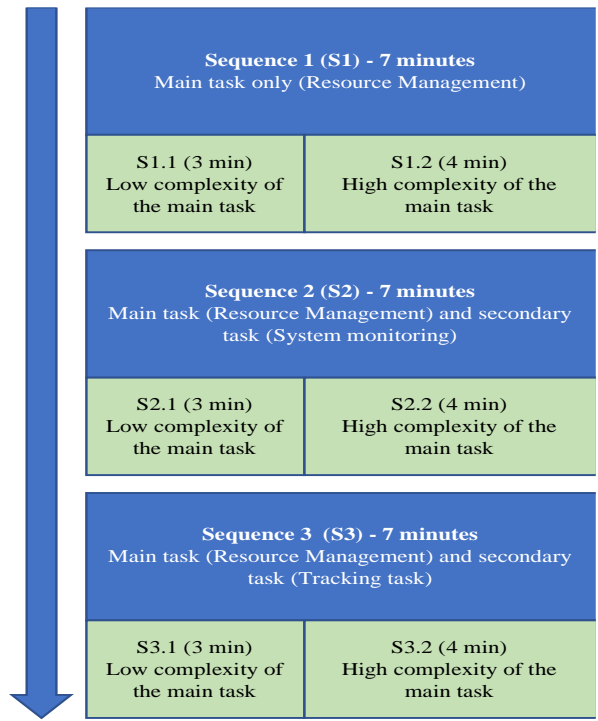

*Figure 32: Three experimental sequences, each with two levels of complexity of the main task*

Thus, the succession of sequences in our scenario results in an increase in difficulty: first there is a continuous task alone, then a continuous task with a discrete task (which generates "discrete" stimuli occasionally disturbing the participant in the main task), and finally two continuous tasks (which requires the control of two processes whose dynamic evolution must be managed).

Furthermore, Figure 32 shows that each sequence itself was broken down into two periods: one 3-minute period during which executing the main task was less complex and a second 4-minute period during which the long breakdown of one pump made the task more complex. At the end of each 7-minute sequence, a NASA-TLX questionnaire was given to the participants through the MATB interface.

### 4.2.2. Measures and coding

The performance of the main task has been coded to identify the modes of cognitive control likely to be adopted by the operator. To do this, we studied whether the participants complied with the instructions for the task (i.e., keeping the level of each of the 2 reservoirs between 2000L and 3000L), and how they managed their "safety margin" compared to the low threshold of 2000L (the emptying of the reservoirs increase when there are more pump failures). Table 7 shows the characteristics of the operations used to operationalize each control mode likely to be used for the main task of fuel tank management. In accordance with Hollnagel's model, the strategic and the tactical modes are associated with a satisfying performance, whereas the opportunistic mode is associated with some errors and the scrambled mode with a poor performance.

*Table 7: Characteristics of the control modes*

| Control mode | Margins |
|---|---|
| Strategic | Complying with instructions and high margins for at least one of the two |
| | tanks (maximum upper value between 2,750 and 3,000) |
| Tactical | Complying with instructions and lower margins for both tanks (values oscillate between 2,000 and 2,750 around the target value of 2,500). |
| Opportunistic | Errors for at least one tank; the participant takes action when the minimal value (between 2,000 and 1,950) is exceeded |
| Scrambled | Serious errors for at least one tank; the minimum value (inferior to 1,950) or the maximum value (superior to 3,050) is exceeded by a large margin when the participant intervenes. |

Two types of physiological data were collected and analysed as indicative of mental strain: cardiac activity with Bioharness 3 belt (Zephyr, Medtronic, Ireland), and oxygenation and deoxygenation of the prefrontal cortex with the 8-channel functional near-infrared spectroscopy (fNIRS) system (Octamon, Artinis Medical, Netherlands). These sensors were especially chosen for their known and robust relationships with mental strain (see Table 8) as well as their ease of implementation in real world application (without too many interferences with ambient factors, such as light variations. The processing of fNIRS data was performed using a bandpass filter (0.01hz-0.09hz). To choose the cut-off frequencies we followed the approach of Pinti et al. (2019), which advocates a low frequency of 0.01hz and a high frequency lower than the Mayer wave frequency (0.1hz).

*Table 8: Relationships between neurophysiological indicators and mental strain*

| | Indicators |
|---|---|
| **Cardiac activity** | Heart rate variability (HRV), computed within time-domain parameters with the standard deviations over 100 successive RR intervals *Relationship with mental strain* Decreases with an increased mental workload (Malik, 1996, Durantin et al., 2014) |
| **Prefrontal Cortex activity** | Concentrations of oxygenated hemoglobin (HBO2) and deoxygenated hemoglobin (HBB) on the 8 optodes of the fNIRS (with T1 and T2 located on the right dorsolateral cortices, T3 to T6 located on orbitofrontal cortices, and T7 and T8 located on the left dorsolateral cortices) *Relationship with mental strain* Neuronal activity is associated with an increase in concentration of oxygenated hemoglobin and a decrease in deoxygenated hemoglobin (Fairclough et al., 2018, Causse et al., 2019) |

### 4.2.3. Data analysis method

We used a two-step methodology to analyse the data. A chi-square analysis was conducted to investigate possible links between mental stress and control modes. Additionally, we used R (R Core Team, 2012), and especially the lme4 package (Bates, Maechler & Bolker, 2012), to perform linear mixed effects analyses. Visual inspection of residual plots did not reveal any obvious deviations from homoscedasticity or normality. As fixed effects, we entered sequence (single or double task), complexity (low or high, corresponding to the fact that the main task has few or many incidents on the pumps) and cognitive control mode (with interaction terms) into the full model. As random effect, we had intercept for participants. Regarding fixed effects, a stepwise model selection by AIC (stepAIC) was conducted. During each step, a new model was fitted, in which one of the model terms was eliminated and tested against the former model.

## 4.3. TRANSITIONS IN MODES AND fNIRS SENSITIVITY

### 4.3.1. Effect of mental stress upon the control modes

First, we conducted a multinomial logistic regression between control modes and the two factors related to mental stress (sequence and complexity). No effects of interaction were found between these two factors.

Then, we examined the effect of the complexity of the main task by comparing the control modes adopted when complexity is low (first period) and when it is more important (second period) as shown in Table 9.

*Table 9: Adoption of the cognitive control modes according to the complexity level*

|  | Scrambled mode | Opport. mode | Tactical mode | Strategic mode |
|---|---|---|---|---|
| Low Complexity | 4 | 3 | 26 | 23 |
| High Complexity | 15 | 10 | 3 | 28 |

We observed that tactical and strategic modes are largely adopted when the task complexity is low. The strategic mode is still used when the complexity increases but the tactical mode disappears. A Bhapkar test revealed that the level of complexity has a significant effect on the control mode regardless of the secondary task: resman ($\chi 2(3,19) = 30.38$, $p < 0.001$), resman with sysmon ($\chi 2(3,19) = 11.08$, $p = 0.01$), and resman with track ($\chi 2(3,18) = 20.79$, $p < 0.001$). McNemar post-hoc tests with Bonferroni adjustment revealed that, for the main task alone (resman), the scrambled mode is significantly more frequent when the level of complexity is high ($p = .03$). Besides, the tactical mode is significantly more frequent in tasks with low complexity level than in tasks with high complexity level: resman ($p <.001$), resman with Sysmon ($p=.043$) and resman with track ($p = .019$). The probability of moving from an X mode when the complexity is low to a Y mode when the complexity is higher was calculated from a transition matrix (see Table 9).

Examining the transitions between the two periods (hence between the two complexity levels) shows (see Figure 33) the stability of the strategic mode (among the 23 participants who adopted the strategic mode in the first period, 19 maintained it in the second one) and the instability of the tactical mode (among the 26 participants who adopted the tactical mode in the first period, only 3 maintained it in the second one).In contrast, the comparisons conducted for each complexity level between sequence 1 (main task alone) and sequences 2 and 3 (main task and secondary tasks of system monitoring and tracking) do not show any negative effect of the secondary task upon the control modes. As a matter of fact, most participants kept the control mode they had adopted for sequence 1 (main task alone), or else they adopted a more effective control mode, which shows the effect of learning.

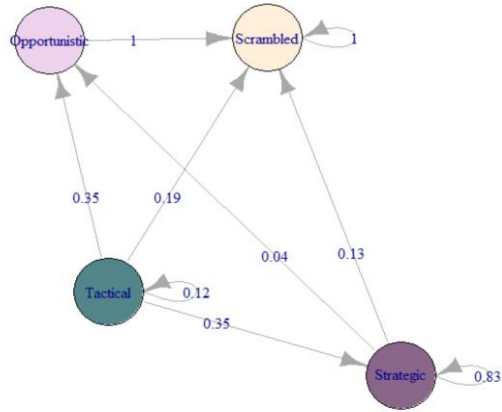

*Figure 33: Transitions between modes between periods of low and greater complexity*

### 4.3.2. Relations between mental stress and mental strain

We conducted different linear mixed-effect analyses to test the effects of task, complexity, and cognitive control modes (CCM) on physiological responses (see Table 4, the figures in bold corresponding to significant effects).

These analyses show that HRV can be explained by mental stress, i.e., by sequence, complexity, and their interactions (see Table 10 and Figure 34).

*Table 10: Estimates of fixed effects from linear mixed-effect model for HB02 and for HRV.*

| | HBO2 T1 ~ CCM + Sequence | HBO2 T2 ~ CCM | HBO2 T3 ~ CCM | HBO2 T4 ~ CCM + Sequence | HBO2 T5 ~ CCM | HBO2 T6 ~ CCM | HBO2 T7 ~ CCM | HRV ~ Complexity * Sequence |
|---|---|---|---|---|---|---|---|---|
| (Intercept) | 0.13 (1.66) | **3.08** (0.95)** | 0.26 (2.56) | 1.39 (1.16) | 0.36 (2.53) | **2.66** (0.95)** | 1.35 (0.94) | 7.25 (4.66) |
| **CCM (reference = Tactical)** | | | | | | | | |
| *Opportunistic* | **0.94** (0.21)*** | **1.11** (0.32)*** | **0.69** (0.30)* | **0.99** (0.34)** | 0.48 (0.29) | **0.97** (0.32)** | **0.59** (0.28)* | |
| *Scrambled* | **1.08** (0.24)*** | **0.71** (0.36)* | **1.26** (0.33)*** | **0.84** (0.39)* | 0.62 (0.33) | **0.85** (0.35)* | **0.71** (0.32)* | |
| *Strategic* | **0.63** (0.18)*** | **0.78** (0.27)** | **0.76** (0.25)** | 0.53 (0.29) | **0.71** (0.25)** | **0.74** (0.27)** | **0.79** (0.24)** | |
| **Sequence (reference = single task resman)** | | | | | | | | |
| *with sysmon* | 0.23 (0.15) | | | 0.30 (0.24) | | | | **-13.72** (3.60)*** |
| *with track* | **0.48** (0.15)** | | | **0.69** (0.24)** | | | | **-12.46** (3.69)*** |
| **Complexity (reference = low complexity)** | | | | | | | | |
| high complexity | | | | | | | | **-24.03** (3.60)*** |
| **Sequence:Complexity** | | | | | | | | |
| *with sysmon : high complexity* | | | | | | | | **13.07** (5.09)* |
| *with track: high complexity* | | | | | | | | **13.24** (5.19)* |
| Num. obs. | 100 | 100 | 100 | 100 | 100 | 100 | 100 | 82 |
| Num. groups: Participant_ | 17 | 17 | 17 | 17 | 17 | 17 | 17 | 14 |
| Var: Participant_ (Intercept) | 46.61 | 14.82 | 110.76 | 21.72 | 108.17 | 14.62 | 14.52 | 212.55 |
| Var: Residual | 0.36 | 0.82 | 0.70 | 0.93 | 0.69 | 0.80 | 0.64 | 90.84 |

*Note.* The fixed factors resulting from the stepwise model selection by AIC are indicated below the response variables; the reference condition is indicated in brackets for each explanatory factor; ***p < 0.001, **p < 0.01, *p < 0.05.

We found a significant main effect of complexity, with HRV more likely to decrease in the high complexity than in the low complexity condition ($\beta$=-24.03, SE=3.60, t(63)=-6.67, p<0.001). Moreover, there is also a significant effect of sequence. We found lower HRV when the main task was carried out with the secondary tracking task ($\beta$=-12.46, SE=3.69, t(63)=-3.38, p<0.01) or with system monitoring task ($\beta$=-13.72, SE=3.60, t(63)=-3.81, p<0.001), than when it was conducted as a single task.

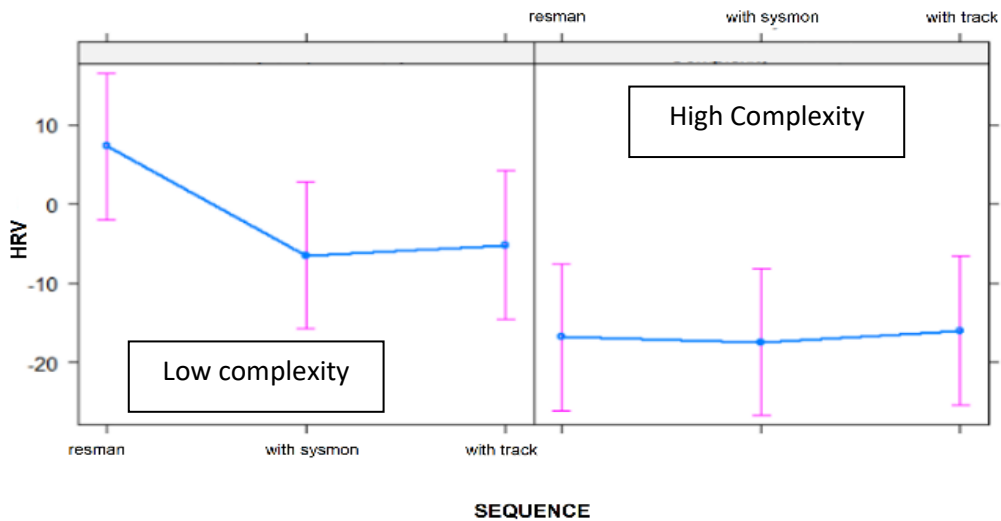

*Figure 34: Interactions of sequence and complexity on HRV*

This effect of sequence upon operator strain is also observed, on the mental demand dimension of the NASA-TLX. A one-way between subjects ANOVA shows that the single task condition involved a significantly lower mental demand than the double task conditions (F(2,48)=4.32, p<0.05). The interaction between complexity and sequence is also found to be significant, with a higher contrast between single task and double task conditions when the complexity is low. Moreover, there is no significant correlation between neurophysiological indicators and NASA-TLX scores.

### 4.3.3.   Relations between control modes and mental strain

The linear mixed-effect analyses also showed a significant effect of the control modes (CCM) upon the concentration in oxy-hemoglobin (HBO2). According to the stepwise model selection by AIC, HBO2 can be explained by control modes only, for optodes T2, T3, T5, T6 and T7, whereas HBO2 for optodes T1 and T4 can be explained by two main fixed effects, CCM and sequence. We followed the same procedure for concentration in deoxy-hemoglobin (HBB), but no significant results were found.

It should be noted that, for all optodes from T1 to T7, the tactical control mode (set as reference condition in the linear mixed model) always produces a significant lower HBO2 concentration, in comparison with the less effective modes (scrambled and opportunistic control) or the more anticipatory one (strategic control).

## 4.4. DISCUSSION

The study findings are both theoretical and methodological in nature.

First, regarding the stress-strain relationship, we observed a significant effect of mental stress on HRV, which is unsurprising. There is a main effect of the complexity of the reservoir management task on cardiac activity. The other constraint factor of sequence (i.e., the addition of a secondary task) also plays a significant but lesser role when the complexity of the main task is low. We should also note that the neurological indicators (fNIRS) are not or not very sensitive to the constraint.

In addition, we investigated, in a more original way, the relation between the cognitive control modes and mental workload, from the perspective of both mental stress and mental strain. Our analyses reveal two main theoretical contributions.

On the one hand, there is a significant effect of task complexity on the adoption and the variation of control modes. Particularly, we found an instability of the tactical mode, showing attraction between this mode and low complexity, and repulsion between this mode and higher complexity. This instability of the tactical mode was also analysed with the finer-grained analysis of transitions between the consecutive periods of low and higher complexity. We observed that an increase in complexity mainly leads to transitions from the tactical mode to a less effective mode (54% of the transitions). In contrast, the strategic and scrambled modes were mostly stable (respectively 83% and 100% of participants in one of this mode remained in the same mode, between low and high complexity periods within a given sequence). Furthermore, and congruent with the study of Stanton et al. (2001), we observed that a major part of the transition is between two "close" modes (70% of transitions from tactical to opportunistic or strategic modes, and 100% of transitions from opportunistic to scrambled modes). This result suggests that people move between control modes in a linear manner.

On the other hand, we found links between the modes of control and operator strain, as it was shown by Cegarra et al. (2017). The present study indicates that the tactical mode is associated with lower mental strain, when considering the HBO2 concentration indicator of mental workload. As stated by Leon-Carrion et al. (2008), "the hemodynamics of inter-individual differences in this region may reflect different cognitive strategies used in task resolution". Our study shows that the tactical mode is the most efficient one since it is associated with a satisfying performance and with the lowest mental strain off all control modes.

As shown in Figure 35, the control modes of the COCOM could therefore be seen as a mechanism for regulating operators' mental workload, which would moderate the stress-strain relationship (Hockey, 1997; [C24]; Cegarra, 2017).

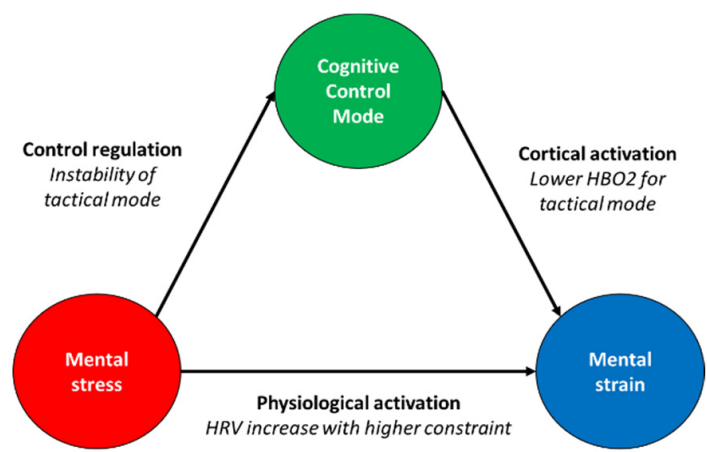

*Figure 35: Towards a moderating effect of control mode on the stress-strain relationship*

This result also calls attention to the advantage of studying brain activity to detect changes in control mode. If, in our study, the cerebral activity seems little correlated with mental stress variations, we nevertheless observe, on almost all the zones of the prefrontal cortex, a significant difference in HBO2 concentration between the tactical mode and other modes.

Hence, an increase in cortical activation could help reveal the shift away from the tactical mode towards less effective and more reactive control (the opportunistic or scrambled mode, where control of the situation is no longer guaranteed) or on the contrary towards more proactive control (strategic mode, requiring more anticipation). This potential detection ability opens new perspectives to design and trigger assistance aimed at keeping operators in the tactical mode, since it appears to be the most efficient one.

Finally, it should be noted this research work has some limitations. The experiment was run with novice participants only, who may be more heterogeneous in terms of cognitive control than an expert population. Therefore, it would be necessary to verify whether the same findings would apply to experts (e.g., a population of aircraft pilots).

In addition, our study, which only involved male participants, may hide gender effects on the adoption of control methods. Moreover, we coded the four control modes of the COCOM according to operators' performance on the main task and not the overall performance in the case of double task situations. When participants had to carry out multiple tasks, there could have been phenomena of focusing on or prioritizing the main task. This focus may have led to the maintenance of an effective cognitive control on the management of the reservoirs, to the detriment of the control of the secondary tasks.

Hence, in future research studies, it would be worth considering cognitive control by adopting an approach modelling operators' multi-task management on MATB-II, as proposed by Gutzwiller et al. (2014).

# Part B. Making autonomy compatible with human activities: Contribution to the adaptation of human-autonomy teaming

# Chapter 5.  Offline design of Human-Autonomy Teaming

*"Perfection is achieved, not when there is nothing more to add, but when there is nothing left to take away."* **– A. de Saint-Exupery**

The Know-How-to-Cooperate of the autonomous agent can be leveraged in the early stage of design. Indeed, it is important to think as soon as possible about how, and in which conditions, the distribution of work between humans and machines could be optimized, by especially determining the trigger for work sharing. Moreover, it is also critical to verify if the operators really belong to the sociotechnical system, and if they can easily cooperate with the designed system in dynamic situation.

This chapter focusses therefore on the development of tools and mechanisms for designing dynamic function allocation (based on CWA method, cf. §5.1), and for verifying the preliminary design of a sociotechnical system and its usability by operators with some simulation techniques (cf. §5.2). This second contribution was a part of the PhD of Sophie Prat (2017).

These different contributions were published in [C14, C15, C18, C19, C21, C23, C27, J2].

## 5.1. A TOOL FOR A PRIORI DESIGNING DYNAMIC FUNCTION ALLOCATION (DFA)

Socio-Technical Systems Engineering (STSE) focuses on the design of complex systems with interconnected human, technical, and organizational components (Baxter & Sommerville, 2011). Especially, this approach raises the issue of the role of operators faced with increasingly autonomous technical systems in dynamic, risky, and sometimes unforeseen situations. The distribution of activities between humans and machines is a central process in Socio-Technical System (STS) design and operation (Challenger et al., 2013). Function allocation and, more precisely, dynamic function allocation (DFA), can help a system maintain a satisfying performance in complex situations. This issue must be considered as early as the preliminary design phase of a project (MOD, 1989; Goom, 1996).

### 5.1.1. CWA as a design framework for considering DFA

**Cognitive Work Analysis.** Several methods have been proposed to design sociotechnical systems: User Centered Design (UCD) approach (Norman and Draper 1986), hierarchical task analysis (Norman and Draper 1986), cognitive task analysis (Chipman, Schraagen, and Shalin 2000), etc. Nevertheless, as stated by Rasmussen (1997), Vicente (1999) or Papantonopoulos (2004), many methods are too normative or too descriptive, unsuitable for designing adaptation, dynamic cooperation and work distribution. On the one hand, normative methods (with ISO standards and ergonomics handbooks) focus too much on the specification of the ideal ways to perform work - and therefore human-system interactions - under certain anticipated conditions, difficult to reproduce in real life within open systems, and with non-expert users that can deviate from the standard procedures. On the other hand, descriptive methods are based on the analysis of the familiar and recurring conditions. As reminded by Romero et al. (2016), the descriptive methods were used to supporting an "Anthropocentric Production Systems" with adjustability to different degrees of user experience or reliability. The resulting design of STS may be more tolerant to the adaptation and the deviation to the rules from the system agents, and it can generate satisfying systems in nominal conditions. Nevertheless, this kind of methods are again limited to consider unforeseen events and novel conditions that should typically occur in the context of dynamic and flexible systems, like autonomous transportation systems or smart factories. Indeed, these two main approaches may forbid the correct evaluation of design choices in terms of human awareness of the situation or mental workload when designing intelligent systems integrating the human.

By contrast, Cognitive Work Analysis (CWA), proposed by Rasmussen (1986), Rasmussen, Pejtersen, and Goodstein (1994) and further developed and codified by Vicente (1999), appears as one of the most comprehensive. It combines the contributions of engineering and human factors to provide designers with a powerful framework for SST design. As depicted on Table 11, it is a formative constraint-based approach, consisting of five successive stages: a) Work Domain Analysis (WDA), b) Control Task Analysis (ConTA), c) Strategies Analysis (StrA), d) Social Organization and Cooperation Analysis (SOCA), and e) Worker Competencies Analysis (WCA). The issue of function allocation is addressed at the fourth stage, namely SOCA. This issue is a crucial one, but the exploration of the social and organisation phase has received less attention than the application of the WDA or ConTA (Jenkins, Stanton, Salmon, Walker, & Young, 2008). SOCA does not deal explicitly with the dynamic distribution of functions between humans and machines (Chauvin & Hoc, 2014). No modelling tools existed for this stage before the recent proposals made by Jenkins et al. (2008) or Stanton and Bessell (2014). In this study, we propose to make up for these weaknesses.

*Table 11: CWA methodology (adapted from Jenkins et al., 2008)*

| WHY | WHAT | HOW | |
|---|---|---|---|
| **CONSTRAINTS IDENTIFICATION** | **PHASES OF CWA** | **DATA ACQUISITION** | **REPRESENTATION** |
| System Functional Constraints | Work Domain Analysis (WCA) | Document Analysis, Reviews by SME | Abstraction Hierarchy (AH), Abstraction Decomposition Space (ADS) |
| Decisional and Situational Constraints | Control Task Analysis (ConTA) | Cognitive Walkthrough, Study of Work Practices | Decision Ladder (DL), Contextual Activity Template (CAT) |
| Strategy Constraints to Achieve System Goals | Strategies Analysis (StrA) | Critical Decision Methods, Interaction Analysis, Verbal Protocol Analysis | Information Flow Map (IFM) |
| Functional Allocation Constraints | Social Organization & Cooperation Analysis (SOCA) | Communication Analysis, Interaction Analysis | Color Codes applied dynamically on all of the above (SOCA-ADSSOCA-CAT, SOCA-DL, SOCA-IFM) |
| Functional Competency Constraints | Worker Competencies Analysis (WCA) | Repertory Grid Analysis, Review of Decision Ladder | Skills Rules Knowledge (SRK), Functional Matrix |

The present work aims at improving the SOCA stage and at integrating explicitly DFA into the CWA framework. It proposes a tool for designers that will enable them to verify that a particular solution will meet the purpose of the system, regardless of the work situation.

**Dynamic Function Allocation.** Dynamic function allocation (DFA) requires knowing the work functions that should be allocated (what), the situations in which they may be allocated (when and where), and the resources that could be associated with a given function (who). Three phases of the CWA (WDA, ConTA, and SOCA) provide these data through two main existing tools: the Abstraction Hierarchy (AH) and the Contextual Activity Template (CAT).

WDA deals with the constraints that are placed on actors by the functional structure of the field or environment in which the work occurs (Naikar, 2013). This phase is associated with a modelling tool, the AH. This tool enables the description of a work domain in terms of five levels of abstraction: functional purpose (the purpose of the work domain, its "raison d'être"), value and priority measures (the criteria ensuring that the system progresses toward the functional purpose), purpose-related functions (the general functions that are performed in order to achieve the functional purpose), object-related processes (processes and capabilities characterising the objects used by the general functions), and physical objects.

ConTA is related to the activity required for achieving a system's purpose with a set of specific resources. Naikar (2013) propose to characterize this activity as a set of recurring work situations, work functions, or control tasks. This author introduces the CAT for modelling activities in work systems. This template highlights the contextual relationships between the various element of ConTA and graphically illustrates all the combinations of work situations, work functions, and control tasks that are possible. She explained that the decomposition of activity into work situations is appropriate in systems where work is segmented according to time and space (in hospitals or schools for example), whereas activity is better characterized by its content, independently of its temporal of spatial characteristics, in other systems. In those cases, it is appropriate to decompose activity into a set of work functions. Work functions are related to functions to be performed in a work system. They are defined at the purpose-related functions level or at the object-related processes level in the AH (Jenkins et al., 2008). In a research laboratory, activity is divided into work functions such as writing papers, conducting experiments, and reading. The CAT is designed to represent activity both in terms of work situations and work functions. A graphical code is used to distinguish work situations in which a work function *can* occur and those in which a work function will *typically* occur. According to Stanton and Bessell (2014) and as depicted in Figure 36, a work function - in a given situation - may be qualified as *expected* (it can occur and typically occurs), *optional* (it can occur but does not typically occur), or *impossible* (it never occurs). The decision ladder is then used to decompose activity into a set of control tasks for each work situation and/or work functions.

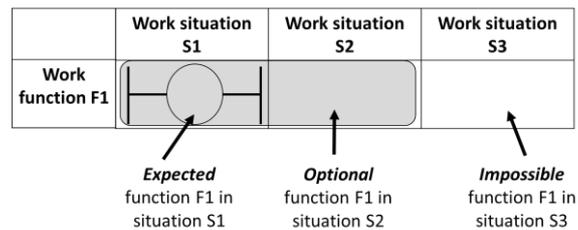

*Figure 36: CAT Layout (from Bessell and Stanton, 2014)*

SOCA addresses the constraints governing how the team communicates and cooperates (Jenkins et al., 2008). It aims to determine the distribution of work, communication, and cooperation amongst actors (i.e., the different resources of the system). Jenkins et al. (2008) propose to map actors (represented by means of a colour code) onto the AH and more precisely onto the functions described at the levels of the purpose-related function and of the object-related processes (SOCA-AH). In the same way, they map actors onto the CAT to show where these can have an influence on the system (SOCA-CAT). They consider, at this stage, the actors' capability to perform a certain work function during a certain work situation. Cells occupied by more than one actor indicate that either or all the identified actors can support the activity. According to these authors, this representation of constraints helps to identify and evaluate potential combinations of working practices to determine optimal practices. This analysis may be carried out according to Rasmussen et al.'s (1994) six criteria: (a) actor competencies, (b) access to information and means of action, (c) coordination demands, (d) workload, (e) safety and reliability, and (f) existing regulations. These criteria are either input data to model DFA problems (e.g., actor competencies) or evaluation criteria to choose allocations (e.g., workload).

The existing tools of the CWA enable the identification of potential allocations of resources to work functions. However, they do not provide the means to evaluate and optimize these according to the work situation characteristics and, most importantly, according to the work situation variations.

**Modelling and solving dynamic function allocation problem with CWA.** Defining a dynamic function allocation entails taking dynamic situations and resource availability into account. For that purpose, designers need a definition of work situations and a modelling of resource constraints adapted to the specific problem, as well as a method used to formalize and to evaluate the STS according to different complex situations.

The notion of work situation seems useful to deal with the question of DFA although its "modern" definition (Jenkins et al., 2008; Stanton & Bessell, 2014) was not originally thought to model this specific problem. The use of this concept for DFA problem raises therefore new questions:

*Are time and location sufficiently detailed to distinguish all the work situations?* According to Naikar, Moylan, and Pearce (2006), work situations are characterized by some absolute or relative constants of time or location (work can occur at a specific place or at a specific distance of a moving position, and work can occur at a specific moment or just before or after a mission phase). For instance, in the context of aircraft system analysis, these authors described five different situations fitting with mission phases ("on ground not in aircraft", "on ground in aircraft", "enroute to station", "on station", "enroute to base"). However, could the situation "flying in bad weather conditions" be considered as a spatio-temporal situation? This kind of work situation can occur at any place and any time, as there is no unit of time and place, or relationship with a mission phase or a moving place. Cuny and Chauvin (2009) remind that "in ergonomic psychology, the situation theoretically includes all variables forming a system of potential interactions with the activity as its operational framework". Work situations can be therefore more generally influenced and characterized by the external and internal conditions of the system (the information level or the nature of the system environment, temporal pressure, etc.). "When" and "where" questions should be thus completed or replaced by the question "In which internal and external conditions does the system operate?" to define work situations.

*Do work situations include incidental or critical situations?* The recent applications of the CAT (Jenkins et al., 2008; Stanton & Bessell, 2014) are centred on nominal phases of the mission. However, the dynamic function allocation could also take some degraded situations into account (the failure of some system components, uncertainty, or absence of knowledge regarding information relative to the mission, etc.).

*Are work situations independent from each other?* The different work situations are independent from each other if they are defined according to time or location. However, if we also consider work situations according to internal and external conditions, the situations « enroute to station » and « flying in bad weather conditions » could occur in parallel.

*What is the granularity of the modelling of work situations?* Naikar et al. (2006) assert that the decomposition of activity into work situations and work functions can be done at different levels of detail or granularity. They provide an example of this granularity issue. Situations such as "On ground not in aircraft", "On ground in aircraft", and the "in the air" situations ("enroute to station" and "enroute to base") are typically the conjunction of two elementary situations, defined by the location of the activity in relation to the plane and to the ground ("on ground" versus "in the air", "in aircraft" versus "not in aircraft"). The different elements of internal and external conditions could therefore be a unit of description of work situations.

*How can forgetting work situations and a combinatorial explosion of situations be avoided?* The number of situations could increase very fast if numerous conditions are considered and combined. For instance, taking into account the weather conditions (cloudy or sunny), tactical conditions (in fight zone, not in fight zone) and system capacity conditions (full tank of fuel, almost empty tank) and the five situations given by Naikar et al. (2006) results in having to consider and model 40 different situations. Complex situations should be considered as the result of the conjunction of several elementary situations that are not always independent from each other.

The modelling of resource availability in dynamic situations is also a crucial question for dealing with DFA issues. Jenkins et al. (2008) and Stanton and Bessell (2014) propose to map resources and actors, especially on the AH and CAT. However, they do not formalize in detail the constraints that can occur between these resources in dynamic situations, which is necessary to define the DFA problem. Selecting to allocate a resource to a function could be dependent on the use of this resource or another one for another function. This relation of dependence among resources can be expressed at the design stage (modelled in SOCA-AH) or in the case of a situation that creates some unavailability or dependence (modelled in SOCA-CAT). The following list represents an attempt to model these constraints: *a) binary constraints:* a resource can be allocated or not to a function; *b) disjunctive constraints:* one or several resources can be allocated to the same function; *c) exclusive constraints:* two resources cannot work in parallel on the same or on different functions; *d) capacity constraints:* the number of functions allocated to one resource or both resources is limited; *e) conditional constraints:* a resource can be allocated to a function only if one or several resources are allocated to one or several functions; *f) antecedence constraints:* this is a special case of conditional constraints to which a temporal dimension has been added; a resource may be allocated to a function only if one or several resources were previously allocated to one or several functions.

Using these elements of detail or adaptation of CWA leads to proposing a method using SOCA-AH and SOCA-CAT models and SOCA criteria to formalize and solve the DFA problem.

*SOCA-AH is centred on the analysis of functions and resources and would be used to assess the choices made by the designers regarding the composition of the system.* The model provides a means of assessing whether a function is statically allocated to a resource (only one resource is planned in the system to carry out the function: there is only one coloured actor in a box of AH) or whether a function is admissible for dynamic allocation to a resource (several resources are planned and some of them could carry out the function: there are at least two coloured actors in a box of AH).

*SOCA-CAT is centred on the analysis of situated functions and resources and would be used to assess the choices made by the designers regarding the functioning of the system in dynamic situations.* The model provides a means to assess the potential risks of the DFA in different complex situations and to find, when it is possible, the best system configuration to deal with situations. The SOCA-CAT is composed of the designers' choices that are represented by different types of functions actionable in a given elementary situation. A function can be "expected" (a function with a bar inside a dotted box can occur and typically occurs), "optional" (a function inside a dotted box without a bar can occur but does not typically occur) or "impossible" (a function outside the dotted box can never occur). Moreover, some functions are designed with different allocation possibilities (different resources or configurations of resources can carry out the function).

When complex situations are considered, namely when the conjunction of several elementary situations is examined, designers should check whether there is no conflict between the choices made for the elementary situations. They must look for functions that are "expected" in an elementary situation and that are "impossible" for all resources of the system in other elementary situations. Let us consider situation S* as the conjunction of elementary situations Si and Sj. SOCA-CAT would be useful to model:

- a minimal configuration list MinConfig(S*) of functions that can be allocated to a resource and are "expected" in a complex situation. The list is composed of the function-resource couples, noted Fi-Ri, that are at least considered once as "expected" in situation Si and Sj.
- a list Pot(S*) representing all the functions that can be allocated to a resource and are "possible" (i.e., "expected" or "optional") in a complex situation. The list is composed of the Fi-Ri couples that are considered in all situations Si and Sj as "expected" or "optional".

From these two lists, the designers could first check whether there are any design conflicts between concurrent elementary situations modelled with SOCA-CAT, i.e., whether MinConf(S*) is included in Pot(S*). Hence, they deal with a decision problem, depicted in Figure 37, which can be written as: $\{MinConf(S*) \subseteq Pot(S*)\} = TRUE$?

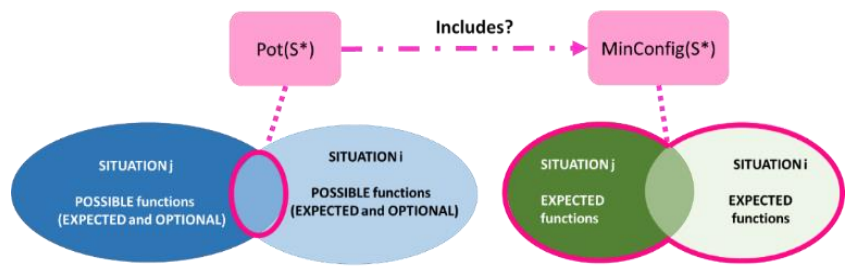

*Figure 37: Decision problem of function allocation*

If the answer to the decision problem is positive, that is, if there is no design conflict, the designers could then deal with an optimization problem. They look for the best configuration in the list Pot(S*) that minimizes a criterion of dynamic function allocation defined in SOCA (e.g., the workload of some resources). This problem can be written as: $\forall S*$, $minimize\ (Solution(S*))$, $with\ MinConf(S*) \subseteq Solution(S*) \subseteq Pot(S*)$. The following section presents an implementation of this method on a case study.

### 5.1.2. Example of application

The proposed CWA-based DFA method was applied on a small human-machine system composed of technical components (an electric pedal-assist bike, a GPS navigation system, physiological sensors, a battery gauge) and a human agent (a cyclist). The system can be considered as an instance of STS. It is both an intentional and a causal system: the system reacts to the variations on the road due to actions of other road users and to the laws of nature. This "simple" case study was chosen to illustrate the method proposed in this work. The application example must therefore be considered as a first proof of concept. It simulates a design problem inspired by the new needs resulting from the recent popularization of pedal-assist bikes and the development and integration of new technologies.

On the one hand, new uses have appeared: cyclists want to avoid daily battery recharging, or they wish to integrate unplanned routes to their usual routine without having to do without power assist on the final slopes before arriving home. Hence, the assist capacities of electric bikes need to be better adapted to the cyclists' individual goals (such as duration and length of trips) and to dynamic situations (ascents, wind, road traffic). Adaptation also involves improved battery use and assistance optimization any time on the route while securing the bikers' safety. On the other hand, new technologies enable adding physical and software devices onto the electric bikes to guarantee bikers a safe and effective ride. Designers thus need to be given a method to evaluate whether the resources and the dynamic function allocation are sufficient to meet these objectives of safety and performance.

**Modeling function allocation with the existing CWA tools.**

CWA modelling tools were used to model the functional constraints (AH), the situational constraints (CAT), and the resource constraints related to the DFA problem.

*Work Domain Analysis (Abstraction Hierarchy).* The functional purpose of the system is to guarantee a safe and effective ride to a destination.

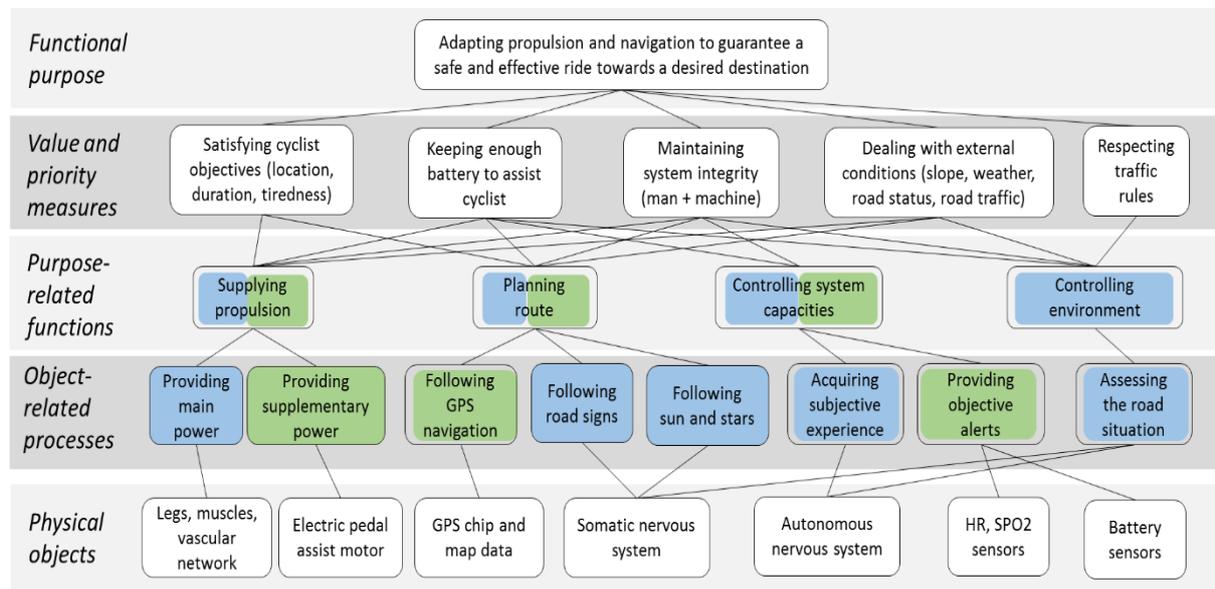

*Figure 38: Abstraction Hierarchy*

Meeting this objective entails that the system must comply with some values and priorities related to performance and safety, shown in Figure 38 from left to right. The safety priorities from the smallest to the largest scale of the system are the following ones: minimal battery level for cyclists' safety; system integrity; adaptation of the system to the road; adaptation of the system to the road management system. The performance priorities are related to the management of the location objective (the system must help the cyclists reach the desired destination), the management of the path duration, and the management of the cyclists' tiredness. Consideration should be given to the human-machine cooperation issue, in terms of the following purpose-related functions, which can be cognitive and motor functions: supplying propulsion to ride the bike and to reach the location objective; route planning means regularly geolocating the system and choosing an adapted path; controlling system capacities to monitor internal conditions (in this case, the energetic states of human and technical components of the system that could result in an accident or underperformance); controlling the environment (i.e. monitoring external conditions such as the weather, road grade and quality, stop signs, etc). The object-related processes and the physical objects are defined in relation to the hybrid nature of the human-machine system. The human actor is situated within the system; hence, the physical objects can be defined in terms of human capacity (e.g., muscles or the nervous system) and technical capacity (e.g. motor or sensors). This dichotomy between human and technical parts will be used in the SOCA phase to categorize the resources to which functions could be allocated.

*Control Task Analysis (Contextual Activity Template).* The work situations were modelled with different internal and external conditions, as indicated in the proposal of the method (see Figure 39). Some situations (S2 to S5) could result from the conjunction of elementary conditions (knowledge level of the cyclists on the path to reach their destination, and difficulty level resulting from the road quality, grade, and traffic). Other situations depend only on one condition (speed, GPS signal access). The distinction between "expected", "optional", and "impossible" situation-related functions was examined in this CWA phase and then refined in SOCA-CAT.

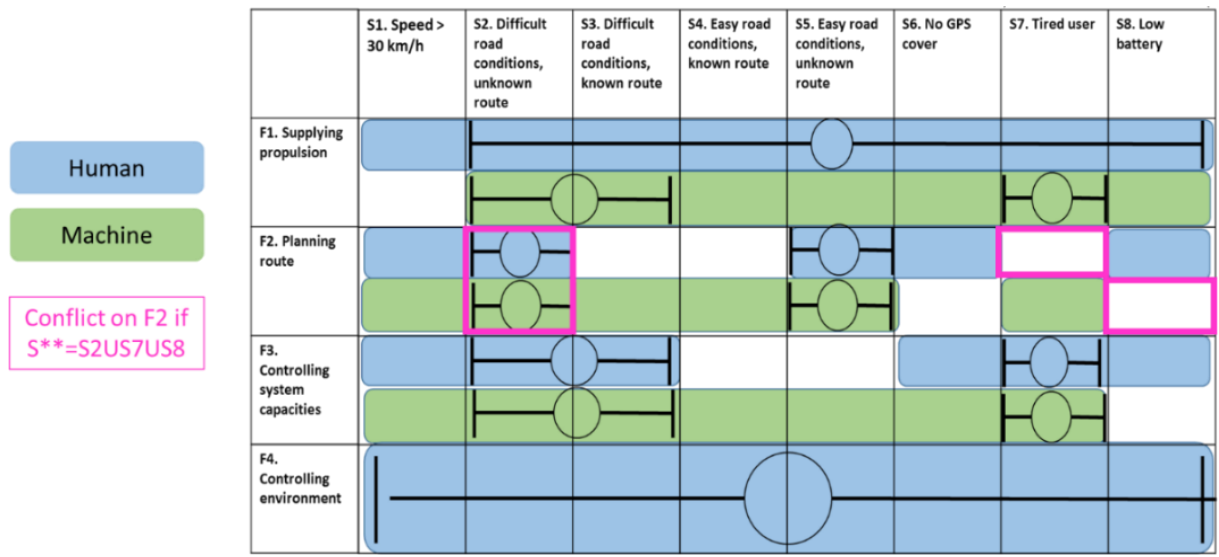

*Figure 39: Contextual Activity Template*

**Solving the dynamic function allocation problem with CWA.**

The use of previous CWA modelling tools would help designers assess whether their function- and situation- dependent choices of resources generate conflicts threatening the safety or the performance of the activity of bike riding and would allow them choosing an optimal situated function allocation when there is no conflict.

In Figure 39 only eight « elementary » or « simple » situations were defined, arising from the consideration of six variables of internal or external conditions (speed, road difficulty, knowledge of route, GPS signal access, user tiredness, and battery level). The proposed method entails verifying whether the function allocation choices onto the eight modelled elementary situations can deal with complex situations (i.e., the different conjunctions of the elementary situations), instead of considering and modelling all the combinations of the six condition variables. In the latter case, if only two modalities were considered for each variable (e.g., difficult or easy road conditions), up to 64 situations should be completely examined and defined by the designers. The proposal seeks to deal with this combinatorial explosion and to reduce this number by stressing the conflictual conjunctions that should be modelled in addition to these eight elementary situations.

Let us consider two cases: S*=S1US4US6, a rather favourable conjunction of elementary situations (Speed>30km/h, known route, easy road conditions, no GPS cover), and S**=S2US7US8, a difficult complex situation (unknown route, difficult road conditions, tired user, and low battery). The notation Fi-H and Fi-M explained in the proposal is kept for dealing with these examples. MinConfig is the list of all the Fi-H and Fi-M that are "expected" (with circle and whiskers) in at least one elementary situation composing the complex situation. For this minimal list, it should be noted that only one resource is sometimes sufficient to allocate to a function like F2 or F3 (this is therefore an exclusive constraint noted XOR). Moreover, for the specific case of F1 (ruled by a conditional constraints of F1-H on F1-M), F1-M is expected as well as F1-H in S2, S3 or S7.

In the complex situation S*, F4-H is expected in S1, both F1-H and F4-H are expected in S4 and in S6, so *MinConf(S*)={F1-H, F4-H}*. In the complex situation S**, all modelled Fi-H and Fi-M of SOCA-CAT are expected in S2. So *MinConf(S**) = {F1-H; F1-M; F2-H XOR F2-M; F3-H XOR F3-M; F4-H}*. Pot is the list of all the Fi-H and Fi-M that are not "impossible" (not outside dotted boxes) in every elementary situation composing the complex situation (i.e., the list of all the Fi-H and Fi-M that are "expected" or "optional" in every considered elementary situation). For this maximal list of potential Fi-H and Fi-M, two resources can be allocated separately or together to the same functions: they are therefore both included in the list. Moreover, the conditional constraints are taken into account (e.g., for the specific case of F1, the constraint will be noted F1-M if F1-H). In the case of S*, F2-M is impossible in S1 and in S6, F2-H and F3-H are both impossible in S4.

So *Pot(S*) = {F1-H; F1-M if F1-H; F3-M; F4-H}*. In the case S**, F2-H is impossible in S7, and both F2-M and F3-M are impossible in S8, whereas all the Fi-H and Fi-M are possible in S2. So *Pot(S**) = {F1-H; F1-M if F1-H; F3-M; F4-H}*.

First the problem decision must be solved: "$\{MinConf(S*) \subset Pot(S*)\} = TRUE$?". If the answer is negative, there exist a design conflict, and there is no admissible solution to the problem of dynamic function allocation in the situation under investigation. This happens in the case S**, where neither F2-H nor F2-M are present in Pot(S**) whereas they were in MinConf(S**) under the form F2-H XOR F2-M: MinConf(S**) is not included in Pot(S**). This design conflict is represented in Figure 39, in purple. If the answer is positive, there is at least one admissible allocation in the complex situation that meets the system functional purpose. This happens in the case S*, where MinConf(S*) is included in Pot(S*).

When possible, an optimization problem can then be solved with DFA criteria. For instance, let us consider the physical and mental workload of the cyclist, or the consumed power of the machine (i.e., the allocation of functions must be minimal respectively on the human or on the machine). The solutions of DFA in S* are thus:

- *Solution(S*, min cyclist workload)={F1-H; F1-M; F4-H}:* electric assist must be implemented to decrease the physical workload, and the automated monitoring of system capacities F3-M is deactivated to avoid an information overload.
- *Solution(S*, min energy consumption)={F1-H ;F4-H}:* the machine can be completely deactivated for propulsion and information processing, so as to keep enough battery to help the cyclist in hard road conditions.

**Findings.**

At the theoretical level, this study proposes a method that follows the formative nature and the focus on constraints modelling of CWA to deal with the DFA issue. This contribution aims at continuing the work made on SOCA around the DFA question (Jenkins et al., 2008; Stanton and Bessell, 2014) by considering: a) SOCA-AH as a means to examine the constraints relative to the design choice of resources in terms of static function allocation (one sole resource for one function) or potential dynamic function allocation (several separate resources for one function); b) SOCA-CAT as a means to examine the constraints relative to the activation of resources in different situations that would influence the possibility and the choice of dynamic function allocation. This last consideration especially involved revisiting the concept of work situation defined by Naikar et al. (2006) relative to the specific question of DFA by characterizing it in terms of external and internal conditions.

At the methodological level, the proposal is intended to help designers deal with the combinatorial explosion resulting from the combination of the different conditions that form complex situations. Instead of modelling all these complex situations, designers would be able to simply add new condition variables to the previously examined elementary situations and to observe the DFA properties of emergent situations. The analysis of the conjunctions of elementary situations in SOCA-CAT would then enable them to detect design conflicts. In this case, these conflictual complex situations should be completely defined and modelled by designers. Otherwise, the dynamic function allocation is functioning in these complex situations, and the DFA problem can be considered as a local optimization problem (the best configuration is looked for in each situation according to specific criteria).

In terms of future perspectives, the proposed method could be further developed by integrating the temporal dimension: in the early design stages, situational constraints could be tested according to baseline scenarios to help designers assess the quality and the influence of their choices on the safety and the performance of the system in realistic situations. The number of design conflicts or the total cost generated from the DFA criteria could then be calculated to assess different design solutions. Considering the temporal dimension is also a way of thinking of a DFA problem not only as a local optimization but also as a global optimization problem (i.e., over entire scenarios).

## 5.2. SIMULATION FLOW GENERATION TO VERIFY SOCIOTECHNICAL SYSTEM DESIGN

In the context of Human-Autonomy Teaming, process control system deserves special attention. This system makes it possible to control the process and adapt its behaviour according to changes in the environment (whether these changes are planned or not). A process control system can be understood through the concept of an open socio-technical system. This concept highlights the fact that the system pursues objectives (which it can achieve in different ways) in a dynamic context to which it adapts (aspect of operational reliability and resilience). It also underlines the fact that technical and human sub-systems are distinct but interdependent, the performance of the system thus based on a joint optimization of these sub-systems. The design of a process control system is therefore part of the design of socio-technical systems. This requires an articulation between technocentric and anthropocentric visions, through a participatory design approach. The principle of this design approach is to integrate the end user (the human operator) as an actor in the design. It is therefore a question of involving him/her in the design activities, from specification to evaluation.

### 5.2.1. A design flow to generate command and interface

The evolution of the complexity in systems to be designed requires adapting the design procedures. Indeed, the traditional sequential approach of design does not allow the system to be understood as a "whole". Design choices at early stages can have a big impact on the entire system and can only be detected at the end of the design (in the integration stages). To understand the system, a simultaneous design approach is therefore necessary. In addition, having short and iterative cycles makes it possible to integrate verification and validation loops, to ensure that the requirements are well defined and that the design solution meets the user needs. With this design approach, corrections could be made sooner, reducing costs and redesign times.

Besides, the design activity relies on the use of models. This notion of a model is important because it helps designers reason and communicate with each other, and with end users. We are thus moving from document-based engineering to model-based engineering. To facilitate exchanges between designers and thus limit misinterpretations, the idea is to re-use representations that have been constructed as part of the design activities. However, each discipline has its own representations (which are not necessarily intelligible to everyone). The use of Model Driven Engineering (MDE) techniques can automate the transition from one model to another (model transformation), at lower costs and by limiting the introduction of errors.

**A framework for automated sociotechnical design flow**

The work previously carried out between Lab-STICC and Segula Technologies (Bignon, 2012; Goubali et al., 2014) resulted in the creation of a design support tool jointly generating a control program and a supervision interface. This tool, named Anaxagore, uses the MDE model transformation techniques to generate the control and the monitoring interface from a business model (provided by a mechanical engineer) and a library of elements. The business model is a P&ID (Piping & Instrumentation Diagram), based on the standard (ANSI / ISA-5.1, 1992), describing the physical system. An element is defined as the constituent unit of the system process. It can relate to equipment (valve, pump, etc.) or to system functions. Each item in the library contains multiple views. The concept of view is here an extension of that used by Lallican et al. (2007), it corresponds to the different points of view of the designers. An element of the library thus contains several views including those relating to supervision, and that relating to the command. Another business model is considered to generate the supervision interface according to ergonomic criteria (Rechard, 2015), with the WDA model from CWA.

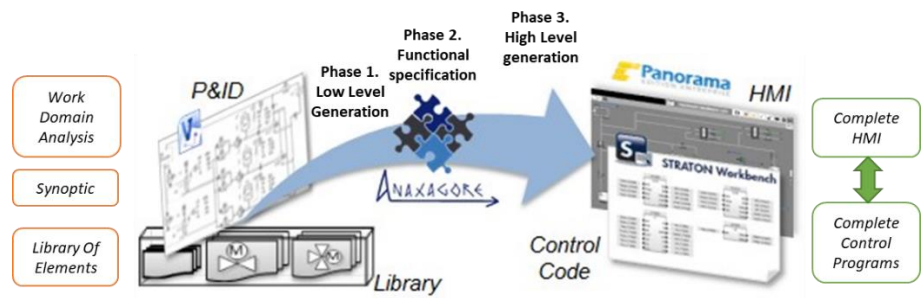

*Figure 40: Anaxagore Principles*

The design flow, making it possible to generate the supervision HMI and the control programs, can be broken down into three phases (Figure 40). The first phase consists of generating the supervision interface and the associated low level control program (Low Level Generation). The control-command chain generated only makes it possible to control the elements of the system one by one, via elementary commands. A valve may, for example, have two basic commands: one corresponding to its opening command, the other to its closing command. The second phase corresponds to the functional specification of global commands (also called high-level commands) (Goubali et al., 2014). To specify these global commands, end-user development (EUD) techniques are used. One advantage is for instance to allow an expert mechanical engineering to directly create high-level programs without having computer skills. To do this, a specification interface (based on the low-level interface) makes it possible to record elementary command sequences. The expert records two examples to perform the function to be specified. The starting and ending points are different from example to example. From the recorded examples, a module makes it possible to find all the configurations associated with the function which is being specified (generalization). Once all the global commands have been specified, we move on to the last phase (High Level Generation). It generates the supervision interface and high-level commands, as well as the part that manages the reconfiguration. The result is the complete control system (from a software point of view).

More operationally, this design flow uses the following transformation: the P&ID entered on Microsoft Visio is transformed into a platform independent model: the synoptic diagram. The generation is performed from this model and a library of standard elements based on the *view* concept. Moreover, a mapping is automatically created through the *list of parts*. This model is composed of the list of all the system elements instances and their associated variables. Furthermore, the *list of parts* contains all the connections between each element which are in the *synoptic*. Thus, the *list of parts* is a synthesis of information contained in the P&ID, enriched with complementary information related to variables exchanged between supervision, control and the physical process. This model is therefore used throughout the design flow for the generation.

The instrumented Anaxagore design flow takes therefore place in the context of simultaneous design and aims to implement a design centered on the end user (participatory design). Syntactic consistency between the control programs and the supervision interface is also ensured, thanks to joint generation. Nevertheless, this approach does not include means to test the generated control/monitoring system, and to verify if the designed system can be easily understandable and usable by end users.

**Needs for verifying the automated design flow**

In the context of manufacturing systems, a stage of virtual tests (so called *virtual commissioning*) can be introduced just before the stage of real tests. Oppelt and Urbas (2014) go further by proposing to integrate the virtual commissioning in the control system design process. According to them, the *integrated virtual commissioning* will permit the automation engineers to test continuously their work during the project. Moreover, it should permit to conduct user tests earlier. As a result, errors should be detected earlier, and these should be easily corrected without adding huge costs. Despite all the research work, *virtual commissioning* has not become a standard in the industry. According to Oppelt et al. (2014), two main causes could be pointed out: virtual commissioning requires significant efforts for modelling and should be supported by the expertise of automation engineers, who must then have a considerable knowledge in simulation techniques. Barth and Fay (2013) underlined that manual development of simulation models is a time-consuming task. Moreover, these models must be updated at each modification in the design steps.

In this context, automatic generation of simulation models from business models could be a solution. For example, Bevan et al. (2012) used Model Driven Engineering (MDE) to generate the simulation model of the operating part, and the control program. Then, the simulation is used to check the control program. However, this work cannot be applied to systems with a strong continuous feature. In this respect, our work focuses on designing control/monitoring system for sub-systems of a ship. These control/monitoring systems are mainly fluid management systems and can be considered as process control systems. Thus, the addition of verification techniques by simulation of the process would make it possible to guarantee the quality of the programs generated as soon as possible, in terms of semantic consistency and taking into account of operating safety. On the other hand, it would make it possible to set up user tests to validate the control-command chain, and thus really place oneself in a context of participatory design.

Li et al. (2013) proposed a classification of simulation modelling methods for complex systems as follows. A first group gathers unified modelling methods which include in particular those using multi-domain concepts or based on multiple formalisms. The use of multi-domain concepts can be done through a language such as *Modelica*. Indeed, *Modelica* allows the description of heterogeneous physical systems with mathematical equations. As a result, this object-oriented language permits the simulation of continuous or hybrid systems. Methods based on multiple formalisms can be those using language such as UML, or DEVS formalism (Zeigler et al., 1995). On opposition, a second group gathers combined modelling methods with co-simulation or multi-modelling. The aim is to compose models with several formalisms and simulate them. The co-simulation consists in using different simulation systems which exchange data whereas multi-modelling concerns the composition of models with different semantics. The last group gathers MDE-based modelling methods. These methods allow formal modelling, independently of the model implementation, and then generation of the code from it. MDE comes from software engineering. More details can be found in (Bezivin, 2005).

Although simulation modelling approaches have been separated in three groups, it is possible to use a combination of methods from these groups. MDE-based modelling methods seem to be appropriate for automatic generation of simulation models. Multi-domain languages such as *Modelica* seem to be appropriate for supporting simulation models all along the design process. This language has been selected in both academia and industrial applications for hybrid simulation of complex systems. Thus, the joint use of a multi-domain language, such as *Modelica*, and a MDE-based modelling method should permit to define a generation flow of simulation models, which could be used throughout the design project.

Oppelt et al. (2014) proposed a mechanism for the simulation models generation from engineering data, in five steps. The first step, which is accomplished by a simulation expert, is the *creation of simulation objects*. This could correspond to the creation of elements simulation models library. The second step is the *extending planning object*, the aim is to extend the engineering model to integrate supplementary data which is relevant for a simulation purpose. In the third step, a *mapping* is done between the elements' simulation models and their equivalent in the engineering model. The fourth step corresponds to the collection and *data exchange* for creating the simulation model. The last step corresponds to the *simulation model generation*. The mechanism presented by Oppelt et al. (2014) is interesting because it could be implemented independently of the target simulation software and/or of the simulation modelling approach. However, the mechanism's formalisation is still at a relatively abstract level.

To support the designers' activities of control/monitoring system, the simulation models generation must be used throughout the design project (concept of integrated virtual commissioning). Consequently, the simulation models of system elements must be extended according to the needs (for example, regarding what must be checked). To succeed, a solution consists in using an object-oriented simulation language which offers a multi-domain modelling (such as *Modelica*). The advantage is that models from existing libraries can be reused, modified, or new ones created. The use of *Modelica* has already been successfully implemented, for instance, by (Barth and Fay, 2013; Arroyo et al., 2016). However, the work on automated generation of *Modelica* model from P&ID lacks formalisation. Using MDE techniques, a generation flow could be defined to fulfil these requirements.

### 5.2.2. Simulation-based verification of the automated design flow

**Proposal of a simulation flow generation**

Following the literature review, the generation of the simulation models was proposed for the target *Modelica* language, from a library of elements and a P&ID, in parallel of automation engineer activities. This proposal is inspired by the Anaxagore tool-supported approach of Bignon et al. (2013). In particular, the synoptic diagram was considered as input model since it is a generic P&ID representation (independently of the platform in which it was entered). The library of standard elements based on the view concept has been extended by a simulation view for storing the elements simulation models.

The simulation model and the related simulation library were both generated (containing all elements' models to be instantiated) in *Modelica* language. This library is required to use the generated simulation model in a *Modelica* software. The advantage is to only import in the *Modelica* software the elements' simulation models that are used.

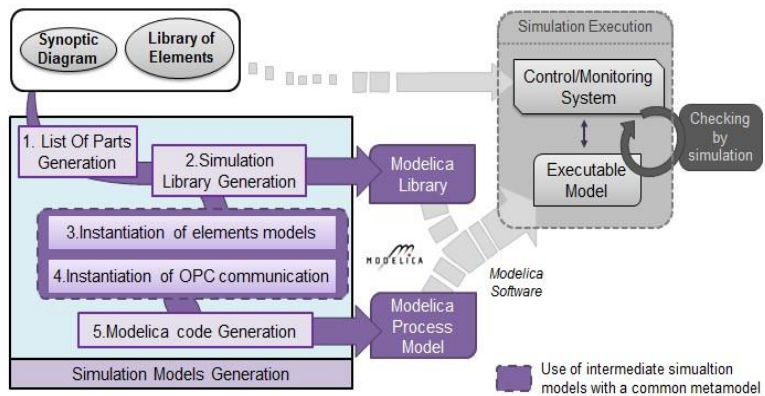

*Figure 41: Overview of the proposal*

To implement the generation of both the simulation model and the associated *Modelica* library, five steps were defined (see Figure 41). The first step, *List Of Parts Generation*, maps the different views with information contained in the Synoptic Diagram. The concept of List Of Parts (Bignon et al., 2013) was used to this purpose. The second step corresponds to the *Simulation Library Generation*, in the *Modelica* language. The third step corresponds to the *Instantiation of elements models* and their related connections, whereas the fourth step concerns the *Instantiation of OPC communication* for exchange with the control system. The last step corresponds to the *Modelica code Generation* of the complete simulation model. These steps are implemented through the generation flow presented in Figure 42.

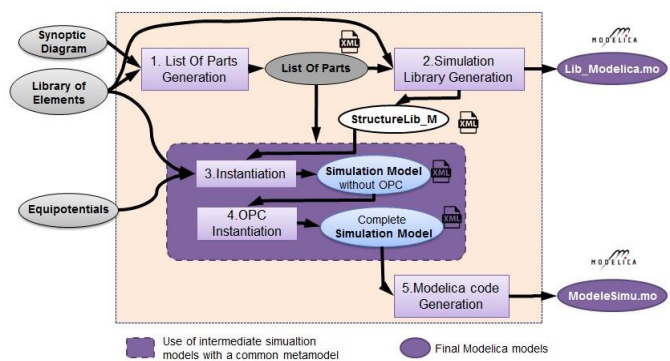

*Figure 42: Generation flow of simulation models*

**Step 1.** From the Library of elements and the Synoptic Diagram, the List of Parts is generated. Besides the mapping between the Library views, the List of Parts contains a synthesis of information related to the elements' instances. Thus, all the variables exchanged between the control/monitoring system and the process are indexed. All the connections (informational and physical flows) between the elements' instances included in the Synoptic Diagram are also expressed. Consequently, this model is used for generating both the *Modelica* library and the simulation model.

**Step 2.** The List Of Parts is then used, with the Library of elements (cf. the metamodel of the Library of elements, Figure 43, left), for the generation of the Modelica Library. This library contains all simulation models of elements that must be instantiated during the generation of the process simulation model. In addition to the Modelica Library, its structure is also generated (*StructureLib M*). It contains all the paths of elements' simulation models for the next step.

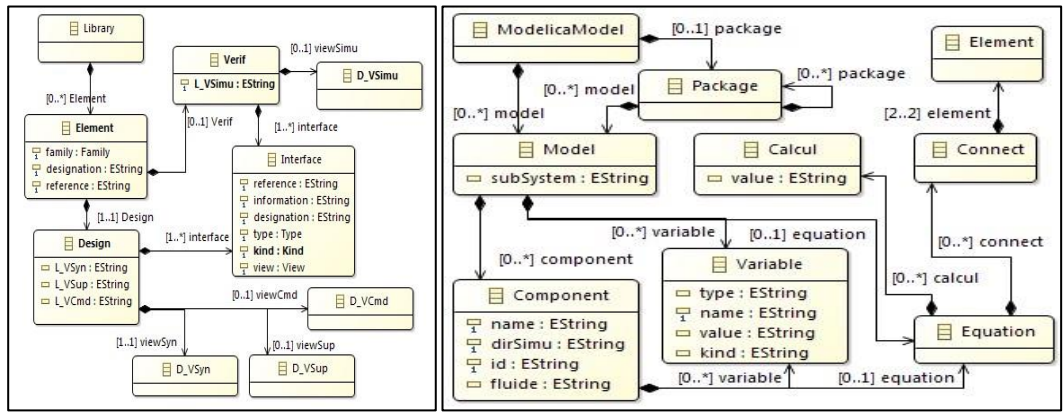

*Figure 43: Metamodels of Library Of Elements (left) and Simulation Model (right)*

The next two steps concern the building of the simulation model. This building phase is carried out using intermediate simulation models in XML format.

**Step 3.** First, the elements models are instantiated with their connections (step 3). From the List Of Parts and the StructureLib M, the elements models are instantiated. Connections between these models are instantiated from the List Of Parts and the generic rules of simulation views. Another input model is used, namely Equipotentials, that models the connection points at a same potential. This model comes from (Bignon, 2012) and can be directly derived from the List Of Parts. At the end of this step, an intermediate simulation model is generated, without the communication for exchanges with the control part (*Simulation Model without OPC*). To this purpose, a common metamodel of the simulation models is introduced. A simulation model (*ModelicaModel* on Figure 43) is composed of a "Model" which corresponds to the model to be executed during the simulation. As the *Modelica* language is hierarchical, this Model can be encapsulated into a Package. The *subSystem* attribute of "Model" refers to the subsystem name which is represented. The "Model" is composed of several "Component". These "Components" correspond to the instantiated simulation model of elements. Each "Component" is referenced by an *id* attribute for the instance name, and a *name* attribute for the element name. This information comes from the List Of Parts. The "Model" can have several "Variable". A "Variable" corresponds to an internal or external variable of simulation. The "Model" can also include an "Equation" section. The "Equation" section can contain two types of equations related to the *Modelica* language. Within *Modelica*, the connections of information flows or physical flows, which are defined by a "connector", can be done using a *connect* equation. Consequently, the "Connect" equation is composed of two "Element" that represent the two variables to be connected. The other type of equation is a classic equation, and it corresponds to the "Calcul" equation. A "Component" can also have several "Variable" which are internal to the simulation, and an "Equation" section. Moreover, "Model" and "Component" can have initial equations.

**Step 4.** To allow communication between the simulated process and the control/monitoring system, the instantiation of the OPC communication is done through step 4. From the List Of Parts, OPC variables are instantiated. Then, these are added to the *Simulation Model without OPC* to obtain the *Complete Simulation Model*.

**Step 5.** Finally, the last step transforms the *Complete Simulation Model* into *Modelica* code (*ModeleSimu.mo*). As a result, the simulation model that targets *Modelica* language is generated.

A prototype of our proposal has been implemented on Eclipse framework and applied for testing the control/monitoring system of a Fresh Water System described by Prat et al. [C19]. SystemModeler software was used as *Modelica* execution environment. The generated simulation model was successfully compiled in this software. Nevertheless, a manual initialisation of the simulation model is necessary to run the tests scenario. Through the presented proposal, an effort of formalisation has been made. This approach could be used in the industrial context and is independent of a particular proprietary software.

**Use of the simulation flow to verify the usability of sociotechnical system**

The simulation models would allow to perform verifications at three levels of the design flow (cf. Figure 40). A first series of verifications can be carried out after the low-level generation step. This involves checking monitoring & control chain of an element (for example, a valve), and checking the proper functioning of the alarm system (related to the instrumentation, for example). A second set of verifications can be performed following the high-level functional specification step. In this case, it is a question of verifying that the specified elementary command sequences allow the process to be brought into a state which delivers the service. Finally, a third series of tests can be performed following the high-level generation. This is to test the implementation of high-level functions, and the reconfigurability of the system.

The use case considered for the implementation of verifications is shown in Figure 44. This system is thus made up of two bunkers St1 and St2 which can contain 70m$^3$. Each has a filling valve (V2VM01 for St1, and V2VM02 for St2) and a drain valve (V2VM03 for St1, and V2VM04 for St2). A level transmitter is associated with each hold (LT0001 for St1, and LT0002 for St2). Different alarm thresholds are associated with the level measurement in the bunkers: very low level (LAL), low level (LIL), high level (LIH) and very high level (LAH). Each bunker also has an overflow detector (LS0001 for St1 and LS0002 for St2), which is integrated into a feedback loop which, if necessary, sends an electrical closing command to the filling valve.

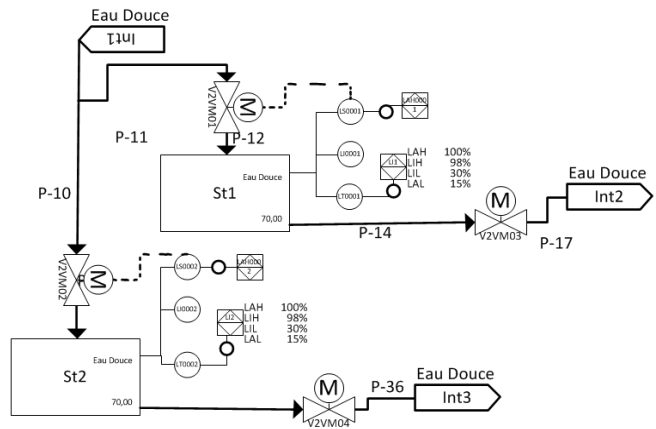

*Figure 44: P&ID use case*

To check the correct functioning of the alarms linked to the instrumentation of the bunkers during filling, a scenario has been designed. The valves are initially closed, bunker St1 contains 10m³ of fresh water and bunker St2 contains 40m³. Noticing the very low-level alarm in the cargo hold St1, the operator should ask for the V2VM01 valve to be opened to fill it. As the cargo compartment fills up, the level transmitter returns information related to the cargo hold percentage. Once 98% full, the operator requests the valve to be closed.

To run the simulation, we import the generated library and the simulation model into the SystemModeler software. The simulation model which comprises 298 equations and 298 variables was compiled in less than a minute, as it is for the generation of the executable. To provide a proof of concept, it was considered the point of view of the automation engineer who wishes to check and correct, iteratively, his program upstream of the implementation phase. The first low level verification is here presented, as an illustration of the approach. For the extended proof of concept at the different phases of the design flow, the author could refer to the PhD dissertation of Sophie Prat (2017).

When the simulation is launched, the supervision interface indicates to the operator that the level in the bunker St1 is low (via a low-level alarm). Since this is not a very low-level alarm in the bunker, the operator does not request the opening of the V2VM01 valve, which is conditioned by the appearance of a very low-level alarm (as initially expected in the scenario).

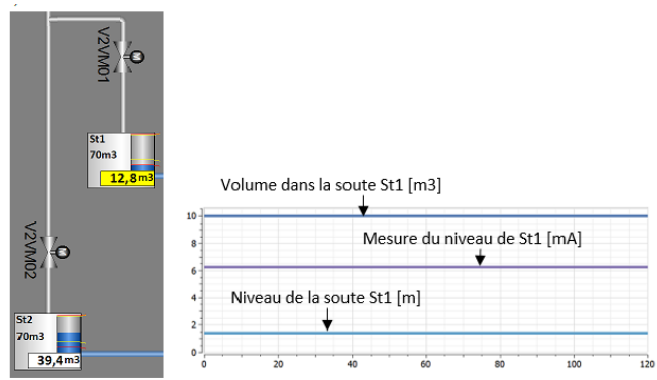

*Figure 45: Monitoring HMI and results of simulation*

In addition, the supervision interface indicates that the bunker St1 contains 12.8m³ and 39.4m³ for the bunker St2 (Figure 45), while the simulation results indicate that St1 contains 10m³ (Figure 46) and 40m³ for St2. We thus note that the displayed volumes of the bunkers do not correspond to those defined, but nevertheless that the low-level alarm is consistent with the displayed volume. The error therefore does not seem to lie with the supervision HMI. Looking at Figure 46, it can be observed the values displayed on the HMI correspond to those sent by the control program. The same goes for the alarm. However, the level transmitter measurement signal (sent and received) is indeed 6.285714 mA, which corresponds to a volume of 10m³.

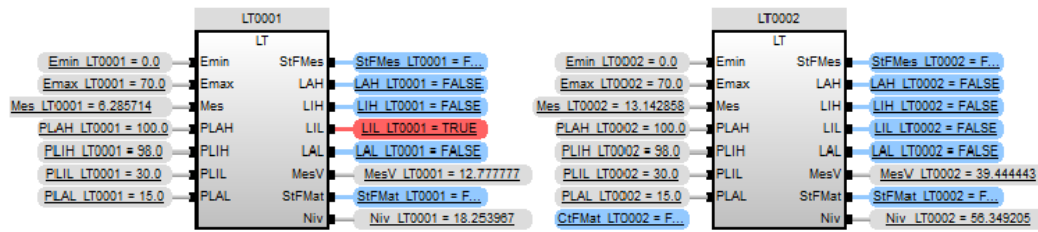

*Figure 46: Extract of the initial state of the control program*

There is therefore a problem in the processing of the measurement signal at the level of the control program. Thus, the implementation of this test allowed to realize that there was a conversion error in the control program, that had so far not been detected.

## 5.3. DISCUSSION

The adaptation mechanisms of HATs must therefore be considered from the preliminary phases of the design, to avoid the occurrence of failures or significant costs at the level of human-system integration, in the phases of deployment or operation of the socio-technical systems.

One of the major challenges here is to consider, in a dynamic way, the distribution of roles and functions between humans and machines. This is indeed necessary for the hybrid teams to remain efficient, by reconfiguring themselves following changes in the work situation or in the capacities of their members.

To this purpose, designers must not only look at approaches that are sometimes too prescriptive (defining an ideal system, but which can only be used by expert operators and in situations foreseen by the designer) and based only on formal proof (with model-checking techniques).

Formative methods, such as CWA, and simulation techniques, can thus help define and verify the functioning of human-machine systems, with a less rigid definition of the space of possibilities that leaves room for more emerging behaviours. Moreover, this type of approach also makes it possible to consider the temporal dynamics and the transitions that HATs may encounter, by identifying very early potential design conflicts.

The contributions presented in this chapter have thus made it possible to explore this path. On the one hand, a tool was proposed for designing the dynamic allocation of functions based on the CWA method and identifying certain design paradoxes in work situations. On the other hand, an approach based on the MDE allowed to generate a simulation flow, to verify the correct functioning of a human-machine system whose control programs and interfaces were generated automatically.

# Chapter 6.  Online adaptation of Human-Autonomy Teaming

*"To improve is to change; to be perfect is to change often"* – **W. Churchill**

If we have proposed to consider and provide tools for elaborating the dynamic function allocation from the preliminary design stage, it is also important to develop capacities to trigger this allocation in real time, with indicators associated to the status of the operator (link with part A). This would indeed improve human-machine cooperation, in the context where the operator and the AI share the same operational and tactical activity. This is what it is developed in section §6.1.

But this chapter also discusses the question of dynamic adaptation of human-autonomy teaming at the upper level of system of systems (SoS), where autonomy could monitor and improve cooperation within a human team (so there must be human-machine cooperation around cooperation human-human). Particularly, we studied this within a patrol of fighter pilots. This research occurred in the ANR TAPAS project, described in section §6.2. The same type of approach has also been used in the domain of cyber defense, with some experimental studies at the cybersecurity center in Vannes.

These are these 2 cooperation configurations, which were illustrated in Figure 16, on which we have proposed and evaluated adaptation mechanisms. These different works have been highlighted in the following publications: [J5, J8, J9, J14, C16, C28, C34].

## 6.1. System adaptation triggered by online mental workload monitoring

During the study carried out on the activity of operating a swarm of drones (Alexandre Kostenko's PhD, 2017), we also sought to implement a dynamic function allocation. This was controlled and triggered by the objective and real-time mental workload indicator that we had synthesized (see §3.3.2). In the following pages, we present the different types of assistance that we have designed, especially for the levels of high workload (level 4) and overload (level 5), then we analyse the impact of these dynamics aids on operators' performance and the cognitive condition.

### 6.1.1.  Design of assistances as a function of operator MWL

With dynamic function allocation - and unlike static allocation - tasks are distributed differently over time and according to various events. According to Inagaki (2003), the distinction between static and dynamic function allocation is based on the idea that a function does not necessarily have to be definitively allocated to an agent and that there are many situations where the machine and the operator can perform the same function in turn satisfactorily. Figure 47 illustrates the distinction between static function allocation and dynamic function allocation for a function f.

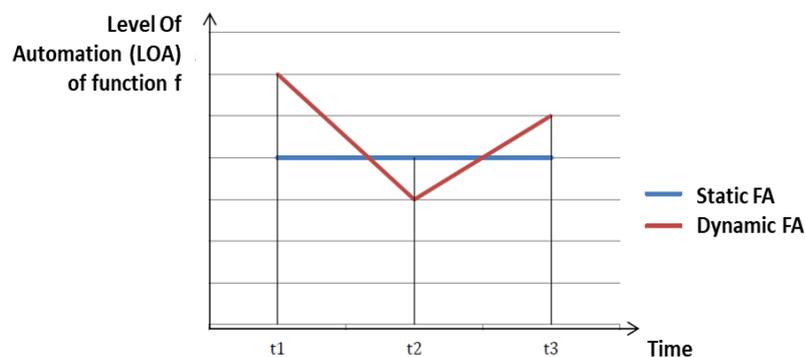

*Figure 47: Difference between static and dynamic function allocation*

**Dynamic function allocation conditions and triggers**

According to Debernard (1993) four conditions must be met to set up a dynamic function allocation, otherwise we must move towards a static allocation:

- The system must be multitasking.
- The level of requirement must vary over time, with phases of underload and overload.
- Tasks usually performed by the operator must be able to be automated.
- The operator must be present within the system.

According to Kaber, Perry, Segall, McClernon and Prinzel (2006), the objective of dynamic function allocation is to moderate the mental workload of the operator, and to support situational awareness, by facilitating the management of the cognitive tradeoff between task demand and cognitive resources. Indeed, according to Parasuraman et al. (1992), dynamic function allocation allows a better combination between the capabilities of the operator and those of the machine.

Kaber and Endsley (2004) especially identified three criteria to drive function allocation:

- *Critical events:* the allocation is driven by the occurrence of a critical event impacting the objectives of the system (Hilburn, Jorna, Byrne & Parasuraman, 1997). These critical events can be breakdowns, malfunctions, etc. and they must be identified during design or by experience feedback.

- *Performance:* function allocation is triggered when performance drops below a certain threshold (Parasuraman et al., 1992).
- *Physiological activity:* the allocation is controlled by the level of physiological activity of the operator.

According to Hoc (2001) the increase in the workload of an agent generates the need to add another agent with the same skills. In such a case, the establishment of an augmentative cooperation - in the sense of Schmidt et al. (1991) – can be justified (cf. §2.2.1). In the scenario developed on SUSIE (cf. §2.3.2), we therefore sought to automate functions initially performed by the operator, so that human and autonomy share some common skills. Considering the four main stages of information processing, Parasuraman et al. (2000) identify four processes - or "functions" - which can be subject to more or less automation (cf. §2.3.2). The assistance solutions we designed were therefore related to each of these "functions", for the different tasks involved in operating a swarm of drones.

- *Information gathering:* according to Inagaki (2003), a strong automation of information gathering consists of highlighting information to attract the attention of the operator. Two such automations are possible in the case of drone swarm operations on SUSIE simulator: one relates to message processing, and the other deals with zones management (cf. §2.3.2). These automations consist in highlighting the unprocessed messages that appeared less than two minutes ago (for the message processing task) as well as the zones without drones (for the zone management task), by flashing them to encourage the operator to treat them. This is a matter of mutual control (Hoc, 2001), in the sense that the system checks that the operator complies with the prescription concerning the processing of messages (Always read a message within two minutes of its appearance), as well as the prescription concerning the treatment of messages (Never leave an area empty, by deleting the zone or transfer drones into it).
- *Information analysis:* according to Parasuraman et al. (2000), a low level of automation involves presenting data in its raw form. At the highest level of automation, the system combines several data into one, to provide a so-called "sophisticated" representation. We can set up such automation for the vehicle search task. This automation consists in analysing and representing in a more elaborate way the information delivered by a message that has not been processed. This processing could consist in machine reading the message with the coordinates of a potential target, then in putting a red cross on the map at these coordinates.
- *Decision making:* this function can especially be automated on SUSIE for the task of zone management. To manage a zone without a drone, the operator must first judge the usefulness of the zone, then transfer drones to the zone if it is useful (with the presence of some detected targets) or delete it if it is not. Automation would be to automatically consider areas without drones to be useful and decide to send drones there.
- *Implementation of action:* wo automations related to the implementation of the solution can be implemented here: one relates to vehicle locking and the other to the task of zone management. To lock a vehicle, the operator must perform an inspection. The execution of this inspection action can be automated. Automation related to the zone management task would consist of automatically sending drones to the zone (i.e., transferring drones to areas without drones).

**Activation of assistances as a function of mental workload level**

<u>Mental Workload level 4 aids:</u> since the operator rather focusses on the processing of vehicles when the task load increases, we have implemented aids to improve their situation awareness. At mental workload level 4 were triggered automation solutions that focus only on information gathering, with a vertical type of cooperation (cf. top of Figure 48).

<u>Mental Workload level 5 aids:</u> at this level of mental workload, the operator is overwhelmed, and many tasks are no longer performed. It is therefore relevant to set up a horizontal type of cooperation, where the "functions" of decision-making and execution of the solution are distributed between human and machine. The machine can therefore make and execute decisions. In addition to the automations made at level 4, we are adding automations for information analysis, decision making and solution implementation.

Figure 48 summarizes the aids and the rules for activating these aids according to the levels of mental workload.

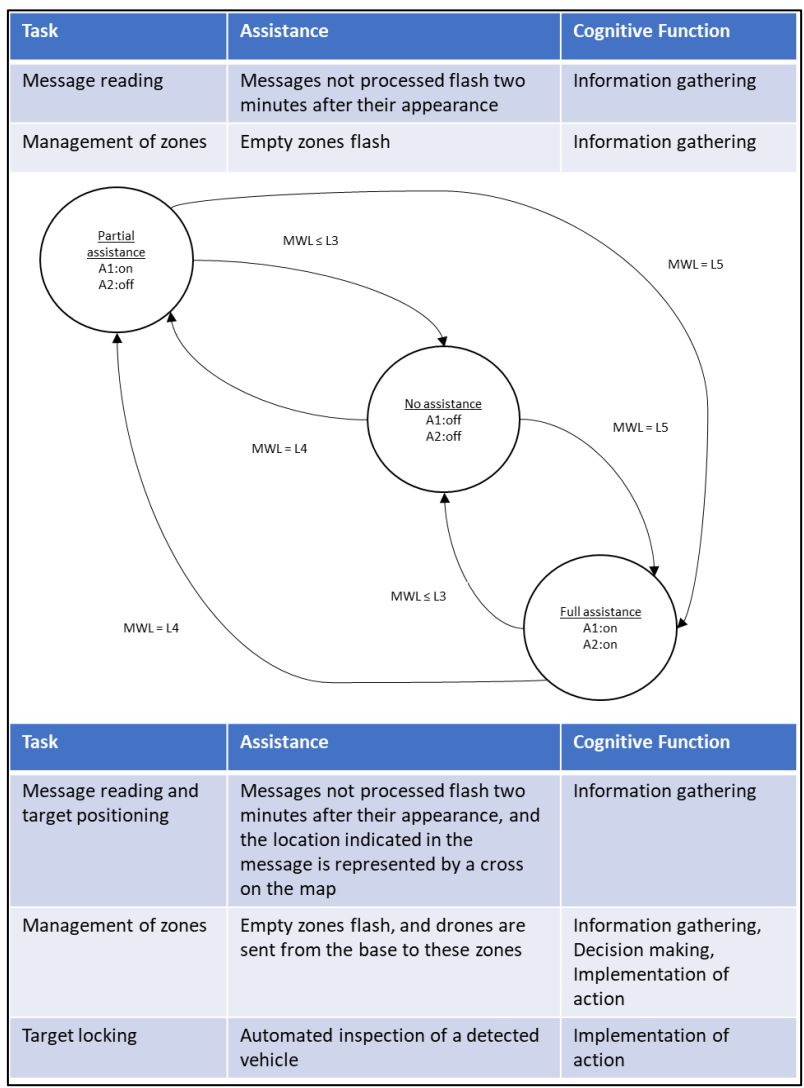

| Task | Assistance | Cognitive Function |
|------|-----------|-------------------|
| Message reading | Messages not processed flash two minutes after their appearance | Information gathering |
| Management of zones | Empty zones flash | Information gathering |

| Task | Assistance | Cognitive Function |
|------|-----------|-------------------|
| Message reading and target positioning | Messages not processed flash two minutes after their appearance, and the location indicated in the message is represented by a cross on the map | Information gathering |
| Management of zones | Empty zones flash, and drones are sent from the base to these zones | Information gathering, Decision making, Implementation of action |
| Target locking | Automated inspection of a detected vehicle | Implementation of action |

*Figure 48: Assistances and activation rules*

### 6.1.2. Effects of assistances on operator state and performance

To study the effects of the aids previously described, the experiment performed in §3.2.2 was reproduced with 24 participants. The scenario was the same, but here the assistances were automatically triggered by the level of mental workload assessed in real time with the objective and synthetic indicator presented in §3.3. We then compared the physiological, behavioural and performance data between the first experiment and this new experiment, to assess the impact of the assistance.

**Effect on operators' performance and regulations**

The results mainly show a beneficial effect of the aid on *participants' performance and on behaviour* (participants better comply with prescribed operating procedures).

A series of Wilcoxon tests was carried out on performance indicators P1 (relating to the treatment and neutralization of vehicles) and P2 (relating to the processing of messages) to be able to compare the mean values obtained with the aid (DFA condition) and without the aid (Control condition). Statistical results revealed a significant increase in P2 with aid (Z (45) = 2.094, p <0.05). However, no difference was found for P1. Furthermore, a Wilcoxon test also revealed that the percentage of vehicles that were detected was significantly higher with the aid (Z (45) = 3.133, p <0.01). Thus, even if the number of neutralized vehicles has not significantly increased, the number of targets identified and therefore the operator's perception of threats has been greatly improved.

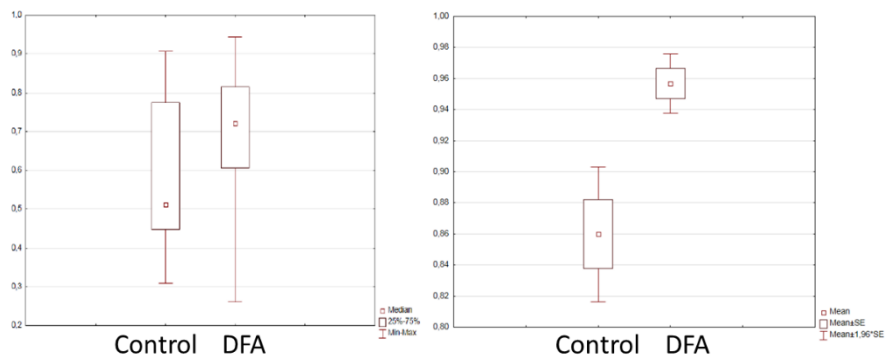

*Figure 49: Effect of DFA aids on the performance (message processing on the left, and % of detected vehicles on the right)*

Besides, a Wilcoxon test on compliance with prescribed operating modes also showed that compliance is significantly better in Experiment 2 (DFA condition) than in Experiment 1 (Control condition), Z (45) = 2.126, p <0.05. In particular, participant behaviour is based more on performance / compliance management and less on priority management when dynamic allocation of functions was implemented.

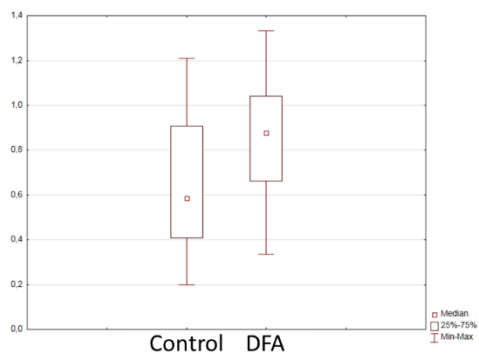

*Figure 50: Effects of DFA aids on the compliance with prescribed operating modes*

**Effect on operators' mental workload**

In addition, we looked at the effect of the aid on the mental workload of the participants. Student's tests for paired data were carried out on the average time spent at high workload and overload levels (respectively level 4 and level 5), to compare the values obtained with and without the aid (cf. Figure 51). Statistical results revealed that participants spent significantly less time at level 4 with assistance (t (45) = 2.22, p <0.05). No significant effect was observed on time spent on level 5.

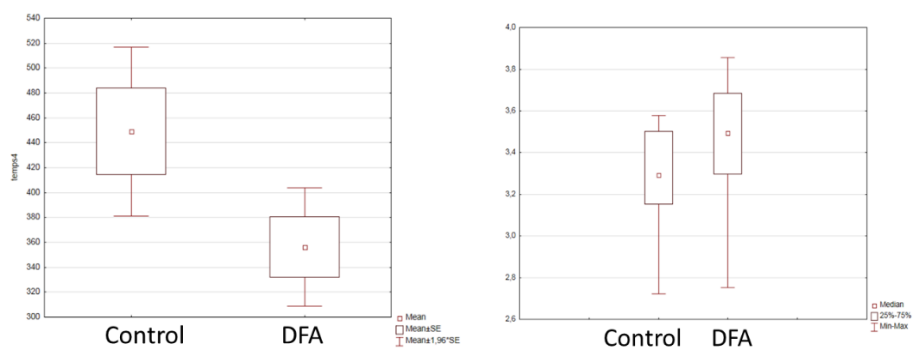

*Figure 51: Effect of DFA aids on mental workload (left: level 4, right: level 5)*

**Findings**

Ultimately, the aid deployed seems to help participants to better regulate their activity, by encouraging them to implement regulations based on performance and on compliance. The automation related to information gathering (for the tasks of processing messages and managing zones) encourages the participants to carry out tasks which had not been carried out in experiment 1. The implementation of horizontal cooperation (at level 5 of mental workload) makes it possible to fight against prioritization, by allocating tasks that become secondary for the operator to the machine, which explains the increase in compliance with operating methods and in performance in the experiment 2.

The aid provided to operators therefore facilitates the management of the cognitive compromise (between performance and cognitive cost) and improves performance. These results are consistent with the literature. Indeed, according to Parasuramam et al. (1992), dynamic function allocation allows a better combination of operator and machine capabilities and, according to Kaber et al. (2006), aid facilitates the trade-off between demand and resources.

These findings are however activity dependent. Indeed, the nature of the activity carried out in SUSIE simulator is very exploratory, a kind of "target search and rescue" task where the task load depends on previous actions achieved by operators. The more participants detected targets and they are proactive in their search, the more they faced stimuli to be processed and they risked being highly loaded. That can explain why the average mental workload did not decrease with the aid. Instead, an increased cognitive efficiency was observed.

Moreover, we considered here operator state monitoring and dynamic adaptation at an individual level. It is another challenge to consider this problem, but at a team level. This is what it is developed in the following section.

## 6.2. AN AGENT FOR MONITORING TEAMS AND COACHING HUMAN DIALOGUE

Collective work between pilots in a fighter squadron is a very complex dynamic process. It requires accurate analysis to understand and anticipate the tasks that could increase the risk to jeopardize mission or pilot's safety. The risk is defined as a probability of damage or loss (e.g., mission abort or accident), caused by external or internal vulnerabilities. It seems therefore particularly important to investigate how the different squadron collective tasks (CTs) can be deteriorated by these vulnerabilities.

As depicted in Fig. 52 (bottom left box), the potential deterioration of collective work can result from team-external factors (the threat status for example, given the situation or the function of the CTs), namely task constraints.

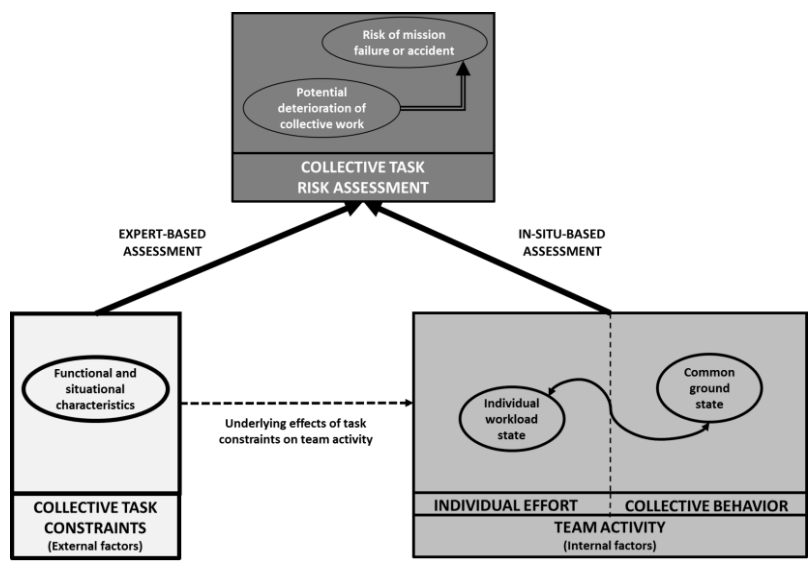

*Figure 52: Two ways to assess collective task criticality: task analysis (external factors) and team activity analysis (internal factors)*

Nevertheless, in a context where pilots are trained to face often very complex and difficult situations, it is also relevant to examine team-internal factors. These are related to the state of the pilots or the state of the team common ground (see Figure 52, light grey box) and can help understand how a patrol reacts to the task constraints. The degradations of collective work can be examined by analysing team activity, at individual and team levels (Cannon-Bowers & Salas, 1997). The individual level involves team members' cognition and can be assessed by measuring their mental workload.

Various studies (Nählinder et al., 2004; Svensson & Wilson, 2002) have shown that an increase in mental workload, reported from both subjective and physiological measures, leads to a decrease in situation awareness. This degradation of the perception and comprehension of environmental information and projection of the situation over time (Endsley, 1995) by one pilot could therefore have an impact on the upper-level team situation awareness (Gorman et al., 2006), due to the quality of interaction interference (Hoc, 2001). From a team-level perspective, the analysis can be made through the study of team communications that constitute a central component of information processing (Salas et al., 2008). Failures in communication (e.g., sharing incorrect or confusing information) could trigger off uncertainty about the common representation of the situation faced by the team (Xiao et al., 1996). Moreover, degraded communications could lead to errors and loss of common ground (Klein et al., 2005), which would limit or disrupt the way in which teams are performing their joint activity (Salas et al., 2008). These two approaches of task risk assessment – the expert assessment of external factors highlighted by a task analysis and the in-situ assessment of internal factors highlighted by a team activity analysis - are synthesized in Figure 52.

This chapter proposes a method for determining the fighter squadron CTs that would be the most critical for mission success and pilots' safety. The COWORK[2] method (COmmunications and WORKload to study COllective WORK) is designed to classify CTs according to their potential risk of deterioration of collective work, at individual and team levels. Therefore, this classification follows the approach based on internal factors, relying on analyses of physiological measures of mental workload and the quality of communication.

The following sections outline the three stages of the COWORK[2] method and illustrate its deployment in a military field-based experiment. This classification based on internal factors are then benchmarked against a classification based on external factors, i.e., the risk level categorization of collective tasks achieved by the subject matter expert. Finally, the results of this application are discussed in terms of the effectiveness of the method to identify the critical collective tasks. This work was developed in the ANR TAPAS project.

### 6.2.1. COWORK[2] method for monitoring common ground

The COWORK[2] method propose a classification based on the in-situ assessment of internal factors. It involves three main stages, as illustrated in Figure 53: (1) a task analysis of the collective work (CW) achieved in military tactical flights to identify CTs and characterize their constraints; (2) an analysis of team activity (individual effort and collective behaviour) for each CT, based on mental workload and communication quality assessment; and (3) the classification of CTs in terms of deterioration risk of individual and team-level activities.

Moreover, as depicted in Figure 53 (dotted arrow), the classification of CTs resulting from the COWORK[2] method should be validated by comparing it with another classification of reference. This "benchmark" classification may rely on an expert-based assessment of external factors, providing a categorization of the CTs in terms of risks level, based on the knowledge of their different situational and functional constraints (defined in stage 1).

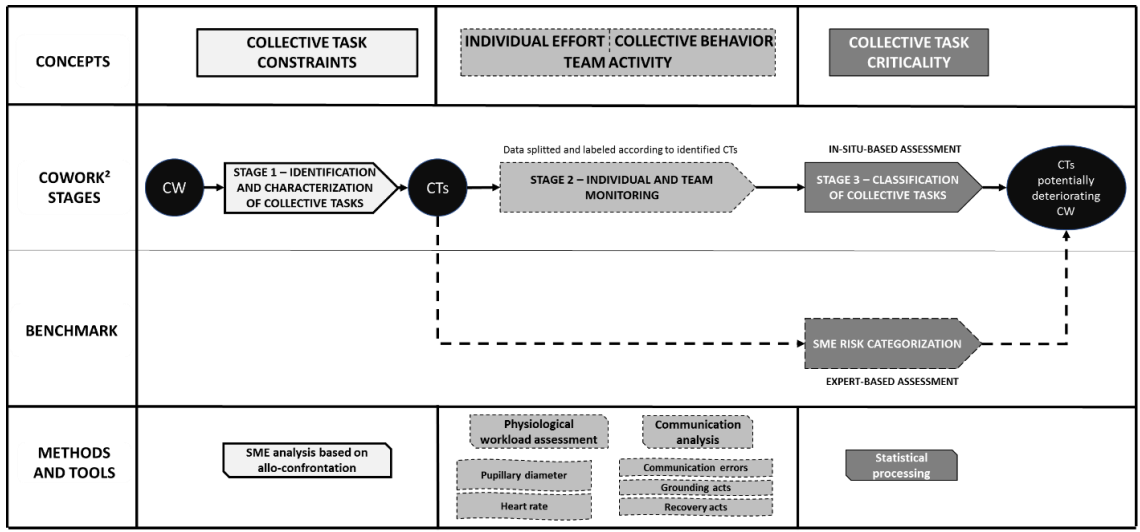

*Figure 53: The 3-stage COWORK² method (CW: Collective Work, CT: Collective Tasks)*

The following sections detail the three stages of the method by explaining and justifying the choice of variables, methods, and tools found in the literature.

**Stage 1: Identification and characterization of collective tasks**

To analyse pilots' collective work, it is necessary to first identify collective flight tasks. The study of CTs in patrols can rely on radio communications. Consequently, intra-patrol radio communications are used to carry out the analysis of CTs (Guerin et al., 2014). A subject matter expert (SME) can help with this identification through an allo-confrontation method (Mollo & Falzon, 2004). Confrontation methods are known to yield good immersion and data close to the deployed activity (Lahlou, 2011). These recorded communications, in addition to providing tactical and operational information about the flight context, are presented to the SME to retrieve a set of communication standards. The expected result is a definition of the tasks composing the collective work. The acquisition of a corpus of CTs - used as a CT coding scheme (the first stage of the method) - is a prerequisite for segmenting the pilots' physiological activity before analysing the workload on the one hand and the communication quality on the other.

**Stage 2: individual and team level monitoring (toward a low-intrusiveness device)**

In this stage, two kinds of assessment are conducted: the individual-workload level assessment and the team-communication level assessment.

At individual level, when pilots are engaged in a cognitive activity, indirect indications of their level of effort may be collected from the variations of different physiological variables. These measures therefore provide information that is not contained in performance or subjective measures (Hankins & Wilson, 1998; Veltman & Gaillard, 1996a; Wilson, 2002a). Numerous studies have used physiological and/or ocular data to examine workload in the field of pilot activities. A higher physiological activation (activation of the sympathetic branch of the autonomous nervous system), namely a higher workload, is then observed between baseline measurement (resting period) and flight activity (Karavidas et al., 2010; Lehrer et al., 2010; Veltman & Gaillard, 1996a, 1996b, 1998; Veltman, 2002; Wilson, 2002a, 2002b; Yao et al., 2008; Ylonen et al., 1997) and during the most difficult flight phases, namely take-off or approach segments with a high informational load (landing, touch and go). This increasing activation is reflected by an increase in heart rate (Hankins & Wilson,

1998; Lee & Liu, 2003; Veltman & Gaillard, 1996a, 1996b; Yao et al., 2008; Ylonen et al., 1997; Wilson, 2002a, 2000b), respiratory rate (Karavidas et al., 2010; Yao et al., 2008), skin conductance (Wilson, 2002a), pupil diameter (Dehais et al., 2008) and a decrease in heart rate variability (Hankins & Wilson, 1998; Veltman & Gaillard, 1996b; Wilson, 2002a; Wilson et al., 1994). Collecting this information, in addition to the assessment of intra-patrol communication quality, is relevant to estimate a potential degradation of collective work during flight, since a high workload at an individual level may have a negative impact on CT performance (one member of the team fails to execute his/her own part of the collective work because of the workload) and on the quality of the common ground (because of the workload, one member of the team makes communication failures that lead to the degradation of the information and knowledge shared by the team).

At team level, the assessment entails the analysis of the radio communications exchanged within the squadron during the flight. Different frameworks exist in the literature for coding communications according to errors, recovery acts, or grounding acts. Gibson et al. (2006) assert that communication errors play an important role in incidents within the context of aviation. They propose a taxonomy of human communication errors, which can be directly identified with a two-dimensional analysis grid: the "level of grammar" dimension and the "external error mode" dimension. Regarding the first dimension, the distinct levels of grammar at which the error arises can be seen from various perspectives: phonology (the sound patterns of language), syntax (the word patterns of language), semantics (the meaning patterns of language) or pragmatics (language in context). The pragmatic level deals with deviations from standard grammar communication required in procedures underpinning the work domain (Corradini & Cacciari, 2002). The second dimension, the external error mode, provides observable behaviours of communication: for instance, the communication can be omitted or repeated or conducted too early or too late. Communication errors can also be identified indirectly from their consequences, by analysing the processes of recovery (Gibson et al., 2006). Furthermore, the team-communication assessment can be supported by the analysis of grounding in communication. This is a well-known process that helps update the common ground between two interlocutors (Clark & Brennan, 1991). Klein et al. (2004) describe several activities to support common ground, by inserting various clarifications and reminders into the communication.

Individual-workload level and team-communication level assessment requires the development of a suitable device. The deployment of an experimental setup to assess individual workload and communication quality requires fulfilling two main conditions: (a) adapting to the simulation environment with high quality data acquisition and (b) not disrupting the usual flying activity (i.e., the device had to be unnoticed) and the preparation of pilots. To this end, Lassalle et al. [C16] propose to implement a low-intrusiveness experimental setup for continuous recording of the patrol leader's physiological activity and intra-patrol communications. Thus, this device enables the acquisition of measures reflecting with some accuracy workload variations that could be associated with communication failures within the team and consequently a degradation of collective work.

**Stage 3: classification of collective tasks**

The last step of the COWORK$^2$ method consists in evaluating the potential degradation of collective work by discriminating CTs with the aid of physiological activity and communication analyses. Different statistical and multivariate exploratory techniques can be used to achieve this purpose, such as hierarchical ascendant classification or K-means. CTs are considered as endangering collective work when an increase of workload and/or failures in communication and grounding acts are detected.

### 6.2.2. Application to the dialogue between two fighter pilots

The COWORK[2] method was applied in the context of tactical flight simulation during actual training sessions of a Rafale (a multi-role combat aircraft for the French Navy and French Air Force) patrol. This training is carried out within a highly realistic simulation environment, and it represents a real stake for pilots who are trying to obtain a higher rank through completing a section leader test. In this evaluation context, flight scenarios were not known in advance because they were adapted to the evaluation needs. Nevertheless, the training flights chosen had a high level of complexity to produce numerous CTs; they involved air-to-air or some air-to-ground tactical missions.

**Stage 1 application: identification and characterization of collective tasks**

Allo-confrontation (Stage 1 of the method for CT analysis) was carried out by a SME (Lt.-Col., French Air Force pilot working with Dassault Aviation) from an audio and video recording of a simulated air-to-air training mission made by an experienced pilot. The CTs were then defined in terms of communication sequences; a communication sequence included at least two transmitted sentences. The mission under investigation was specifically chosen to cover a maximum of typical CTs. The SME was given all the available information regarding the pilot during the mission (i.e., visual displays of aircraft outside, the cockpit, and the pilot's actions in the cockpit). A summary of this corpus of collective tasks is given in Table 12 (left side).

*Table 12: Final corpus of collective tasks resulting from stage 1 (left side) and benchmark classification (right side)*

| STAGE 1: IDENTIFICATION AND CHARACTERIZATION OF COLLECTIVE TASKS | | BENCHMARK | |
|---|---|---|---|
| COLLECTIVE TASK CORPUS | TASK CHARACTERISTICS | TASK RISK | TASK CATEGORIES |
| Take-off | Basic management tasks (patrol and aircraft administration) → There is no threat to manage | Low risk | Operational flight management |
| Fence-in | | | |
| Join up | | | |
| Visual/Blind report | | | |
| Weapon status check | | | |
| Airfield weather and status report | | | |
| Handover | | | |
| Radio frequency change | | | |
| Basic flying | | | |
| Formation change | | | |
| Mission management | | | |
| In/Out check | | | |
| Oil check | | | |
| 9-line recap (recap of some mission information) | | | |
| Landing | | | |
| Engagement | Shot status checking tasks (verification of the target status after a shot) → The threat is processed | | High tactical fight management |
| Monitoring | Target approach tasks (Decision making and implementation of the strategy to deal with the identified threat) | | |
| Commit | | | |
| Threat | | | |
| Interception | → **The threat is identified and located and the patrol attempts to have a window of opportunity** | | |
| Target | | | |
| Situation awareness report | **Target search tasks** (proactive search of target) | | |
| Contact radar faded report | → **The threat could be in the battlefield and is looked for** | | |
| Delouse (management of unknown platform trailing friendly platform) | **Protection manoeuvres tasks** | | |
| Self-protection | → **The patrol is attacked by the threat** | | |
| System breakdown management | **Survival management tasks** → **The aircraft integrity is endangered** | **High Risk** | |

Moreover, the SME was asked to provide a risk level assessment and an expert categorization of the CTs defined in this stage 1. This categorization (see Table 12, right side) would be used later as a benchmark to discuss the quality of the classification obtained by the COWORK² method. From the SME's point of view, CTs can be examined according to a risk level related to task purpose, depending on the status of the threat (see the columns "Task Characteristics"). From this standpoint, two main categories of CTs can be distinguished: (a) the basic CT routines involved in operational flight management (e.g., weather management, basic flying), (b) the high tactical CTs applied in fight situations (e.g., engagement of a target, management of an aircraft system breakdown).

**Stage 2 application: individual and team level monitoring**

Population. Five male pilots, all French Navy fighter pilots ranging from 29 to 32 years old, participated in the experiments. Participants had an average of 870 hours of flying experience on all platforms (ranging from 700 to 1100 hours) and 338 hours on Rafale aircraft specifically (ranging from 150 to 500 hours). The experiments involved 22 training missions, five of which were run as pre-tests. However, the data obtained for two pilots was rejected due to technical problems (data synchronization difficulty) or poor quality of the signals. Thus, analyses were conducted using data collected from three pilots over nine missions. Experiments were conducted during tactical flight training sessions in a highly realistic simulation environment.

Simulator. Experiments were conducted in a Rafale simulator located in the Rafale Simulation Centre of Landivisiau Navy Air Base, France. The simulator provides a very high realistic environment (see Figure 54) with the appearance and functions of a real Rafale aircraft cockpit with real flight instruments and G-seat. During the training sessions, radio communications occurred between the lead pilot, his wingman, stationed in a similar simulator in a side room, and controllers located in the instructor's room. The ambient temperature of the simulator was regulated between 20 and 21°C with a humidity rate set at 40%.

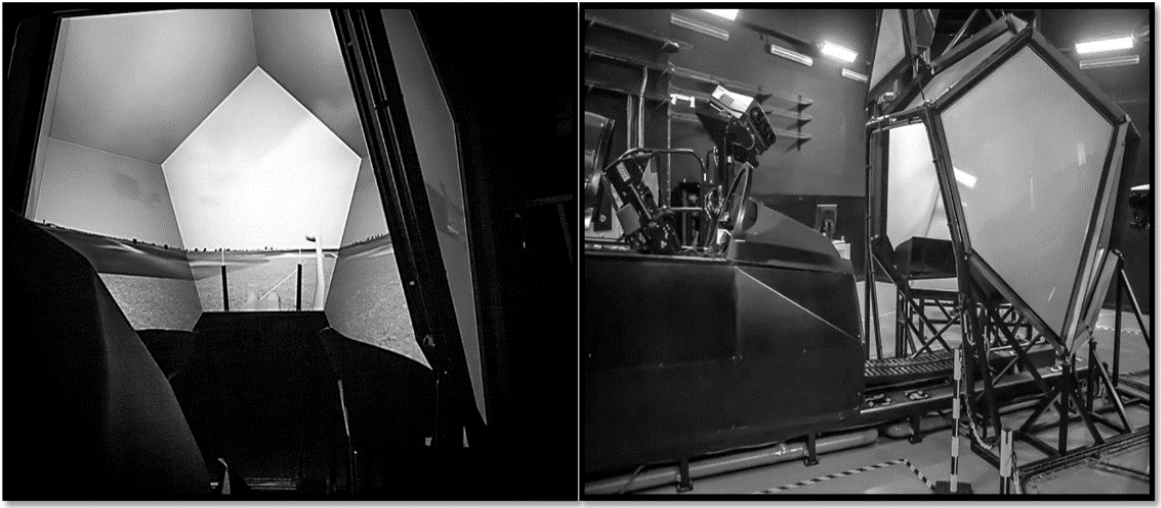

*Figure 54. Rafale simulator presentation.*

<u>Experimental setup.</u> The individual workload of the lead pilot was measured from several physiological indicators during each flight training session. Cardiac, electrodermal, and pupil signals were continuously recorded, as well as respiratory signals (examined as a control variable for cardiac signals). Cardiac and respiratory activities were measured from a BioHarness3™ electrocardiogram belt worn directly on the skin with an adjustable elastic strap. Electrodermal activity was recorded with two Ag/AgCl electrodes kept at the wrist with a band (Affectiva™). A Double-Tracking Device (DTD, see [C16], for a more detailed explanation) was used to record pupil diameter information. The DTD entailed coupling two FaceLab™ eye trackers fixed on a quick-release dedicated support. This device would guarantee maximal tracking coverage during the flight despite the pilots' head movements. All the sensors (belt, wristband, and DTD) were chosen to comply with experimental field constraints to reduce intrusiveness that would bring discomfort to the participants, and they were integrated into the simulation environment without impairing the pilots' activity.

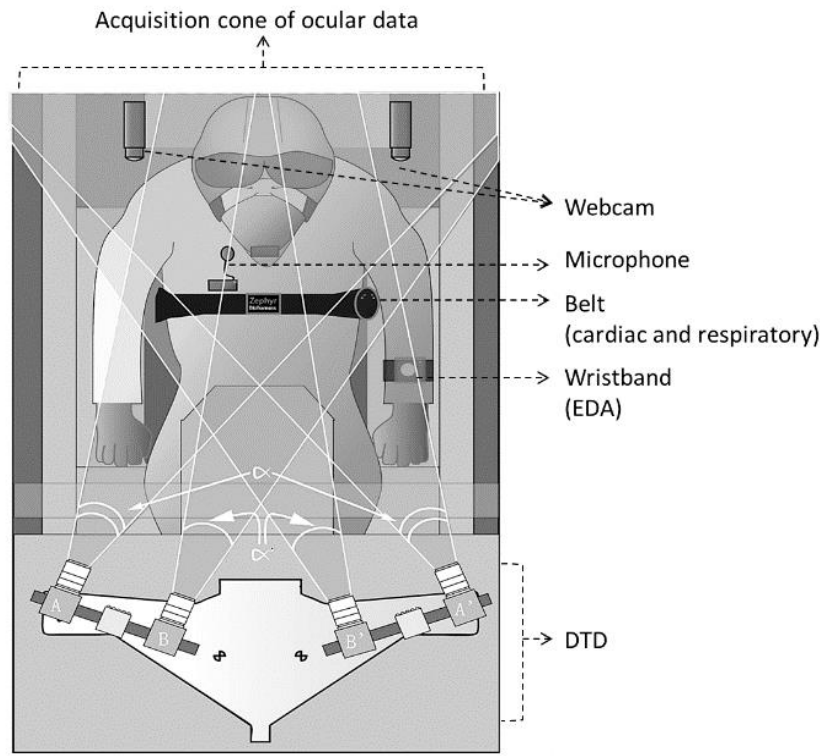

*Figure 55: Experimental setup for individual workload and collective work assessment*

The physiological data were posteriori synchronized with the training session timeline by means of online audio-visual information recordings. To this end, a microphone and two webcams were fixed in the cockpit. Intra-patrol communications were recorded and archived at the Rafale Simulation Centre from an internal device and then posteriori analysed to label the team activity according to the corpus of CTs defined in stage 1. Figure 55 shows the experimental setup.

<u>Physiological activity monitoring and communication coding.</u> Throughout the simulated training sessions, heart rate (HR), respiratory rate (RR – as a control variable of the HR), skin conductance (SC) and pupil diameter (PD) indicators were recorded. The sampling frequency for each index was respectively 250 Hz, 25 Hz, 32 Hz, and 60 Hz. Communications were analysed at a high level by using the literature frameworks presented in section 2.2. Six indicators were used, and the number of occurrences for each indicator was coded for each communication. The communication transcripts cannot be disclosed, due to confidentiality reasons. Communication analysis was therefore conducted at the naval air station, and audio recordings could only be listened and coded on site, with the help of a subject matter. The following indicators were considered:

- *Communication errors:* (a) Non-procedural silences (corresponding to an error at the pragmatic level with the external mode "*omitted*" according to Gibson's typology); (b) Simultaneous speech (when a pilot speaks even though the other pilot has not finished his communication, that can be considered as a pragmatic level error with the external mode *"too early too late"*); (c) Hesitations during the communication, defined by Boomer (1965) as consisting of both filled ("ah") pauses and intra-clause silent pauses of at least 200 msec. duration (they could therefore be viewed as a lower level error related to phonology).
- *Recovery acts:* Self-corrections (when a pilot changes his speech during the communication).
- *Grounding acts:* (a) Repeat information request (as in "*say again*" when one pilot queries a

response that was not received); (b) Situation report request ("*to report*", which reflects the need for clarification).

<u>Procedure.</u> The objectives and the experimental device were presented to the pilots before the training day. Each pilot also signed a written consent form before the data collection. Sensors (belt and wristband) and the microphone were fixed on pilots when they arrived. Then, the pilots sat down in the cockpit and carried out the various adjustments needed (e.g., seat height). Calibration and creation of the individual head model were then realized for each pair of the double eye tracker. This stage took about five minutes. As reported by Gerin, Pieper et Pickering (1994), this 5-minutes-calibration phase would give enough time for the physiological measures to stabilize (reduction of the effect of indoor/outdoor temperature difference, stress, nicotine, caffeine, etc.). Usually, the mission started shortly after the calibration process. The exact start time was recorded for the subsequent synchronization using an independent trigger module.

<u>Signal processing and data reduction.</u> All radio communications during flight sessions were transcribed and coded based on the previously defined CT coding scheme. Physiological and communication data were then divided and processed according to this segmentation. As a second step, the physiological data were processed to:

- remove outliers and singular steps out of the domain of physiological validity (Sami et al., 2004; Storm et al., 2000),
- remove the signal noise: for instance, the pupillary diameter signal was cleansed of all light reflexes by using the routine proposed by Marshall (2002) to keep only the cognitive dimension of the pupillary response,
- normalize the interpretable data to erase the differences between subjects (z-normalization for PD, mean-normalization for the other data),
- calculate the temporal mean scores (tonic value) for each physiological variable and for each segment of communication sequences. Mean scores were computed taking into account a time window including five seconds before and after segments to cover residual or anticipation effects on measures.

<u>Experimental device validity.</u> Lassalle et al. [C16] indicate that the whole setup was successfully deployed and offered a non-invasive solution to collect high quality data in a realistic flight environment. Nevertheless, examining the signal quality acquisition also revealed a very noisy signal regarding skin conductance, probably due to artefact movements. Consequently, SC measures were not included in the dataset used for analyses.

**Stage 3 application: classification of collective tasks**

<u>Individual-workload level classification.</u> After processing and normalizing the physiological data, a hierarchical ascendant clustering (HAC) was used to identify the different classes of CTs according to HR and PD pattern variations. The partition obtained was then consolidated by K-means clustering. Finally, a Friedman analysis of variance (ANOVA) was performed to observe between-classes significant differences.

As shown in Figure 56, the HAC revealed five different classes of CTs. The analysis of K-mean scores highlighted the same five classes, hence showing the robustness of the HAC results. Particularly, results showed the isolation of the landing CT (Class 2) and the grouping of CTs corresponding to tactical or combat phases (Class 1) such as the shooting or self-protection ones. Thus, CTs belonging to Class 1 share the feature of involving the pilots' survival or the mission achievement. Results indicated that both the HR and the PD were jointly highest in Class 1 (delouse, threat, interception, engagement, 9-line recap, monitoring, self-protection, targeting, contact radar faded report, situation awareness report). However, the HR was lowest in Class 5, and the PD was lowest in Class 3 (airfield weather, fence-in, handover, and radio frequency change) and Class 2. Class 2, corresponding to the landing CT, is explained by a larger decrease of the PD, compared to all other tasks. The right side of Figure 56 shows the final centres derived from K-means clustering.

Previous results are confirmed by a Kruskall-Wallis ANOVA (for non-parametric data because they were not all normally distributed) conducted on HR and PD and examining the fixed factor "CT classes". The analysis revealed a significant effect of classes on both the HR (H (4, N= 2302) =23.38, $p$<0.001) and the PD (H (4, N= 2234) =123.77, $p$<0.001). A test for multiple comparisons of mean ranks mainly revealed a significant increase between Class 1 and all the other classes for the PD ($p$ <0.001) and between Class 1 and Class 3 ($p$ <0.001) and Class 4 ($p$ <0.05) for the HR. These results confirmed physiological activation during the CTs belonging to Class 1 (characterized by a significant increase of both HR and PD). It should also be noticed that the HR was not influenced by the RR. The absence of correlation between HR and RR (as a control variable) was controlled through a Spearman rank-order correlation. Hence, as shown in Figure 56 (right side), CTs included in Class 1 could be considered as generating a high mental workload for the leader pilot, Class 2 would represent a low level of mental workload, and Classes 3, 4, and 5 could indicate a medium level of mental workload.

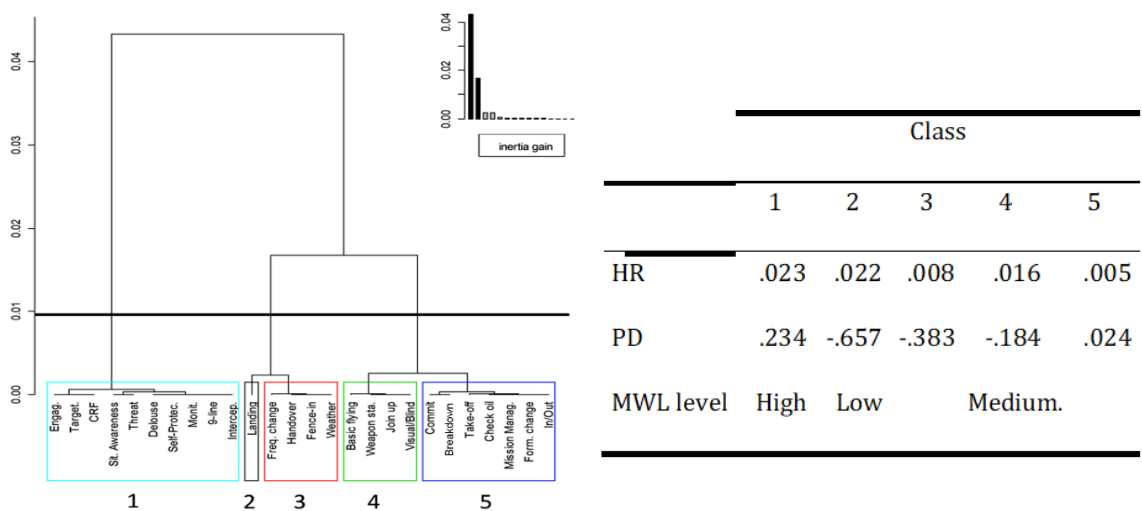

*Figure 56: Hierarchical clustering of the physiological indicators, with final centres from K-means clustering performed on physiological normalized data (MWL means Mental WorkLoad)*

<u>Team-communication level classification.</u> Team-level assessment entailed a communication analysis using six different indicators. First, a principal component analysis (PCA) was conducted to detect patterns of contributory factors potentially involved in deteriorating collective work. As before, it was completed with HAC designed to create different classes of CTs. An analysis by K-means and a Kruskall-Wallis ANOVA were then carried out. Figure 57 (left side) shows the projection of the communication-based indicators on the factorial design.

A PCA was conducted with the six communication-based indicators: non-procedural silences, hesitations, self-corrections, simultaneous speech, repeat information request, and situation report request. Axis 1 explains 36.54% of the inertia. It is determined by the following indicators: non-procedural silences, simultaneous speech, and self-correction. All these indicators can be viewed as communication errors. Axis 2 describes 20.33% of the inertia. It is determined by grounding acts indicators: report situation and repeat information. These two indicators reflect the need to clarify the situation. Axis 3 explains 15.76% of the inertia and is mainly determined by hesitations, namely a deviation of communication standard at phonologic level. Together, these three axes explain over 72% of the data variability. Fig. 56 (left side) shows Axes 1 and 2 of the PCA. The HAC was carried out from the three PCA axes. The analysis shows four different classes of CTs (see Figure 57 right side and Figure 58).

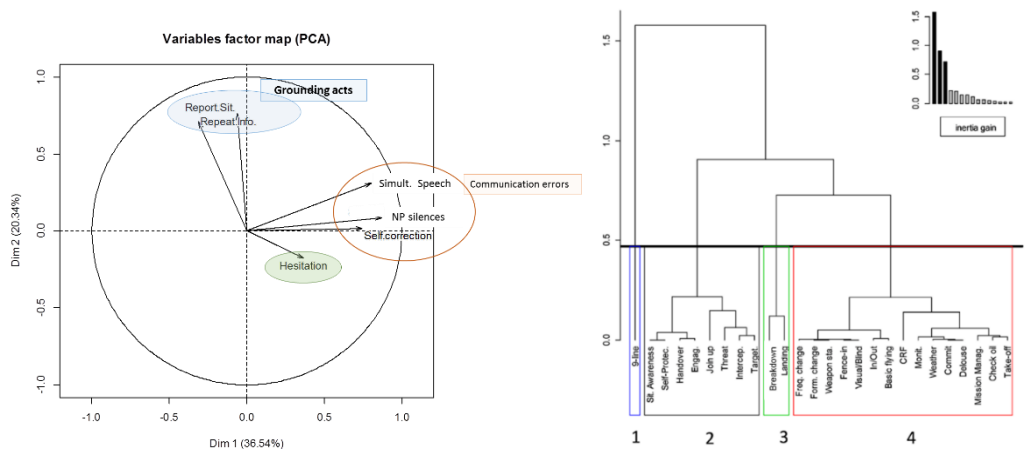

*Figure 57: PCA factor map and hierarchical clustering on communication-based indicators*

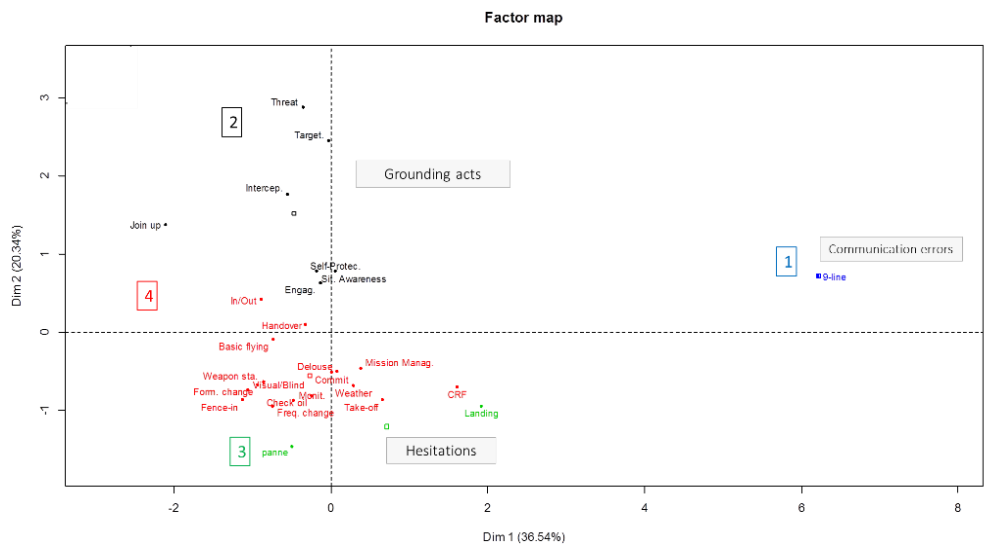

*Figure 58: Clusters projected on the factorial plan.*

Class 1 (Communication errors) refers to the 9-line recap CT for which an increase of these kinds of errors is observed. This task, particularly long, has thus been the cause of many communication errors. Class 2 (Grounding acts) included CTs for which an increase of clarification requests is observed. This class includes delicate manoeuvres (join up, threat, in/out check) and shooting (engagement) CTs on the one hand, and tasks for which the mission success or the pilots' survival are potentially threatened (e.g., self-protection) on the other. Another noteworthy result is that, except for the join up CT, all the tasks belonging to this class correspond to the tasks having triggered an increase in both heart rate and pupil size (Class 1 of the individual-workload level). Class 3 (Hesitations) is explained by an increase of the number of pilot hesitations. The most noticeably landing CT belongs to this class. Class 4 includes the tasks characterized by a significant decrease of hesitation, report situation, and repeat information.

The K-means analysis failed to find the same classes as those obtained during the HAC (except for Class 1). Finally, a Kruskal-Wallis ANOVA performed for each of the six communication-based indicators as the fixed factor class revealed a significant effect of the class on hesitation with H (4, N = 2301) = 13.76 p <0.01, non-procedural silences with H (4, N = 2301) = 42.68 p <0.001, and simultaneous speech H (4, N = 2301) = 39,004 p <0.001. However, no significant difference between classes was shown from post-hoc test of multiple comparisons by means rank.

<u>Analysis of the classifications of collective tasks.</u> The results of classification, summarized in Table 3, show the dichotomy of CTs between operational flight management and high tactical fight management. This classification matches therefore the characterization of CTs of the SME (see Table 12) that distinguishes two categories of tasks according to the status level of threat.

*Table 13: Individual and team characterization of collective tasks*

| TASK CATEGORIES | COLLECTIVE TASKS | INDIVIDUAL MENTAL WORKLOAD LEVEL | TEAM COMMUNICATION CHARACTERISTICS |
|---|---|---|---|
| Operational flight management | Take-off | Medium | |
| | Fence-in | Medium | |
| | Join up | Medium | Grounding acts |
| | Visual/Blind report | Medium | |
| | Weapon status check | Medium | |
| | Airfield weather and status report | Medium | |
| | Handover | Medium | |
| | Radio frequency change | Medium | |
| | Basic flying | Medium | |
| | Formation change | Medium | |
| | Mission management | Medium | |
| | In/Out check | Medium | |
| | Oil check | Medium | |
| | 9-line recap | Medium | Communication errors |
| | Landing | Low | Hesitations |
| High tactical fight management | Engagement | High | Grounding acts |
| | Monitoring | High | |
| | Commit | Medium | |
| | Threat | High | Grounding acts |
| | Interception | High | Grounding acts |
| | Target | High | Grounding acts |
| | Situation awareness report | High | Grounding acts |
| | Contact radar faded report | High | |
| | Delouse | High | |
| | Self-protection | High | Grounding acts |
| | System breakdown management | Medium | Hesitations |

The results indicate two main findings characterizing the CTs. In comparison with operational flight management CTs, the CTs related to high tactical fight management trigger off (1) an increase of the patrol leader's mental workload at individual level, very often categorized as involving a high level of mental workload, and (2) a difficulty to maintain the patrol common ground at the team level, especially observed with grounding acts densification.

Individual characteristics of collective tasks. At the individual-workload level, results showed that pupil diameter and heart rate tend to increase during the CTs related to tactical phases (interception, engagement, self-protection, etc.) compared to more routine tasks of collective work such as airfield weather management or radio frequency change. This implies an increase in mental workload and then a potential deterioration of the collective work in the tactical phases. It seems therefore necessary to support collective work during these specific tasks for which the collective work could be weakened by an increased workload that could be detrimental to information acquisition and shared situation awareness. The routine tasks could be considered as highly automated and therefore very inexpensive for the pilot. Mouala et al. (2001) report that if the negative effects of a high workload on performance are well documented, a low workload could also affect performance because of the diminution of the pilots' vigilance. The decrease in pupil diameter observed for the landing task may indicate a state of fatigue. This last finding is at variance with the literature data that generally report an increase in mental workload (inter-alia reflected by a PD increase) during landing, which is one of the most dangerous flight segments (Kasarskis et al., 2001). However, the literature data are collected during either civil flights or military flights (with predefined scenarios) but not necessarily tactical flights or flights carried out in truly immersive conditions. The training flights used in this study were tactical military missions with undefined scenarios performed on a highly realistic simulator. Moreover, the flights under investigation were integrated into the pilots' training course and efforts to reach a higher rank. The challenges of training were therefore important. Thus, the landing task that marks the end of the mission (i.e., the end of both the combat critical period performed in a highly realistic simulated environment and the evaluation time) could reflect a state of fatigue for the lead pilot, expressed through physiological indicators. Results showed that collective work presents a higher risk of degradation during high tactical tasks, based on the finding of an increased individual workload. The observation of a fatigue state during the landing task, however, could also lead to degraded collective work during this flight task.

Team characteristics of collective tasks. At the team-communication level, results first showed that the communication procedures were generally followed. Communication errors occurred most noticeably during 9-line recap collective task. This is a long and tedious task during which the pilots must memorize and exchange information (about targets, weapons, etc.) given by the controller in one single communication. This step is particularly crucial for good team situation assessment and good common ground sharing about the mission and its objectives. Results showed that communication quality was degraded during this CT and could affect collective work. Otherwise, communication quality decreased during the landing CT, as shown by the increase of pilots' hesitations. This can be explained by the fatigue occurring during this final phase, also noticed in the individual-workload level assessment. According to Prinzo (2001), hesitation pauses could reflect cognitive demands related to a reduced workload (operators switching from an automatic and repetitive processes to more deliberate thought). This observation confirms that deteriorated collective work could occur during landing and when there is a low level of workload. Finally, the team-level analysis also showed an increase of the clarification requests (report situation and information repetition requests) during high tactical CTs, namely those for which mission success or survival is directly engaged. In other words, this observation confirms the presence of uncertainty in the team, which the pilots need to reduce to perform tactical tasks. It should be noted that the CTs characterized by an increase of clarification requests are also (except one) characterized by an increased workload. These tasks, crucial for mission success, can thus be classified as endangering collective work from the perspective of both induced individual workload and the quality of the team interactions.

### 6.2.3. Towards a coaching system for team dialogue

Overall, the application of the COWORK² method has revealed the critical tasks of patrol collective work, i.e., for which collective work could be weakened by a workload increase and/or a team interaction quality decrease. Furthermore, this classification was corroborated by the SME risk level categorization of CTs based on the threat status. Consequently, this method could be used to evaluate, in a posteriori or real-time way, the risk of collective work degradation. The application of the method showed the relevance to detect the CTs which present a risk of deterioration of cognitive activities at individual and team levels. Indeed, the critical CTs, identified with the COWORK² method, can potentially be detrimental to the mission success or the patrol survival, and should therefore require more vigilance. The combination of the two complementary analyses, related to workload and communication, provides a reliable multicriteria way for a risk-based classification of collective tasks.

Moreover, the statistical analysis of communication indicators used at team level in the experiment indicates two different indices of collective work risk degradation. On the one hand, the clarification requests could be interpreted as direct measures of collective work degradation, when pilots explicitly express that they face uncertainty and need to upgrade their situation representation with the model of their patrol partner. On the other hand, the communication errors and the hesitations could also represent indirect measures of collective work degradation. Indeed, the weakened information sharing could generate uncertainty and misunderstanding resulting in degraded team situation awareness. Both classes of indicators could reveal difficulties in the construction of mutual understanding.

As practical perspectives, the evaluation obtained from the proposed method could therefore help design a new collective work support system for human-human or hybrid patrols. This would enable a control entity (human operator or artificial intelligence) to maintain or strengthen collective work in the patrol. The identified "collective work endangered" CTs (obtained in Stage 3 of the method) and the real-time monitoring of physiological activity and communication (Stage 2) could provide dynamic indicators alerting pilots about the state of the collective work and helping them to reconfigure and improve their common ground (see Figure 59). This module could simply alert them (so that they react to prevent deleterious effects) or assist them in dialogue management, dynamic functions allocation and control interface adaptation.

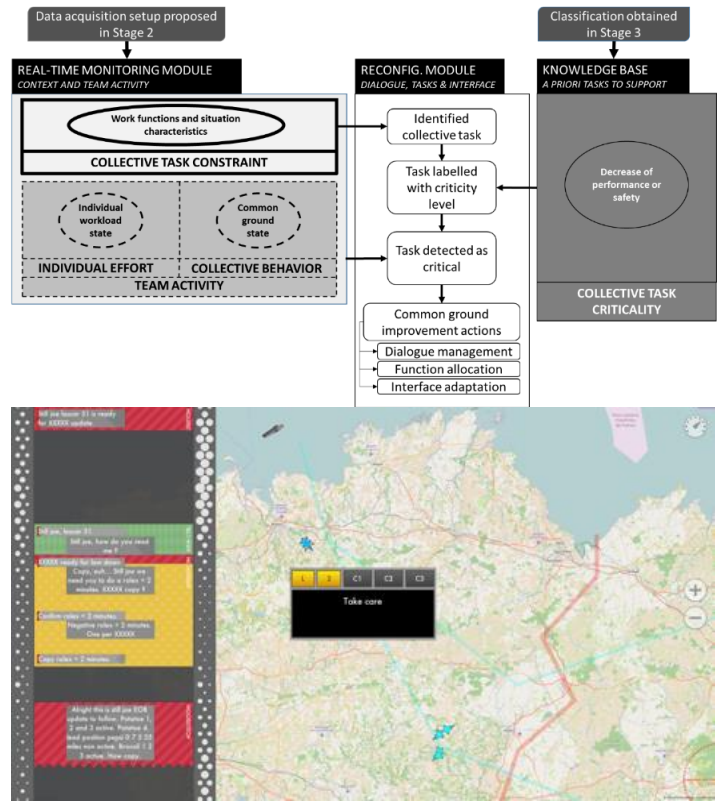

*Figure 59: Use of COWORK² for designing a system enhancing pilots' common ground*

In TAPAS project, a proof of concept was developed to illustrate this kind of team assistance, where the collective where pre-labeled according to the classification produced in Table 13, and where there is a real-time monitoring of each pilot workload as well as the intra-patrol communications as well as their communication errors. The entry in a given collective task and in a given situation (here situated according to a tactical map), or the gap between observed strain on the teams and a priori collective task stress, can trigger some system actions. For instance, system can request to repeat or clarify an information from a pilot to the other one.

## 6.3. DISCUSSION

In this chapter we sought to define and implement different types of adaptation mechanisms for HATs, in different configurations. We can especially distinguish:

- a "HMS configuration", where autonomous system and human are at the same level of Human Machine System, they are in the same team. Here, the autonomy plays the role of a teammate, and like in soccer, it can be a pivot player with a tactical contribution to the team, or it can carry out more operational actions.

- In a "SoS configuration" (System of Systems) where the autonomy is over and external to a team (as a soccer coach can be, advising team players about new strategies).

In the HMS configuration (§6.1), we designed and observed the effect of some adaptation mechanisms. Different findings can be highlighted here. Especially, there is a kind of authority paradigm shift in HAT, from (a) a situation where human operator is only at a tactical level, supervising and planning the actions that the autonomous system achieves at a lower level; to (b) a situation where autonomy can additionally monitor human performance, and may be substituted for the operator when this one is experiencing high or very level of workload. This authority shift corresponds also to a change in cooperation structure and form, since situation (a) is rather a vertical cooperation where the skills of human and AI are integrated (operator has the overview and plans the zone to be inspected, and the AI handles the operational drone swarm moves), whereas situation (b) is more a horizontal cooperation with a debative form (human supervise what AI is doing with drone swarm and can correct some actions, but meanwhile AI can also supervise human behaviour and cognitive effort to potentially take over in case of overload). Moreover, the proposed mechanisms allowed to practically demonstrate the use of mental workload, as an OFS based trigger for balanced automation and adaptive HMI.

In the SoS configuration (§6.2), we considered an autonomous system that can be over and external a human team. The autonomy was thought to support, advise, and guide the team. That results in a kind of {HT}AT, a Human Team-Autonomy Teaming. In this case, the monitoring of human operators becomes more complex, with multi-level aspects. It is necessary to integrate different indicators for the common ground, measured at individual level (e.g., mental workload of the leader) and at the collective level (communication errors). Finally, this projection of what a hybrid team could be in a fighter jet squadron showed that cooperation is not only a question of responsibility and authority. Indeed, an AI may also take a role of facilitator, using a "soft power" to minimally request team member to maintain the common ground. However, defining this minimal action of AI on the dialogue is by itself a challenge, since too much or not enough actions could have a counterproductive effect disturbing the team.

# *Discussion and perspectives*

# Scientific outlooks

*"Future is a direction, not a destination."* – **E. Catmull**

Human-Autonomy Teaming is an emerging paradigm. It of course inherits years of research and debate on humans and automation interactions, but it must also consider the singularity of autonomous systems and their raising power of action. Exponential progress in AI and robotics poses many new challenges, not only on technical aspects, but also on societal and ethical criteria.

**A dystopian vision at the antipodes of the human-autonomy relationship**

Job destruction. As Harari (2018) prophesies, "Machine learning and robotics will change virtually everything from making yogurt to teaching yoga." This transformation of the role of humans in the society is likely to increase as technological advances does not compete only with humans in the realm of physical capacities, but now in that of cognition. Indeed, the speed of calculation of algorithms, the capacity to process tons of data, as well as the properties of connectivity (networking) and actualization (considering new data in inference models) of AI, allow decisions to be made with a level of confidence and with a speed that human experience or intuition can no longer match. Certain cognitive activities have, for example, become very hermetic to humans: computers have made the financial system so complicated and transactions so fast that few humans can handle it.

The end of individual liberties. Human freedom is also a critical question with the raise of autonomous systems in our society. Gaspard Koenig (2018) thus points to the risk of the loss of the notion of individuals, since the authority of Big Data algorithms in the AI era could supplant the human authority promoted by the liberal philosophers of the Enlightenment. Indeed, with the invention of the biometric sensor (neurophysiological or behavioral data), machine learning can understand, or even hack, the desires, decisions, and opinions of humans (who often do not themselves very well). We could therefore be in a society where autonomy decides when the human is in or outside the control loop, and where the "nudge" (i.e., a subtle and incentive mechanism) guides the human in greedy exploitation of certain information, to the detriment of human exploration capacities.

The loss of human autonomy. Contentment and complacency phenomena may finally emerge, as the experience of using autonomous systems incite humans to trust them on an ever-increasing number of topics. For instance, there have been various incidents of drivers sinking into a lake, or falling from a demolished bridge, apparently following GPS instructions. Another cause of the loss of autonomy is the amount of information that humans must process. Whether in the operation of complex systems or in everyday life, we must absorb and process countless bits of data. As Harari (2018) points out, while we are overwhelmed by the mass of information, how can we find a sense in our activity or in our life, and how to situate our data bits in relation to the bits produced by billions of other humans and computers?

**Enhancing Human-Autonomy Teaming to avoid the dystopia**

This very dystopian vision is not the only possible future, however. Aside "Black Mirror" and "West World" perspectives exists also some successful examples of HAT, making the most of human and machine. For example, human-AI teams, known as "centaurs," outperform both men and computers at chess game. In this manuscript, I have also attempted to propose solutions that would better achieve the positive interdependence between human and machine, rather than moving towards sterile codependence. But the two axes on OFS monitoring and HAT dynamic adaptation, that we have developed in the previous chapters, are still to be investigated or even to be extended. Whether this about machine understanding of human, cooperative structure, training capacity, or human-machine communication, there is still a long way to go. We define in the following paragraphs 4 axes which seem interesting to study in the future (cf. Figure 60).

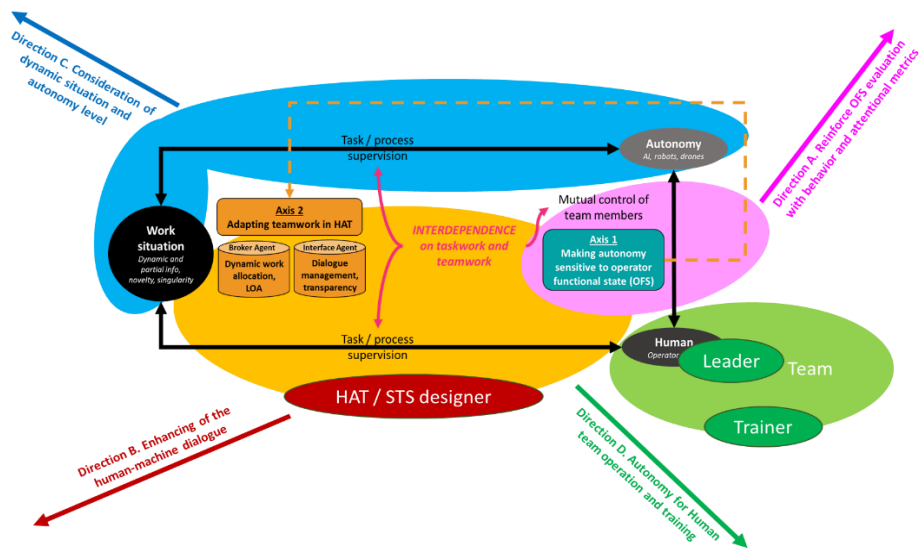

*Figure 60: 4 directions to further explore Human-Autonomy Teaming*

## DIRECTION A. A MULTILEVEL AND NEUROERGONOMICS APPROACH FOR OFS MONITORING

The contributions presented in chapters 3 and 4 focused on the development of tools and indicators to monitor the OFS. The concept of MWL was chosen as a first approximation, because it makes it possible to explore the integration of physio-behavioral measures, as well as the consideration of adaptation and regulation mechanisms (which we more particularly investigated with the notion of cognitive control). However, the issue of OFS monitoring could be further explored and refined, by adopting a multilevel and neuroergonomic approach.

**A1. Multilevel and temporal considerations.**

A first avenue would be to integrate different cognitive states and to take into account the temporal dimension in a more important way in the modeling of the OFS.

Hence, we could integrate the states of sustained attention and mental fatigue over the long term, or that of physiological stress during more sudden and intense events. This would allow to produce a cascadic classification of the OFS (see Figure 61), an approach that we recently sought to implement within the framework of the PEA MMT PRECOGS project [R10].

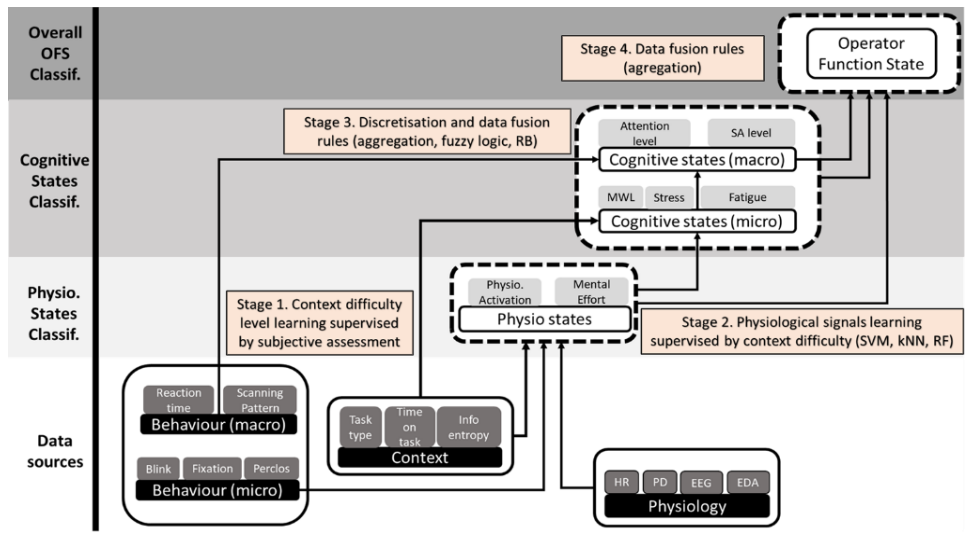

*Figure 61: A mutlilevel approach for OFS classification*

On the temporal aspect, it would also be interesting to see how the mental workload can be anticipated (cf. Hancock, 2017) when the changes in task demand are predictable (as on the experimental situation of TAPAS, where pilots know when they will enter a risky battle space).

### A2. Further integration of behavioural and neurophysiological metrics

A second avenue would be to improve the evaluation of the OFS by merging other indicators which can be very sensitive to the mental activity of the operator.

We can move towards the integration of much more behavioral metrics, for example related to the mechanisms of visual attention (on the detection of new stimuli, the visuospatial search patterns on an interface, the velocity of saccades, etc.), or towards the categorization of motor actions (for example to determine how risky a driver behavior is). I began to explore the use of this type of indicators, in particular through collaborations with Assaf Botzer ([C22]) and with Anna Ma-Wyatt (who leads the Active Vision Lab at the University of Adelaide) on attentional measures, and with AXA and Renault IRT System X for profiling risky driving behavior ([J10, J11]).

In addition, in line with recent work on neuroergonomics (Parasaruman, 2003), and with what some colleagues are currently exploring (in particular Julien Cegarra and Frédéric Dehais, members of the jury for the PhD defences of Alexandre Kostenko and Amine Laouar), the integration of neurological measurements related to the central nervous system (EEG or fNIRS) would be a plus to better discriminate cognitive states, compared to what we can already obtain from physiological measurements related to the peripheral nervous system (Pupillary diameter, HRV, EDA). I have already started to take this turn, especially during Amine Laouar's PhD, with a first use and exploitation of fNIRS signals.

**A3. Towards a finer-grained OFS classification**

Finally, it would be interesting to move towards a finer categorization of OFS, in order to better adapt the response to be provided when the operator is no longer functional.

Additional work should therefore be carried out on classification techniques, whether in supervised or unsupervised learning, and with more discriminating approaches. Following the TRM104 project [R7] that we briefly presented in section §3.3, we sought to improve the classification of mental effort using a neural network during the PRECOGS project [R11].

Moreover, to move towards a finer and more usable measurement of the OFS, a final avenue would be to combine the detection of different cognitive states with modes of cognitive control. We could thus imagine measuring the energetic (physiological activation) and behavioral variations (in particular the strategies implemented in terms of visual attention), then classifying these data in terms of modes of cognitive control, as we propose in a recent paper [C35].

# DIRECTION B. A CONSTRUCTIVE HUMAN-AUTONOMY DIALOGUE

Keeping humans in the loop of control becomes a crucial issue in HATs: we must avoid confining them to the role of the teammates who stay on the sidelines, or that of the passenger who sleeps in the back seat. For this purpose, we must fight against the complacency phenomenon ("I blindly trust autonomy because previous experiences have been conclusive"), and loss of SA ("I no longer understand the situation, so I trust autonomy through default"). Conversely, it is important to prevent humans from being too suspicious of machines and turn away from them.

In this quest for the right balance, autonomy must therefore be seen not as a means of comforting humans in their habits or of orienting - with good or bad intentions - their decisions, but rather as a kind of Jiminy Cricket that challenges humans. This would notably help to improve the sensemaking capacities of humans (Klein et al., 2007) by helping to detect potential discrepancies between the inference models they used and new singular data they processed. This would also make it possible to take better advantage of the dual-system models of reasoning (the automatic intuition of System 1 versus the controlled logic of System 2, cf. Kanheman, 2011), with the help of a machine capable of identifying our personal needs while escaping our cognitive illusions.

## B1. Informing human operator about their OFS and their limits

A first cognitive illusion in humans could be knowledge of oneself and one's own limits. By equipping operators with physiological sensors and by classifying in real time their level of mental workload, their level of fatigue or the cognitive control mode they instantiate, we can help these operators to better detect possible faults, and to trigger appropriate countermeasures.

However, communicating this new information is a sensitive issue. What the AI "mirror" shows is sometimes difficult for humans to accept: they may think that the "mirror" is distorted ("the AI tells me that I am overloaded, but I think that making more effort will pay off to overcome the present demand"), or have difficulty facing reality (« machine said I am not functional… How dares it? »). Furthermore, providing reflective feedback to humans on their own state could help better understand certain decisions and actions of the autonomous system (« IA takes this task over control whereas I have the skills to do it! … Ah, ok, this is because I am too busy on another task or overloaded by the current activity »).

We must therefore find the right mode of communication to adapt to the individual and the operational context. Should we choose explicit or subliminal communication? Should we try to convince humans with purely rational arguments, or rely on more emotional and relational dimensions to get the message across? What etiquette or politeness (Miller, 2005) should the autonomy adopt to announce unpleasant news to humans, and with which role (Jamieson, 2014; Flemisch et al., 2003) the system should exert the control on humans: butler, coach, life partner, horseback rider?

## B2. Agent Transparency for trust calibration and repair

Another essential aspect is the transparency of the autonomous agent, on its perception and understanding of the situation, as well as on the actions and intentions it may have. This notion of transparency, which we presented in paragraph §2.3.3., is a key element in creating an effective human-autonomy team, to establish a minimum of trust and common reference between human and machine. Several questions can be considered here.

First, we can think about the dimensions on which the autonomous agent must be transparent. As suggested by Lyons (2013) and then Chen et al. (2016), the adjustment of the level of transparency must not only relate to the "Robot-TO-Human" models (how the system explains to humans what it is doing), but also to the "Robot-OF-Human" (how the system communicates on what it perceives of the state and actions of the human operator or the team, in connection with point B1). Based on recent works we have been doing in ANR Humanism project, it would be interesting to see how the transparency settings on these 2 dimensions can be combined, and whether there are beneficial interaction effects or not on the performance of the HAT and on the trust-in-robot. (cf. Figure 62). This could help designers to calibrate the level of operator's trust in the system.

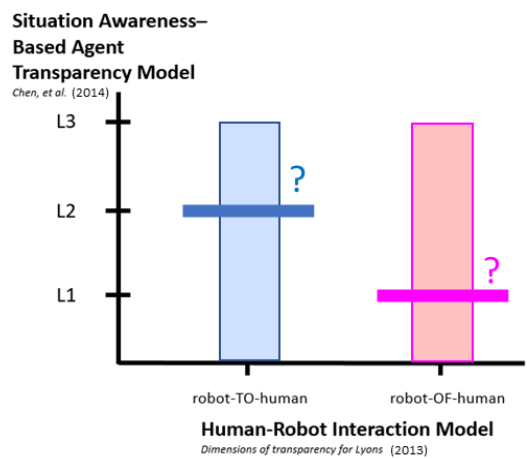

*Figure 62: What is the best combination of robot-TO and robot-OF-human transparency?*

Moreover, transparency is also a point to be explored to repair or tune human trust in the autonomous agent. Phenomena of over-trust or under-trust can indeed appear, either because of past experiences with AI or the robot, or because of the consequences that this can have of trusting or not trusting the robot. Transparency could therefore be seen as a dynamic and online adjustment which would allow to get out of overtrust or mistrust in the machine (for example due to irrational decision biases linked to loss aversion and risk aversion, cf. Prospect Theory Tversky & Kahneman, 1992, cf. top of Figure 63). Let suppose an autonomous system which communicates about the action it is taking, with a transparency level SAT1 (according to the transparency levels taxonomy of Chen et al., 2014). We could then imagine that the addition of transparency on information related loss or gain values could help correctly re-anchor human decision to trust the autonomy (SAT3a, bottom of Figure 63). We could also work on how we could improve the perception of the probability that autonomy is right or not in the information it is sharing (by adding qualitative rationales explaining the robot's reasoning or a probabilistic projection of the results of the robot's action, see respectively SAT2 and SAT3b on bottom of Figure 63).

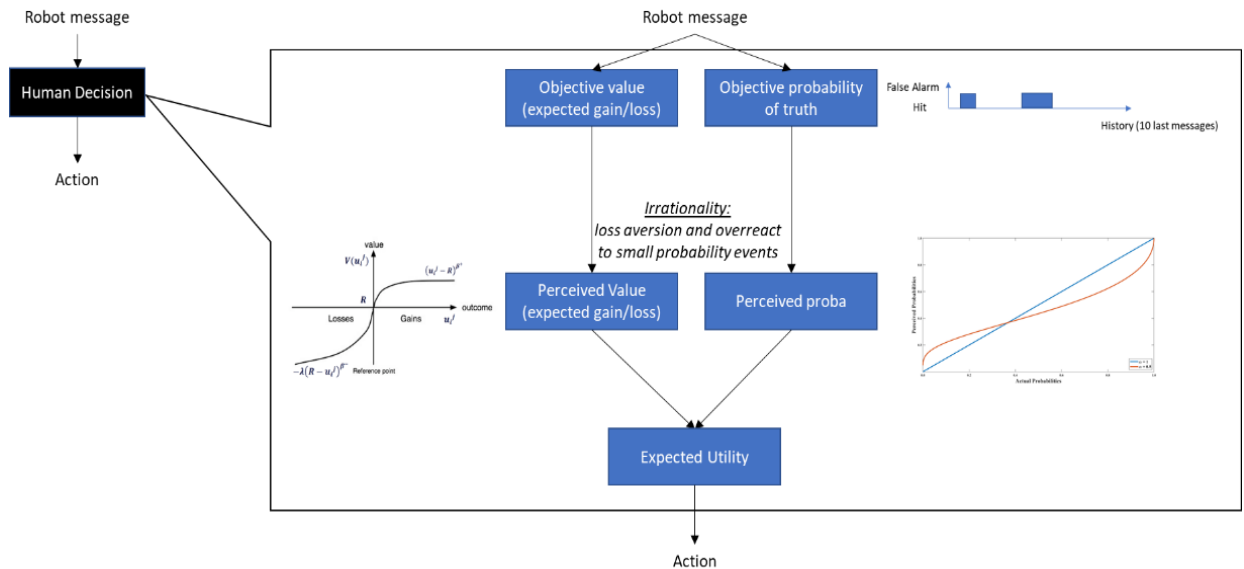

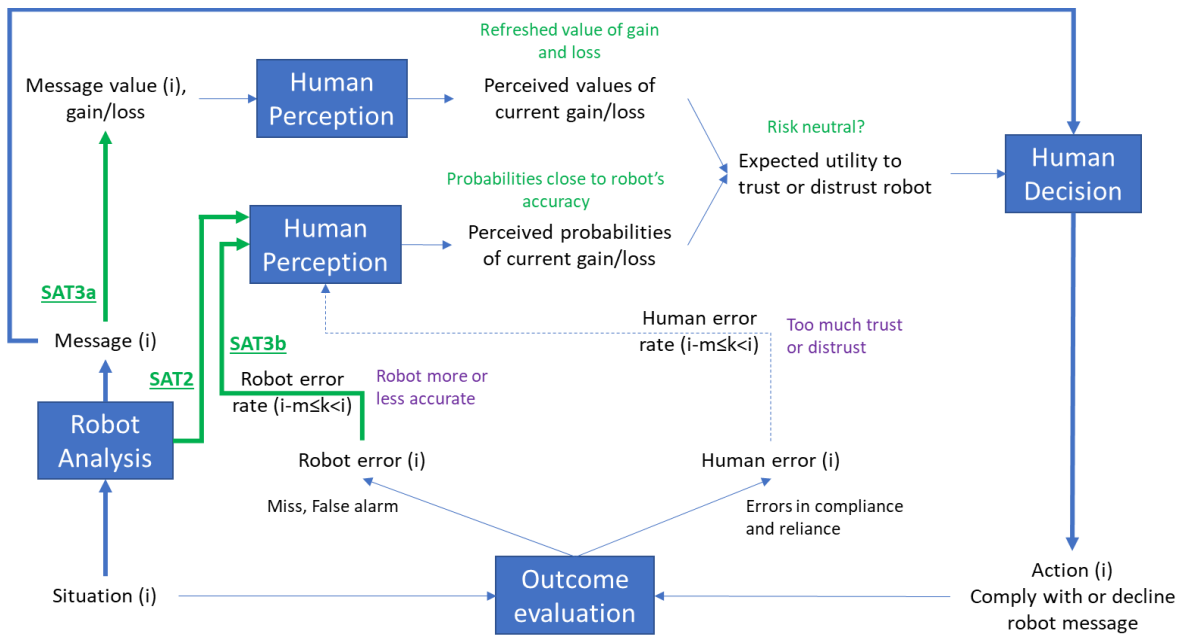

*Figure 63: Agent transparency as a mechanism for trust repair and irrationality correction?*

## DIRECTION C. ABOUT TRANSITION AND LONGITUDINAL RELATIONSHIPS IN HAT

### C1. Detecting system mode/state awareness in transition to fight OOTL syndrome

If humans can change their mode of cognitive control (cf. Chapter 4), autonomous system can also adopt different behaviour according to the situation. It is what we have recently explored in a collaboration with Renault System X ([J10], [J11]), studying the transitions between automated driving and manual driving. These transitions in mode may not only cause some difficulty in recovering a sufficient situation awareness and keeping human in the loop (of the road scene in the case of car driving), and it sometimes also results in loss of mode or state awareness. This problem with mode or state awareness and OOTL syndrome (Out-Of-The-Loop) is well known in aviation (that can explain automation surprise, Sarter, 1997) and it is also involved in maritime accidents (especially during the manoeuvres when there is a switch between two control boxes).

It could be interesting to study these phenomena by considering transitions over long period of activity. It is indeed difficult to sustain high attention over a long period of time, which sometimes results in a tendency of the human brain to disconnect from the task at hand and to focus on more personal matters (Mind Wandering effect). We could look at how time on task changes the way in which the human operators perceive changes in the situation or in the supervised system, and how and how fast they regain control in the case of transitions in system modes.

Following recent studies (Hartwich et al., 2018), another key aspect to be addressed could also be the discomfort caused by the gaps between the human cognitive control preferences (like driving style in cars), and the dynamic behaviour implemented by autonomous system. We have started to explore this question through the master internship of Mathieu Legendre.

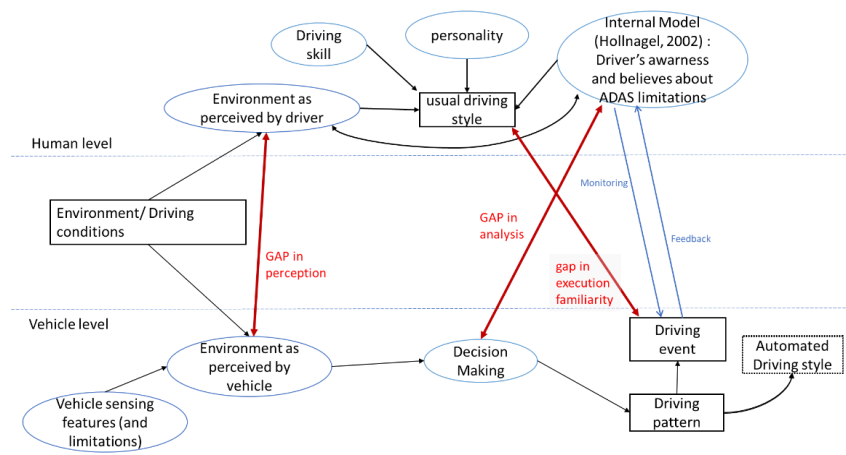

*Figure 64: Discomfort and gaps between human driving style and system control mode*

**C2. Longitudinal considerations: Meta-cooperation and long-term trust**

In the longer term, the question of meta-cooperation (Hoc, 2001) between human and autonomy remains a major challenge in the HAT paradigm.

Building a common frame of reference and instilling mutual trust is often a long process when it comes to the relationship between two humans. This is based on experiences of cooperative activities where each individual observes and learns from the other (and sometimes also, as Harari (2016) points out, on the ability of humans to gossip so as to assess the reliability of a member from the community!). However, trust and meta-cooperation between humans are based on a certain stability and a certain inertia of individual knowledge and skills. How can we imagine the evolution of the relationship between human and autonomy, with an autonomous system that learns at a crazy speed, and whose performance and skills can change in a very short time?

For Harari (2018), this relationship will be a standoff between man and machine, rather than a lifelong partnership in which hierarchies and routines can form and last for decades. In other words, the symbiosis would then only be an unstable equilibrium, and not a global optimum that could be reached. For example, the chess centaur teams progress through a process of cross learning, where the human challenges the machine (by giving him new objectives or by guiding him with his know-how), and where the machine challenges the human (using counterfactual reasoning to explore new avenues).

We should therefore be able to study in a longitudinal way how trust and meta-cooperation are created in such teams, in particular by looking at the effect of the differences in learning dynamics between humans and machines. Can the autonomous system help humans to move from the novice to the expert? How to adapt the assistance of the autonomous agent accordingly, as a function of human progress? How to get a kind of mutual sensemaking, where the actions of each member of the team can be seen as a new "data" to question, confirm or break the "frame" used for the inferences performed by the human or the machine?

# DIRECTION D. A SCALE-UP CHALLENGE: AUTONOMY FOR AND WITH HUMAN TEAMS

We mainly explored the question of HAT in the simplest configuration: a human and an autonomous agent, forming a human-machine system. In this case, the symbiosis results in a kind of "augmented operator" (Romero et al., 2018), where the monitoring of the human operator help adjust the level of assistance delivered by the autonomy. In the last chapter (§6.2), we also investigated another configuration, where autonomy plays the role of a mediator between two humans. However, it remains much work to do to answer to this scale-up challenge, by considering all the complexity and the emergent dynamics of team level.

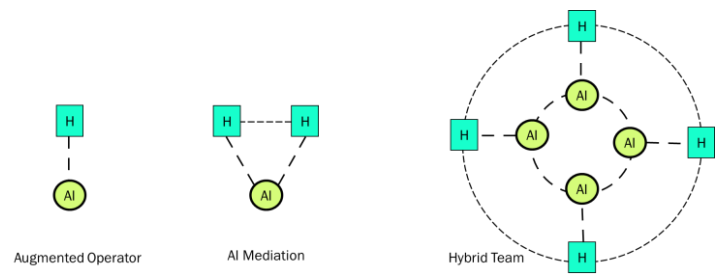

*Figure 65: Team configurations to be explored*

### D1. From individual OFS monitoring to team monitoring

A first challenge is to study the relationship between individual performance and teamwork. This is especially a crucial question to understand how interindividual differences in terms of mental workload or mental fatigue, may influence team cognitive processes, like leadership, social support, or mutual performance monitoring (cf. Salas et al., 2005). Modelling the relationship between OFS and team performance would further help predict degradation of some part of collective work and help to advise or assist the team in recovering some resilience mechanisms. For instance, we recently observed in a study on a cyber team [J9] that a high workload resulted in less emotional or appreciative support, but that the technical support does not change.

### D2. Connecting operations, training, and design: OFS monitoring as a service?

Finally, it would be interesting to think OFS monitoring as a service, that could be helpful for different kind of users. We have already tested how OFS can provide feedback and alert to individuals about their own states and how it helps reconfigure the human-machine interaction (at the operational level). But we could also imagine how OFS monitoring could be used by a team leader to reinforce crew resource management (again at the operational level), how it could help for improving operators' training with the aid of data collected during operational situation (at the organizational level), and finally how system designers could use the collected OFS data to redesign the workspace and the work distribution (at the system level).

# A roadmap between Lorient and Adelaide

*"It's a long way to the top if you wanna rock 'n' roll"* — **AC/DC (CROSSING motto** 😊**)**

*"Coming together is a beginning. Keeping together is progress. Working together is success."* - **Henry Ford**

To support these research directions, I could rely on different ecosystems and networks that I should strengthen in the coming years. Figure 66 shows that the research directions have been starting from previous projects, and they are supported with new and ongoing collaborations.

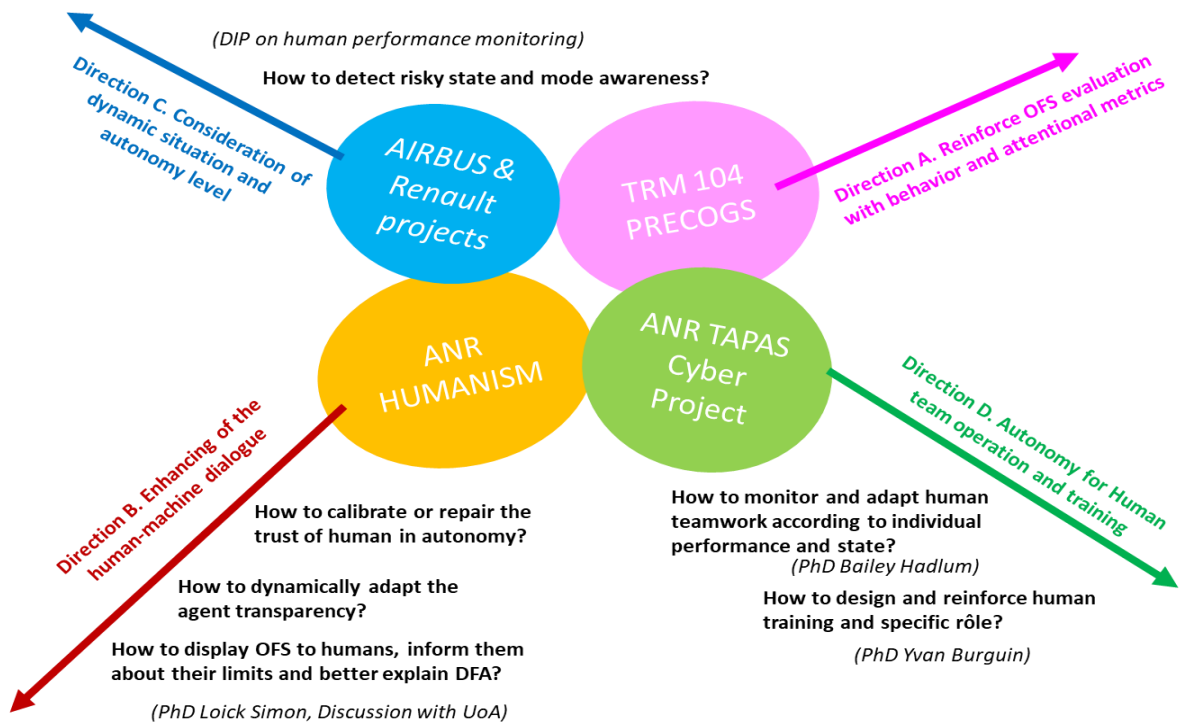

*Figure 66: Collaborative roadmap for supporting the research directions*

## SCIENTIFIC INITIATIVES IN FRANCE

**Perspectives within Lab-STICC and FHOOX team.** The Lab-STICCis a research unit historically recognized in Brittany and in France in the field of ICTS (Information & Communication Technology and Science). The Lab-STICC's motto "From sensors to knowledge: communicating and deciding" initially underlined the intention to give "meaning" to the "sign", in terms of interpretations and increased value with respect to treatments based on a signal in all its forms, from lower to upper layers (up to Human-Complex Systems cooperation). The upper layer (left tip of the Lab-STICC triangle, cf. Figure 67) is especially addressed by SHM Department (or HMS in English for Human Machine System), devoted to the modeling, analysis and design of hybrid systems gathering human users (individuals or teams) and adaptive artefacts. In this department, FHOOX is a multidisciplinary team whose I took over the scientific management in 2020. Composed of 7 permanent researchers in cognitive ergonomics, social psychology, and automatics, it intends to optimize human-system cooperation in complex systems, such as transportation systems (vessels, aircraft, autonomous vehicles), smart factories and crisis management system, by focusing on the modelling and the monitoring of cognitive processes and operator states, as well as by proposing new mechanisms of adaptation at individual, collective and organizational levels.

These axes are noticeably well aligned with the Innovation and Research Strategy of Brittany Region (S3), where the integration of human appears as crucial for maritime and industrial economies (smart factory and cyber crisis), and that could open positive opportunities.

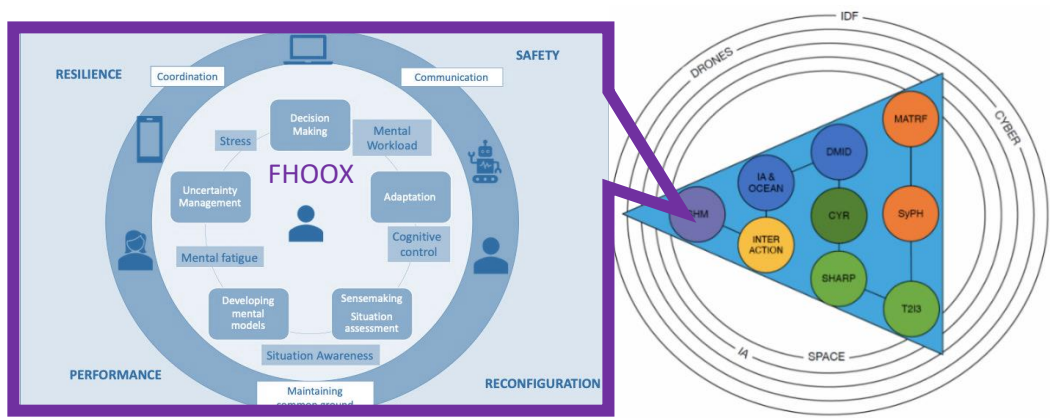

*Figure 67: FHOOX team and Lab-STICC organization*

Moreover, this institutional framework also allows the synergy with other research teams and departments. In 2021, we will especially start a PhD with IRIS team, from CyR department, to explore the directions A and D that I have previously defined.

<u>Yvan Burguin's PhD (Started in October 2021):</u> This PhD topic focuses on improving and customizing cyberdefense training. This involves identifying the tasks and work situations that generate stress or mental workload on cybersecurity operators, based on machine learning of physiological data. This identification will then be used to automatically generate training scenarios adapted to the operators. It would be a means to study how we can extend and integrate the contributions developed in Chapters 3 and 6.

**Opportunities for building a research platform.** To support the research directions, as well as the activities of FHOOX team, we could also develop and enrich an experimental platform or a network of platforms to evaluate human-machine cooperation and human performance monitoring. To this purpose, FHOOX team will contribute to the CPER program about Smart Factories between 2021 and 2027 (National/regional funding of about 100k€). In this global program (more than 2M€ of funding overall), we especially aim at developing new capabilities to study human factors in realistic and naturalistic settings, and by also joining them to the existing assets from Lab-STICC (like the SCAP platform in Lorient developed by Eric Martin, and the Loustic-Brest platform develop by Gilles Coppin, on which we have already made different studies).

**Industrial and academic partnerships in France.** Future works could also rely on strong relationships I have established in France, and that I would like to continue to strengthen.

On the application domain of Industry of the Future, I look for continuing collaborative works, following the ANR Humanism Project (2018-2021) we carried out with colleagues from Valenciennes (LAMIH) and Reims (CRESTIC). We applied in 2021 for a new ANR project (RODIC). We have also been participating to project SEANATIC since June 2020, funded by AMI ADEME and gathering a local consortium in Brittany (Piriou, Thalos, IoT.bzh, Azimut).

<u>Loïck Simon's PhD (started in June 2020):</u> In this project, we are working with Azimut on the definition of new interfaces that can help maintenance officer in vessels to understand and assess the benefits and the risks provided by a maritime predictive maintenance solution, by further exploring the questions of ecological interface design and transparency (links with direction B).

On the application domain of transportation systems, I will continue to strengthen the existing partnerships with Renault, Airbus, Dassault Aviation and Thales, following the recent projects with us (Amine Laouar's PhD, TRM104 and PRECOGS projects). We have also a privileged relation with ENSM (French Merchant Navy School in Le Havre) that could help us to study decision making and human-machine cooperation with end-users and realistic simulators.

Finally, scientific interactions and new projects could also emerge from scientific networks: I especially participated to the workgroup on Human Machine Systems and Automatics of the GDR MACS and the GIS GRAISyHM.

## SCIENTIFIC INITIATIVES ABROAD

The "Australian Experience" in 2021, as well as some previous and starting collaborations with colleagues in other countries, will also help to support my scientific outlooks.

Hence, I hope to continue to work with my Israel colleague Assaf Botzer (he recently shared experimental data on a study about vigilance in driving, and we started to analyze them). Some other actions could be also reinforced, at the European level, following some discussions we had with Aschaffenburg to build a European project on human-cooperation in autonomous car.

**Continuation of the collaboration with CROSSING, UoA and UniSA in Adelaide**

Within the framework of IRL CROSSING, very focused on the question of Human-Autonomy Teaming, a promising collaboration have started with UniSA on research direction D.

Bailey Hadlum PhD (started in July 2021): Based on previous collaborative work between UniSA and Lab-STICC (cf. [34]), this IRL Crossing PhD aims at combining methods and expertise from Naval Group on naval operators and submariners' activities and environment, from Lab-STICC on operator states and team performance, and from UniSA on sleep and circadian rhythms. The collaboration will allow for innovative and enhanced ways of quantifying operator and team fatigue states, whilst taking time of day and fatigue into account. Especially, it is expected to model and experimentally evaluate the effect of fatigue on different dimensions of Team Performance (cohesion, communication loops typology and quality, supportive behaviour), in line with previous studies exploring the influence of various factors on the quality of Team Processes (Salas et al., 2005). To this purpose, a laboratory-based sleep deprivation study at UniSA will be conducted, by using the Sleep and Chronobiology laboratory of Pr. Siobhan Banks.

Moreover, this year in Australia is also an opportunity to build new projects with colleagues from UoA. I have applied for a DIP project (Defense Innovation Partnership in South Australia) led by Pr. Anna Ma-Wyatt and in collaboration with Naval Group, on the question of human performance monitoring in transitional situations (especially to study the awareness of system modes/states, cf. Direction C). I have also started some workshops with Pr. Langford White and Pr. Anna Ma-Wyatt, to explore the question of transparency control as a mechanism for trust adjustment and repair (cf. direction B). All these actions constitute a good basis to support a long-term collaboration, and to come back regularly in Adelaide!

# *References*

# Table of figures

# Table of tables

# Bibliography

Abbass, H. A., Scholz, J., & Reid, D. J. (2018). Foundations of trusted autonomy. *Springer Nature.*

Adam, M., Cardin, O., Berruet, P., and Castagna, P. (2012). Data processing from manufacturing systems to decision support systems: propositions of alternative design approaches. *IFAC Proceedings Volumes*, 45(6), 1129–1134.

Akash, K., McMahon, G., Reid, T., & Jain, N. (2020). Human Trust-based Feedback Control: Dynamically varying automation transparency to optimize human-machine interactions. *IEEE Control Systems Magazine, 40(6), 98-116.*

ANSI/ISA-5.1 (1992). 5.1 instrumentation symbols and identification.

Arad, B.A. (1964). Note on the measurement of control load and sector design in the enroute environment, *Federal aviation administration Washington DC systems research and development service.*

Aranbarri Artetxe, J., Da Cunha, C., Vergara, V., & Rauffet, P. (2009). Vertical integration: modeling and evaluation, *13th IFAC Symposium on Information Control Problems in Manufacturing (INCOM 2009)*, Moscow (Russia).

Aranbarri Artetxe, J., Da Cunha, C., Vergara, V., & Rauffet, P. (2009). Vertical integration: modeling and evaluation, *13th IFAC Symposium on Information Control Problems in Manufacturing (INCOM 2009)*, Moscow (Russia).

Arroyo, E., Hoernicke, M., Rodr´ıguez, P., and Fay, A. (2016). Automatic derivation of qualitative plant simulation models from legacy piping and instrumentation diagrams. *Computers & Chemical Engineering*, 92, 112– 132.

Atoui, M. A., Cohen, A., Rauffet, P., & Berruet P. (2019). Fault Diagnosis by Bayesian Network Classifiers with a Distance Rejection Criterion. In 16th International Conference on Informatics in Control, Automation and Robotics, ICINCO (pp. 463-468).

Atoui, M. A., Cohen, A., Rauffet, P., & Berruet P. (2019). Fault Diagnosis by Bayesian Network Classifiers with a Distance Rejection Criterion. In 16th International Conference on Informatics in Control, Automation and Robotics, ICINCO (pp. 463-468).

Bainbridge, L. (1983). Ironies of automation. In *Analysis, design, and evaluation of man–machine systems (pp. 129-135).* Pergamon.

Barth, M. and Fay, A. (2013). Automated generation of simulation models for control code tests. *Control Engineering Practice*, 21(2), 218–230.

Bates, D.M., Maechler, M., Bolker, B., (2012). lme4: Linear mixed effects using S4 classes. *R package* version 0.999999

Bauer, R., & Gharabaghi, A. (2015). Estimating cognitive load during self-regulation of brain activity and neurofeedback with therapeutic brain-computer interfaces. *Frontiers in behavioral neuroscience, 9, 21.*

Baxter, G. & Sommerville, I. (2011). Socio-technical systems: From design methods to systems engineering. *Interacting with Computers, 23, 4-17.*

Beatty, J. (1982). Task-Evoked Pupillary Responses, Processing Load, and the Structure of Processing Resources. *Psychological Bulletin, 91(2), 276-292.*

Bergasa, L.M., Nuevo, J., Sotelo, M.A., Barea, R., & Lopez, M.E. (2006). Real-time system for monitoring driver vigilance. *IEEE Trans. Intell. Transport. Syst., 7, 63–77, 2006*

---

Bernard, A., Rauffet, P., & Da Cunha, C. (2011). Organizational capability management for improving performance of global production networks, *44th CIRP Conference on Manufacturing Systems*, madison (USA).

Bernard, A., Rauffet, P., & Da Cunha, C. (2011). Organizational capability management for improving performance of global production networks, *44th CIRP Conference on Manufacturing Systems*, madison (USA).

Berruet, P. (1998). Contribution au recouvrement des SFPM: analyse de la tolérance et reconfiguration. *PhD dissertation. Ecole Centrale Lille.*

Bevan, R., Berruet, P., de Lamotte, F., Adam, M., Cardin, O., and Castagna, P. (2012). Generation of multiplatform control for transitic systems using a componentbased approach. In *Emerging Technologies & Factory Automation (ETFA), 2012 IEEE 17th Conference on*, 1–8. IEEE.

Bezivin, J. (2005). On the unification power of models. *Software & Systems Modeling*, 4(2), 171–188.

Bhaskara, A. Skinner, M., & et Loft, S. (2020). Agent Transparency: A Review of Current Theory and Evidence, *IEEE Transactions on Human-Machine Systems, vol. 50, no 3, p. 215-224*, doi: 10.1109/THMS.2020.2965529.

Bignon, A. (2012). *G´en´eration conjointe de commandes et d'interfaces de supervision pour syst`emes sociotechniques reconfigurables*. Ph.D. thesis, Universit´e de Bretagne Sud.

Bignon, A., Rossi, A., and Berruet, P. (2013). An integrated design flow for the joint generation of control and interfaces from a business model. *Computers in industry*, 64(6), 634–649.

Boomer, D. S. (1965). Hesitation and grammatical encoding. *Language and Speech*, 8(148).

Borghini, G., Aricò, P., Di Flumeri, G., Cartocci, G., Colosimo, A., Bonelli, S., ... & Pozzi, S., (2017). EEG-based cognitive control behaviour assessment: an ecological study with professional air traffic controllers. *Scientific reports, 7(1), 1-16.*

Boy, G. A. (2014). Requirements for Single Pilot Operations in Commercial Aviation: A First High-Level Cognitive Function Analysis. *In CSDM (Posters)* (pp. 227-234).

Braver, T. S., (2012). The variable nature of cognitive control: a dual mechanisms framework. *Trends in cognitive sciences, 16(2), 106-113.*

Cannon-Bowers, J. A., & Salas, E. (1997). A framework for developing team performance measures in training. In *Team performance assessment and measurement: Theory, methods, and applications* (pp. 45-62). Mahwah, NJ: LEA.

Causse, M., Chua, Z. K., & Rémy, F., (2019). Influences of age, mental workload, and flight experience on cognitive performance and prefrontal activity in private pilots: a fNIRS study. *Scientific reports, 9(1), 1-12.*

Causse, M., Chua, Z., Peysakhovich, V., Del Campo, N., & Matton, N., (2017). Mental workload and neural efficiency quantified in the prefrontal cortex using fNIRS. *Scientific reports, 7(1), 1-15.*

Cegarra, J. and Chevalier, A. (2008). The use of Tholos software for combining measures of mental workload: Toward theoretical and methodological improvements. Behavior Research Methods, 40(4), 988–1000.

Cegarra, J., Baracat, B., Calmettes, C., Matton, N., Capa R.L., (2017). A neuroergonomics perspective on mental workload predictions in Jens Rasmussen's SRK framework, *Le travail humain, 80, 7-22.*

Challenger, R., Clegg, C. W., & Shepherd, C. (2013). Function allocation in complex systems: reframing an old problem. *Ergonomics, 56:7, 1051-1069.*

Chapanis, A. (1965). Words, words, words. *Human Factors, 7, 1-17.*

Chauvin, C., & Hoc, J. M. (2014). Integration of Ergonomics in the Design of Human–Machine Systems. in P. Millot (Ed.) *Designing Human-Machine Cooperation Systems, 43-86.*

Chauvin, C., Coppin, G., Rauffet, P. (2019). From the characterization of the operator cognitive states to the design of assistance system. *1st German French Workshop on New paradigms for Autonomous Driving*. TUM, Munich (Germany)

Chauvin, C., Coppin, G., Rauffet, P. (2019). From the characterization of the operator cognitive states to the design of assistance system. *1st German French Workshop on New paradigms for Autonomous Driving*. TUM, Munich (Germany)

Chauvin, C., Rauffet, P., Guennal, L., & Clostermann, J.-P. (2017). AERONAV: A new concept for small HSC Implications for bridge equipment and decision support, FAST - *International Conference on Fast Sea Transportation*, September 2017, Nantes (France).

Chauvin, C., Rauffet, P., Guennal, L., & Clostermann, J.-P. (2017). AERONAV: A new concept for small HSC Implications for bridge equipment and decision support, FAST - *International Conference on Fast Sea Transportation*, September 2017, Nantes (France).

Chauvin, C., Rauffet, P., Tréhin, M., Berruet, P., & Lassalle, J. (2015). Using Cognitive Work Analysis to design smart grid interfaces, Human Factors & User Experience in everyday life, medicine, and work, October 2015, *HFES European Chapter Conference,* Groningen (The Netherlands).

Chauvin, C., Rauffet, P., Tréhin, M., Berruet, P., & Lassalle, J. (2015). Using Cognitive Work Analysis to design smart grid interfaces, Human Factors & User Experience in everyday life, medicine, and work, October 2015, *HFES European Chapter Conference,* Groningen (The Netherlands).

Chauvin, C., Said, F., Langlois, S., (2019). Does the Type of Visualization Influence the Mode of Cognitive Control in a Dynamic System? in *International Conference on Intelligent Human Systems Integration, (pp. 751-757).* Springer, Cham.

Chauvin, C., Saïd, F., Rauffet, P. (2017). Intérêt de la réalité augmentée pour la reprise en main d'un véhicule fortementautomatisé. *Research report in collaboration with Renault IRT System X*

Chauvin, C., Saïd, F., Rauffet, P. (2017). Intérêt de la réalité augmentée pour la reprise en main d'un véhicule fortementautomatisé. *Research report in collaboration with Renault IRT System X*

Chauvin, C., Said, F., Rauffet, P., Langlois, S. (2020). Analyzing the take-over process in a highly automated vehicle: From drivers' behaviour to cognitive control modes. *Le Travail Humain, 83(4), 379-405.* https://doi.org/10.3917/th.834.0379

Chauvin, C., Said, F., Rauffet, P., Langlois, S. (2020). Analyzing the take-over process in a highly automated vehicle: From drivers' behaviour to cognitive control modes. *Le Travail Humain, 83(4), 379-405.* https://doi.org/10.3917/th.834.0379

Chauvin, C., Thiébaut-Rizzoni, T., Clostermann, J.P., Rauffet, P. (2021). Teamwork in Pilotage operations: creating and maintaining a COFOR between pilots and masters. *Le Travail Humain*

Chauvin, C., Thiébaut-Rizzoni, T., Clostermann, J.P., Rauffet, P. (2021, submitted in January). Teamwork in Pilotage operations: creating and maintaining a COFOR between pilots and masters. *Le Travail Humain*

Chen, J. Y., Barnes, M. J., Selkowitz, A. R., Stowers, K., Lakhmani, S. G., Kasdaglis, N. (2016). Human-autonomy teaming and agent transparency. *Companion Publication of the 21st International Conference on Intelligent User Interfaces (IUI) (pp. 28–31).*

Chen, J., Procci, K., Boyce, M., Wright, J., Garcia, A., & Barnes, M. (2014). Situation Awareness-Based Agent Transparency, *ARMY RESEARCH LAB ABERDEEN. ARL-TR-6905.*

Chipman, S. F., J. M. Schraagen, and V. L. Shalin. 2000. "Introduction to Cognitive Task Analysis." in *Cognitive Task Analysis*, edited by J. M. Schraagen, S. F. Chipman, and V. L. Shalin. Mahwah, NJ.

Clark, H.H., & Brennan, S.E. (1991). Grounding in Communication. In L. B. Resnick, J. M. Levine, S. D. Teasley (Eds.). *Perspectives on socially shared cognition* (pp. 127-149). Washington, DC, US: American Psychological Association.

Coppin, G. and Legras, F. (2012). Controlling Swarms of Unmanned Vehicles through User-Centered Commands. In *AAAI Fall Symposium: Human Control of Bioinspired Swarms.*

Coppin, G., Rauffet, P. Kostenko, A. (2017). Recommandations sur les modes de reconfiguration système / IHM. *Research report of project TRM104 in collaboration with Dassault Aviation*

Coppin, G., Rauffet, P. Kostenko, A. (2017). Recommandations sur les modes de reconfiguration système / IHM. *Research report of project TRM104 in collaboration with Dassault Aviation*

Corradini, P., & Cacciari, C. (2002). The effect of workload and workshift on air traffic control: a taxonomy of communicative problems. *Cognition, Technology & Work*, *4*(4), 229-239.

Cuny, X., & Chauvin, C. (2009). Decision-making in controlling development of driving/piloting situations. *Safety Science, 47(9), 1201-1204.*

D'Orazio, T., Leo, M., Guaragnella, C., & Distante, A. (2007). A visual approach for driver inattention detection. *Pattern Recog., 40, 2341–2355, 2007*

Daaboul, J., Xu, Y., Vergara, V., Le Duigou, J., Chevallereau, B., Rauffet, P., Laroche, F., Da Cunha, C., & Bernard, A. (2009). Amélioration de la performance industrielle par l'ingénierie numérique, *11ème Colloque National AIP-PRIMECA* La Plagne (France).

Daaboul, J., Xu, Y., Vergara, V., Le Duigou, J., Chevallereau, B., Rauffet, P., Laroche, F., Da Cunha, C., & Bernard, A. (2009). Amélioration de la performance industrielle par l'ingénierie numérique, *11ème Colloque National AIP-PRIMECA* La Plagne (France).

D'Angelo, M., Cervero, R., Durning, S., & Varpio, L. (2019). The teamwork model: Proposing a model for studying interprofessional healthcare teams. *MedEdPublish, 8.*

De Waard, D. and Studiecentrum, V. (1996). The measurement of drivers' mental workload. *Groningen University, Traffic Research Center Netherlands.*

Dearden, A., Harrison, M., & Wright, P. (2000). Allocation of function: scénarios, context and the economics of effort. *International Journal of Human-Computer Studies, 52, 289–318.*

Debernard, S. (1993). Contribution à la répartition dynamique de tâches entre opérateur et système automatisé : application au contrôle du trafic aérien. *PhD dissertation, Valenciennes*

Dehais, F., Causse, M., & Pastor, J. (2008). Embedded eye tracker in a real aircraft: new perspectives on pilot/aircraft interaction monitoring. In *3rd International Conference on Research in Air Transportation Proceedings*. Fairfax, USA: Federal Aviation Administration.

Deline, S., Guillet, L., Guerin, C., & Rauffet, P. (2017). Assessment of stress sources and moderators among analysts in a cyber-attacks simulation context, *HFES European Chapter Conference*, Roma (Italy).

Deline, S., Guillet, L., Guerin, C., & Rauffet, P. (2017). Assessment of stress sources and moderators among analysts in a cyber-attacks simulation context, *HFES European Chapter Conference*, Roma (Italy).

Deline, S., Guillet, L., Rauffet, P., & Guerin, C. (2016). Le stress dans un contexte de cyberdéfense : relations entre mesures subjectives et physiologiques, *Congrès Annuel de la Société Française de Psychologie*, Paris (France).

Deline, S., Guillet, L., Rauffet, P., & Guerin, C. (2016). Le stress dans un contexte de cyberdéfense : relations entre mesures subjectives et physiologiques, *Congrès Annuel de la Société Française de Psychologie*, Paris (France).

Deline, S., Guillet, L., Rauffet, P., Guérin, C. (2019). Team cognition in a cyber defense context: Focus on social support behaviors. *Cognition, Technology and Work*. https://doi.org/10.1007/s10111-019-00614-y

Deline, S., Guillet, L., Rauffet, P., Guérin, C. (2019). Team cognition in a cyber defense context: Focus on social support behaviors. *Cognition, Technology and Work*. https://doi.org/10.1007/s10111-019-00614-y

Demir, M., McNeese, N. J., Cooke, N. J. (2016). Team communication behaviors of the human-automation teaming [Conference session]. *IEEE International Multi-Disciplinary Conference on Cognitive Methods in Situation Awareness and Decision Support (CogSIMA) (pp. 28–34).*

Diallo, L.M., Kostenko, A., Rauffet, P., Moga, S., Coppin, G. (2021). PRECOGS : Identification et classification des états cognitifs à partir d'indicateurs neurophysiologiques. Application à la situation expérimentale MATB-II. *Research report of project PEA MMT Precogs*

Diallo, L.M., Kostenko, A., Rauffet, P., Moga, S., Coppin, G. (2021). PRECOGS : Identification et classification des états cognitifs à partir d'indicateurs neurophysiologiques. Application à la situation expérimentale MATB-II. *Research report of project PEA MMT Precogs*

Diederichs, F., Bischoff, S., Widlroither, H., Reilhac, P., Hottelart, K., & Moizard, J. (2015). Smartphone integration and SAE L3 car automation: a new cockpit concept and its evaluation in a car simulator. Der Fahrer im 21, 63-78.

DIN EN ISO 10075-1, (2018). Ergonomic principles related to mental workload—Part 1: general issues and concepts, terms and definitions *(ISO 10075-1:2017).*

Durantin, G., Gagnon, J. F., Tremblay, S., & Dehais, F., (2014). Using near infrared spectroscopy and heart rate variability to detect mental overload. *Behavioural brain research, 259, 16-23.*

Durkee, K. T., Pappada, S. M., Ortiz, A. E., Feeney, J. J., & Galster, S. M. (2015, March). System decision framework for augmenting human performance using real-time workload classifiers. *In IEEE International Multi-Disciplinary Conference on Cognitive Methods in Situation Awareness and Decision (pp. 8-13).*

Endsley, M. R. (1995). Toward a theory of situation awareness in dynamic systems. *Human Factors, 37*(1), 32-64.

Endsley, M. R. (2020). The divergence of objective and subjective situation awareness: A meta-analysis. *Journal of cognitive engineering and decision making, 14(1), 34-53.*

Endsley, M. R., & Kaber, D. B. (1999). Level of automation effects on performance, situation awareness and workload in a dynamic control task, *Ergonomics, 42, 462–492.*

Eriksson, A., Stanton, N.A., (2017). Driving performance after self-regulated control transitions in highly automated vehicles, *Human Factors, 59, pp. 1233-1248.*

Fairclough, S. H., Burns, C., & Kreplin, U., (2018). FNIRS activity in the prefrontal cortex and motivational intensity: impact of working memory load, financial reward, and correlation-based signal improvement. *Neurophotonics, 5(3), 035001.*

Fall, I., & Rauffet, P. (2008). Analyse de l'utilisation du systeme de gestion des capacité s organisationnelles au sein du groupe Valeo. *Research report of project ANR Pilot 2.0, in collaboration with MNM Consulting*

Fall, I., & Rauffet, P. (2008). Analyse de l'utilisation du systeme de gestion des capacité s organisationnelles au sein du groupe Valeo. *Research report of project ANR Pilot 2.0, in collaboration with MNM Consulting*

Fantini P., Pinzone M., Taisch M. (2018). Placing the operator at the centre of Industry 4.0 design: Modelling and assessing human activities within cyber-physical systems, *Computers & Industrial Engineering, 139, 105-128.*

Fitts, P. M. (1951). Human engineering for an effective air navigation and traffic control system. *Ohio state University Foundation Report*, Columbus, OH.

Flemisch F., M. Heesen, T. Hesse, J. Kelsch, A. Schieben, and J. Beller, (2012): Towards a dynamic balance between humans and automation: authority, ability, responsibility and control in shared and cooperative control situations, *Cognition, Technology & Work, vol. 14, no. 1, pp. 3–18.*

Flemisch, F. O., Adams, C. A., Conway, S. R., Goodrich, K. H., Palmer, M. T., & Schutte, P. C. (2003). The H-Metaphor as a guideline for vehicle automation and interaction. *NASA report*

Gagnon, O., Parizeau, M., Lafond, D., & Gagnon, J. F. (2016, March). Comparing methods for assessing operator functional state. *In IEEE International Multi-Disciplinary Conference on Cognitive Methods in Situation Awareness and Decision Support (CogSIMA), (pp. 88-92).*

Gerin, W., Pieper, C., & Pickering, T. G. (1994). Anticipatory and residual effects of an active coping task on pre-and post-stress baselines. *Journal of psychosomatic research, 38(2), 139-149.*

Gibson, W., Megaw, E., Young, M., & Lowe, E. (2006). A taxonomy of human communication errors and application to railway track maintenance. *Cognition, Technology & Work*, 8(1), 57-66.

Gilani, M. (2016), Machine Learning Classifiers for Critical Cardiac Conditions, *Master of Engineering and Applied Science, University of Ontario Institute of Technology*

Goom, M. K. (1996). Improving function allocation for integrated systems design, pp. 45-61. *Technical Report CSERIAC SOAR 96-01*. Crew Systems Ergonomics Information Analysis Centre, Wright-Patterson Airforce Base, OH, USA.

Gorman, J. C., Cooke, N. J., & Winner, J. L. (2006). Measuring team situation awareness in decentralized command and control environments. *Ergonomics, 49*(12-13), 1312-1325.

Grimm, D., Demir, M., Gorman, J. C., Cooke, N. J. (2018). Team situation awareness in Human-Autonomy Teaming: A systems level approach. Proceedings of the *Human Factors and Ergonomics Society Annual Meeting, 62, 149–149.* doi:10.1177/1541931218621034

Groom, V., Nass, C. (2007). Can robots be teammates? Benchmarks in human–robot teams. *Interaction studies, 8, 483–500.*

Grote, G., Weik, S., Wäfler, T. & Zölch, M. (1995). Criteria for the complementary allocation of functions in automated work systems and their use in simultaneous engineering projects. *International Journal of Industrial Ergonomics, 16, 367-382.*

Guerin, C., Leroy, B., Chauvin, C. & Coppin, G. (2014). Task analysis from the expert point of view: a prerequisite condition to analyse physiological activity of fighter pilot aircraft. Poster presented at *HFES European Chapter.* Lisbon, Portugal.

Guérin, C., Rauffet, P., Chauvin, C. (2021). Human operator in smart factory: towards a symbiotic cognitive system. *UBS Showcase/ Trends on Digital Twins*

Guérin, C., Rauffet, P., Chauvin, C. (2021). Human operator in smart factory: towards a symbiotic cognitive system. *UBS Showcase/ Trends on Digital Twins*

Guerin, C., Rauffet, P., Chauvin, C., & Martin, E. (2019). Toward production operator 4.0: modelling Human-Machine Cooperation in Industry 4.0 with Cognitive Work Analysis. *IFAC HMS.* Tallin (Estonie)

Guerin, C., Rauffet, P., Chauvin, C., & Martin, E. (2019). Toward an order picker 4.0: modelling operator 4.0 in the perspective of joint cognitive and physical production system. *INAIT.* Cambridge (UK)

Guerin, C., Rauffet, P., Chauvin, C., & Martin, E. (2019). Toward production operator 4.0: modelling Human-Machine Cooperation in Industry 4.0 with Cognitive Work Analysis. *IFAC HMS.* Tallin (Estonie)

Guerin, C., Rauffet, P., Chauvin, C., & Martin, E. (2019). Toward an order picker 4.0: modelling operator 4.0 in the perspective of joint cognitive and physical production system. *INAIT.* Cambridge (UK)

Guérin, C., Rauffet, P., Chauvin, C., Martin, E. (2018). L'opérateur 4.0 en symbiose avec les technologies de la smart factory. *Journées de l'automatique JAMACS.* Nantes (France)

Guérin, C., Rauffet, P., Chauvin, C., Martin, E. (2018). L'opérateur 4.0 en symbiose avec les technologies de la smart factory. *Journées de l'automatique JAMACS.* Nantes (France)

Gupta, C., Centofanti, S., Rauffet, P., Banks, S., Coppin, G., & Chauvin, C. (2019). Framework and Metrics for Online Fatigue Monitoring Within Teams Workin in 24/7 Environments. *IFAC HMS*, Tallin (Estonia)

Gupta, C., Centofanti, S., Rauffet, P., Banks, S., Coppin, G., & Chauvin, C. (2019). Framework and Metrics for Online Fatigue Monitoring Within Teams Workin in 24/7 Environments. *IFAC HMS*, Tallin (Estonia)

Gutzwiller, R. S., Wickens, C. D., & Clegg, B. A., (2014). Workload overload modeling: An experiment with MATB II to inform a computational model of task management. In Proceedings of the *Human Factors and Ergonomics Society Annual Meeting, Vol. 58, No. 1, (pp. 849-853).* Sage CA: Los Angeles.

Hancock, P. A. (2017, June). Whither workload? Mapping a path for its future development. In *International Symposium on Human Mental Workload: Models and Applications* (pp. 3-17). Springer, Cham.

Hancock, P. A., & Matthews, G. (2019). Workload and performance: Associations, insensitivities, and dissociations. *Human factors, 61(3), 374-392.*

Hankins, T.C., & Wilson, G.F. (1998). A comparison of heart rate, eye activity, EEG and subjective measures of pilot mental workload during flight. *Aviation, Space, and Environmental Medicine, 69*, 360-367.

Harari, Y. N. (2016). Sapiens. *Bazarforlag AS.*

Harari, Y. N. (2018). 21 Lessons for the 21st Century. *Random House.*

Hart, S.G. and Staveland, L.E. (1988). Development of NASA-TLX (Task Load Index): Results of Empirical and Theoretical Research. In Peter A. Hancock and Najmedin Meshkati (ed.), *Advances in Psychology, Human Mental Workload. North-Holland, pp. 139–183.*

Hartwich, F., Beggiato, M., & Krems, J. F. (2018). Driving comfort, enjoyment and acceptance of automated driving–effects of drivers' age and driving style familiarity. *Ergonomics, 61(8), 1017-1032.*

Hilburn, B. (2004). Cognitive complexity in air traffic control: A literature review. *EEC note 4.*

Hilburn, B., Jorna, P. G., Byrne, E. A., & Parasuraman, R. (1997). The effect of adaptive air traffic control (ATC) decision aiding on controller mental workload. *Human-automation interaction: Research and practice, 84-91.*

Hirsch-Kreinsen, H., (2014), "Wandel von Produktionsarbeit – Industrie 4.0". *Soziologisches Arbeitspapier, 38.* Dortmund: Wirtschafts- und Sozialwissenschaftliche Fakultät, Technische Universität Dortmund.

Hoc J.M., Amalberti, R., (2007). Cognitive control dynamics for reaching a satisfying performance in complex dynamic situations, *Journal of Cognitive Engineering and Decision Making, 1, 22–55.*

Hoc, J. M. (2001). Towards a cognitive approach to human–machine cooperation in dynamic situations. *International journal of human-computer studies, 54(4),* 509-540.

Hoc, J. M., (2000). La relation homme-machine en situation dynamique. *Revue d'intelligence artificielle, 14(1), 55-71*

Hoc, J.M., (1996). Supervision et contrôle de processus : la cognition en situation dynamique. *Presses universitaires de Grenoble.* Grenoble.

Hockey, G. R. J., (1997). Compensatory control in the regulation of human performance under stress and high workload: A cognitive-energetical framework. *Biological psychology, 45(1-3), 73-93.*

Hollnagel, E., (1993). Human Reliability Analysis – Context and Control. *Academic Press.* London.

Hollnagel, E., (2002). Time and time again. *Theoretical Issues in Ergonomics Science*, 3, 143-158.

Hozdić, E. (2015). Smart factory for industry 4.0: A review. *International Journal of Modern Manufacturing Technologies, 2(1), 2067-3604.*

Hu, S., Zheng, G. (2009). Driver drowsiness detection with eyelid related parameters by support vector machine. *Exp. Syst. Appl., 36, 7651–7658*

Inagaki, T. (2003). Adaptive Automation: Sharing and Trading of Control. In: E. Hollnagel (Ed.), *Handbook of Cognitive Task Design. (pp. 147-169),* Lawrence Erlbaum Associates. Mahwah, NJ.

Jenkins, D.P., Stanton, N.A., Salmon, P.M., Walker, G.H., Young, M.S. (2008). Using cognitive work analysis to explore activity allocation within military domains. *Ergonomics, 51(6), 798-815.*

Ji, Q., Lan, P., & Looney, C. (2006). A probabilistic framework for modeling and real-time monitoring human fatigue. *IEEE Transactions on systems, man, and cybernetics-Part A: Systems and humans, 36(5), 862-875.*

Johnson, M., Bradshaw, J. M., Feltovich, P., Jonker, C., van Riemsdijk, B., Sierhuis, M. (2011). The fundamental principle of coactive design: Interdependence must shape autonomy. In De Vos, M. (Ed.), *Coin international workshops, LNAI (Vol. 6541, pp. 172–191)*.

Jordan, C.S. (1992). Experimental study of the effects of an instantaneous self-assessment workload recorder on task performance. *Report No. DRA/TM (CAD5)/92011*. Farnborough: Defence Evaluation & Research Agency.

Judas, S., Morel, G., Rauffet, P., Chauvin, C., Berruet, P. (2012). Le concept d'opération au centre d'un modèle de conception et d'analyse des systèmes sociotechniques, *Séminaire PERGO*, Vannes (France)

Judas, S., Morel, G., Rauffet, P., Chauvin, C., Berruet, P. (2012). Le concept d'opération au centre d'un modèle de conception et d'analyse des systèmes sociotechniques, *Séminaire PERGO*, Vannes (France)

Judas, S., Rauffet, P., Morel, G., Chauvin, C., & Berruet, P. (2012). Test d'une interface écologique appliquée au pilotage de sous-marin, ErgoIHM, Bidart (France). *24th French Speaking Conference on Human-Computer Interaction*, Bidart (France)

Judas, S., Rauffet, P., Morel, G., Chauvin, C., & Berruet, P. (2012). Test d'une interface écologique appliquée au pilotage de sous-marin, ErgoIHM, Bidart (France). *24th French Speaking Conference on Human-Computer Interaction*, Bidart (France)

Kaber, D. B., & Endsley, M. R. (2004). The effects of level of automation and adaptive automation on human performance, situation awareness and workload in a dynamic control task. *Theoretical Issues in Ergonomics Science, 5(2), 113-153.*

Kaber, D. B., Perry, C. M., Segall, N., McClernon, C. K., & Prinzel III, L. J. (2006). Situation awareness implications of adaptive automation for information processing in an air traffic control-related task. *International Journal of Industrial Ergonomics, 36(5), 447-462.*

Kahneman, D. (1973). Attention and Effort. *Englewood Cliffs*, NJ: Prentice-Hall.

Kahneman, D. (2011). Thinking, fast and slow. *Macmillan.*

Kampik, T., Nieves, J. C., & Lindgren, H. (2019, May). Explaining sympathetic actions of rational agents. In *International Workshop on Explainable, Transparent Autonomous Agents and Multi-Agent Systems (pp. 59-76).* Springer, Cham.

Karavidas, M.K., Lehrer, P.M., Lu, S.E., Vaschillo, E., Vaschillo, B., & Cheng, A. (2010). The effects of workload on respiratory variables in simulated flight: A preliminary study. *Biological Psychology, 84*, 157-160.

Kasarskis, P., Stehwien, J., Hickox, J., Aretz, A., & Wickens, C. (2001). Comparison of expert and novice scan behaviors during VFR flight. Paper presented at the *11th International Symposium on Aviation Psychology,* Columbus, OH: The Ohio State University.

Kavitha, R., & Christopher, T. (2016). Classification of Heart Rate data using BFO-KFCM clustering and improved extreme learning machine classifier. *In IEEE Computer Communication and Informatics (ICCCI), (pp. 1-6).*

Khushaba, R.N., Kodagoda, S., Lal, S., & Dissanayake, G. (2011). Driver drowsiness classification using fuzzy wavelet-packet-based feature-extraction algorithm. *IEEE Trans. Biomed. Eng., 58, 121–131*

Klein, G., Feltovich, P. J., Bradshaw, J. M., & Woods, D. D. (2005). Common ground and coordination in joint activity. Organizational simulation, 53.

Klein, G., Feltovich, P.J., & Woods, D.D. (2004). Common Ground and Coordination in Joint Activity. In W.R. Rouse & K.B. Boff (Eds.), *Organizational simulation.* New York, US: Wiley.

Klein, G., Phillips, J. K., Rall, E. L., & Peluso, D. A. (2007). A data–frame theory of sensemaking. *In Expertise out of context (pp. 118-160).* Psychology Press.

Knowles, W.B. (1963). Operator Loading Tasks. *Human Factors: The Journal of the Human Factors and Ergonomics Society, 5(2), 155–161.*

Koenig, G. (2019). La fin de l'individu. Voyage d'un philosophe au pays de l'intelligence artificielle. *Éditions de l'Observatoire.*

Kostenko, A. S. (2017). Évaluation multidimensionnelle et dynamique de la maitrise de la situation par l'opérateur : création d'un indicateur temps réel de charge mentale pour l'activité de supervision de drones. *PhD dissertation, Université de Bretagne Sud.*

Kostenko, A., Coppin., Rauffet, P. (2017). État de l'art sur les techniques d'apprentissage et de classification appliquées à l'analyse des états d'un opérateur en interactions avec un système critique & Spécifications du module logiciel de classification. *Research report of project TRM104 in collaboration with Dassault Aviation*

Kostenko, A., Coppin., Rauffet, P. (2017). État de l'art sur les techniques d'apprentissage et de classification appliquées à l'analyse des états d'un opérateur en interactions avec un système critique & Spécifications du module logiciel de classification. *Research report of project TRM104 in collaboration with Dassault Aviation*

Kostenko, A., Rauffet, P., Chauvin, C., & Coppin, G. (2015). La question de la maîtrise de la situation dans la supervision de systèmes autonomes : vers des indicateurs temps réel de la charge mentale, *JDJN MACS*, Bourges (France).

Kostenko, A., Rauffet, P., Chauvin, C., & Coppin, G. (2015). La question de la maîtrise de la situation dans la supervision de systèmes autonomes : vers des indicateurs temps réel de la charge mentale, *JDJN MACS*, Bourges (France).

Kostenko, A., Rauffet, P., Chauvin, C., & Coppin, G. (2016). A dynamic closed-looped and multidimensional model for Mental Workload evaluation, HMS 2016: *13th IFAC/IFIP/IFORS/IEA Symposium on Analysis, Design, and Evaluation of Human-Machine Systems*, September, Kyoto (Japan).

Kostenko, A., Rauffet, P., Chauvin, C., & Coppin, G. (2016). A dynamic closed-looped and multidimensional model for Mental Workload evaluation, HMS 2016: *13th IFAC/IFIP/IFORS/IEA Symposium on Analysis, Design, and Evaluation of Human-Machine Systems*, September, Kyoto (Japan).

Kostenko, A., Rauffet, P., Chauvin, C., Coppin, G. (2015). Création d'indicateurs synthétiques temps réel de la charge mentale, *Journée du GT ASHM du GDR MACS*, Paris (France)

Kostenko, A., Rauffet, P., Chauvin, C., Coppin, G. (2015). Création d'indicateurs synthétiques temps réel de la charge mentale, *Journée du GT ASHM du GDR MACS*, Paris (France)

Kostenko, A., Rauffet, P., Moga, S., & Coppin, G. (2019). Operator Functional State: measure it with attention intensity and selectivity, explain it with cognitive control. 3[rd] conference on *Human Workload*, Roma (Italy)

Kostenko, A., Rauffet, P., Moga, S., & Coppin, G. (2019). Operator Functional State: measure it with attention intensity and selectivity, explain it with cognitive control. 3[rd] conference on *Human Workload*, Roma (Italy)

Kostenko, A., Rauffet, P., Moga, S., Coppin, G. (2019). Etat de l'art sur la classification des états fonctionnels de l'opérateur et proposition de mise en œuvre. *Research report of project PEA MMT Precogs*

Kostenko, A., Rauffet, P., Moga, S., Coppin, G. (2019). Etat de l'art sur la classification des états fonctionnels de l'opérateur et proposition de mise en œuvre. *Research report of project PEA MMT Precogs*

Kostenko, A., Rauffet, P., Moga, S., Coppin, G. (2019). Operator Functional State: measure it with attention intensity and selectivity, explain it with cognitive control. *Human Mental Workload: Models and Applications, Revised Selected Papers*, Editors: Longo, Luca, Leva, M. Chiara (Eds.)

Kostenko, A., Rauffet, P., Moga, S., Coppin, G. (2019). Operator Functional State: measure it with attention intensity and selectivity, explain it with cognitive control. *Human Mental Workload: Models and Applications, Revised Selected Papers*, Editors: Longo, Luca, Leva, M. Chiara (Eds.)

Kotsiantis, S. B., Zaharakis, I., & Pintelas, P. (2007). Supervised machine learning: A review of classification techniques. *Emerging artificial intelligence applications in computer engineering, 160, 3-24.*

Kramer, A.F. (1991). Physiological metrics of mental workload: A review of recent progress. In Damos, D.L. (ed.), *Multiple-task performance. London: Taylor & Francis, 279-328.*

Lahlou, S. (2011). How can we capture the subject's perspective? An evidence-based approach for the social scientist. *Social Science Information, 50*, 607-655.

Lallican, J.L., Berruet, P., Rossi, A., and Philippe, J.L. (2007). A component-based approach for conveying systems control design. In *4th International Conference on Informatics in Control, Automation and Robotics ICINCO 2007*, 329–336.

Laouar, A. (2021). Etude de la prise de décision chez les pilotes d'aviation commerciale : relation entre le contrôle cognitif et la charge mentale. *PhD Dissertation, Université Bretagne Sud, Lorient.*

Laouar, A., Rauffet, P., Chauvin, C., & Bressolle, M. C. (2018). A step toward decision-making assistance: Understanding pilots' decision making and workload impact, 2nd conference on *Human Workload*, Amsterdam (The Netherlands).

Laouar, A., Rauffet, P., Chauvin, C., & Bressolle, M. C. (2018). A step toward decision-making assistance: Understanding pilots' decision making and workload impact, 2nd conference on *Human Workload*, Amsterdam (The Netherlands).

Larson, L., DeChurch, L. A. (2020). Leading teams in the digital age: Four perspectives on technology and what they mean for leading teams. *The Leadership Quarterly, 31 doi:10.1016/j.leaqua.2019.101377*

Lassalle, J., Leroy B., Rauffet, P., Guerin, C., Chauvin, C., Coppin, C. (2015) Technique d'Analyse pour le Partage d'Autorité dans les Systèmes de systèmes : méthode COWORKS². *Research report of project ANR ASTRID TAPAS*

Lassalle, J., Leroy B., Rauffet, P., Guerin, C., Chauvin, C., Coppin, C. (2015) Technique d'Analyse pour le Partage d'Autorité dans les Systèmes de systèmes : méthode COWORKS². *Research report of project ANR ASTRID TAPAS*

Lassalle, J., Rauffet, P., Leroy, B., Guérin, C., Chauvin, C., Coppin, G., Saïd, F. (2017). COmmunication and WORKload analyses to study the COllective WORK of fighter pilots: the COWORK2 method. *Cognition, Technology and Work*, *19, 477–491.* https://doi.org/10.1007/s10111-017-0420-8

Lassalle, J., Rauffet, P., Leroy, B., Guérin, C., Chauvin, C., Coppin, G., Saïd, F. (2017). COmmunication and WORKload analyses to study the COllective WORK of fighter pilots: the COWORK2 method. *Cognition, Technology and Work*, *19, 477–491*. https://doi.org/10.1007/s10111-017-0420-8

Lassalle, J., Rauffet, P., Leroy, B., Guillet, L., Chauvin, C., & Coppin, G. (2014). Innovative multi-sensors device deployment for fighter pilots' activity study in a highly realistic Rafale simulator, HFES, Lisboa (Portugal).

Lassalle, J., Rauffet, P., Leroy, B., Guillet, L., Chauvin, C., & Coppin, G. (2014). Innovative multi-sensors device deployment for fighter pilots' activity study in a highly realistic Rafale simulator, HFES, Lisboa (Portugal).

Lassalle, J., Simon, L., Rauffet, P. (2021). Spécifications des besoins - recommandations psycho-ergonomiques et cognitives pour la conception des IHS. *Research project of project AMI ADEME Seanatic.*

Lassalle, J., Simon, L., Rauffet, P. (2021). Spécifications des besoins - recommandations psycho-ergonomiques et cognitives pour la conception des IHS. *Research project of project AMI ADEME Seanatic.*

Law, T., & Scheutz, M. (2021). Trust: Recent concepts and evaluations in human-robot interaction. In *Trust in Human-Robot Interaction (pp. 27-57)*. Academic Press.

Lee, J.D., & See, J.K (2004). Trust in Automation: Designing for Appropriate Reliance. *Human Factors, vol. 46, no 1, p. 50-80*. doi: 10.1518/hfes.46.1.50_30392.

Lee, Y.-H., & Liu, B.-S. (2003). Inflight workload assessment: Comparison of subjective and physiological measurements. *Aviation, Space, and Environmental Medicine*, *74*(10), 1078-1084.

Lehrer, P., Karavidas, M., Lu, S.E., Vaschillo, E., Vaschillo, B., & Cheng, A. (2010). Cardiac data increase association between self-report and both expert ratings of task load and task performance in flight simulator tasks: An exploratory study. *International Journal of Psychophysiology, 76*, 80-87.

Lemoine M-P., Debernard S., Crévits I., Millot P. (1996): Cooperation between humans and machines: first results of an experimentation of a multi-level cooperative organization in air traffic control. *Computer Supported Cooperative Work, 5, pp. 299-321.*

León-Carrion, J., Damas-López, J., Martín-Rodríguez, J. F., Domínguez-Roldán, J. M., Murillo-Cabezas, F., Martin, J. M. B., & Domínguez-Morales, M. R., (2008). The hemodynamics of cognitive control: the level of concentration of oxygenated hemoglobin in the superior prefrontal cortex varies as a function of performance in a modified Stroop task. *Behavioural brain research, 193(2), 248-256.*

Leplat, J. (2006). La notion de régulation dans l'analyse de l'activité. *Perspectives interdisciplinaires sur le travail et la santé, 8(1), 1-25.*

Lewis, M., Sycara, K., & Walker, P. (2018). The role of trust in human-robot interaction. *In Foundations of trusted autonomy (pp. 135-159)*. Springer, Cham.

Li, X., Lei, Y., Wang, W., Wang, W., & Zhu, Y. (2013). A dsm-based multi-paradigm simulation modeling approach for complex systems. In *Proceedings of the 2013 Winter Simulation Conference: Simulation: Making Decisions in a Complex World*, 1179–1190. IEEE Press.

Liu, J., Zhang, C., & Zheng, C (2010). EEG-based estimation of mental fatigue by using KPCA-HMM and complexity parameters. Biomed. *Signal Process. Contr, 5, 124–130*

Longo, F., Nicoletti, L., & Padovano, A. (2017). Smart operators in industry 4.0: A human-centered approach to enhance operators' capabilities and competencies within the new smart factory context. *Computers & Industrial Engineering, 113, 144-159.*

Longo, L. (2017, September). Subjective usability, mental workload assessments and their impact on objective human performance. In *IFIP Conference on Human-Computer Interaction (pp. 202-223).* Springer, Cham.

Lyons, J. B., Mahoney, S., Wynne, K. T., & Roebke, M. A. (2018). Viewing machines as teammates: A qualitative study. *2018 AAAI Spring Symposium Series*

Lyons, J.B. (2013) Being transparent about transparency: A model for human-robot interaction. In *2013 AAAI Spring Symposium Series.*

Mäki-Marttunen, V., Hagen, T., & Espeseth, T., (2019a). Task context load induces reactive cognitive control: An fMRI study on cortical and brain stem activity. *Cognitive, Affective, & Behavioral Neuroscience, 19(4), 945-965.*

Mäki-Marttunen, V., Hagen, T., & Espeseth, T., (2019b). Proactive and reactive modes of cognitive control can operate independently and simultaneously. *Acta psychologica, 199, 102891.*

Malik, M., (1996). Heart rate variability: Standards of measurement, physiological interpretation, and clinical use: Task force of the European Society of Cardiology and the North American Society for Pacing and Electrophysiology. *Annals of Noninvasive Electrocardiology, 1(2), 151-181.*

Marshall S.P. (2002). Method and apparatus for eye tracking and monitoring pupil dilation to evaluate cognitive activity. Patent US 6090051A. Washington, DC: U.S. Patent and Trademark Office.

McNeese, N. J., Demir, M., Cooke, N. J., Myers, C. (2018). Teaming with a synthetic teammate: Insights into human-autonomy teaming. Human Factors: *The Journal of the Human Factors and Ergonomics Society, 60, 262–273.* doi:10.1177/0018720817743223

Miller, C. A. (2005, July). Trust in adaptive automation: the role of etiquette in tuning trust via analogic and affective methods. *In Proceedings of the 1st international conference on augmented cognition (pp. 22-27).*

Millot P., & Pacaux-lemoine M. P. (2013). A common Work Space for a mutual enrichment of Human-machine Cooperation and Team-Situation Awareness. *12th IFAC/IFIP/IFORS/IEA Symposium on Analysis, Design, and Evaluation of Human-Machine Systems,* Las Vegas, Nevada, USA, August.

Millot P., Mandiau R. (1995). Men-Machine Cooperative Organizations: Formal and Pragmatic implementation methods. In J.M. Hoc, P.C. Cacciabue, E. Hollnagel Editors, *Expertise and Technology: Cognition Computer Cooperation*, Lawrence Erlbraum Associates, New Jersey, chap. 13, pp. 213-228, 199.

Millot, P., & Lemoine, M. -P. (1998). An attempt for generic concepts toward Human-Machine Cooperation. In *IEEE international conference on systems, man and cybernetics*. San Diego, USA.

MOD (1989). Defence Standard 00-25. Human Factors for Designers of Equipment. UK Ministry of Defence.

Mollo, V., & Falzon, P. (2004). Auto- and allo-confrontation as tools for reflective activities. *Applied Ergonomics*, *35*(6), 531-540.

Mouloua, M., Gilson, R., Kring, J., & Hancock, P. (2001). *Workload, situation awareness, and teaming issues for UAV/UCAV operations.* Proceedings of the 45th Human Factors and Ergonomics Society (HFES), 162-165.

Myers, C., Ball, J., Cooke, N., Freiman, M., Caisse, M., Rodgers, S., Demir, M., & McNeese, N. (2019). Autonomous intelligent agents for team training. *IEEE Intelligent Systems, 34, 3–14.* doi:10.1109/MIS.2018.2886670

Nählinder, S., Berggren, P., & Svensson, E. (2004). *Reoccurring LISREL patterns describing mental workload, situation awareness and performance.* Proceedings of the 48th Human Factors and Ergonomics Society (HFES).

Naikar, N. (2013). Work domain analysis: Concepts, guidelines, and cases. *CRC Press.*

Naikar, N., Moylan, A., & Pearce, B. (2006). Analysing activity in complex systems with cognitive work analysis: concepts, guidelines and case study for control task analysis. *Theoretical Issues in Ergonomics Science, 7(4), 371-394.*

Nhim, V., Rauffet, P., Da Cunha, C., Bernard, A., Uys, W., Louw, L., & Du Preez, N. (2009). Controlling ideas and concepts generation stage for a better management of innovation portfolio: analysis of the Fugle methodology's funnel, *International Conference on Industrial Engineering and Systems Management.* Montreal (Canada).

Nhim, V., Rauffet, P., Da Cunha, C., Bernard, A., Uys, W., Louw, L., & Du Preez, N. (2009). Controlling ideas and concepts generation stage for a better management of innovation portfolio: analysis of the Fugle methodology's funnel, *International Conference on Industrial Engineering and Systems Management.* Montreal (Canada).

Noel, J. B., Bauer, K. W., & Lanning, J. W. (2005). Improving pilot mental workload classification through feature exploitation and combination: a feasibility study. *Computers & operations research, 32(10), 2713-2730*

Norman, D. A. and S. W. Draper. 1986. User Centered Design. edited by D. A. Norman and S. W. Draper. *Hillsdale, NJ: Lawrence Erlbaum Associates.*

O'Neill, T., McNeese, N., Barron, A., & Schelble, B. (2020). Human–Autonomy Teaming: A Review and Analysis of the Empirical Literature. *Human Factor*s, 0018720820960865.

Oppelt, M. & Urbas, L. (2014). Integrated virtual commissioning an essential activity in the automation engineering process: From virtual commissioning to simulation supported engineering. In *Industrial Electronics Society, IECON 2014-40th Annual Conference of the IEEE*, 2564–2570. IEEE.

Oppelt, M., Wolf, G., Drumm, O., Lutz, B., Stöß, M., & Urbas, L. (2014). Automatic model generation for virtual commissioning based on plant engineering data. In *Proceedings of the 19th World Congress of the International Federation of Automatic Control*.

Pacaux-lemoine M. P., & Debernard S. (2002). A Common Work Space to support the Air Traffic Control. *Control Engineering Practice, A Journal of IFAC, 10 (5), pp. 571-576.*

Pacaux-Lemoine, M. P. (2020). HUMAN-MACHINE COOPERATION: Adaptablity of shared functions between Humans and Machines-Design and evaluation aspects. *Accreditation to direct research, Université Polytechnique Hauts-de-France.*

Pacaux-Lemoine, M. P., & Flemisch F., (2018). Layers of Shared and Cooperative Control, assistance and automation, *Cognition, Technology & Work,* https://doi.org/10.1007/s10111-018-0537-4,

Pacaux-Lemoine, M. P., & Loiselet, A. (2002). A Common Work Space to support the cooperation into the cockpit of a two-seater fighter aircraft. *Cooperative Systems Design: A Challenge of Mobility Age, 157–172.*

Pacaux-Lemoine, M.-P., Berdal, Q., Guérin, C., Rauffet, P., Chauvin, C., Trentesaux, D. (2021). Designing Human-Systems Cooperation in Industry 4.0 with Cognitive Work Analysis: a first evaluation. *Cognition, Technology & Work.* https://doi.org/10.1007/s10111-021-00667-y

Pacaux-Lemoine, M.-P., Berdal, Q., Guérin, C., Rauffet, P., Chauvin, C., Trentesaux, D. (2021). Designing Human-Systems Cooperation in Industry 4.0 with Cognitive Work

Analysis: a first evaluation. *Cognition, Technology & Work.* https://doi.org/10.1007/s10111-021-00667-y

Papantonopoulos, S. (2004). "System Design in Normative and Actual Practice: A Comparative Study of Cognitive Task Allocation in Advanced Manufacturing Systems." *Human Factors and Ergonomics in Manufacturing 14(2):181–96.*

Parasuraman R., & Sheridan T.B., & Wickens C.D. (2000): Models for Types and Levels of Human Interaction with Automation, *IEEE Transactions on Systems, Man, and Cybernetics, Vol. 30, No 3, May.*

Parasuraman, R. (2003). Neuroergonomics: Research and practice. *Theoretical issues in ergonomics science, 4(1-2), 5-20.*

Parasuraman, R., & Hancock, P. A. (2001). Adaptive control of mental workload.

Parasuraman, R., Bahri, T., Deaton, J. E., Morrison, J. G., & Barnes, M. (1992). Theory and design of adaptive automation in aviation systems. *Catholic University of America Washington DC cognitive science lab.*

Phillips, C. A., Repperger, D. W., Kinsler, R., Bharwani, G., & Kender, D. (2007). A quantitative model of the human–machine interaction and multi-task performance: A strategy function and the unity model paradigm. *Computers in biology and medicine, 37(9), 1259-1271.*

Pinti, P., Scholkmann, F., Hamilton, A., Burgess, P., & Tachtsidis, I. (2019). Current status and issues regarding pre-processing of fNIRS neuroimaging data: an investigation of diverse signal filtering methods within a general linear model framework. *Frontiers in human neuroscience, 12, 505.*

Prat, S. (2017). Intégration de techniques de vérification par simulation dans un processus de conception automatisée de contrôle commande. *PhD Dissertation, Université Bretagne Sud, Lorient.*

Prat, S., Cavron, J., Kesraoui, D., Rauffet, P., Berruet, P., & Bignon, A. (2017). An Automated Generation Approach of Simulation Models for Checking Control/Monitoring System, *IFAC World Congress*, Toulouse (France).

Prat, S., Cavron, J., Kesraoui, D., Rauffet, P., Berruet, P., & Bignon, A. (2017). An Automated Generation Approach of Simulation Models for Checking Control/Monitoring System, *IFAC World Congress*, Toulouse (France).

Prat, S., Rauffet, P., Berruet, P., & Bignon, A. (2016). A multi-level requirements modeling for sociotechnical system simulation-based checking, *International Conference on System Man and Cybernetics (IEEE SMC)*, Budapest (Hungary).

Prat, S., Rauffet, P., Berruet, P., & Bignon, A. (2016). A multi-level requirements modeling for sociotechnical system simulation-based checking, *International Conference on System Man and Cybernetics (IEEE SMC)*, Budapest (Hungary).

Prat, S., Rauffet, P., Bignon, A., & Berruet, P. (2015). Vers l'intégration d'une approche de génération automatique de mod ele de simulation dans un flot de conception de contrôle - commande, *JDJN MACS*, Bourges (France).

Prat, S., Rauffet, P., Bignon, A., & Berruet, P. (2015). Vers l'intégration d'une approche de génération automatique de mod ele de simulation dans un flot de conception de contrôle - commande, *JDJN MACS*, Bourges (France).

Prinzo, O. V. (2001). *Data-linked pilot reply time on controller workload and communication in a simulated terminal option* (technical report DOT/FAA/AM-01/8). Civil Aeromedical Institute, Office of Aviation Medicine Federal Aviation Administration.

R Core Team, (2012). R: A language and environment for statistical computing. *Foundation for statistical Computing.*

Rachedi, N. D. E. (2015). Modélisation et surveillance de systèmes Homme-Machine : application à la conduite ferroviaire *(PhD Dissertation, Valenciennes).*

Rajabiyazdi F., & Jamieson, A. (2020). A Review of Transparency (seeing-into) Models, in *2020 IEEE International Conference on Systems, Man, and Cybernetics (SMC), oct. 2020, p. 302-308,* doi: 10.1109/SMC42975.2020.9282970.

Rao, A.S., & Georgeff, M.P. (1995). BDI agents: From theory to practice. in *Proc. 1st Int. Conf. Multi-Agent Syst., pp. 312–319.*

Rasmussen, J. (1986). Information Processing and Human–machine Interaction. Amsterdam, North Holland: Elsevier.

Rasmussen, J., (1983). Skills, rules, and knowledge - signals, signs, and symbols, and other distinctions in human performance models. In *IEEE Transactions on Systems, Man, and Cybernetics, vol. 13, pp. 257-266.*

Rasmussen, J., Pejtersen, A.M., & Goodstein, L.P. (1994). *Cognitive Systems engineering.* New York: Wiley.

Rauffet, P, Coppin, G., Chauvin, C. (2019). *Towards detection and mitigation of fatigue risk in submarine teams*, invited talk, UniSA (Pr. Siobhan Banks) au *Workshop Human Factor Research for Navy*, Adelaide (Australia)

Rauffet, P, Coppin, G., Chauvin, C. (2019). *Towards detection and mitigation of fatigue risk in submarine teams*, invited talk, UniSA (Pr. Siobhan Banks) au *Workshop Human Factor Research for Navy*, Adelaide (Australia)

Rauffet, P. (2007). A methodology for the application of an automated and interactive reification process in a virtual Community of Practice, Master dissertation, Centrale Nantes, Nantes (France)

Rauffet, P. (2007). A methodology for the application of an automated and interactive reification process in a virtual Community of Practice, Master dissertation, Centrale Nantes, Nantes (France)

Rauffet, P. (2010). Prise en compte des facteurs formels et contextuels dans la gestion des capacités organisationnelles. Application aux organisations matricielles. PhD dissertation, Centrale Nantes. Nantes (France)

Rauffet, P. (2010). Prise en compte des facteurs formels et contextuels dans la gestion des capacités organisationnelles. Application aux organisations matricielles. PhD dissertation, Centrale Nantes. Nantes (France)

Rauffet, P. (2017). Initiative vigilance au volant, définition des facteurs de vigilance & choix des signaux et des capteurs. *Research report in collaboration with AXA*

Rauffet, P. (2017). Initiative vigilance au volant, définition des facteurs de vigilance & choix des signaux et des capteurs. *Research report in collaboration with AXA*

Rauffet, P. (2021). *Operator state monitoring and teamwork adaptation: tools for Human-Autonomy Teaming.* Seminar series, School of Psychology, University of Adelaide (Australia)

Rauffet, P. (2021). *Operator state monitoring and teamwork adaptation: tools for Human-Autonomy Teaming.* Seminar series, School of Psychology, University of Adelaide (Australia)

Rauffet, P., Bernard, A., & Da Cunha, C. (2009). Le pilotage d'entreprise par les capacités organisationnelles : étude de l'approche roadmapping, *JDJN MACS*, 17-18 mars 2009, Angers (France).

Rauffet, P., Bernard, A., & Da Cunha, C. (2009). Le pilotage d'entreprise par les capacités organisationnelles : étude de l'approche roadmapping, *JDJN MACS*, 17-18 mars 2009, Angers (France).

Rauffet, P., Bernard, A., Da Cunha, C., & Labrousse, M. (2009). Progress Management in Performance-Driven Systems: Study of the 5Steps Roadmapping, a Solution for Managing Organizational Capabilities and Their Learning Curves., *13th IFAC Symposium on Information Control Problems in Manufacturing (INCOM),* Moscow (Russie).

Rauffet, P., Bernard, A., Da Cunha, C., & Labrousse, M. (2009). Progress Management in Performance-Driven Systems: Study of the 5Steps Roadmapping, a Solution for Managing Organizational Capabilities and Their Learning Curves., *13th IFAC Symposium on Information Control Problems in Manufacturing (INCOM),* Moscow (Russie).

Rauffet, P., Bernard, A., Da, Cunha C., Du Preez, N., Louw, L., & Uys, W. (2008). Assessed, interactive and automated reification in a virtual community of practice, *International Symposium on Tools and Methods of Competitive Engineering.* Izmir (Turkey).

Rauffet, P., Bernard, A., Da, Cunha C., Du Preez, N., Louw, L., & Uys, W. (2008). Assessed, interactive and automated reification in a virtual community of practice, *International Symposium on Tools and Methods of Competitive Engineering.* Izmir (Turkey).

Rauffet, P., Botzer, A., Chauvin, C., Said, F. (2020). Effects of Engagement in a Non-Driving Task and Type of Display on Driver Behavior when Taking Control from an Automated Vehicle. *Cognition, Technology and Work 22, 721–731.* https://doi.org/10.1007/s10111-019-00611-1

Rauffet, P., Botzer, A., Chauvin, C., Said, F. (2020). Effects of Engagement in a Non-Driving Task and Type of Display on Driver Behavior when Taking Control from an Automated Vehicle. *Cognition, Technology and Work 22, 721–731.* https://doi.org/10.1007/s10111-019-00611-1

Rauffet, P., Botzer, A., Kostenko, A., Chauvin, C., & Coppin, G. (2016). Eye activity measures as indicators of drone operators' workload and task completion strategies, *HFES European Chapter Conference*, Praha (Czech Republic).

Rauffet, P., Botzer, A., Kostenko, A., Chauvin, C., & Coppin, G. (2016). Eye activity measures as indicators of drone operators' workload and task completion strategies, *HFES European Chapter Conference*, Praha (Czech Republic).

Rauffet, P., Chauvin, C. (2018). L'analyse des processus collectifs et l'évaluation des états cognitifs de l'opérateur pour améliorer la prise de décision, invited talk, Human Factors department, Airbus, Toulouse (France)

Rauffet, P., Chauvin, C. (2018). L'analyse des processus collectifs et l'évaluation des états cognitifs de l'opérateur pour améliorer la prise de décision, invited talk, Human Factors department, Airbus, Toulouse (France)

Rauffet, P., Chauvin, C., Morel, G., & Berruet, P. (2015). Designing sociotechnical systems: a CWA-based method for dynamic function allocation, 33rd European Conference on Cognitive Ergonomics, 1-3 juillet 2015, Warsaw (Poland). *European Conference on Cognitive Ergonomics,* Warsaw (Poland)

Rauffet, P., Chauvin, C., Morel, G., & Berruet, P. (2015). Designing sociotechnical systems: a CWA-based method for dynamic function allocation, 33rd European Conference on Cognitive Ergonomics, 1-3 juillet 2015, Warsaw (Poland). *European Conference on Cognitive Ergonomics,* Warsaw (Poland)

Rauffet, P., Chauvin, C., Nistico, C., Judas, S., & Toumelin, N. (2017). Effect of control-display compatibility on the mental workload of submarine helmsmen, *1st conference on Human Workload*, Dublin (Ireland).

Rauffet, P., Chauvin, C., Nistico, C., Judas, S., & Toumelin, N. (2017). Effect of control-display compatibility on the mental workload of submarine helmsmen, *1st conference on Human Workload*, Dublin (Ireland).

Rauffet, P., Chauvin, C., Nistico, C., Judas, S., Toumelin, N. (2017). Effect of control-display compatibility on the mental workload of submarine helmsmen," pp. 198-212 *Human Mental Workload: Models and Applications, Revised Selected Papers*, Editors: Longo, Luca, Leva, M. Chiara (Eds.), ISBN: 978-3-319-61061-0

Rauffet, P., Chauvin, C., Nistico, C., Judas, S., Toumelin, N. (2016). Analysis of submarine steering: effects of cognitive and perceptual–motor requirements on the mental workload and performance of helmsmen. *Cognition, Technology and Work, 18, 657–672.* https://doi.org/10.1007/s10111-016-0384-0

Rauffet, P., Chauvin, C., Nistico, C., Judas, S., Toumelin, N. (2016). Analysis of submarine steering: effects of cognitive and perceptual–motor requirements on the mental workload and performance of helmsmen. *Cognition, Technology and Work, 18, 657–672.* https://doi.org/10.1007/s10111-016-0384-0

Rauffet, P., Chauvin, C., Nistico, C., Judas, S., Toumelin, N. (2017). Effect of control-display compatibility on the mental workload of submarine helmsmen," pp. 198-212 *Human Mental Workload: Models and Applications, Revised Selected Papers*, Editors: Longo, Luca, Leva, M. Chiara (Eds.), ISBN: 978-3-319-61061-0

Rauffet, P., Coppin, G. (2018). Online classification of cognitive states and perspectives for dialogue management with autonomous systems. Invited talk, University of Adelaide (Pr. Anna Ma-Wyatt), *Workshop Human and robotic vision: from scene understanding to augmentation*, Adelaide (Australia)

Rauffet, P., Coppin, G. (2018). Online classification of cognitive states and perspectives for dialogue management with autonomous systems. Invited talk, University of Adelaide (Pr. Anna Ma-Wyatt), *Workshop Human and robotic vision: from scene understanding to augmentation*, Adelaide (Australia)

Rauffet, P., Coppin, G. Kostenko, A. (2016). État de l'art sur la classification des états fonctionnels de l'opérateur et propositions de mise en œuvre. *Research report of project TRM104 in collaboration with Dassault Aviation*

Rauffet, P., Coppin, G. Kostenko, A. (2016). État de l'art sur la classification des états fonctionnels de l'opérateur et propositions de mise en œuvre. *Research report of project TRM104 in collaboration with Dassault Aviation*

Rauffet, P., Coppin, G., Chauvin, C. (2017). Modeling of sociotechnical systems: analysis of macro-cognitive processes and assessment of operator functional states to improve human-machine cooperation, invited talk by UniSA (Pr. Siobhan Banks). *Workshop Team Cohesion, Teamwork and Decision Making in Maritime Environments*, Adelaide (Australia)

Rauffet, P., Coppin, G., Chauvin, C. (2017). Modeling of sociotechnical systems: analysis of macro-cognitive processes and assessment of operator functional states to improve human-machine cooperation, invited talk by UniSA (Pr. Siobhan Banks). *Workshop Team Cohesion, Teamwork and Decision Making in Maritime Environments*, Adelaide (Australia)

Rauffet, P., Da Cunha C., Bernard A. (2016). Managing resource learning in distributed organisations with the organisational capability approach. *International Journal of Technology Management*, 70(4) https://doi.org/10.1504/IJTM.2016.075902

Rauffet, P., Da Cunha C., Bernard A. (2016). Managing resource learning in distributed organisations with the organisational capability approach. *International Journal of Technology Management*, 70(4) https://doi.org/10.1504/IJTM.2016.075902

Rauffet, P., Da Cunha, C. & Bernard, A. (2010). Knowledge Sharing and Communities of Practices for Intra-organizational Interoperability, pp. 397-406 *Enterprise Interoperability IV*. Huang, George and Mak, K. and Maropoulos, Paul-Springer London. ISBN 978-1-84996-257-5

Rauffet, P., Da Cunha, C. & Bernard, A. (2010). Knowledge Sharing and Communities of Practices for Intra-organizational Interoperability, pp. 397-406 *Enterprise Interoperability IV*. Huang, George and Mak, K. and Maropoulos, Paul-Springer London. ISBN 978-1-84996-257-5

Rauffet, P., Da Cunha, C., & Bernard, A. (2008). Pilotage de la performance organisationnelle : D'un modèle intuitif à un modele inductif des capacités organisationnelles, *Colloque Recherche Inter Ecoles Centrales - CRIEC*, Nantes (France)

Rauffet, P., Da Cunha, C., & Bernard, A. (2008). Pilotage de la performance organisationnelle : D'un modèle intuitif à un modele inductif des capacités organisationnelles, *Colloque Recherche Inter Ecoles Centrales - CRIEC*, Nantes (France)

Rauffet, P., Da Cunha, C., & Bernard, A. (2009). Designing and managing Organizational Interoperability with organizational capabilities and roadmaps, *5th International Conference on Interoperability for Enteprise, Software and Appliccations (I-ESA)*, 20-23 April 2009, Beijin (Chine).

Rauffet, P., Da Cunha, C., & Bernard, A. (2009). Designing and managing Organizational Interoperability with organizational capabilities and roadmaps, *5th International Conference on Interoperability for Enteprise, Software and Appliccations (I-ESA)*, 20-23 April 2009, Beijin (Chine).

Rauffet, P., Da Cunha, C., & Bernard, A. (2009). Le pilotage d'entreprise par les compétences organisationnelles : étude de la méthode 5 steps, *8e Congrès International de Génie Industriel (CIGI)*, June, Tarbes (France).

Rauffet, P., Da Cunha, C., & Bernard, A. (2009). Le pilotage d'entreprise par les compétences organisationnelles : étude de la méthode 5 steps, *8e Congrès International de Génie Industriel (CIGI)*, June, Tarbes (France).

Rauffet, P., Da Cunha, C., & Bernard, A. (2010). Knowledge sharing and communities of practices for intra-organizational interoperability, *6th International Conference Interoperability for Enterprise Software and Applications* (UK).

Rauffet, P., Da Cunha, C., & Bernard, A. (2010). Knowledge sharing and communities of practices for intra-organizational interoperability, *6th International Conference Interoperability for Enterprise Software and Applications* (UK).

Rauffet, P., Da Cunha, C., & Bernard, A. (2011). Vers un apprentissage organisationnel durable dans le contexte de groupe d'entreprises : comparaison du progrès fonctionnel et de la performance opérationnelle, *Colloque National AIP PRIMECA,* Mont Dore (France)

Rauffet, P., Da Cunha, C., & Bernard, A. (2011). Vers un apprentissage organisationnel durable dans le contexte de groupe d'entreprises : comparaison du progrès fonctionnel et de la performance opérationnelle, *Colloque National AIP PRIMECA,* Mont Dore (France)

Rauffet, P., Da Cunha, C., Bernard, A. (2008). Intégration de la conduite du progrès dans le pilotage de la performance, *Journées STP GDR MACS*, Metz (France)

Rauffet, P., Da Cunha, C., Bernard, A. (2008). Intégration de la conduite du progrès dans le pilotage de la performance, *Journées STP GDR MACS*, Metz (France)

Rauffet, P., Da Cunha, C., Bernard, A. (2009). Pilotage des capacités organisationnelles, PGSO *Interop VLAB*, Tarbes (France)

Rauffet, P., Da Cunha, C., Bernard, A. (2009). Pilotage des capacités organisationnelles, PGSO *Interop VLAB*, Tarbes (France)

Rauffet, P., Da Cunha, C., Bernard, A. (2012). Conceptual model and IT system for Organizational Capability management. *Computers in Industry*, 63(7): 706-722. https://doi.org/10.1016/j.compind.2012.05.004

Rauffet, P., Da Cunha, C., Bernard, A. (2012). Conceptual model and IT system for Organizational Capability management. *Computers in Industry*, 63(7): 706-722. https://doi.org/10.1016/j.compind.2012.05.004

Rauffet, P., Da Cunha, C., Bernard, A. (2014). A double loop learning system for knowledge transfer and reuse in groups: application of a roadmapping approach. *International Journal of Knowledge and Learning*, 9(1). https://doi.org/10.1504/IJKL.2014.067171

Rauffet, P., Da Cunha, C., Bernard, A. (2014). A double loop learning system for knowledge transfer and reuse in groups: application of a roadmapping approach. *International Journal of Knowledge and Learning*, 9(1). https://doi.org/10.1504/IJKL.2014.067171

Rauffet, P., Da Cunha, C., Bernard, A. (2014). A dynamic methodology and associated tools to assess organizational capabilities. *Computers in Industry*, 65(1):158-174. https://doi.org/10.1016/j.compind.2013.08.006

Rauffet, P., Da Cunha, C., Bernard, A. (2014). A dynamic methodology and associated tools to assess organizational capabilities. *Computers in Industry*, 65(1):158-174. https://doi.org/10.1016/j.compind.2013.08.006

Rauffet, P., Guennal, L., Chauvin, C., Clostermann, J.P. (2017). Projet AERONAV : Passerelle pour les Navires Grande Vitesse. *Research report of project ADEME AERONAV, in collaboration with ENSM*

Rauffet, P., Guennal, L., Chauvin, C., Clostermann, J.P. (2017). Projet AERONAV : Passerelle pour les Navires Grande Vitesse. *Research report of project ADEME AERONAV, in collaboration with ENSM*

Rauffet, P., Guerin, C., Chauvin, C., & Martin, E. (2018). Contribution of Industry 4.0 to the emergence of a joint cognitive and physical production system, *HFES European Chapter Conference*, 8 octobre 2018, Berlin (Germany).

Rauffet, P., Guerin, C., Chauvin, C., & Martin, E. (2018). Contribution of Industry 4.0 to the emergence of a joint cognitive and physical production system, *HFES European Chapter Conference*, 8 octobre 2018, Berlin (Germany).

Rauffet, P., Guerin, C., Chauvin, C., Martin, E. (2019). Function allocation rules and agent transparency levels: principles for designing cognitive assistances. *Workshop GRAISyHM Human & Industry 4.0.* Valenciennes (France)

Rauffet, P., Guerin, C., Chauvin, C., Martin, E. (2019). Function allocation rules and agent transparency levels: principles for designing cognitive assistances. *Workshop GRAISyHM Human & Industry 4.0.* Valenciennes (France)

Rauffet, P., Guérin, C., Chauvin, C., Martin, E. (2019). Lot 2 : état de l'art sur les facteurs humains dans l'industrie 4.0. *Research report of project ANR HUMANISM*

Rauffet, P., Guérin, C., Chauvin, C., Martin, E. (2019). Lot 2 : état de l'art sur les facteurs humains dans l'industrie 4.0. *Research report of project ANR HUMANISM*

Rauffet, P., Labrousse, M., Da Cunha, C., & Bernard, A. (2009). Organizational learning in group: A digital double-loop system based on knowledge maturity and performance assessment, *6th International Conference on Digital Enterprise Technology.* Hong Kong (China).

Rauffet, P., Labrousse, M., Da Cunha, C., & Bernard, A. (2009). Organizational learning in group: A digital double-loop system based on knowledge maturity and performance assessment, *6th International Conference on Digital Enterprise Technology.* Hong Kong (China).

Rauffet, P., Labrousse, M., Da Cunha, C., Bernard, A. (2010). Sustainable Organizational Learning in Group: A Digital Double-Loop System Based on Knowledge Maturity and Performance Assessment. pp. 1769-1786 *Advances in Soft Computing*. Popplewell, Keith and Harding, Jenny and Poler, Raul and Chalmeta, Ricardo-Springer Berlin. ISBN 978-3-642-10429-9

Rauffet, P., Labrousse, M., Da Cunha, C., Bernard, A. (2010). Sustainable Organizational Learning in Group: A Digital Double-Loop System Based on Knowledge Maturity and Performance Assessment. pp. 1769-1786 *Advances in Soft Computing*. Popplewell, Keith and Harding, Jenny and Poler, Raul and Chalmeta, Ricardo-Springer Berlin. ISBN 978-3-642-10429-9

Rauffet, P., Lassalle, J., Leroy, B., Coppin, G., & Chauvin, C. (2015). Tapas project: facilitating cooperation in hybrid combat air patrols including autonomous UCAV, *Applied Human Factors and Ergonomics* (AHFE), Las Vegas (USA).

Rauffet, P., Lassalle, J., Leroy, B., Coppin, G., & Chauvin, C. (2015). Tapas project: facilitating cooperation in hybrid combat air patrols including autonomous UCAV, *Applied Human Factors and Ergonomics* (AHFE), Las Vegas (USA).

Rauffet, P., Lassalle, J., Leroy, B., Coppin, G., Chauvin, C. (2015). The TAPAS Project: Facilitating Cooperation in Hybrid Combat Air Patrols Including Autonomous UCAVs. *Procedia Manuf.*, 3,974-981. https://doi.org/10.1016/j.promfg.2015.07.152

Rauffet, P., Lassalle, J., Leroy, B., Coppin, G., Chauvin, C. (2015). The TAPAS Project: Facilitating Cooperation in Hybrid Combat Air Patrols Including Autonomous UCAVs. *Procedia Manuf.*, 3,974-981. https://doi.org/10.1016/j.promfg.2015.07.152

Rauffet, P., Morel, G., Berruet, P., Chauvin, C. (2013). Approche pluridisciplinaire pour la conception des systèmes sociotechniques. *Journal national de la recherche en IUT*, 4:221-232.

Rauffet, P., Morel, G., Berruet, P., Chauvin, C. (2013). Approche pluridisciplinaire pour la conception des systèmes sociotechniques. *Journal national de la recherche en IUT*, 4:221-232.

Rauffet, P., Morel, G., Chauvin, C. & Berruet, P. (2012). Approche pluridisciplinaire pour la conception et le pilotage de systèmes sociotechniques résilients, *Colloque de la Recherche en IUT - CNRIUT*, Tours, (France).

Rauffet, P., Morel, G., Chauvin, C. & Berruet, P. (2012). Approche pluridisciplinaire pour la conception et le pilotage de systèmes sociotechniques résilients, *Colloque de la Recherche en IUT - CNRIUT*, Tours, (France).

Rauffet, P., Saïd, F., Laouar, A., Chauvin, C., & Bressolle, M. C. (2021). Cognitive control: transitions in modes and fNIRS sensitivity. *CCIS series*

Rauffet, P., Saïd, F., Laouar, A., Chauvin, C., & Bressolle, M. C. (2020, November). Cognitive Control Modes and Mental Workload: An Experimental Approach. *In 4th International Conference on Computer-Human Interaction Research and Applications (pp. 17-26).* SCITEPRESS-Science and Technology Publications.

Rauffet, P., Saïd, F., Laouar, A., Chauvin, C., & Bressolle, M. C. (2020, November). Cognitive Control Modes and Mental Workload: An Experimental Approach. *In 4th International Conference on Computer-Human Interaction Research and Applications (pp. 17-26).* SCITEPRESS-Science and Technology Publications.

Rauffet, P., Saïd, F., Laouar, A., Chauvin, C., & Bressolle, M. C. (2021). Cognitive control: transitions in modes and fNIRS sensitivity. *CCIS series*

Reilhac, P., Hottelart, K., Diederichs, F., & Nowakowski, C. (2017). User Experience with Increasing Levels of Vehicle Automation: Overview of the Challenges and Opportunities as Vehicles Progress from Partial to High Automation. *In Automotive User Interfaces. 457-482.* Springer, Cham.

Romero, D., Bernus, P., Noran, O., Stahre, J., Fast-Berglund, Å. (2016, September) The Operator 4.0: Human Cyber-Physical Systems & Adaptive Automation towards Human-Automation Symbiosis Work Systems

Romero, D., Stahre, J., Wuest, T., Noran, O., Bernus, P., Fast-Berglund, Å., & Gorecky, D. (2016, October). Towards an operator 4.0 typology: a human-centric perspective on the fourth industrial revolution technologies. *In International Conference on Computers and Industrial Engineering (CIE)*

Romero, D., Wuest, T., Stahre, J., & Gorecky, D. (2017, September). Social Factory Architecture: Social Networking Services and Production Scenarios Through the Social Internet of Things, Services and People for the Social Operator 4.0. In *IFIP International Conference on Advances in Production Management Systems (pp. 265-273).* Springer, Cham.

Russell, S (2019). Human-Compatible Artificial Intelligence. *Penguin.*

Rüßmann, M., Lorenz, M., Gerbert, P., Waldner, M., Justus, J., Engel, P., & Harnisch, M. (2015). Industry 4.0: The future of productivity and growth in manufacturing industries. Boston Consulting Group.

Ryman, S. G., El Shaikh, A. A., Shaff, N. A., Hanlon, F. M., Dodd, A. B., Wertz, C. J., ... & Abrams, S., (2019). Proactive and reactive cognitive control rely on flexible use of the ventrolateral prefrontal cortex. *Human brain mapping, 40(3), 955-966.*

SAE International. (2018). Taxonomy and definitions for terms related to driving automation systems for on-road motor vehicles *(J3016_201806).*

Salas, E., & Fiore, S. M. (2004). *Team cognition: Understanding the factors that drive process and performance*. Washington, DC: American Psychological Association.

Salas, E., Cooke, N. J., & Rosen, M. A. (2008). On teams, teamwork, and team performance: Discoveries and developments. *Human Factors, 50*(3), 540-547.

Salas, E., Rosen, M. A., Burke, C. S., Nicholson, D., & Howse, W. R. (2007). Markers for enhancing team cognition in complex environments: the power of team performance diagnosis. *Aviation, Space, and Environmental Medicine, 78*(1).

Sami S., Seppänen M., & Kuusela A. (2004). Artefact correction for heartbeat interval data. Proceedings of 1st Probisi.

Santiago-Espada, Y., Myer, R. R., Latorella, K. A., & Comstock Jr, J. R., (2011). The multi-attribute task battery ii (matb-ii) software for human performance and workload research: A user's guide

Sarter, N. B., Woods, D. D., & Billings, C. E. (1997). Automation surprises. *Handbook of human factors and ergonomics, 2, 1926-1943.*

Scallen, S.F. and Hancock, P.A., (2001). Implementing adaptive function allocation. *The International Journal of Aviation. Psychology, 11, 197–221.*

Schmidt, K. (1991). Cooperative Work: A Conceptual Framework. In Rasmussen, J., Brehmer, B and Leplat, J. (eds.), Distributed Decision Making: Cognitive Models for Cooperative Work, John Wiley & Sons, Chichester, pp. 75-109.

Schmidt, K. (1994). Cooperative work and its articulation: requirements for computer support. *Le travail humain, 345-366.*

Schulte, A., Donath, D., & Honecker, F. (2015, October). Human-system interaction analysis for military pilot activity and mental workload determination. In *IEEE International Conference on Systems, Man, and Cybernetics (pp. 1375-1380).*

Schwerd, S., & Schulte, A. (2020, July). Experimental Validation of an Eye-Tracking-Based Computational Method for Continuous Situation Awareness Assessment in an Aircraft Cockpit. In *International Conference on Human-Computer Interaction (pp. 412-425). Springer, Cham.*

Seeber, I., Bittner, E., Briggs, R. O., de Vreede, T., de Vreede, G.-J., Elkins, A., Maier, R., Merz, A. B., Oeste-Reiß, S., Randrup, N., Schwabe, G., Söllner, M. (2020). Machines as teammates: A research agenda on AI in team collaboration. *Information & Management, 57*, 103174.doi:10.1016/j.im.2019.103174

Sheridan, T.B. and Stassen, H.G. (1979). Definitions, Models and Measures of Human Workload. In Moray, N. (ed.), *Mental Workload: its theory and measurement*. Springer US, 219–233.

Simon, L., Rauffet, P. & Guérin, C (2021). Développement de règles de conception et des mécanismes adaptatifs des interfaces dans le dialogue humain-machine - Application à l'industrie 4.0. *11th EPIQUE conference*, Lille (France)

Simon, L., Rauffet, P. & Guérin, C (2021). Développement de règles de conception et des mécanismes adaptatifs des interfaces dans le dialogue humain-machine - Application à l'industrie 4.0. *11th EPIQUE conference*, Lille (France)

Simon, L., Rauffet, P. Guérin, C., & Lassalle, J. (2021). Exploiter la méthode CWA pour le design écologique d'une interface de maintenance 4.0. *11th EPIQUE conference*, Lille (France)

Simon, L., Rauffet, P. Guérin, C., & Lassalle, J. (2021). Exploiter la méthode CWA pour le design écologique d'une interface de maintenance 4.0. *11th EPIQUE conference*, Lille (France)

Simon, L., Rauffet, P. Guérin, C., & Lassalle, J. (2021). Using cognitive work analysis to develop predictive maintenance tool for vessels. *31st European Safety and Reliability Conference*. Angers (France)

Simon, L., Rauffet, P. Guérin, C., & Lassalle, J. (2021). Using cognitive work analysis to develop predictive maintenance tool for vessels. *31st European Safety and Reliability Conference*. Angers (France)

Sperandio, J.C. (1971). Variation of Operator's Strategies and Regulating Effects on Workload. *Ergonomics, 14(5), 571–577.*

Stanton, N. A., & Bessell, K. (2014). How a submarine returns to periscope depth: Analysing complex socio-technical systems using Cognitive Work Analysis. *Applied ergonomics, 45(1), 110-125.*

Stanton, N.A, Ashleigh, M.J., Roberts, A.D., & Xu F., (2001). Testing Hollnagel's contextual control model: Assessing team behaviour in a human supervisory control task, *Journal of Cognitive Ergonomics, 5, 21-33*

Storm, H., Fremming, A., Odegaard, S., Martinsen, O., & Morkrid, L. (2000). The development of a software program for analyzing spontaneous and externally elicited skin conductance changes in infants and adults. *Clinical Neurophysiology*, *111*, 1889-1898.

Svensson, E. A., & Wilson, G. F. (2002). Psychological and psychophysiological models of pilot performance for systems development and mission evaluation. *The International Journal of Aviation Psychology, 12*(1), 95-110.

Tulli, S., Correia, F., Mascarenhas, S., Gomes, S., Melo, F. S., & Paiva, A. (2019, May). Effects of agents' transparency on teamwork. In *International Workshop on Explainable, Transparent Autonomous Agents and Multi-Agent Systems (pp. 22-37)*. Springer, Cham.

Tversky, A., & Kahneman, D. (1992). Advances in prospect theory: Cumulative representation of uncertainty. *Journal of Risk and uncertainty, 5(4), 297-323.*

Vanderhaegen, F., Wolff, M., & Mollard, R. (2019). Synchronization of stimuli with heart rate: a new challenge to control attentional dissonances. *Automation Challenges of Socio-technical Systems, 1-28.*

Veltman, J.A. (2002). A comparative study of psychophysiological reactions during simulator and real flight. *The International Journal of Aviation Psychology, 12*, 33-48.

Veltman, J.A., & Gaillard, A.W.K. (1996a). Physiological indices of workload in a simulated flight task. *Biological Psychology, 42*, 323-342.

Veltman, J.A., & Gaillard, A.W.K. (1996b). Pilot workload evaluated with subjective and physiological measures. In K. Brookhuis, C. Weikert, J. Moraal, and D. de Waards (Eds.), *Aging and Human Factors, Proceedings of the Europe Chapter of the Ergonomics Society.* The Netherlands: Traffic Research Center, University of Groningen.

Veltman, J.A., & Gaillard, A.W.K. (1998). Physiological workload reactions to increasing levels of task difficulty. *Ergonomics, 41*, 656-669.

Vicente, K.J. (1999). Cognitive work analysis: Toward safe, productive, and healthy computer-based work. Mahwah, NJ: Erlbaum.

Walliser, J. C., Mead, P. R., Shaw, T. H. (2017). The perception of teamwork with an autonomous agent enhances affect and performance outcomes. Proceedings of the *Human Factors and Ergonomics Society Annual Meeting, 61, 231–235.* doi:10.1177/1541931213601541

Wang, Z., Hope, R. M., Wang, Z., Ji, Q., & Gray, W. D. Cross-subject workload classification with a hierarchical Bayes model. *NeuroImage, 59(1), 64-69, 2012*

Wickens, C., Brookhuis, K., Longo, L., & Sharples, S. (2017). The future of mental workload research and practice. *Panel Discussion in 1ˢᵗ Symposium on Human Mental Workload, Dublin (Ireland).*

Wilson, D., & Sperber, D. (2002). Truthfulness and relevance. *Mind, 111(443), 583-632.*

Wilson, G.F. (2002a). An Analysis of Mental Workload in Pilots During Flight Using Multiple Psychophysiological Measures. *The International Journal of Aviation Psychology, 12*, 3-18.

Wilson, G.F. (2002b). A comparison of three cardiac ambulatory recorders using flight data. *The International Journal of Aviation Psychology, 12*, 111-119.

Wilson, G.F., Fullenkamp, P., & Davis, I. (1994). Evoked potential, cardiac, blink, and respiration measures of pilot workload in air-to-ground missions. *Aviation, Space, and Environmental Medicine, 65*, 100-105.

Woods, D. (1996). Decomposing automation: Apparent simplicity, real complexity. Theory and applications.

Wynne, K. T., & Lyons, J. B. (2018). An integrative model of autonomous agent teammate-likeness. *Theoretical Issues in Ergonomics Science, 19*, 353–374.doi:10.1080/1463922X.2016.1260181

Xiao, Y., Hunter, W. A., Mackenzie, C. F., Jefferies, N. J., & Horst, R. L. (1996). Task complexity in emergency medical care and its implications for team coordination. *Human Factors, 38*, 636–645.

Yang, S. I., & Cho, S. B. (2009). Recognizing human activities from accelerometer and physiological sensors. In Multisensor Fusion and Integration for Intelligent Systems (pp. 187-199). Springer, Berlin, Heidelberg.

Yao, Y.J., Chang, Y.M., Xie, X.P., Cao, X.S., Sun, X.Q., & Wu, Y.H. (2008). Heart rate and respiration responses to real traffic pattern flight. *Applied psychophysiology and biofeedback, 33*, 203-209.

Yin, Z., & Zhang, J. (2014). Operator functional state classification using least-square support vector machine based recursive feature elimination technique. *Computer methods and programs in biomedicine, 113(1), 101-115.*

Ylonen, H., Lyytinen, H., Leino, T., Leppaluoto, J., & Kuronen, P. (1997). Heart rate responses to real and simulated BA Hawk MK 51 flight. *Aviation, Space, and Environmental Medicine, 68*, 601-605.

Young, M.S., Brookhuis, K.A., Wickens, C.D., and Hancock, P.A. (2015). State of science: mental workload in ergonomics. *Ergonomics, 58(1), 1–17.*

Zeeb, K., Buchner, A., & Schrauf, M. (2016). Is take-over time all that matters? The impact of visual-cognitive load on driver take-over quality after conditionally automated driving. *Accident Analysis & Prevention, 92, 230-239*

Zeigler, B.P., Song, H.S., Kim, T.G., & Praehofer, H. (1995). Devs framework for modelling, simulation, analysis, and design of hybrid systems. In *Hybrid Systems II*, 529–551. Springer.

Zhang, J., Yin, Z., & Wang, R. (2015). Recognition of mental workload levels under complex human–machine collaboration by using physiological features and adaptive support vector machines. *IEEE Transactions on Human-Machine Systems, 45(2), 200-214.*

Zhang, Z. & Zhang, J. (2010). A new real-time eye tracking based on nonlinear unscented Kalman filter for monitoring driver fatigue. *J. Contr. Theor. Appl. 8, 181–188, 2010*

Zhao, C., Zhao, M., Liu, J., & Zheng, C. (2012). Electroencephalogram and electrocardiograph assessment of mental fatigue in a driving simulator. *Accident Analysis & Prevention, 45, 83-90.*